\documentclass[twocolumn]{aastex701}
\usepackage{xcolor}
\usepackage{amsmath}
\usepackage{lipsum}
\usepackage{comment}
\usepackage{makecell}
\usepackage{csvsimple}
\usepackage{xspace}
\usepackage{hyperref}
\usepackage{longtable,array,booktabs}
\usepackage{ulem}                          
\usepackage{graphicx}
\usepackage{tikz}
\usepackage{soul}

\extrafloats{100}
\newcommand{\msun}{\,\mathrm{M}_\odot}

\newcommand{\asymnsmaller}[3]{%
  \ensuremath{#1^{\scalebox{0.8}{$+#2$}}}_{\scalebox{0.8}{$-#3$}}}

\newcommand{\hunt}{H\&R23\xspace}

\received{July 9, 2025}
\revised{September 29, 2025}
\accepted{October 27, 2025}

\submitjournal{The Astrophysical Journal}

\begin{document}

\title{The White Dwarf Initial-Final Mass Relation from Open Clusters in Gaia DR3}

\correspondingauthor{David R. Miller}
\email{drmiller@phas.ubc.ca}

\author[0000-0002-4591-1903]{David R. Miller}
\affiliation{Department of Physics and Astronomy, University of British Columbia, Vancouver, BC V6T 1Z1, Canada}
\email{drmiller@phas.ubc.ca}

\author[0000-0002-4770-5388]{Ilaria Caiazzo}
\affiliation{Institute of Science and Technology Austria, Am Campus 1, Klosterneuburg, 3400, Austria}
\affiliation{TAPIR, Walter Burke Institute for Theoretical Physics, Mail Code 350-17, Caltech, Pasadena, CA 91125, USA}
\email{Ilaria.Caiazzo@ist.ac.at}

\author[0000-0001-9739-367X]{Jeremy Heyl}
\affiliation{Department of Physics and Astronomy, University of British Columbia, Vancouver, BC V6T 1Z1, Canada}
\email{heyl@phas.ubc.ca}

\author[0000-0001-9002-8178]{Harvey B. Richer}
\affiliation{Department of Physics and Astronomy, University of British Columbia, Vancouver, BC V6T 1Z1, Canada}
\email{richer@astro.ubc.ca}

\author[0000-0003-0089-2080]{Mark A. Hollands}
\affiliation{Department of Physics, University of Warwick, Coventry, CV4 7AL, UK}
\email{Mark.Hollands@warwick.ac.uk}

\author[0000-0001-9873-0121]{Pier-Emmanuel Tremblay}
\affiliation{Department of Physics, University of Warwick, Coventry, CV4 7AL, UK}
\email{P.Tremblay@warwick.ac.uk}

\author[0000-0002-6871-1752]{Kareem El-Badry}
\affiliation{Division of Physics, Mathematics and Astronomy, California Institute of Technology, Pasadena, CA 91125, USA}
\email{kelbadry@caltech.edu}

\author[0000-0003-4189-9668]{Antonio C. Rodriguez}
\affiliation{Division of Physics, Mathematics and Astronomy, California Institute of Technology, Pasadena, CA 91125, USA}
\email{tonycuevas98@gmail.com}

\author[0000-0002-0853-3464]{Zachary P. Vanderbosch}
\affiliation{Division of Physics, Mathematics and Astronomy, California Institute of Technology, Pasadena, CA 91125, USA}
\email{zvanderb@caltech.edu}


\begin{abstract}

The initial-final mass relation (IFMR) links a star's birth mass to the mass of its white dwarf (WD) remnant, providing key constraints on stellar evolution. Open clusters offer the most straightforward way to empirically determine the IFMR, as their well-defined ages allow for direct progenitor lifetime estimates. We construct the most comprehensive open cluster WD IFMR to date by combining new spectroscopy of 22 WDs with an extensive literature review of WDs with strong cluster associations. To minimize systematics, we restrict our analysis to spectroscopically confirmed hydrogen-atmosphere (DA) WDs consistent with single-stellar origins. We separately analyze a subset with reliable Gaia-based astrometric membership assessments, as well as a full sample that adds WDs with strong cluster associations whose membership cannot be reliably assessed with Gaia. The Gaia-based sample includes 69 spectroscopically confirmed DA WDs, more than doubling the sample size of previous Gaia-based open cluster IFMRs. The full sample, which includes 53 additional literature WDs, increases the total number of cluster WDs by over $50\%$ relative to earlier works. We provide functional forms for both the Gaia-based and full-sample IFMRs. The Gaia-based result useful for $M_i\geq2.67\msun$ is
$$M_f = \left[0.179 - 0.100 H(M_i-3.84\msun) \right ] \times (M_i-3.84\msun)+0.628 \msun$$
where $H(x)$ is the Heaviside step function. Comparing our IFMR to recent literature, we identify significant deviations from best-fit IFMRs derived from both Gaia-based volume limited samples of field WDs and double WD binaries, with the largest discrepancy occurring for initial masses of about $5\msun$.

\end{abstract}



\section{Introduction}
\label{sec:intro}

When a star exhausts its nuclear fuel, it ends its life as a stellar remnant. More massive stars collapse into neutron stars or black holes, while lower-mass stars form white dwarfs (WDs). WDs consist of a degenerate electron core surrounded by thin layers of lighter elements. As they no longer sustain fusion, they gradually cool, radiating away residual thermal energy and becoming increasingly faint. Their low luminosities make them difficult to identify, but in recent years, the Gaia observatory \citep{2016A&A...595A...1G,2018A&A...616A...1G,2021A&A...649A...1G,2023A&A...674A...1G} has expanded the known WD population by nearly an order of magnitude, significantly improving the ability to study these objects.

A fundamental tool for understanding WDs is the initial-final mass relation (IFMR), which links a star's birth mass to the mass of its WD remnant. The IFMR not only enables the characterization of an individual WD progenitor but also provides critical insight into the transition mass between WD and neutron star formation. It is also a key ingredient in all population synthesis codes that model stellar populations. WDs are supported by electron degeneracy pressure, while stars massive enough to undergo neon burning eventually overcome this pressure in their cores and collapse, leading to a supernova. Constraining this threshold is essential for refining stellar evolution models, understanding heavy element enrichment rates, tracing galactic star formation histories, and improving predictions of supernova rates and compact object formation.

Several approaches exist for studying the IFMR. One method that has gained traction in recent years involves wide binaries, where both components share a total age but are assumed to have evolved independently. This technique has been applied to WD-main sequence pairs (e.g., \citealt{2008A&A...477..213C}; \citealt{2012ApJ...746..144Z}), WD-turnoff/subgiant pairs (e.g., \citealt{2021ApJ...923..181B}), and double WD systems (e.g., \citealt{2015ApJ...815...63A}; \citealt{2024MNRAS.527.9061H}). While these methods have provided useful constraints, they primarily probe low-mass progenitors and include relatively few high-mass WDs, limiting their ability to constrain the upper end of the IFMR. Another approach relies on more general field WD populations (e.g., \citealt{2018ApJ...860L..17E}; \citealt{2024MNRAS.527.3602C}). This approach benefits from large sample sizes, particularly with the precision and breadth of Gaia astrometry. However, field WDs lack well-constrained total ages, requiring indirect inferences of progenitor masses based on assumptions such as the form of the initial mass function (IMF) and star formation history, among others. Additionally, this method is poorly suited to constraining the high-mass end of the IFMR, as high-mass WDs from single-star evolution are rare, and contamination from merger remnants cannot be reliably identified in field populations. Moreover, the field WD population cannot constrain the maximum progenitor mass, which must be set using models or another IFMR. 

The most direct and widely used approach for studying the IFMR involves identifying WDs born in star clusters (e.g., \citealt{2005MNRAS.361.1131F}; \citealt{2009ApJ...692.1013S}; \citealt{2009ApJ...693..355W}; \citealt{2009MNRAS.395.1795C}; \citealt{2009MNRAS.395.2248D}; \citealt{2015ApJ...807...90C}; \citealt{2016ApJ...818...84C}; \citealt{2018ApJ...866...21C}; \citealt{2021ApJ...912..165R}; \citealt{2020NatAs...4.1102M}; \citealt{2022ApJ...926..132H}; \citealt{2022ApJ...926L..24M}). Cluster membership provides a crucial advantage, as all stars in a cluster are thought to have formed simultaneously. By determining the cluster's age, typically from the main-sequence turnoff (MSTO), and subtracting a WD's cooling age, one can infer the progenitor's lifetime and, consequently, its mass. Though Gaia has dramatically increased the number of known WDs, IFMRs based on open clusters remain sparsely populated, particularly in the high-mass regime.

Beyond increasing the known WD population, Gaia has also significantly improved the census of Milky Way star clusters. In this work, we use the \citet{2023A&A...673A.114H} Milky Way Cluster catalog, which represents the most complete single Milky Way cluster catalog to date, to identify previously unrecognized cluster member WDs. We combine these new identifications with an extensive literature review of known cluster WDs to refine the open cluster WD IFMR. We restrict our IFMR analysis to spectroscopically confirmed hydrogen-atmosphere (DA) WDs. This choice is motivated by the limited number of non-DA cluster WDs with well-determined atmospheric and cooling parameters. As such, their inclusion would provide only a minor increase in the sample while potentially introducing systematic errors, as it remains unclear whether the IFMR follows the same form for DA and non-DA WDs. \citet{2021AJ....162..162B} found that non-DAs appear to have $\approx0.07\msun$ lower final masses than DAs at a given initial mass. However, it is unclear whether this discrepancy reflects a physical difference, small-number statistics, or systematic issues with fitting non-DA atmospheres (see, e.g., \citealt{2019ApJ...876...67B}). Our DA IFMR considers two samples: one restricted to sources with Gaia-based astrometric cluster membership determinations and another that also includes sources for which Gaia cannot fairly assess membership.

We describe our sample selection in Section~\ref{sec:sample}, outline analysis techniques for WDs and their host clusters in Section~\ref{sec:analysis}, present our final sample and analyze the resulting IFMR in Section~\ref{sec:results_and_discussion}, and summarize our overall findings in Section~\ref{sec:conclusions}.


\section{Sample}
\label{sec:sample}


To develop a sample of candidate open cluster member WDs, we use the Milky Way cluster catalog from \citet{2023A&A...673A.114H} (hereafter \hunt), which contains 7,167 clusters comprising 121 globular clusters, 6,818 open clusters, and 228 moving groups. \hunt employed the Hierarchical Density-Based Spatial Clustering of Applications with Noise (HDBSCAN) algorithm, identified in earlier work as the most sensitive for recovering clusters in Gaia data \citep{2021A&A...646A.104H}. While HDBSCAN delivers substantial completeness, it has a high false positive rate. To address this, \hunt performed a Cluster Significance Test (CST) using nearest-neighbor distributions of cluster stars versus field stars. A cluster must pass a CST threshold of at least $3\sigma$ to be considered valid, with those exceeding $5\sigma$ classified as high-probability clusters.

They further classified clusters photometrically using a machine learning method that relies on simulated open clusters generated using SPISEA \citep{2020AJ....160..143H}. A sample of 10,000 simulated clusters was used to train their neural network to classify clusters, along with a sample of 10,000 false-positive fake clusters. Using this simulated training dataset, they employed a convolutional neural network to assign a homogeneity parameter to each cluster identified by HDBSCAN. They recommend using a median homogeneity classification cutoff $>0.5$. For each HDBSCAN cluster, they used 1,000 samples of the neural network to estimate each cluster's age, extinction, differential extinction, and distance modulus for any cluster within 15~kpc. Only stars given a $>50\%$ probability of cluster membership in HDBSCAN were used to determine these parameters. They compared their photometric parameters to other literature and found that they were generally inconsistent outside the most well-behaved clusters. We will consider some of these parameters in the initial analysis but, as required, will follow up with detailed examinations.

\begin{figure}[ht]
    \centering
    \includegraphics[width=1.0\columnwidth]{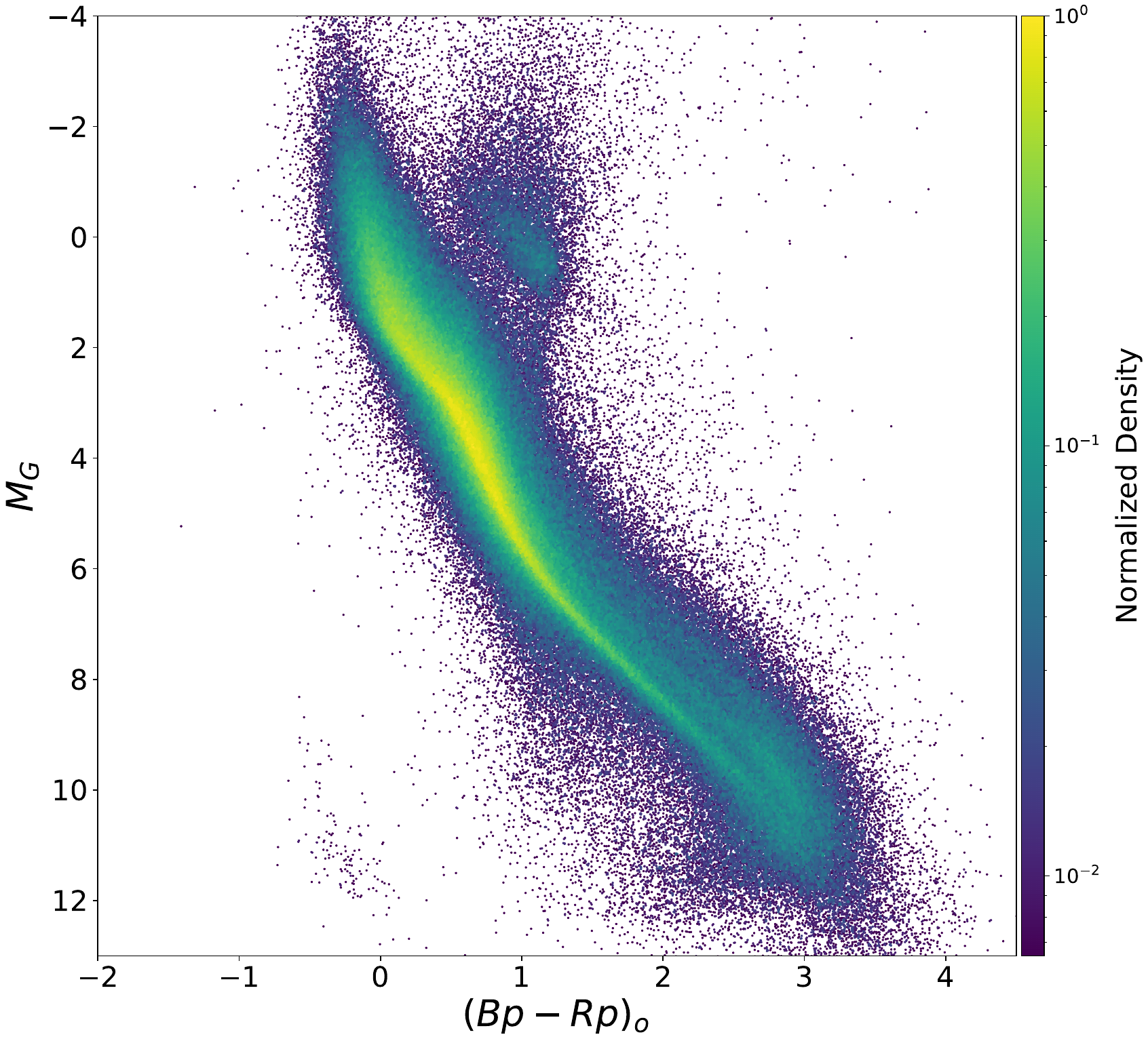}
    \caption{Dereddened Gaia EDR3 CMD of all candidate members of the 4,004 high-probability open clusters in \hunt. Contains 560,436 total member candidates. $(\mathrm{Bp{-}Rp})_\mathrm{o}$ denotes the dereddened Gaia color, and $M_G$ is the extinction-corrected absolute Gaia $G$-band magnitude.}
    \label{fig:all_clusters_CMD}
\end{figure}

Applying their recommended high-probability cluster cuts on median homogeneity parameter ($\mathrm{class}\_50$ in \hunt) of $>0.5$, and CST $>5\sigma$, and selecting only open clusters reduces the sample to 4,004 clusters, which we adopt for this work. Each cluster star is assigned a probability of cluster membership by HDBSCAN, which is based on the star's proximity to the majority of cluster members in 6D phase space. Given the tendency for WDs to be found far from their birthplaces (e.g., \citealt{2022ApJ...926..132H}; \citealt{2022ApJ...926L..24M}; \citealt{2023ApJ...956L..41M}), potentially as a result of receiving a small natal velocity kick (e.g., \citealt{2003ApJ...595L..53F}; \citealt{2007MNRAS.382..915H}; \citealt{2009ApJ...695L..20F}; \citealt{2018MNRAS.480.4884E}), we opted against performing a specific cut on this probability. While some cluster member candidates will be interlopers, cluster membership can be further examined with follow-up spectroscopy. It turns out that many heavily studied WDs that are long-thought members of the most well-studied clusters are given reasonably low membership probabilities by HDBSCAN, supporting our decision not to cut on this parameter. Fig.~\ref{fig:all_clusters_CMD} shows the candidate member stars from this subset of clusters, where each member has been dereddened using the mean cluster $A_v$ reddening in \hunt.

\begin{figure}[ht]
    \centering
    \includegraphics[width=1.0\columnwidth]{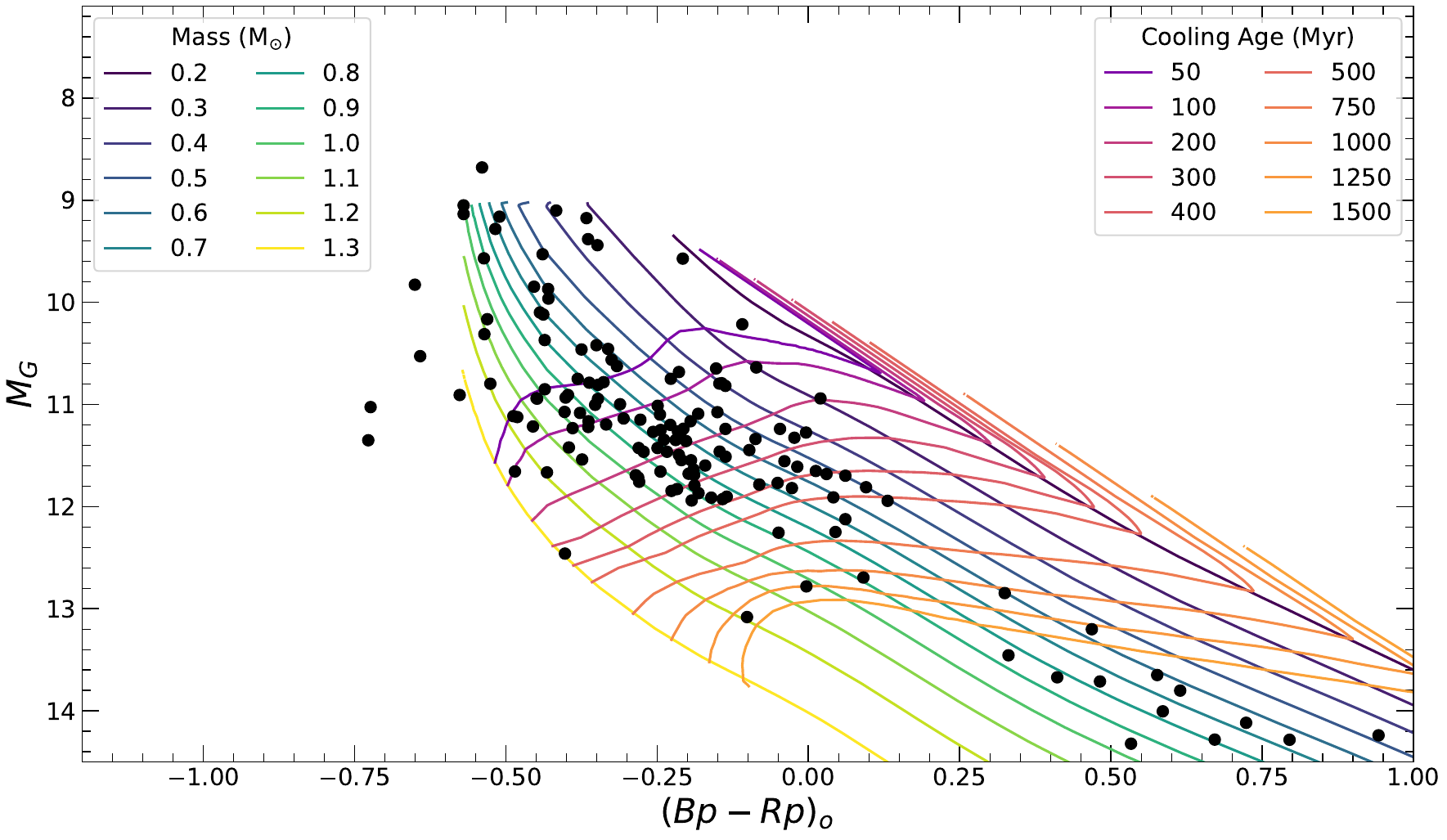}
    \caption{Candidate cluster member white dwarfs with each source dereddened using cluster mean $A_v$ from associated cluster in \hunt. Contours of constant age shown from 50 to 1,500~Myrs (top to bottom), along with contours of constant mass from 0.2 to 1.3~$M_\odot$ (right to left).}
    \label{fig:WDs_cluster_reddening}
\end{figure}

\begin{figure*}[ht]
    \centering
    \includegraphics[width=0.9\textwidth]{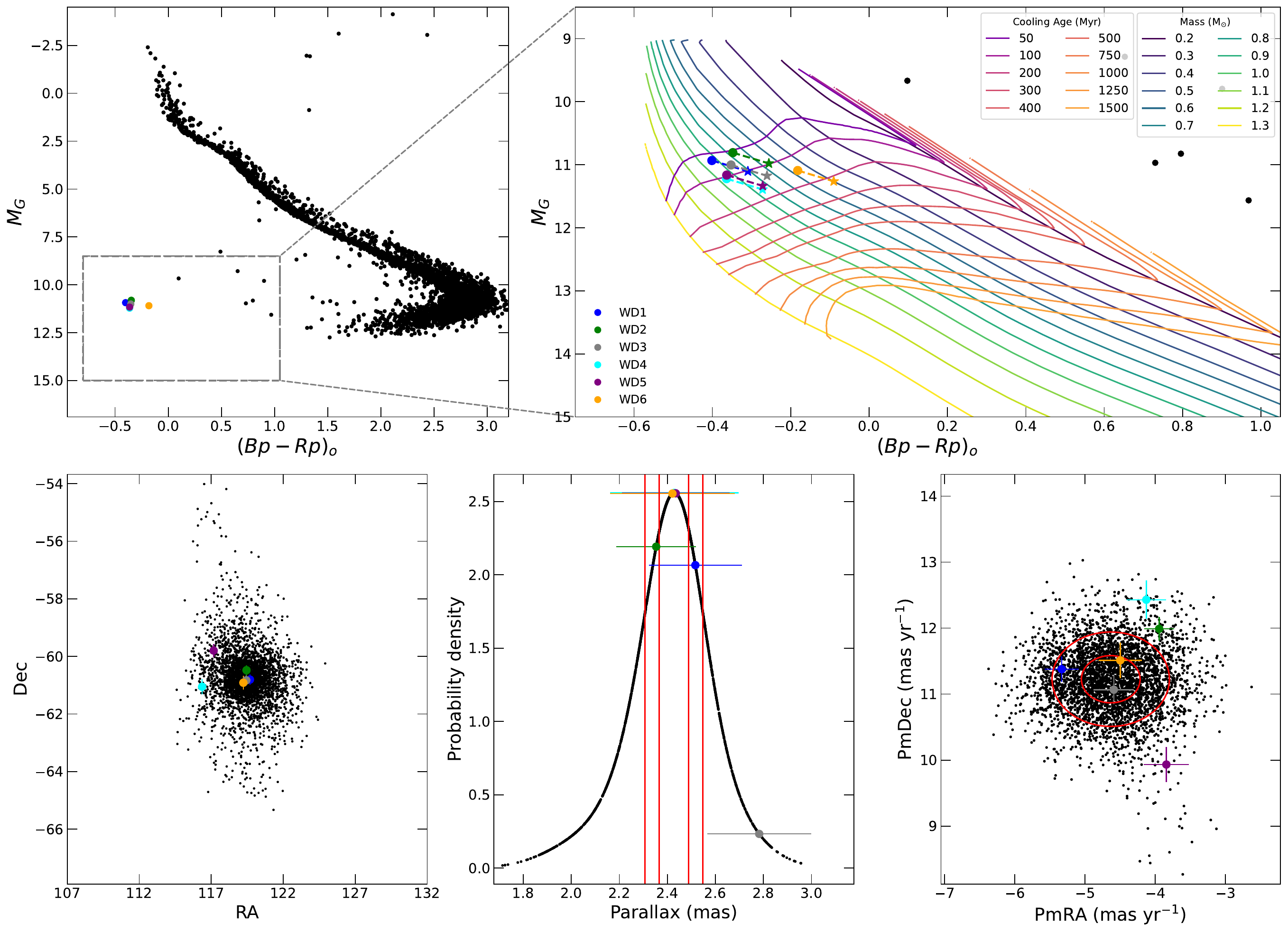} 
    \caption{\textit{Top left}: Dereddened cluster CMD for NGC 2516 cluster using \hunt catalog cluster member candidates and the catalog's mean $A_v$.WD member candidates are shown as colored circles in all panels; colors are consistent across panels. \textit{Top right}: Zoom in on the WD sequence, with WD cooling models overlaid. The star symbols show WD candidates without any applied dereddening. The dashed lines connect the \hunt catalog mean reddening and no reddening cases. \textit{Bottom}: Distribution for astrometric data of cluster member candidates in NGC 2516, showing the position (left), kernel density estimation of parallax (center), and proper motion (right). Solid red lines are 1 and $2\sigma$ regions relative to cluster mean values, as calculated in \hunt. See Appendix~\ref{sec:appendix-figsandtables} Figs.~\ref{fig:CMD_alessi_13} through \ref{fig:CMD_upk_624}
    for similar plots of all other examined clusters.}
    \label{fig:CMD_NGC2516}
\end{figure*}

To identify WD candidates, we use the Gaia EDR3 WD catalog from \citet{2021MNRAS.508.3877G}, which includes approximately 359,000 high-probability WDs. By crossmatching the high-probability open cluster member candidates from \hunt with the $P_\mathrm{WD} > 0.7$ sample of \citet{2021MNRAS.508.3877G}, we identify 144 candidate cluster member WDs across 65 total clusters. The WD sequence for these candidates is shown in Fig.~\ref{fig:WDs_cluster_reddening}, where sources are dereddened using \hunt cluster reddening values and carbon-oxygen (CO) core, thick hydrogen envelope cooling models from \cite{2020ApJ...901...93B}\footnote{\url{http://www.astro.umontreal.ca/~bergeron/CoolingModels}} are overlaid. An example cluster color-magnitude diagram (CMD) is shown in Fig.~\ref{fig:CMD_NGC2516}, with the full set of CMDs for all 65 clusters available in the online journal. These figures additionally show the distribution of stars in position, parallax, and proper motion space. Select \hunt data for these clusters is provided in Table~\ref{tab:candidate_clusters} of the appendix. Throughout this work, we refer to most clusters primarily by the primary cluster names in \hunt; the final column of Table~\ref{tab:candidate_clusters} lists commonly used alternative names for some well-known clusters. For the particularly well-studied Alpha Persei, Hyades, Pleiades, and Praesepe clusters, we use these names instead of the corresponding \hunt names (Melotte 20, Melotte 25, Melotte 22, and NGC 2632, respectively). 

In work published after our initial candidate selection and most of our follow-up observations (see Section~\ref{sub_sec:candidate selection}), \citet{2024A&A...686A..42H} updated the \hunt catalog by incorporating derived masses and Jacobi radii to better distinguish open clusters from moving groups. Their revision was motivated by the observation that many groups classified as clusters based on CMD classifiers exhibited unusually flat stellar distributions, which are atypical of true clusters. While they initially considered using velocity dispersions as a distinguishing factor, they found this unreliable for the \hunt sample due to binary stars, tidal tails, and other dynamical features.

To differentiate between open clusters and moving groups, they calculated the Roche surface for each candidate based on its mass and Jacobi radius. If a candidate had no radius at which a Roche sphere could form, it is likely a moving group. They calculate the presence of a valid Jacobi radius as a function of distance from the cluster center and use this to estimate the probability that a given cluster has a valid Jacobi radius. To be considered an open cluster, a candidate required a probability of at least 0.5 and a total mass of at least 40~$M_\odot$ within its best-fitting Jacobi radius. This mass threshold addressed limitations of their method when applied to low-mass clusters, particularly the assumption of a Kroupa IMF \citep{2001MNRAS.322..231K} and spherical symmetry, both of which can break down in smaller groups. Additionally, they required at least 10 member stars within the best-fitting Jacobi radius.

Thirty clusters from our candidate list were reclassified as moving groups in \citet{2024A&A...686A..42H}, as indicated in Appendix~\ref{sec:appendix-figsandtables} Table~\ref{tab:candidate_clusters}. Although this represents nearly half of the clusters with candidate WDs, most of these WDs had already been assigned low priority due to assumed WD properties or poorly defined CMDs. Of the WDs we selected for follow-up observations, five were in clusters that were reclassified as moving groups. We discuss the inclusion of these moving group sources in the following section. 


\subsection{Candidate Selection}
\label{sub_sec:candidate selection}

Our crossmatch between the \citet{2021MNRAS.508.3877G} and \hunt catalogs yielded 144 WD candidate cluster members. Although these sources are candidate members based on their Gaia astrometry, we expect some to be interlopers or false candidates. Additionally, some of the candidates may be in unresolved binaries, which can significantly impact the IFMR due to the effects of mass transfer on WD evolution. To identify likely interlopers and binaries and reduce the sample to a more manageable subset for spectroscopic follow-up, we examine Gaia photometry to estimate each WD's cooling age and mass and, from that, its progenitor lifetime. Summing the cooling age and progenitor lifetime gives an estimate of the WD's total age. In the case of unresolved binaries, the presence of two unresolved stars makes the WDs appear overluminous, leading to underestimated masses and overestimated total ages when interpreted with single-star evolutionary models. We employed these total age estimates to rank the candidate WDs in terms of priority for follow-up spectroscopy, emphasizing the ones that are more likely to be members and not interlopers or unresolved binaries.

As most WDs are classified as DA WDs \citep{2023MNRAS.518.5106J}, we estimate the cooling age and mass of each source by linearly interpolating the \cite{2020ApJ...901...93B} DA WD cooling model grid, which assumes a CO core composition. While most ultramassive WDs ($M > 1.05\msun$) are expected to have oxygen-neon (ONe) cores, the assumption of a CO core does not significantly affect the resulting estimates. Additionally, because ultramassive WDs are high-priority targets for constraining the upper end of the IFMR, they were naturally emphasized in our follow-up selection.

The mass derived from the cooling model fit is used to estimate the progenitor mass by interpolating an IFMR constructed from \cite{2022ApJ...926L..24M}
and \citet{2020NatAs...4.1102M}, with the latter included to better populate the lower-mass region that is not well sampled by Gaia-based IFMRs. To address scatter in the combined IFMR, we employ isotonic regression combined with a smoothed univariate spline to derive robust progenitor mass estimates. Using these progenitor masses, we estimate progenitor lifetimes with PARSEC isochrones \citep{2012MNRAS.427..127B,2014MNRAS.445.4287T,2014MNRAS.444.2525C,2018MNRAS.476..496F,2022A&A...665A.126N} at solar metallicity. By combining the cooling age and progenitor lifetime, we calculate the total age of each WD based on Gaia DR3 data. Because these total age estimates were used only as order-of-magnitude checks to help guide candidate selection, uncertainties in the assumed IFMR do not significantly impact the selection.

We estimate progenitor masses only for WDs $\geq0.51\msun$, the lower limit of the combined IFMR data range, and do not extrapolate below this, as such sources are unlikely to be the result of single-star evolution and thus are not useful for constraining the IFMR, even in cases where they are true cluster members. For WDs exceeding the upper mass limit (1.2~$M_\odot$) of the combined IFMR, we fix the mass at this limit. This underestimates progenitor mass and overestimates lifetime for very massive WDs, but has little impact on selection given our prioritization of high-mass WDs.

There are several limitations to this methodology. Cooling model fits rely on assumed reddening, taken as the mean cluster reddening from \hunt. Progenitor lifetimes are computed using solar-metallicity isochrones, and also depend the adopted IFMR. Additionally, optical photometry is insufficient to characterize the parameters of hot, blue WDs, as colors become rather insensitive to $T_\textrm{eff}$. Because of the uncertainty in total age estimates, we used them only as rough guides for candidate prioritization, comparing each source total age to the cluster age from \hunt, with cluster ages redetermined later in this work. We also prioritized higher-mass WDs, candidates with low Gaia astrometric uncertainties, and those from clusters with well-defined CMDs that allow for precise age determination.

Based on these prioritization criteria, our final sample was divided as follows. We selected 22 WDs for spectroscopic follow-up, as shown in Table~\ref{tab:candidates_observed}. An additional 38 candidates were found to have existing literature spectra and did not require follow-up; these are listed in Appendix~\ref{sec:appendix-figsandtables} Table~\ref{tab:candidates_literature}. This left 84 candidates without spectroscopic follow-up (Appendix~\ref{sec:appendix-figsandtables} Table~\ref{tab:candidates_notobserved}). Many of these lower-priority candidates are extreme outliers, with photometrically determined WD parameters suggesting total ages much older than their cluster ages. Others are from clusters with poor-quality CMDs that preclude precise age estimates. Among the candidates with high-quality CMDs and consistent WD parameters, a few showed very high astrometric uncertainty and were deemed unreliable as a result. Some of the remaining candidates were simply given lower priority due to low expected mass, as we focused primarily on higher-mass WDs to better populate the high-mass regime of the IFMR. The primary reason each source was not selected for follow-up is provided in Appendix~\ref{sec:appendix-figsandtables} Table~\ref{tab:candidates_notobserved}. We make one exception to this prioritization by observing a candidate from RSG 5. While its inferred total age is much greater than the cluster age, its potential association with an exceptionally young cluster with a poorly populated upper main sequence (see Appendix~\ref{sec:appendix-figsandtables} Fig.~\ref{fig:CMD_rsg_5}) warrants spectroscopic follow-up. This source had also been independently identified as a candidate member of RSG 5 in prior work by \citet{2022AJ....164..215B}.

\begin{table*}
\tiny
\centering
\caption{Candidate white dwarf cluster members for which we obtained new spectroscopy. Initial parameter estimates are derived from Gaia photometry. No extrapolation is applied for sources that fall outside the \citet{2020ApJ...901...93B} DA model grids. \((t_{\mathrm{total}}-\text{Age}_{\mathrm{cl}})/\sigma_{\text{Age,\,cl}}\) is the difference between the total WD age and the cluster mean age, normalized by the $1\sigma$ age range in \hunt; their lower bound is used for sources that appear younger than their birth cluster and the upper bound for those that appear older. $\dagger$ indicates the source is in a cluster that was reclassified as a moving group by \citet{2024A&A...686A..42H}. $\ddagger$ RSG 5 WD candidate is included despite a total age greatly exceeding the cluster age, due to its potential high mass and association with a very young cluster with a poorly constrained age.}
\label{tab:candidates_observed}
\begin{tabular}{lccccccccc}
\toprule
Name & 
Gaia DR3 Source ID &
RA & 
Dec & 
$G_\mathrm{obs}$ & 
$t_\mathrm{cool}$ &
$M_\mathrm{WD}$ &
$M_\mathrm{prog}$ &
$t_\mathrm{total}$ & 
$\tfrac{t_\mathrm{total}-\mathrm{Age}_\mathrm{cl}}{\sigma_\mathrm{Age,cl}}$ \\
      & 
      & 
[deg] & 
[deg] & 
[mag] & 
[Myr] & 
[$M_\odot$] & 
[$M_\odot$] & 
[Myr] & 
[$\sigma_\mathrm{Age,cl}$] \\
\midrule
BH 99 $\dagger$ & 5350664076917572352 & 159.691 & -58.840 & 19.74 & - & $>1.3$ & - & - & - \\
HSC 601 $\dagger$ & 2060960191793682176 & 305.163 & 37.224 & 19.65 & - & $>1.3$ & - & - & - \\
NGC 2516 WD4 & 5289447182180342016 & 116.506 & -61.054 & 19.46 & 105 & 0.95 & 4.72 & 229 & 0.8 \\
NGC 2516 WD5 & 5294015515555860608 & 117.278 & -59.798 & 19.40 & 97 & 0.93 & 4.44 & 245 & 0.9 \\
NGC 3532 WD2 & 5340165355769599744 & 166.416 & -58.594 & 20.16 & 150 & 1.19 & 7.59 & 191 & -0.5 \\
NGC 3532 WD4 & 5339402672685054720 & 167.391 & -58.658 & 19.93 & 83 & 1.13 & 6.29 & 145 & -1.0 \\
NGC 3532 WD9 & 5340288226187644160 & 167.302 & -57.916 & 19.84 & - & $>1.3$ & - & - & - \\
NGC 3532 WD10 & 5340530320614001792 & 164.660 & -57.385 & 19.55 & 6 & 1.19 & 7.69 & 45 & -2.1 \\
RSG 5$\ddagger$ & 2082008971824158720 & 302.137 & 44.828 & 18.82 & 79 & 0.98 & 5.06 & 184 & 12.3 \\
Stock 1 & 2021451685334449152 & 294.343 & 25.148 & 20.01 & 150 & 1.07 & 5.92 & 220 & -0.4 \\
Stock 2 WD2 & 507105143670906624 & 31.857 & 59.412 & 20.17 & 228 & 0.99 & 5.19 & 327 & 0.2 \\
Stock 2 WD3 & 507054806657042944 & 31.334 & 59.354 & 19.78 & 54 & 1.17 & 6.93 & 103 & -1.5 \\
Stock 2 WD4 & 507119265523387136 & 32.419 & 59.530 & 20.18 & 169 & 0.88 & 3.77 & 388 & 0.6 \\
Stock 2 WD5 & 459270649787942784 & 37.112 & 59.540 & 20.00 & 126 & 1.07 & 5.93 & 196 & -0.7 \\
Stock 2 WD6 & 511159317926025600 & 27.028 & 60.532 & 20.13 & 179 & 0.88 & 3.77 & 407 & 0.7 \\
Stock 2 WD7 & 507128332197081344 & 32.705 & 59.755 & 19.95 & 221 & 0.99 & 5.16 & 321 & 0.2 \\
Stock 2 WD8 & 458778927573447168 & 34.654 & 58.412 & 19.30 & 84 & 0.93 & 4.42 & 232 & -0.5 \\
Stock 2 WD9 & 508276703371724928 & 28.558 & 61.039 & 19.69 & 46 & 0.82 & 3.24 & 392 & 0.6 \\
Stock 2 WD10 & 507899399087944320 & 29.790 & 60.108 & 20.14 & 236 & 0.91 & 4.07 & 416 & 0.8 \\
Theia 248 $\dagger$ & 2174431990805230208 & 323.962 & 54.108 & 19.11 & 1 & 1.06 & 5.88 & 73 & -0.9 \\
Theia 817 WD1 $\dagger$ & 4530122390454022272 & 276.033 & 22.785 & 19.47 & 104 & 1.26 & 7.97 & 141 & -0.5 \\
UPK 303 $\dagger$ & 337013370615249408 & 39.883 & 41.974 & 17.74 & 57 & 1.03 & 5.69 & 134 & 0.2 \\
\bottomrule
\end{tabular}
\end{table*}

As previously mentioned, dereddened CMDs for all 65 clusters with candidate WDs are included in the online figure set (see Fig.~\ref{fig:CMD_NGC2516} for an example). Notably, the CMD for IC 4756 includes one additional candidate WD member, Gaia DR3 4283590687858436608, which was not part of our initial sample as it is not in the \citet{2021MNRAS.508.3877G} WD catalog. This object, the bluest source in the cluster, was identified during visual inspection of the CMD, where it clearly lies on the WD sequence. It was considered for follow-up but ultimately excluded due to its Gaia DR3 parallax over error being $<3$. No additional WD candidates were identified through CMD inspection in any of the other clusters with crossmatched WD cluster member candidates. Some clusters (e.g., NGC 2099, NGC 2168, and NGC 2323) show unusually broad CMDs, likely due to differential reddening or high rotation; relevant clusters are discussed in detail in Appendix~\ref{sec:appendix-clusters}.

Continuing our discussion of the updated cluster classification by \citet{2024A&A...686A..42H}, we find that 37 WDs identified in our crossmatch belong to clusters that have been reclassified as moving groups. Of these, 32 are among the candidates we did not prioritize for follow-up (see Appendix~\ref{sec:appendix-figsandtables} Table~\ref{tab:candidates_notobserved}). While many were given low priority due to either advanced cooling ages or low inferred WD masses, visual inspection of CMD quality was applied only as an additional check after these primary criteria. Many of these WDs would have been deprioritized based solely on poor CMD quality, even if Gaia-based WD parameters better supported membership.

The five remaining WDs, each from reclassified clusters, were all observed with spectroscopy prior to the reclassification by \citet{2024A&A...686A..42H}: BH 99, HSC 601, Theia 248, Theia 817 WD1, and UPK 303. While moving groups are no longer gravitationally bound, they typically consist of stars with a common origin that remain largely kinematically consistent. We do not target moving groups due to the increased difficulty in establishing the membership of candidates, but much of this challenge stems from uncertainty in the moving group's age determination. In each case where we followed up a WD candidate from a later reclassified cluster, the moving groups had well-defined main sequences that allowed for reliable age estimates (CMDs for these clusters are included in the online figure set). Candidates from clusters where a reliable age determination was not possible were given low priority. As such, we retain each of these followed-up candidates and do not exclude them based on the reclassification by \citet{2024A&A...686A..42H}.


\subsection{Spectroscopic follow up}
\label{sub_sec:spectroscopy}

We obtained spectra for twenty-two sources as part of this work. Fourteen sources were observed using the Gemini Multi-Object Spectrographs (GMOS; \citealt{2004PASP..116..425H,2016SPIE.9908E..2SG}) on the twin Gemini Telescopes in Hawaii (GMOS-N) and Chile (GMOS-S). GMOS was configured in long-slit mode with a 1" focal plane mask and 2x2 binning for both spatial and spectral directions. We used the B480 grating for all observations where it was available, as it provides improved sensitivity and a broader wavelength range compared to the older B600 grating. For NGC 2516 WD5 and BH 99, which were observed prior to the release of B480, we used the B600 grating. All observations were performed without a filter, with the central wavelength set to 520~nm, giving a wavelength coverage of approximately 320--720~nm. Spectra were reduced with the DRAGONS data reduction software \citep{2023RNAAS...7..214L}.

Additionally, we observed eight sources using the Low-Resolution Imaging Spectrometer (LRIS; \citealt{1995PASP..107..375O,1998SPIE.3355...81M}) on the Keck I Telescope (Hawaii). We used the R600/4000 grism ($R$\,$\approx$\,$1100$) for the blue arm and either the R400 grating ($R$\,$\approx$\,$1000$, for Stock 2 WD2, Stock 2 WD3, and UPK 303) or the R600/7500 grating ($R$\,$\approx$\,$1800$, for HSC 601, RSG 5, Stock 1, Theia 248, and Theia 817 WD1) for the red arm, covering a wavelength range of approximately 3200--10000\,\AA\ or 3200--9000\,\AA\ respectively. A standard long-slit data reduction procedure was performed with the \texttt{Lpipe} pipeline\footnote{\url{http://www.astro.caltech.edu/~dperley/programs/lpipe.html}} \citep{2019PASP..131h4503P}. Analysis of the spectra obtained from both Gemini and Keck is presented in Section~\ref{subsec:spectro_analysis}.


\subsection{Additional Literature Sources}

To construct a comprehensive DA WD IFMR, we supplemented our newly obtained WD spectra with an extensive literature review to identify reliable cluster member WDs. While we prioritize sources with Gaia-based astrometric membership determinations, we acknowledge that many well-studied WDs belong to clusters whose WDs are either too faint for Gaia or are close to the magnitude limit and thus poorly constrained by Gaia astrometry. Consequently, we use Gaia as the preferred but not the sole criterion for membership. We prioritize Gaia astrometry over other membership assessments when it is available and sufficiently precise to evaluate cluster membership.

We include only sources confirmed as DA WDs via spectroscopy and do not re-derive spectroscopic fit parameters for literature spectra. Since most literature $T_\textrm{eff}$ and $\log g$ values are based on fits using the same models and Balmer-line fitting technique employed in this work, this does not introduce any notable inconsistency. We use these parameters to recompute other WD properties following the procedures outlined in Section~\ref{sec:analysis}. Specific sources included in our analysis are detailed in Appendix~\ref{sec:appendix-clusters}. When evaluating the IFMR in Section~\ref{sec:ifmr}, we will separately consider all supported WD cluster members and a subset consisting only of those with Gaia-supported astrometric membership.


\section{Analysis}
\label{sec:analysis}


\subsection{Spectroscopic Parameters}
\label{subsec:spectro_analysis}

All of the newly observed WDs are consistent with DA spectral types, except for RSG 5 and Stock 2 WD6. RSG 5 shows Zeeman split magnetic hydrogen absorption and hydrogen emission, and a measured rotation period of approximately 6.6 minutes. Because of its short rotation period and high magnetic field, the object is likely a double white dwarf merger remnant and we therefore exclude it from our IFMR analysis. A detailed study of this source is presented in \citet{cristea2025}, where it is found to have a cooling age of $60\pm10\,\mathrm{Myr}$. In contrast, MSTO-based isochrone fitting yields a cluster age of $45^{+8}_{-11}\,\mathrm{Myr}$ for RSG 5, consistent with analysis of the lower main sequence. This age is also supported by the independent pre-main sequence estimate from \citet{2022AJ....164..215B}. Given a likely progenitor lifetime of at least 40~Myr, if not significantly longer, this source is found to be a non-member of the RSG 5 cluster.

\begin{figure}[ht]
    \centering
    \includegraphics[width=1.0\columnwidth]{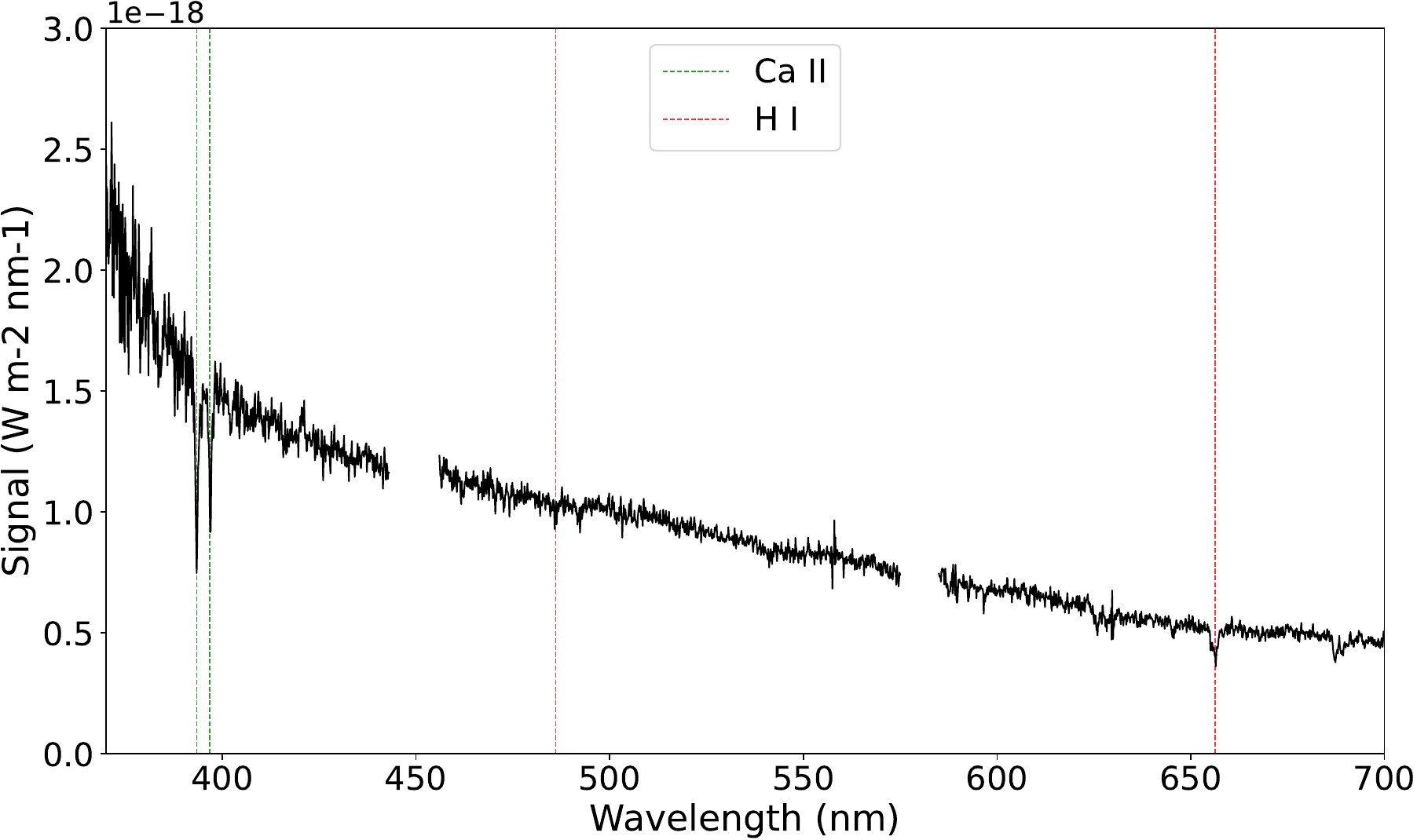}
    \caption{Gemini GMOS spectrum of Stock 2 WD6, highlighting the Ca II H $\&$ K lines (green) and hydrogen Balmer lines H$\alpha$ and H$\beta$ (red). Data near 450 and 580~nm was removed due to artifacts from the chip gaps.}
    \label{fig:Stock_2_WD6}
\end{figure}

Stock 2 WD6 exhibits strong Ca II features, as well as weak, non-broadened H$\alpha$ and H$\beta$ features, suggesting the presence of a thin hydrogen layer. Based on these characteristics, we classify it as a metal-polluted DZA WD. Such objects are thought to have accreted material from disrupted planetary systems \citep{2003ApJ...596..477Z,2014A&A...566A..34K}. The absence of He features suggests a temperature below about $12{,}000\,\mathrm{K}$ \citep{2017MNRAS.467.4970H}. To test for cluster membership, we fit the LRIS spectrum using the WD atmosphere models of \citet{2010MmSAI..81..921K}. The effective temperature ($T_\textrm{eff}$), surface gravity ($\log g$), and abundances of hydrogen and calcium relative to helium were treated as free parameters. Abundances of other metals, such as magnesium and iron, were held at bulk Earth ratios relative to calcium. A least-squares fit was performed using a Levenberg-Marquardt approach, normalizing the model continuum against the data at each step with a low-order polynomial. The resulting best-fit indicates an effective temperature of approximately $10{,}300\,\mathrm{K}$ and $\log g = 7.93$. Applying a mass-radius relation \citep{2020ApJ...901...93B}, we found Gaia absolute magnitudes of about 12.1 in all three filters. Even without considering reddening, this suggests an average mass evolved WD whose total age greatly exceeds the cluster's. When accounting for reddening, the excess age is more pronounced. As such, we definitively rule out Stock 2 WD6 as a candidate cluster member. The Gemini GMOS spectrum for this source is shown in Fig.~\ref{fig:Stock_2_WD6}.

We derived atmospheric parameters for the newly obtained DA WD spectra using non-local thermodynamic equilibrium (NLTE) pure hydrogen atmosphere models from \citet{2011ApJ...730..128T}. The fitting routine follows an approach similar to that of \citet{2005ApJS..156...47L}. We first fit the observed spectrum to a grid of synthetic models combined with a polynomial of up to ninth order in wavelength to correct for continuum calibration errors. After normalizing the spectrum using fixed points outside the Balmer lines, we fit the Balmer line profiles to the synthetic model spectra. This process yields the best-fit values for $T_\textrm{eff}$ and $\log g$. Throughout the fitting procedure we employ the Levenberg-Marquardt nonlinear least-squares minimization method to optimize the fit.

\begin{figure}[ht]
    \centering
    \includegraphics[width=0.6\columnwidth]{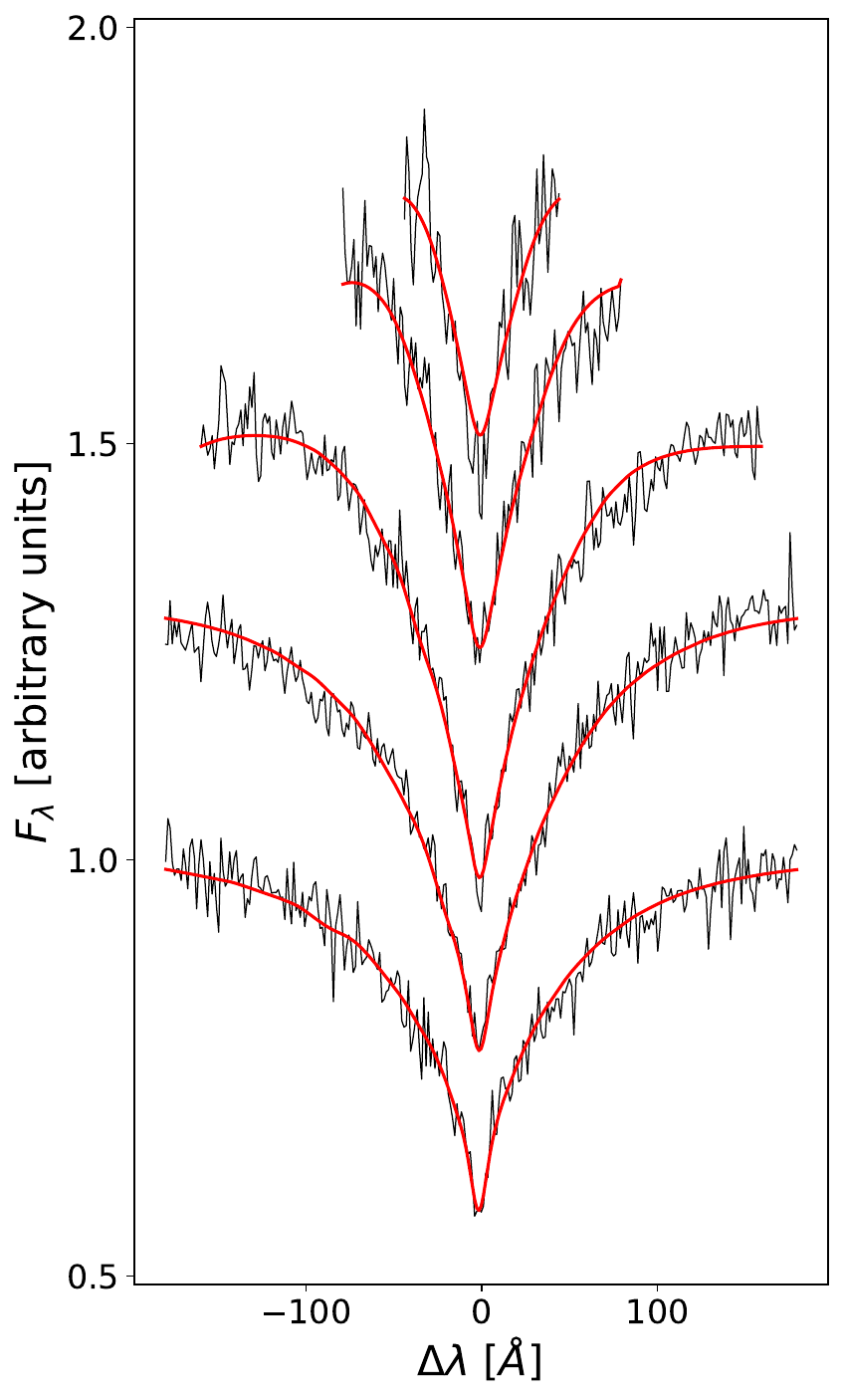}
    \caption{Balmer series from H$\alpha$ to H$\epsilon$ for Stock 2 WD10. Spectrum obtained with Gemini GMOS-N; the best-fitting hydrogen model is superimposed in red. Fits for the remaining DA WD spectra are provided in Figs.~\ref{fig:spectrafits_a} and \ref{fig:spectrafits_b} in the Appendix.}
    \label{fig:spectrafit_stock2wd10}
\end{figure}

An example of the extracted Balmer lines and their corresponding simultaneous fit is presented in Fig.~\ref{fig:spectrafit_stock2wd10}, while the remaining fits are provided in Figs.~\ref{fig:spectrafits_a} and \ref{fig:spectrafits_b} in the Appendix. Note that while the wavelength coverage in all cases includes H$\alpha$, it was excluded in instances where the SNR was too low to reliably fit the line. Among the DA WDs, Stock 2 WD2 exhibits Zeeman splitting, indicating the presence of a magnetic field of about 3~MG. No other WD in the new sample shows the presence of magnetic fields on their surface. In magnetic WDs, degeneracies between magnetic broadening and surface gravity often make $\log g$ estimates particularly uncertain; for this reason, we do not deem this method appropriate for Stock 2 WDs (the only magnetic WD). Additionally, the signal-to-noise in its spectrum is very low. We therefore exclude Stock 2 WD2 from our IFMR analysis.

Instrument and exposure time details, along with the resulting $\log g$ and $T_\textrm{eff}$ fits (where applicable), are summarized in Table~\ref{tab:wd_results_new}. Spectroscopic parameters for literature sources are not rederived, the literature values used in our IFMR analysis are given in Tables~\ref{tab:wd_results_litsources} and \ref{tab:wd_results_nongaia} in Appendix~\ref{sec:appendix-figsandtables}.

\begin{table*}
\tiny
\caption{Spectroscopic follow-up observations of candidate white dwarf members, with best-fit atmospheric parameters for non-magnetic DA WDs. }
\label{tab:wd_results_new}
\centering
\begin{tabular}{cclcccl}
\toprule
Name &
Instrument &
Gemini Program &
Exposure time & 
\multicolumn{1}{c}{$\log g$} &
\multicolumn{1}{c}{$T_\textrm{eff}$} &
Comments \\
  &
  &
  &
\multicolumn{1}{c}{[s]} & 
\multicolumn{1}{c}{$[\mathrm{cm}\,\mathrm{s}^{-2}]$} &
\multicolumn{1}{c}{[K]} & 

\\
\midrule
BH 99         & GMOS-S & GS-2022A-FT-104 & $14{,}320$ & $8.61\pm0.07$   & $52{,}900\pm1{,}100^\dagger$  & $\dagger$error underestimated \\
HSC 601       & LRIS   & -               & $1{,}200$  & $8.41\pm0.04$   & $32{,}900\pm200$ &  \\
NGC 2516 WD4  & GMOS-S & GS-2024A-Q-227 & $3{,}520$ & $8.58\pm0.03$   & $28{,}780\pm130$  &  \\
NGC 2516 WD5  & GMOS-S & GS-2021A-Q-236 &$3{,}200$ & $9.02\pm0.07$   & $32{,}900\pm180$  &  \\
NGC 3532 WD2  & GMOS-S & GS-2024A-Q-227, WD1 in program &$12{,}610$ & $8.01\pm0.06$   & $19{,}500\pm300$ &  \\
NGC 3532 WD4  & GMOS-S & GS-2024A-Q-227, WD2 in program & $7{,}840$ & $8.43\pm0.03$   & $21{,}140\pm190$  &   \\
NGC 3532 WD9  & GMOS-S & GS-2024A-Q-227, WD3 in program & $6{,}640$ & $8.38\pm0.03$   & $23{,}000\pm200$  &   \\
NGC 3532 WD10 & GMOS-S & GS-2024A-Q-227, WD4 in program & $4{,}080$ & $7.83\pm0.02$   & $24{,}240\pm160$  &  \\
RSG 5         & LRIS   & - & $900$ & -               & -               & peculiar magnetic WD \\  
Stock 1       & LRIS   & -& $1{,}200$ & $8.52\pm0.04$   & $18{,}900\pm200$  &  \\
Stock 2 WD2   & LRIS   & -&  $1{,}800$ & -   & -                           & magnetic DA  \\
Stock 2 WD3   & LRIS   & - & $1{,}800$ & $8.40\pm0.03$   & $20{,}200\pm200$  &  \\
Stock 2 WD4   & GMOS-N & GN-2024B-Q-133, WD1 in program & $8{,}320$ & $8.50\pm0.02$   & $19{,}270\pm150$  &  \\
Stock 2 WD5   & GMOS-N & GN-2024B-Q-229, WD2 in program &  $5{,}600$ & $8.527\pm0.018     $ & $22{,}710\pm170$  &  \\
Stock 2 WD6   & GMOS-N & GN-2024B-Q-133, WD3 in program & $6{,}880$ & -               & -               &  DZA \\
Stock 2 WD7   & GMOS-N & GN-2024B-Q-133, WD4 in program & $4{,}800$ & $8.21\pm0.03$   & $20{,}000\pm200$  &  \\
Stock 2 WD8   & GMOS-N & GN-2024B-Q-229, WD5 in program & $2{,}200$ & $8.34\pm0.02$   & $24{,}190\pm180$  &  \\
Stock 2 WD9   & GMOS-N & GN-2024B-Q-133, WD6 in program & $3{,}560$ & $8.355\pm0.019$ & $20{,}550\pm170$  &  \\
Stock 2 WD10  & GMOS-N & GN-2024B-Q-229, WD7 in program & $6{,}800$ & $8.58\pm0.03$   & $19{,}900\pm200$  &  \\
Theia 248     & LRIS   & - &  $900$ & $8.21\pm0.03$   & $27{,}250\pm170$    & \\
Theia 817 WD1 & LRIS   & - & $1{,}200$ & $8.55\pm0.03$   & $27{,}650\pm170$  &  \\
UPK 303       & LRIS   & - & $700$ & $8.71\pm0.02$   & $31{,}530\pm100$    &  \\
\bottomrule
\end{tabular}
\end{table*}


\subsection{White Dwarf Cooling Parameters}
\label{sub_sec: cooling params}

Ultramassive WDs ($M>1.05\msun$) are generally expected to harbour oxygen-neon (ONe) cores, formed when core conditions in the late stages of stellar evolution are sufficient for off-center carbon ignition \citep{2014MNRAS.440.1274C}. While pathways exist for CO WDs to reach this mass range \citep{2022MNRAS.511.5198C,2022MNRAS.512.2972W}, the majority of ultramassive WDs are expected to form with ONe cores as a result of typical stellar evolution processes in progenitors capable of producing such high-mass remnants. Given that core composition is not directly discernible due to the opacity of the outer layers \citep{2008ApJ...683..978D}, and CO and ONe WDs are indistinguishable at the surface, we adopt the practical assumption that WDs exceeding $1.05\msun$ have ONe cores.

From the effective temperatures and surface gravities obtained from spectroscopy, we estimate cooling ages, masses, and radii for all white dwarfs by linearly interpolating a combined grid constructed from three sets of theoretical models. This single grid incorporates CO core models from \citet{2020ApJ...901...93B} for masses between $0.2$ and $1.05\msun$, ONe core models from \citet{2019A&A...625A..87C} for masses between $1.10$ and $1.23\msun$, and ONe core models from \citet{2022A&A...668A..58A} for masses between $1.29$ and $1.369\msun$. The latter models include full general relativistic effects on WD structure and cooling, which become increasingly significant for WDs above $1.29\msun$. All models assume thick hydrogen envelopes. Linear interpolation is performed across the entire mass range, including small gaps between grids (e.g., $1.05$--$1.10$ and $1.23$--$1.29\msun$). These gaps are comparable to the mass grid spacing used within each model set, so no special treatment is required. Any added uncertainty in derived cooling parameters for WDs near the boundaries is modest and does not substantially impact the overall analysis. Derived cooling parameters for newly obtained spectra are given in Table~\ref{tab:wd_results_new_derived}, with rederived parameters for literature spectra given in Appendix~\ref{sec:appendix-figsandtables} Tables~\ref{tab:wd_results_litsources_derived} and \ref{tab:wd_results_nongaia_derived}.

While some cluster member DB WDs are known (e.g., \citealt{2019ApJ...880...75R}; \citealt{2019ApJ...882..106G}; \citealt{2020NatAs...4.1102M}; \citealt{2021AJ....161..169C}), there is a notable scarcity of DB WDs compared to the field population (e.g., \citealt{2005ApJ...618L.129K}). Although DB models can reliably describe their atmospheric and cooling parameters, allowing their inclusion in the IFMR, we find that the significant dearth of DB WDs offers minimal benefit to advancing the understanding of the IFMR. Additionally, their inclusion may introduce potential systematics if the relationship between DB and DA WDs and their progenitor masses differs for a given mass, leading us to exclude DB WDs from our IFMR. We similarly exclude more exotic WDs, which often additionally lack reliable modeling, resulting in a focus purely on DA WDs for this analysis.


\subsection{Cluster Ages}
\label{sub_sec:cluster ages}

For relevant clusters included in \hunt, we estimate the cluster age using PARSEC 2.0 isochrones \citep{2012MNRAS.427..127B,2022A&A...665A.126N} using stars with at least a $50\%$ chance of being members according to \hunt. We adopt a heuristic, iterative approach that combines manual visual fitting with a weighted $\chi^2$ calculation to identify the best-fit age and its associated uncertainties. The weighted $\chi^2$ approach incorporates observational Gaia photometric errors and assigns different importance to selected stellar groups based on their relevance to age determination. Stars near the MSTO are the most critical and are given the highest weighting, comprising $60\%$ of the $\chi^2$ calculation. Giant stars contribute $30\%$, while main sequence stars closely aligned with the isochrone, selected using a narrow cut in color and magnitude to avoid unresolved binaries, make up the remaining $10\%$. These default weightings are sometimes adjusted for clusters with sparse MSTOs or a lack of giant stars. In cases where a cluster has no giant stars in our age sample, the weight is assigned proportionally to the MSTO and MS stars.

Using the mean cluster age and reddening from \hunt as a starting point, we assume a solar metallicity of $Z = 0.015$. We then iteratively adjust both reddening ($A_v$) and metallicity until we achieve a satisfactory fit to the CMD. Reddening is first adjusted to match the central part of the main sequence, which is relatively insensitive to metallicity. Metallicity is then modified to improve the fit to the upper main sequence and turnoff, where it has only a minor effect on color but significantly alters the isochrone shape. Because changes in metallicity also shift the central main sequence, we re-adjust reddening and repeat as needed until both regions are well matched. We also consider the lower main sequence in this process, though it is secondary given our MSTO-based approach to age fitting and the larger photometric uncertainties in this regime. Once the best estimates for reddening and metallicity are obtained, we adjust the cluster age to visually select the best-fit isochrone. While the weighted $\chi^2$ values guide this process, the final fit is determined by careful visual inspection, particularly focusing on the MSTO region.

After manually identifying the best-fit age, we automatically iterate the age in both directions until the weighted $\chi^2$ increases by $30\%$ to determine the upper and lower $1\sigma$ bounds. Greater asymmetry in these bounds is often seen when the $\chi^2$ minimum occurs at an age that differs significantly from the visually determined best fit. In select cases, particularly for clusters with sparse MSTO stars and minimal giants, a larger $\chi^2$ increase threshold was used to provide realistic error bars. We discuss specific cluster fits in the corresponding cluster subsection of Appendix~\ref{sec:appendix-clusters}, along with any required deviations from default parameters. We include an example isochrone fit for the ASCC 113 cluster in Fig.~\ref{fig:ASCC_113_age}, with the remaining fits shown in Appendix~\ref{sec:appendix-figsandtables} Figs.~\ref{fig:cluster_ages_1}, and \ref{fig:cluster_ages_2}.

Manual isochrone fitting is a common practice in open cluster analysis and typically outperforms automated methods for MSTO fitting, as unresolved binaries, blue stragglers, and sparse turnoffs can bias automated approaches. Our method blends the strengths of visual fitting in handling these complexities with a reproducible $\chi^2$ calculation to define error bars that capture both data quality and uncertainty in the fit. Compared to \hunt (see Table~\ref{tab:candidate_clusters}), our results are generally more reliable and have smaller, more realistic uncertainties, owing to a more informed selection of true MSTO stars during visual fitting.

\begin{figure}[ht]
    \centering
    \includegraphics[width=0.8\columnwidth]{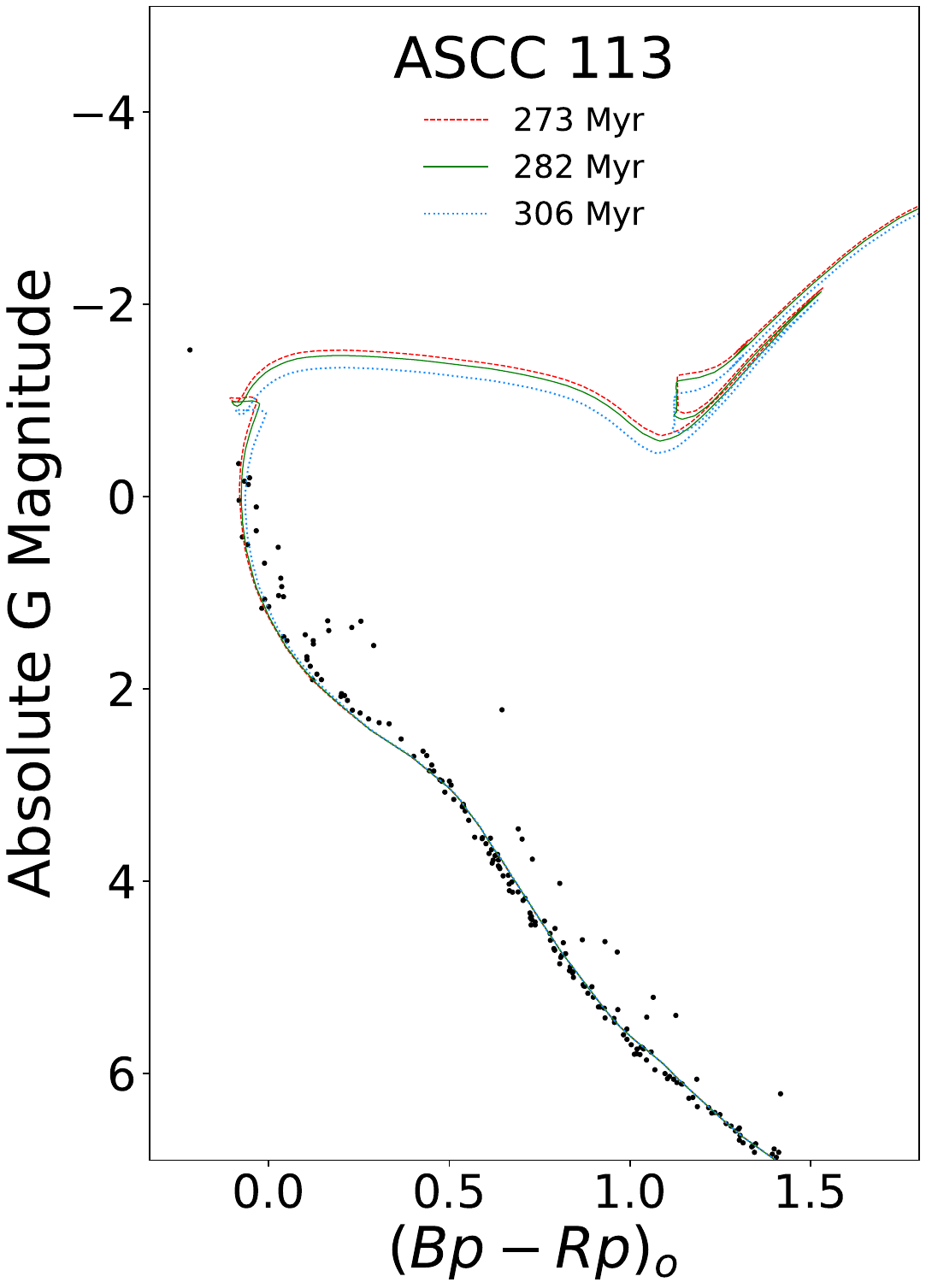}
    \caption{Cluster CMD using only members with $>50\%$ membership probability in \hunt for ASCC 113 cluster. Best-fit (green) and $1\sigma$ error (red and blue) PARSEC isochrones from heuristic fitting method are overlaid.}
    \label{fig:ASCC_113_age}
\end{figure}

Despite their inclusion in the \hunt catalog, we do not fit isochrones to the Pleiades or Alpha Persei clusters, instead adopting the kinematic ages from \citet{2021arXiv211004296H}. These ages are broadly consistent with those found via isochrone fitting, but offer significantly smaller uncertainties and are independent of stellar evolution models, reducing model-dependent systematics and providing a more precise reference age. For some older clusters, while they are present in the \hunt catalog, we found their faint nature resulted in Gaia CMDs that were not well suited for precise age determination. For these cases, we adopt the isochrone ages from \citet{2018ApJ...866...21C}, which also provided the spectral analysis for the relevant cluster WDs. For ASCC 47 and the AB Doradus Moving Group, both of which are not identified by \hunt, we adopt literature-based ages from \citet{2021ApJ...912..165R} and \citet{2018ApJ...861L..13G}, respectively. We summarize determined ages in Appendix~\ref{sec:appendix-figsandtables} Table~\ref{tab:cluster_ages_final}, reddening values are only included if the age was determined using our heuristic isochrone fitting method.


\subsection{Progenitor Masses}

To find the progenitor mass, we calculate the progenitor lifetime from the difference between the cluster age and the WD cooling age and then use PARSEC isochrone tables at the cluster's metallicity to identify the mass of a star with that lifetime. The progenitor mass is taken as the initial mass of the lowest-mass star that has reached the asymptotic giant branch phase in the selected PARSEC isochrone table. We selected this approach, which neglects the duration of the AGB phase, because this phase is not well modeled in these isochrones and, in very young clusters, later stages of stellar evolution are not reliably reached in PARSEC. However, the AGB phase comprises only a tiny fraction of a star's lifetime, typically under a few Myr \citep{2014ApJ...782...17K}, and thus does not significantly impact results. Using the onset of the AGB phase for all sources ensures consistency across the analysis and mitigates discrepancies in the literature, where different studies have used varying cutoff phases, including the start of the AGB, the thermally pulsing AGB, the horizontal branch, or even the main sequence. Progenitor masses for newly obtained spectra are given in Table~\ref{tab:wd_results_new_derived}, while those for literature spectra are provided in Appendix Tables~\ref{tab:wd_results_litsources_derived} and \ref{tab:wd_results_nongaia_derived}.


\section{Results and Discussion}
\label{sec:results_and_discussion}


\subsection{Final White Dwarf Sample}
\label{sec:results_wdsample}

As described earlier, our white dwarf sample was constructed using a twofold approach. First, we developed a candidate list by crossmatching the \hunt Milky Way open cluster catalog with the \citet{2021MNRAS.508.3877G} Gaia EDR3 WD catalog. From this list, we selected high-priority targets for new spectroscopic follow-up, while others already had suitable literature spectra and required no additional observations. Second, we conducted an extensive literature review to identify additional spectroscopically confirmed DA WDs with strong cluster associations that were not part of the Gaia-based crossmatch or could not be reliably assessed using Gaia astrometry. Appendix~\ref{sec:appendix-clusters} provides detailed discussions of clusters with newly observed and literature WD members. We do not discuss clusters where all literature WD members are non-DA, or where we identified candidates but did not follow them up and no prior spectra exist. While we made extensive efforts to identify all spectroscopically confirmed DA WDs with strong cluster associations, we cannot rule out the possibility that some have been missed. 

Appendix~\ref{sec:appendix-figsandtables} includes many tables and figures central to the WD cluster sample, including cluster properties, WD candidate details, spectroscopic fits and parameters, cluster isochrone fits and properties, derived WD and progenitor parameters, and a comparison with the IFMR sample of \citet{2018ApJ...866...21C}. Additional materials include corner plots for IFMR fits and CMDs for all clusters with WDs in the initial crossmatch.

\begin{table*}[ht]
\scriptsize
\caption{Derived parameters for newly followed-up white dwarf candidate cluster members.}
\label{tab:wd_results_new_derived}
\centering
\begin{tabular}{cccccl}
\toprule
Name &
Mass & 
$t_\textrm{cool}$ & 
Radius & 
Initial Mass &
\multicolumn{1}{c}{Comments} \\
  &
[$\msun$] & 
$[\mathrm{Myr}]$ & 
$[\mathrm{km}]$ &
[$\msun$] &
\\
\midrule
BH 99 & $1.03^{+0.03}_{-0.04}$ & $1.46^{+0.15}_{-0.12}$ & $5{,}800\pm400$ & $5.9^{+0.9}_{-0.3}$ & temperature error underestimated \\
HSC 601 & $0.89\pm0.02$ & $16^{+4}_{-3}$ & $6{,}800\pm200$ & $5.1^{+0.4}_{-0.6}$ &  \\
NGC 2516 WD4 & $0.990\pm0.018$ & $71\pm6$ & $5{,}880\pm150$ & $4.67^{+0.17}_{-0.4}$ & \\
NGC 2516 WD5 & $1.19\pm0.03$ & $120\pm20$ & $3{,}900\pm300$ & $5.6^{+0.7}_{-0.8}$ & \\
NGC 3532 WD2 & $0.63\pm0.03$ & $82^{+10}_{-14}$ & $9{,}000\pm400$ & $3.41^{+0.13}_{-0.17}$ & questionable member \\
NGC 3532 WD4 & $0.890\pm0.019$ & $151^{+9}_{-8}$ & $6{,}620\pm160$ & $3.7\pm0.2$ & \\
NGC 3532 WD9 & $0.861\pm0.019$ & $101^{+8}_{-7}$ & $6{,}900\pm160$ & $3.49^{+0.15}_{-0.18}$ & \\
NGC 3532 WD10 & $0.545\pm0.010$ & $20.4^{+1.5}_{-1.0}$ & $10{,}340\pm140$ & $3.19^{+0.10}_{-0.13}$ & non-member \\
Stock 1 & $0.94\pm0.03$ & $253^{+19}_{-17}$ & $6{,}200\pm200$ & $4.05^{+0.3}_{-0.19}$ & \\
Stock 2 WD3 & $0.869\pm0.019$ & $168^{+9}_{-11}$ & $6{,}780\pm160$ & $3.57^{+0.09}_{-0.11}$ & \\
Stock 2 WD4 & $0.931\pm0.013$ & $231^{+7}_{-8}$ & $6{,}250\pm100$ & $3.94^{+0.12}_{-0.16}$ & \\
Stock 2 WD5 & $0.952\pm0.011$ & $146^{+6}_{-5}$ & $6{,}130\pm90$ & $3.48^{+0.07}_{-0.10}$ & \\
Stock 2 WD7 & $0.749\pm0.019$ & $114^{+10}_{-4}$ & $7{,}830\pm170$ & $3.34^{+0.07}_{-0.08}$ & \\
Stock 2 WD8 & $0.837\pm0.013$ & $74\pm4$ & $7{,}130\pm110$ & $3.21^{+0.05}_{-0.07}$ & \\
Stock 2 WD9 & $0.841\pm0.012$ & $143\pm5$ & $7{,}020\pm100$ & $3.46^{+0.07}_{-0.09}$ & \\
Stock 2 WD10 & $0.981\pm0.018$ & $243^{+12}_{-13}$ & $5{,}860\pm150$ & $4.04^{+0.16}_{-0.19}$ & \\
Theia 248 & $0.762\pm0.018$ & $24\pm3$ & $7{,}900\pm180$ & $3.01^{+0.03}_{-0.04}$ & \\
Theia 817 & $0.971\pm0.018$ & $77^{+6}_{-7}$ & $6{,}030\pm150$ & $3.87^{+0.16}_{-0.10}$ & \\
UPK 303 & $1.060\pm0.007$ & $67.4^{+2.0}_{-1.9}$ & $5{,}250\pm100$ & $4.02^{+0.12}_{-0.21}$ & \\
\bottomrule
\end{tabular}
\end{table*}


\subsection{Initial-Final Mass Relation}
\label{sec:ifmr}

\begin{figure*}[ht]
\centering
\includegraphics[width=0.7\textwidth]{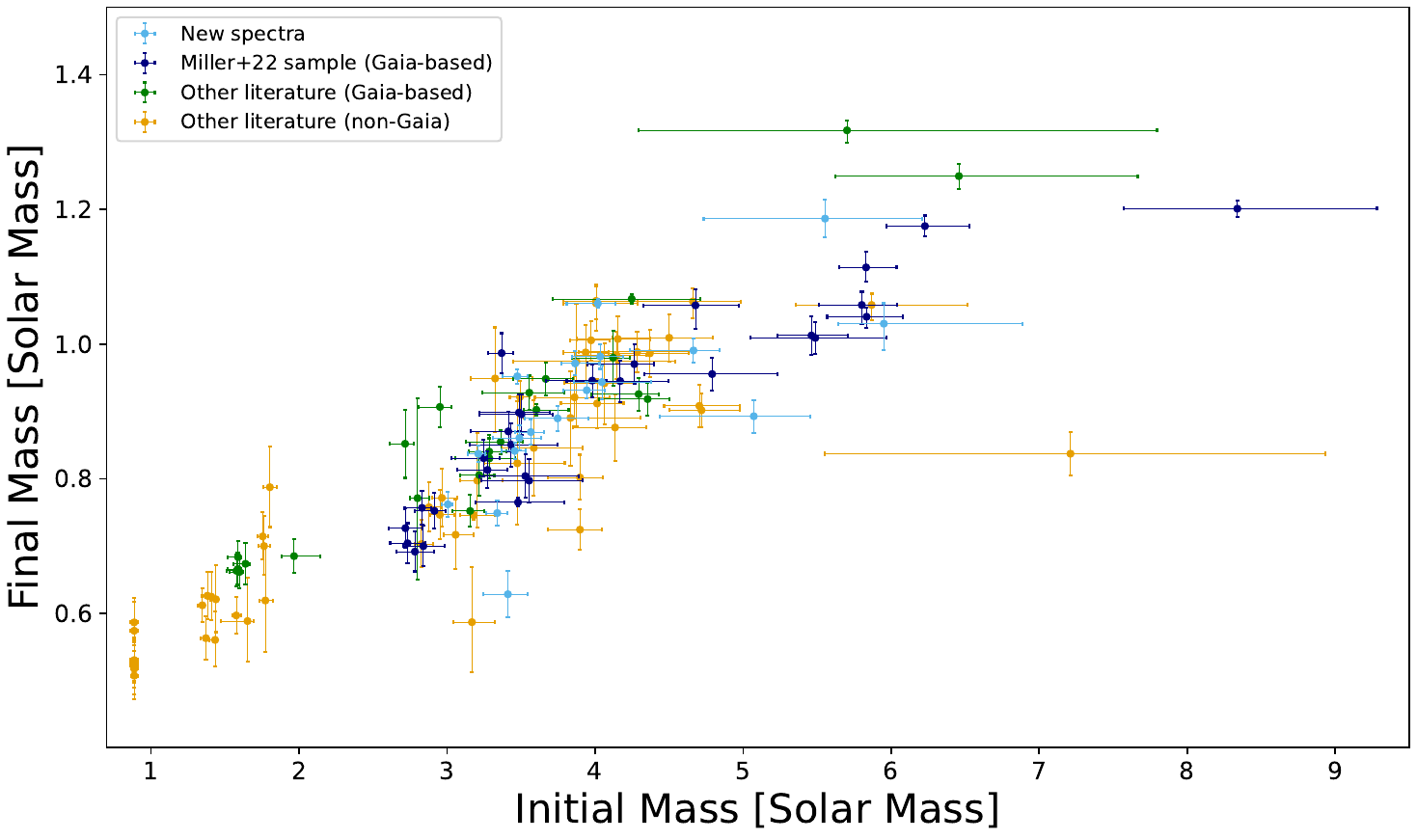} \\
\caption{Initial and final masses for the final sample of white dwarfs included in this study.}
\label{fig:unfit_ifmr}
\end{figure*}

We combine both the newly obtained spectroscopic results and rederived parameters from the literature sample to construct the WD IFMR for our sample, as shown in Fig.~\ref{fig:unfit_ifmr}. In the figure, the newly analyzed candidates are shown in light blue, while those from the Gaia-based \citet{2022ApJ...926L..24M} IFMR are indicated in dark blue. The other sources from the literature are divided into Gaia-based (green) and non-Gaia-based (orange). 

To parametrize the shape of the IFMR, we model it with a continuous piecewise linear function using a set number of flexible breakpoints. The free parameters are taken as the initial mass at each breakpoint, the final mass at each breakpoint and the endpoints, as well as an intrinsic scatter term ($\sigma_\mathrm{int}$), meant to capture additional scatter not reflected in the reported mass uncertainties. The initial mass endpoints are fixed at the minimum and maximum of the data range $\pm0.05\msun$. This results in a piecewise fit function with $2N+3$ free parameters, where $N$ is the number of breakpoints. The models ensures continuity by explicitly defining each linear segment to connect to the next. The final mass at every breakpoint and endpoint are treated as free parameters, so adjacent segments always meet, resulting in an IFMR that remains continuous across the full initial mass range.

We employ physically motivated Gaussian priors on the initial masses at the breakpoints, with Gaussian widths reflecting our confidence in the physical motivation. We tested prior widths as low as $0.1\msun$ and as high as $0.5\msun$, ultimately adopting $0.3\msun$ as a balance between flexibility and physical plausibility. Breakpoint final masses are given broad, noninformative priors. In addition to the physically motivated breakpoints, we also test models that include an additional breakpoint with a broad uniform prior on the initial mass, allowing us to evaluate whether the data supports any additional transitions beyond those expected from stellar evolution models.

We consider several potential breakpoints motivated by stellar evolution models and supported in past studies. One well-motivated breakpoint is at $\approx3.7\msun$, corresponding to the transition where stars experience the second dredge-up at the onset of the asymptotic giant branch (e.g., \citealt{1999ApJ...524..226D}; \citealt{2007A&A...469..239M}; \citealt{2016ApJ...823..102C}). Another breakpoint, predicted at $\approx2.3\msun$, arises from differing helium ignition processes at the tip of the red giant branch, with more massive stars igniting helium smoothly and lower-mass stars undergoing a helium flash. These breakpoints have been included in many past IFMRs (e.g., \citealt{2018ApJ...866...21C}; \citealt{2018ApJ...860L..17E}; \citealt{2024MNRAS.527.3602C}). At lower masses, recent work by \citet{2020NatAs...4.1102M} suggests a kink around $1.8\msun$, potentially linked to carbon star formation. Additional breakpoints at higher masses ($>5\msun$) have been proposed (e.g., \citealt{2018ApJ...860L..17E}; \citealt{2024MNRAS.527.3602C}), though these appear to be primarily empirically motivated and lack strong theoretical justification. While physically motivated transitions guide our fitting, we also consider breakpoints suggested by trends in the observed data.

We combine the Gaussian prior with a Gaussian likelihood to compute the posterior. The likelihood accounts for uncertainties in both initial and final masses, as well the intrinsic scatter $\sigma_{\mathrm{int}}$. The log-likelihood takes the form:

\vspace{-10pt}
{\setlength{\abovedisplayskip}{2pt}
\setlength{\belowdisplayskip}{2pt} 
\begin{equation}
\ln \mathcal{L} = -\frac{1}{2} \sum_i \left[ \frac{\left(y_i - f(x_i)\right)^2}{\sigma_i^2} + \ln\left(2\pi \sigma_i^2\right) \right],
\end{equation}
where
\begin{equation}
\sigma_i^2 = s_k^2 \, \sigma_{x_i}^2 + \sigma_{y_i}^2 + \sigma_{\mathrm{int}}^2.
\end{equation}}
\noindent 
Here, $f(x_i)$ is the predicted final mass for progenitor mass $x_i$, $s_k$ is the slope of the segment containing $x_i$, and $\sigma_{x_i}$ and $\sigma_{y_i}$ are the measurement uncertainties on the initial and final masses, respectively.

We evaluate model quality using the Bayesian Information Criterion (BIC), computed at the best-fit solution. This allows us to assess whether models that include additional breakpoints, with broad uniform priors on the initial mass and no physical motivation, are statistically preferred. As is typical in the literature, a difference of $\Delta\mathrm{BIC}>8$ is considered strong evidence for adopting a more complex model. We apply this threshold when evaluating whether to include such additional breakpoints. In contrast, we do not automatically adopt simpler models with fewer breakpoints, even if they result in a slightly better or similar BIC. Given our emphasis on physically motivated priors, we retain more complex models that better reflect stellar evolution, unless a substantially lower BIC clearly favors a simpler alternative.

To explore the full posterior distribution, we use an ensemble MCMC sampler via the \texttt{emcee} package \citep{2013PASP..125..306F}, initialized near a numerically optimized solution using the L-BFGS-B algorithm. From the resulting samples, we derive medians and $1\sigma$ and $2\sigma$ credible intervals for all parameters. This process yields a robust and physically motivated IFMR. The remainder of this section presents the results from fitting both the Gaia-based and full samples using this approach. Where relevant, we compare to literature IFMRs and discuss breakpoints used in our fitting.


\subsubsection{Gaia-based IFMR}
\label{sec:gaiaifmr}

In our sample of sources supported as cluster members by Gaia astrometry, we have data from only two clusters below an initial mass of $2.7\msun$: Ruprecht 147, where initial masses are approximately $1.6\msun$, and NGC 752, which contributes only a single WD at an initial mass of around $1.97\msun$. Given this significant lack of data, we only fit the IFMR in the mass range above $2.7\msun$ for our Gaia-based sample.

We fit a single breakpoint, with a prior at $3.7\msun$, as both stellar evolution models and the data suggest a change in slope around this mass. While additional breakpoints have been proposed at lower and higher masses, we omit the lower breakpoints entirely, given our Gaia-based IFMR fit only includes data above $2.7\msun$. At higher masses, we find no compelling physical motivation for a breakpoint above $5\msun$, and the data in this regime of our cluster sample is too sparse and scattered to justify its inclusion. To assess whether the data support an additional transition, we also fit a two breakpoint model in which the second breakpoint is assigned a broad uniform prior between $2.7$ and $6.5\msun$, deliberately excluding higher masses given the scarcity of data in that regime. This model yields a BIC of $-135$, notably worse than the BIC of $-143$ for the single breakpoint fit. We therefore clearly favor the simpler model with one physically motivated breakpoint, and adopt it as our best-fit IFMR for the Gaia-based sample.

\begin{figure*}[ht]
\centering
\begin{minipage}{0.9\textwidth}
    \centering
    \includegraphics[width=0.7\textwidth]{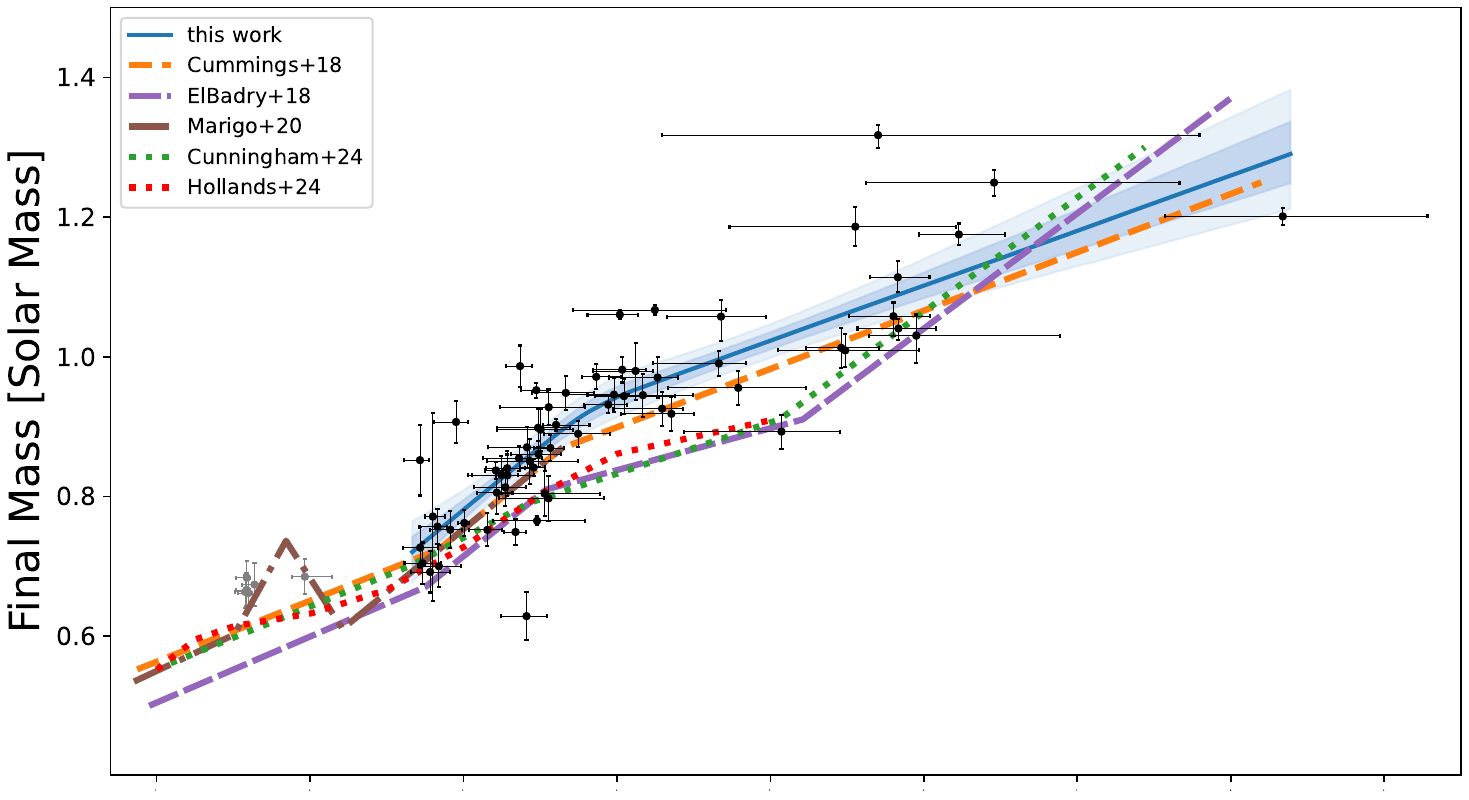} \\[-1.22em]
    \includegraphics[width=0.7\textwidth]{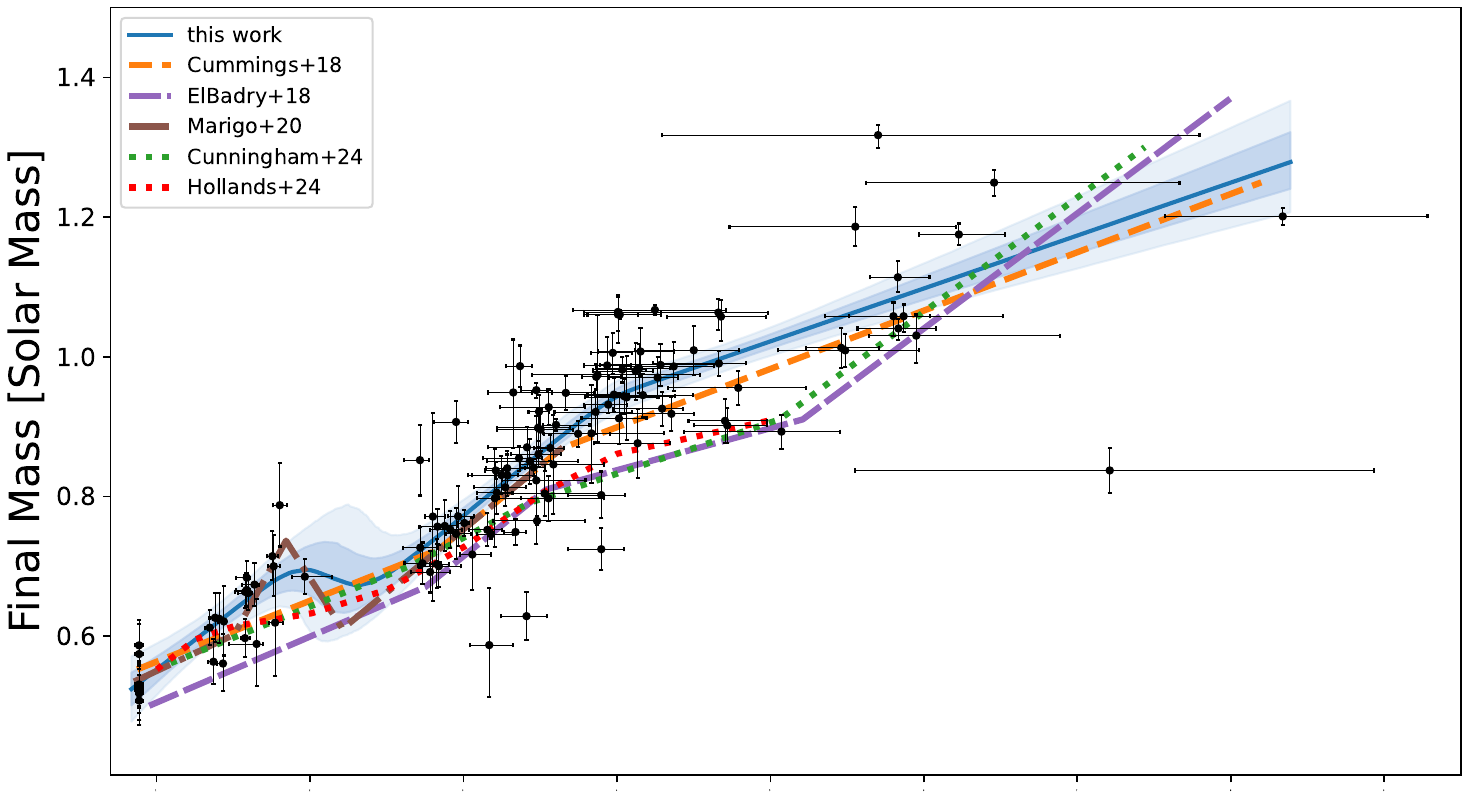} \\[-1.22em]
    \includegraphics[width=0.701\textwidth]{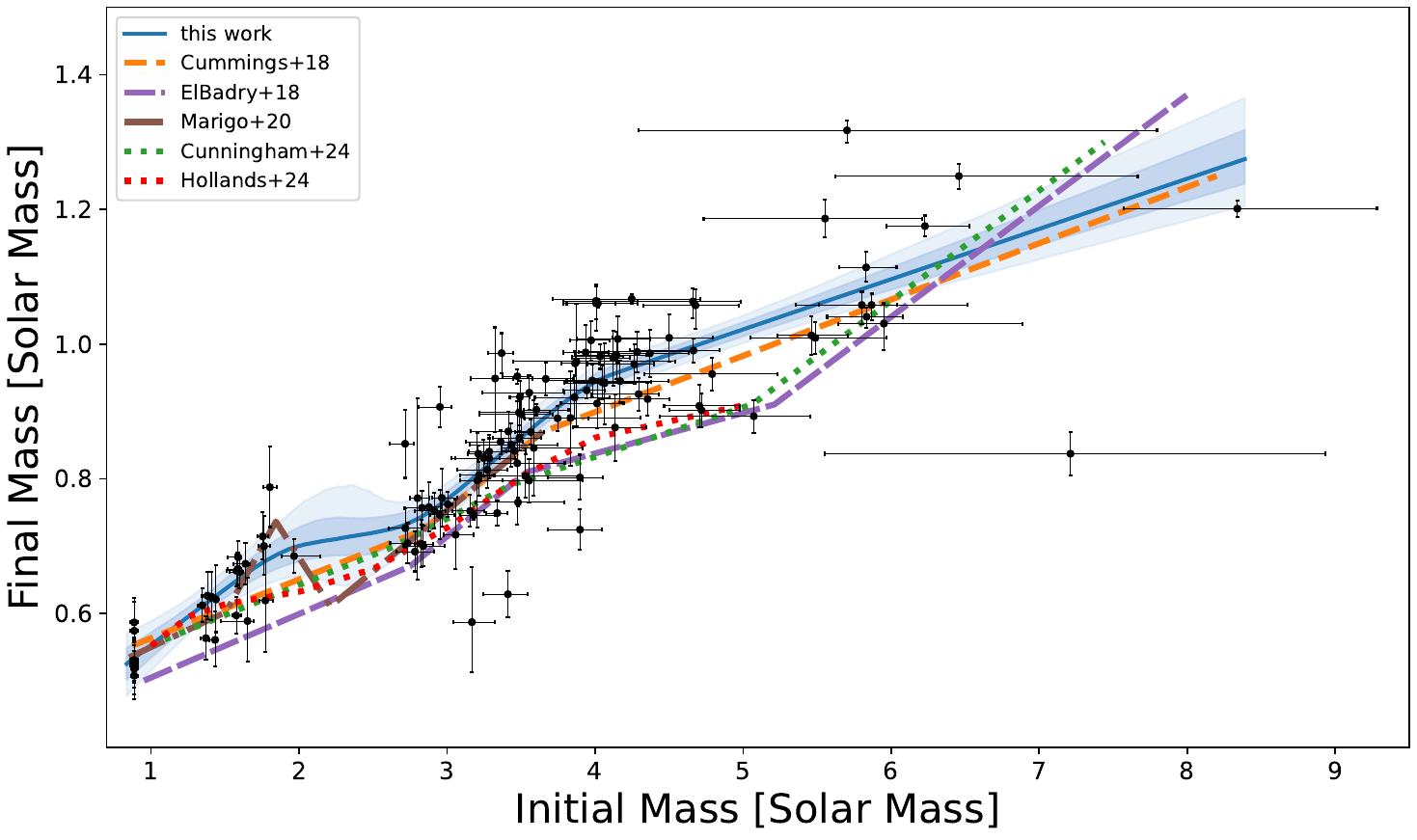}
\end{minipage}
\caption{{\textit{Top}: Derived IFMR for sources supported as members by Gaia astrometry. Solid blue curve shows the posterior predictive median, with the shaded regions indicating the $68\%$ and $95\%$ credible bands, with  select best-fit literature IFMRs overlaid for comparison. WDs in NGC 752 and Ruprecht 147 shown in gray excluded from the fit. \textit{Middle}: Full Sample Fit 1 to the IFMR, including both Gaia-supported sources and those which cannot be fairly assessed with Gaia astrometry. Curves and bands are as in the top panel. \textit{Bottom}: As in the middle panel, but for Full Sample Fit 2.}}
\label{fig:IFMR}
\end{figure*}

The top panel of Fig.~\ref{fig:IFMR} shows the median IFMR for the Gaia-based sample, with shaded bands indicating the $68\%$ and $95\%$ credible regions around the median predictive curve; the corresponding median breakpoint positions and uncertainties are given in Table~\ref{tab:ifmr_fit}. The corner plot showing the posterior distributions of the fit parameters is shown in Appendix~\ref{sec:appendix-figsandtables} Fig.~\ref{fig:corner_gaia_ifmr}. For comparison, we overlaid several literature IFMRs, including those derived from cluster studies \citep{2018ApJ...866...21C, 2020NatAs...4.1102M}, double WD binaries \citep{2024MNRAS.527.9061H}, and Gaia field star populations \citep{2018ApJ...860L..17E, 2024MNRAS.527.3602C}. While we include the \citet{2020NatAs...4.1102M} IFMR in our plot for comparison, we defer discussion of it to the IFMR incorporating non-Gaia-based astrometric sources, which extends to lower masses.

Our IFMR shows significant tension with the cluster-based IFMR of \citet{2018ApJ...866...21C}. Their best-fit falls largely outside our $2\sigma$ confidence region across most of the overlapping mass range, with the greatest discrepancy occurring near $4.7\msun$. The two relations converge slightly above $\sim6\msun$, though both are poorly constrained in this regime due to the small number of high-mass WDs. Our sample benefits from Gaia-based astrometric membership evaluations, whereas the \citet{2018ApJ...866...21C} sample likely includes more interlopers. Of the 43 DA WDs from their sample included in Gaia DR3 and consistent with single-stellar evolution, 28 have data suitable for Gaia astrometric membership assessments. Of these, four are inconsistent with cluster membership, implying a contamination rate of $\sim14\%$. A similar contamination rate may apply to the rest of their sample, including sources with poor or missing Gaia astrometry, though this is only a rough estimate. In contrast, our Gaia-based sample includes only sources deemed likely members using Gaia astrometry.  See Appendix~\ref{sec:appendix-figsandtables} Table~\ref{tab:cummings_2018} for details on the sources from \citet{2018ApJ...866...21C} compared to our sample. For the subset of WDs shared between the two samples, the inferred IFMRs are broadly consistent, with differences primarily driven by updated cooling models and revised cluster ages based on Gaia DR3. This supports the idea that sample contamination likely explains much of the discrepancy between the two IFMRs. 

\begin{table*}[ht]
\small
\caption{IFMR model fit parameters.}
\setlength{\tabcolsep}{16pt}
\begin{tabular}{cc|cc|cc}
\toprule
\multicolumn{2}{c}{\textbf{Gaia-based Sample Fit}} &
\multicolumn{2}{c}{\textbf{Full Sample Fit 1}} & 
\multicolumn{2}{c}{\textbf{Full Sample Fit 2}} \\
\hline
$M_i$ & 
$M_f$ &
$M_i$ & 
$M_f$ &
$M_i$ & 
$M_f$ \\ 
{}[$\msun$] &
[$\msun$] &
[$\msun$] &
[$\msun$] &
[$\msun$] &
[$\msun$] \\
\hline
-               & - &
$0.84$         & $0.52^{+0.02}_{-0.02}$ &
$0.84$         & $0.53^{+0.02}_{-0.02}$ \\
-               & - & 
$1.98^{+0.25}_{-0.23}$   & $0.70^{+0.05}_{-0.05}$ & 
$1.91^{+0.27}_{-0.23}$   & $0.70^{+0.05}_{-0.04}$ \\
$2.67$             & $0.72^{+0.02}_{-0.02}$  &
$2.33^{+0.28}_{-0.25}$   & $0.68^{+0.05}_{-0.05}$ &
$2.78^{+0.23}_{-0.28}$   & $0.73^{+0.04}_{-0.05}$ \\
$3.84^{+0.18}_{-0.20}$ & $0.93^{+0.03}_{-0.03}$ &
$3.98^{+0.09}_{-0.23}$   & $0.95^{+0.02}_{-0.03}$ &
$3.96^{+0.10}_{-0.26}$   & $0.94^{+0.02}_{-0.03}$ \\ 
$8.39$             & $1.29^{+0.05}_{-0.04}$ &
$8.39$         & $1.28^{+0.04}_{-0.04}$ &
$8.39$         & $1.28^{+0.04}_{-0.04}$ \\
\hline
\multicolumn{2}{c|}{$\sigma_{\rm int} = 0.052^{+0.008}_{-0.007}\,\mathrm{M_\odot}$} &
\multicolumn{2}{c|}{$\sigma_{\rm int} = 0.049^{+0.006}_{-0.005}\,\mathrm{M_\odot}$} &
\multicolumn{2}{c}{$\sigma_{\rm int} = 0.049^{+0.006}_{-0.005}\,\mathrm{M_\odot}$} \\
\hline
\end{tabular}
\label{tab:ifmr_fit}
\end{table*}

A striking feature of our results is the significant divergence between our IFMR and Gaia-based field WD IFMRs from \citet{2018ApJ...860L..17E} and \citet{2024MNRAS.527.3602C}, as well as the double WD IFMR of \citet{2024MNRAS.527.9061H}. Across nearly the entire mass range of our fit, the median fits of these relations lie outside our $2\sigma$ confidence region, with the largest deviation occurring near $5\msun$. While this comparison is based on their median relations alone, including their $95\%$ confidence intervals brings agreement within errors below $\sim3.7\msun$, However, the field WD relations diverge substantially at higher masses, falling outside the combined $2\sigma$ confidence regions from approximately $4$--$5\msun$. Some convergence reappears at higher masses, though this is not particularly meaningful given the large uncertainties in all relations in that regime and the highly disparate slopes. A $\sim2\%$ systematic offset between Gaia-derived and spectroscopic WD masses \citep{2019MNRAS.482.5222T} may contribute marginally to this discrepancy, but it cannot account for the full tension. Notably, both the above Gaia-based field WD IFMRs predict a breakpoint slightly above $5\msun$, though we are not aware of any physical motivation for it. The double WD IFMR similarly predicts lower WD masses throughout its recommended range ($1$-$5\msun$), placing it consistently below our relation. 

These discrepancies are not unexpected given the differing methodologies. However, the tension has grown significantly compared to pre-Gaia open cluster IFMRs such as \citet{2018ApJ...866...21C}, highlighting how recent field WD approaches, though benefiting from large sample sizes, may be more susceptible to systematics. As field WD studies infer progenitor masses indirectly, they are especially vulnerable to contamination from merger remnants and unknown atmospheric compositions. In contrast, our cluster-based sample benefits from Gaia-supported membership, direct age estimates, and a clean DA WD sample, reducing contamination and ensuring more reliable progenitor mass determinations. As such, our results offer a valuable external benchmark for calibrating field-based IFMRs, particularly at intermediate progenitor masses where our fit is best constrained. 

We provide the functional form of our Gaia-based IFMR:

\medskip
\noindent\textbf{Gaia-based Sample Fit}
{\small
\setlength{\abovedisplayskip}{2pt}
\setlength{\belowdisplayskip}{2pt} 
\begin{multline}
(2.67 \leq M_i/\msun < 3.84): \\ 
M_f = (0.179 \pm 0.031) \times M_i + (0.244 \pm 0.092) \msun 
\end{multline}
\begin{multline}
(3.84 \leq M_i/\msun < 8.39): \\ 
M_f = (0.079 \pm 0.012) \times M_i + (0.628 \pm 0.057) \msun
\end{multline}}

\noindent
The sheer breadth of data supporting this IFMR fit represents a significant increase over previous spectroscopically supported samples, more than doubling the Gaia-based IFMR sample of \citet{2022ApJ...926L..24M}. Future Gaia data releases will enable even larger and more precise samples, providing the opportunity to further refine the form of the IFMR. Given its clean Gaia-based membership support, which helps minimize systematics, we adopt this as our preferred IFMR for progenitor masses $\geq2.67\msun$. For convenience in modeling, we also express this IFMR in compact form, valid over the same range ($2.67\leq M_i / \msun < 8.39$):

\vspace{-12pt}
{\setlength{\abovedisplayskip}{2pt}
\setlength{\belowdisplayskip}{2pt}
\begin{multline}
    M_f = \left[0.179 - 0.100 H(M_i-3.84\msun) \right ] \times \\ (M_i-3.84\msun)+0.628 \msun
\end{multline}}

\noindent
where $H(x)$ is the Heaviside step function. For plotting purposes, we retain additional significant figures in the coefficients to ensure continuity at the breakpoints.


\subsubsection{General IFMR}
\label{sec:generalifmr}

While the self-consistency of the Gaia-based IFMR sample is valuable, its ability to probe the low-mass regime is limited by the lack of nearby old clusters. In the high-probability open cluster sample from \hunt used in this work, there is only one well-populated cluster ($N > 100$) older than 1~Gyr within 800~pc. Raising the threshold to 2~Gyr yields just one cluster within 1.8~kpc. Even lowering the member threshold to just 30 stars results in only four clusters older than 1~Gyr within 800~pc and only two older than 2~Gyr within 1.8~kpc. Low-mass WDs in more distant clusters typically fall below Gaia's magnitude limit ($\approx20.7\,\mathrm{mag}$), leaving very few clusters capable of hosting detectable low-mass WDs with Gaia.

To better explore this region, we construct an expanded sample that supplements the Gaia-based WDs with literature sources that have strong cluster associations but cannot be fairly assessed using Gaia astrometry. This combined fit retains all 69 WDs from the Gaia-based sample, including the Ruprecht 147 and NGC 752 WDs, which were excluded from the Gaia-based fit due to the lack of sufficient low-mass data below $2.7\msun$, and adds 53 additional WDs, including 20 with progenitor masses below $2\msun$.

Although the inclusion of non-Gaia-astrometric sources provides additional constraints at the low-mass end, there remains a significant gap, with no WDs having progenitor masses between 2 and $2.7\msun$. The IFMR suggested by the WDs below this gap differs from those above, and reconciling these regimes presents a significant challenge. We consider three possibilities: first, that the IFMR flattens between $\approx1.9$ and $2.7\msun$; second, that the IFMR exhibits a non-monotonic behavior in this region, as first suggested by \citet{2020NatAs...4.1102M}; and third, that a single IFMR cannot describe both regimes, potentially due to differences between the clusters involved.

If the IFMR at low masses does not align with the higher-mass regime, the question arises as to what distinguishing feature of these clusters could lead to a different progenitor mass relationship for their member WDs. The low-mass sources originate from six clusters, each at least 1.5~Gyr old, whereas the oldest cluster in the higher-mass regime is roughly half that age. Among these six older clusters, three (NGC 2682, NGC 6121, and NGC 7789) are notably subsolar in metallicity, while the others (NGC 752, NGC 6819, and Ruprecht 147) are near-solar or even supersolar. This raises the question of whether lower metallicity contributes to higher WD masses for a given progenitor mass. However, the discrepancy across the mass gap remains evident even when comparing only clusters with near-solar metallicity. As shown in Table~\ref{tab:cluster_ages_final}, our sample includes both subsolar and supersolar clusters on each side of the gap. While metallicity likely influences the IFMR, increasing the scatter in the data, it cannot explain the significant shift in the observed trend on each side of the $2$--$2.7\msun$ gap.

Similarly, another potential factor is the location of the clusters with respect to the Galactic plane. NGC 752, NGC 7789, and Ruprecht 147 all lie within 200~pc of the Galactic plane in the thin disk, while NGC 2682, NGC 6121, and NGC 6819 are each between 300 and 600~pc above the plane, placing them in the thick disk. These are the most distant clusters from the Galactic plane in our sample. However, as with the metallicity analysis, even if we only consider the clusters close to the galactic plane with low-mass WD members, the discrepancy between the low-mass and higher-mass IFMR remains. Having not attributed the difference to the cluster's metallicity nor the position with respect to the Galactic plane, we find no compelling reason for a different IFMR behavior at low masses due to cluster properties.

The possibility of a non-monotonic feature in the IFMR has received significant recent attention in the literature. This feature was first identified and discussed by \citet{2020NatAs...4.1102M}, who proposed a kink in the IFMR with a peak WD mass of $\approx0.70$ to $0.75\msun$ for progenitor masses between $\approx1.8$ and $1.9\msun$. They interpreted this kink as a consequence of the third dredge-up during the thermally pulsing AGB phase, where carbon and heavier elements are brought to the surface through convection. Stars that surpass a critical C/O ratio become carbon stars, and this threshold corresponds to the beginning of their proposed kink. They hypothesize that the increased WD mass for a given progenitor mass happens due to the carbon excess remaining too low for efficient formation of dust grains and the sustenance of dust-driven stellar winds, which drive mass loss in more massive AGB stars \citep{2010A&A...509A..14M,2019A&A...623A.119B}. This could extend the thermally pulsing AGB phase, allowing for more core growth than predicted by standard stellar evolution models. Further evidence supporting this possibility was presented by \citet{2022ApJS..258...43M}, who analyzed probable cluster-member AGB stars using Gaia data. They characterized these stars, including their initial and core masses, and identified carbon stars with unusually high core masses occupying a similar region of the IFMR kink, supporting the interpretation of \citet{2020NatAs...4.1102M}. They also explored whether these stars could have instead evolved from blue stragglers but concluded this was unlikely.

\citet{2024ApJ...964...51A} followed up on this work by attempting to reproduce the observed kink using PARSEC evolutionary calculations supported by COLIBRI calculations \citep{2013MNRAS.434..488M}. These models simulate thermally pulsing AGB phase evolution through total envelope loss, allowing the determination of final core masses. To reproduce the kink, they examined convective core efficiency and the impact of mass loss at both the base of the convective envelope and the pulse-driven convective boundary. They found that the IFMR kink could be reproduced in some cases, depending on the degree of convective core overshooting. While their results provide additional support for non-monotonic IFMR behavior, further theoretical and observational work is needed to characterize this phenomenon and determine if it is a genuine feature of the IFMR.

Motivated by this, we fit a piecewise linear IFMR that incorporates non-monotonic behavior at the observed kink, which is thought to be a signature of carbon star formation. While \citet{2020NatAs...4.1102M} included multiple breakpoints in this region, we adopt a simpler form with priors on three breakpoints: one near the observed local peak at $1.8\msun$, one at $2.3\msun$ corresponding to the helium ignition transition at the tip of the red giant branch, and one at $3.7\msun$ marking the onset of the second-dredge up. The middle panel of Fig.~\ref{fig:IFMR} shows the resulting IFMR with $68\%$ and $95\%$ credible intervals; the corresponding breakpoint values are listed in Table~\ref{tab:ifmr_fit}. The posterior distributions are shown in the corner plot in Appendix~\ref{sec:appendix-figsandtables} Fig.~\ref{fig:corner_full_fit1_ifmr}. We tested the addition of a forth breakpoint drawn from a broad uniform prior between $1.0$ and $6.5\msun$, but this yielded a worse BIC ($-263$) than the baseline three breakpoint model ($-269$) and was therefore discarded. Removing the breakpoint in the gap resulted in a similar BIC ($-271$), suggesting that it is not statistically required; however, we retain it given the strong physical motivation associated with the red giant branch transition.

At intermediate and high masses, the IFMR fit remains largely unchanged from the Gaia-based fit and continues to show significant disagreement with IFMRs derived from double white dwarf binaries and Gaia field WD populations, as discussed in Section~\ref{sec:gaiaifmr}. With a breakpoint in the middle of the mass gap, the fit allows for a non-monotonic trend similar to that proposed by \citet{2020NatAs...4.1102M}, but the feature is far less pronounced than in their best-fit model. That said, their kink remains largely within our $2\sigma$ credible region, only slightly exceeding it at the peak. Replicating their stronger non-monotonic behaviour would require either a tight prior on the final mass at the breakpoint or, ideally, additional data to directly constrain the IFMR behaviour across the gap. More generally, the results in the lower-mass regime show better agreement with several literature IFMRs compared to the higher-mass regime, although this is not unexpected given the larger credible intervals. These intervals are especially wide and poorly constrained across the $2$ to $2.7\msun$ mass gap.

While this fit provides a reasonable description of the data, we also examine the third possibility discussed earlier: that the IFMR flattens across the low-mass gap. Although we lack a strong physical motivation for this behavior, it offers a simpler empirical description of the observed trend across the gap. 

In this alternative fit, we place the middle breakpoint just beyond the gap rather than within it. We adopt the same priors for the breakpoints at $1.8\msun$ and $3.7\msun$, but now place the middle breakpoint prior at $2.8\msun$, enabling a smoother transition across the mass gap. The bottom panel of Fig.~\ref{fig:IFMR} shows the resulting IFMR. As with the previous fit, full results are given in Table~\ref{tab:ifmr_fit}, and the corresponding corner plot is shown in Appendix~\ref{sec:appendix-figsandtables} Fig.~\ref{fig:corner_full_fit2_ifmr}. The overall behavior is very similar, though the credible intervals across the gap are slightly narrower.

While the inclusion of the $2.8\msun$ breakpoint yields a BIC ($-270$) similar to the simpler two breakpoint model ($-271$), we prefer the extra breakpoint because the simpler models fails to capture the elevated WD masses relative to their progenitors seen just below the gap compared to those just above. This alternative fit allows us to explore how the IFMR might behave if the relation flattens across the gap due to an unknown physical mechanism. A four breakpoint model with the additional breakpoint drawn from a broad uniform prior between $1.0$ and $6.5\msun$ yielded a worse BIC ($-261$) and no improvement in fit quality. We also tested a four breakpoint model using Gaussian priors at $1.8$, $2.3$, $2.8$, and $3.7\msun$, but simultaneously including multiple breakpoints near the observational gap led to unstable fits and was ultimately abandoned.

We provide the functional form of our general sample IFMRs:

\noindent\textbf{Full Sample Fit 1}
{\small
\setlength{\abovedisplayskip}{4pt}
\setlength{\belowdisplayskip}{4pt} 
\begin{multline}
(0.84 \leq M_i/\msun < 1.98): \\ 
M_f = (0.160 \pm 0.045) \times M_i + (0.394 \pm 0.063) \msun
\end{multline}
\begin{multline}
(1.98 \leq M_i/\msun < 2.33): \\ 
M_f = (-0.155 \pm 0.171) \times M_i + (1.015 \pm 1.644) \msun
\end{multline}
\begin{multline}
(2.33 \leq M_i/\msun < 3.98): \\ 
M_f = (0.174 \pm 0.026) \times M_i + (0.251 \pm 0.183) \msun
\end{multline}
\begin{multline}
(3.98 \leq M_i/\msun < 8.39): \\ 
M_f = (0.075 \pm 0.011) \times M_i + (0.644 \pm 0.048) \msun
\end{multline}}

\medskip
\noindent\textbf{Full Sample Fit 2}
{\small
\setlength{\abovedisplayskip}{4pt}
\setlength{\belowdisplayskip}{4pt} 
\begin{multline}
(0.84 \leq M_i/\msun < 1.91): \\
M_f = (0.168 \pm 0.040) \times M_i + (0.384 \pm 0.053) \msun
\end{multline}
\begin{multline}
(1.91 \leq M_i/\msun < 2.78): \\
M_f = (0.021 \pm 0.051) \times M_i + (0.666 \pm 0.310) \msun
\end{multline}
\begin{multline}
(2.78 \leq M_i/\msun < 3.96): \\
M_f = (0.187 \pm 0.033) \times M_i + (0.207 \pm 0.078) \msun
\end{multline}
\begin{multline}
(3.96 \leq M_i/\msun < 8.39): \\
M_f = (0.075 \pm 0.011) \times M_i + (0.646 \pm 0.049) \msun
\end{multline}}

\noindent
This extended IFMR incorporates the full cluster WD sample and includes two separate fits: one with a breakpoint placed within the low-mass gap (Full Sample Fit 1), which allows for non-monotonic behavior, and another with a smoother transition across the gap via a breakpoint placed just beyond it (Full Sample Fit 2). While we do not strongly prefer one fit over the other, Full Sample Fit 1 is modestly favored, given physical support for the possibility of the observed non-monotonic behaviour. For progenitor masses above $2.7\msun$, we recommend the Gaia-based IFMR due to its cleaner sample and reduced systematics. As in the Gaia-based sample, we retain additional significant figures in the coefficients to ensure continuity at the breakpoints. Future work should prioritize identifying and characterizing WDs in clusters with turnoff masses between 2 and $2.7\msun$ to better constrain the low-mass end of the IFMR and determine whether the apparent non-monotonic behavior is a genuine physical feature. 


\subsubsection{Mass loss}

\begin{figure*}[ht]
\centering
\begin{minipage}{0.9\textwidth}
    \centering
    \includegraphics[width=0.7\textwidth]{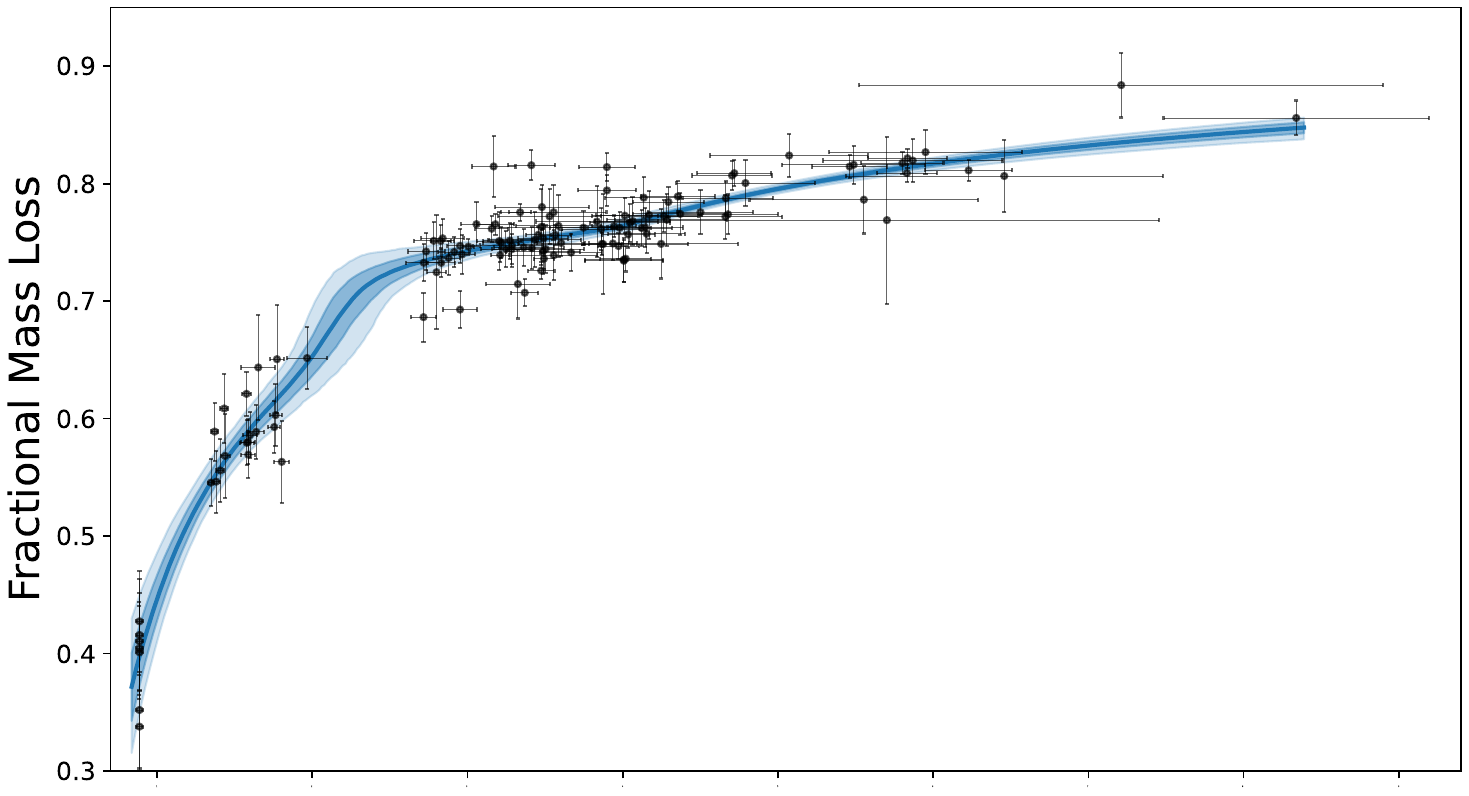} \\[-1.22em]
    \includegraphics[width=0.7\textwidth]{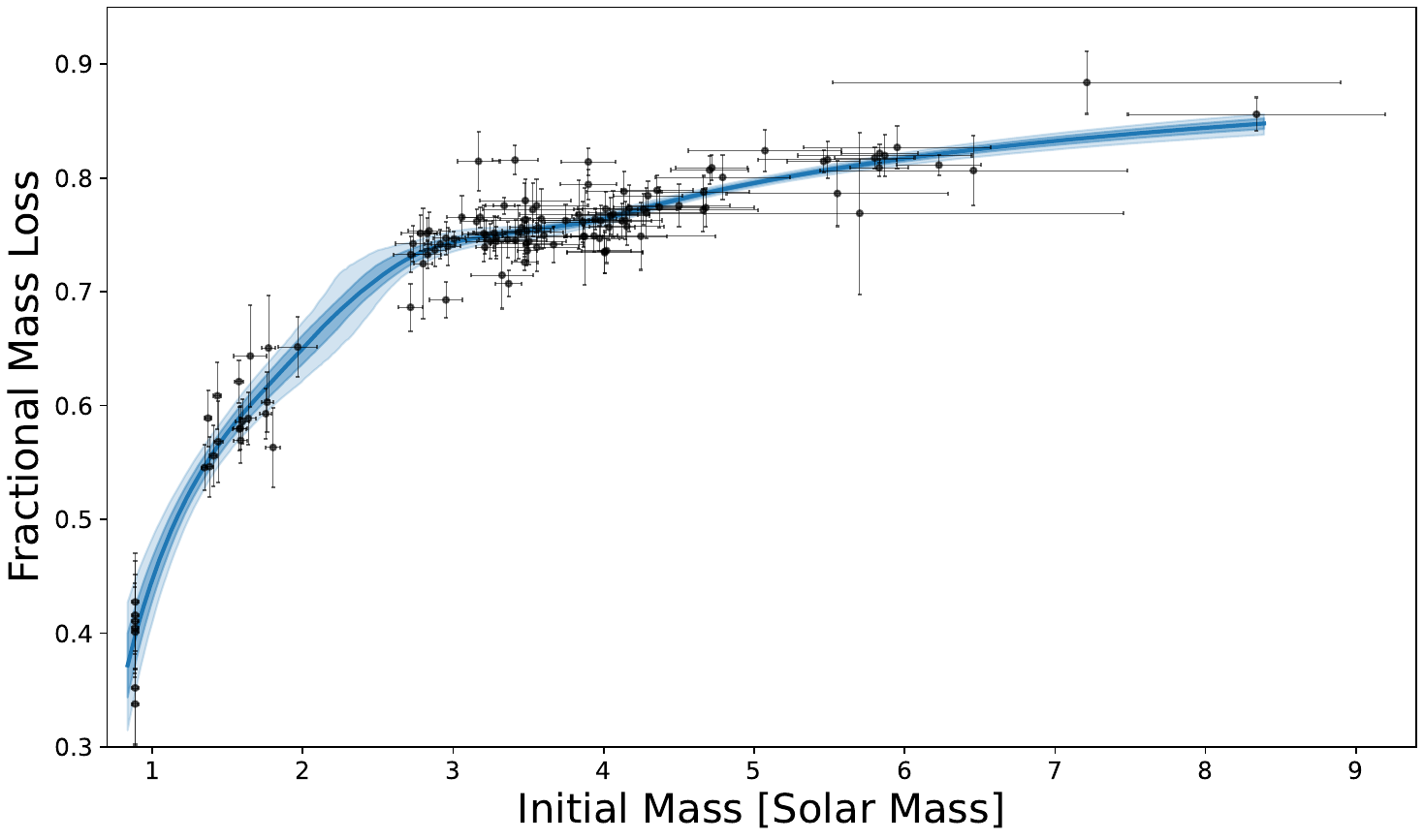}
\end{minipage}
\caption{{\textit{Top}: Fractional mass loss from our Full Sample Fit 1 IFMR. The solid curve is the median predictive relation and the shaded bands are the $68\%$ and $95\%$ credible regions directly computed from the MCMC posterior. \textit{Bottom}: Same as the top panel, but for Full Sample Fit 2.}}
\label{fig:massloss}
\end{figure*}

In Fig.~\ref{fig:massloss}, we present the fractional mass loss, derived from the MCMC posterior distributions of our two full sample IFMR fits. The top panel shows the relation for Full Sample Fit 1, while the bottom shows Fit 2. In both cases, the solid curve represents the median posterior prediction, and the shaded bands show the $68\%$ and $95\%$ credible intervals. 

At the low mass endpoint ($M_i = 0.84\msun$), the inferred mass-loss fraction is $37.2\pm2.8\%$ in Fit 1 and $37.3\pm2.8\%$ in Fit 2. In both cases, the mass-loss fraction rises steeply with progenitor mass and begins to level off toward the end of the mass gap, increasing more gradually at higher masses. The maximum value is reached at the high mass endpoint ($M_i = 8.39\msun$), where both fits converge on $84.8^{+0.4}_{-0.5}\%$.

\section{Summary and Conclusions}
\label{sec:conclusions}

This work presents the most comprehensive and robust open cluster WD IFMR to date, combining newly obtained spectroscopically confirmed cluster member WDs with an extensive literature review. By including only spectroscopically confirmed DA WDs, we reduce potential systematics, distinguishing our approach from many previous IFMR studies that also included DB WDs. We introduce a heuristic cluster age determination method that combines the advantages of manual isochrone fitting for the best fit and automatic statistical methods to assess uncertainties. By rederiving WD cooling parameters and progenitor masses for the full sample, we ensure a fully self-consistent methodology that further minimizes potential systematics. 

Our IFMR fitting approach employs a physically motivated, piecewise linear model, using MCMC methods to determine best fits and credible intervals. We separately fit the IFMR for a Gaia-based sample of 69 WDs with reliable Gaia astrometric cluster membership, and for a full sample that includes an additional 53 literature WDs with strong cluster associations that cannot be fairly assessed with Gaia. For the full sample, we present two fits. One places a breakpoint within the $2$--$2.7\msun$ progenitor mass gap, allowing for a non-monotonic trend consistent with the feature proposed by \citet{2020NatAs...4.1102M}. The other assumes a flatter IFMR across this gap. A key result from both samples is a significant deviation from recent IFMRs derived from Gaia field WDs and double WD binaries, with the strongest discrepancy near $5\msun$. 

Despite these advances, the IFMR remains poorly constrained at the high-mass end and in the low-mass gap, which contains no confirmed cluster-member WDs. Future efforts should focus on identifying and characterizing very high-mass cluster WDs to better constrain the maximum progenitor mass, as well as probing clusters with turnoff masses in the $2$--$2.7\msun$ range to test whether the observed non-monotonic behavior is an intrinsic feature of the IFMR. Overall, our results establish the strongest open cluster constraints on the IFMR to date and reinforce the critical role of open clusters in constraining the IFMR to better understand the evolution of white dwarfs and their progenitors.


\section*{Acknowledgments}

The authors would like to thank the anonymous referee for their constructive feedback, which helped improve the clarity of the manuscript. This work was supported in part by the Natural Sciences and Engineering Research Council of Canada Discovery Grants DG-RGPIN-2022-03051 and DG-RGPIN-2023-04486. This research received funding from the European Research Council under the European Union's Horizon 2020 research and innovation program number 101002408 (MOS100PC). This work includes results based on observations obtained at the international Gemini Observatory, a program of NSF’s NOIRLab, which is managed by the Association of Universities for Research in Astronomy (AURA) under a cooperative agreement with the National Science Foundation on behalf of the Gemini Observatory partnership: the National Science Foundation (United States), National Research Council (Canada), Agencia Nacional de Investigaci\'{o}n y Desarrollo (Chile), Ministerio de Ciencia, Tecnolog\'{i}a e Innovaci\'{o}n (Argentina), Minist\'{e}rio da Ci\^{e}ncia, Tecnologia, Inova\c{c}\~{o}es e Comunica\c{c}\~{o}es (Brazil), and Korea Astronomy and Space Science Institute (Republic of Korea). This work has made use of data from the European Space Agency (ESA) mission {\it Gaia} (\url{https://www.cosmos.esa.int/gaia)}, processed by the {\it Gaia} Data Processing and Analysis Consortium (DPAC, \url{https://www.cosmos.esa.int/web/gaia/dpac/consortium}). Funding for the DPAC has been provided by national institutions, in particular the institutions participating in the {\it Gaia} Multilateral Agreement. Some of the data presented herein were obtained at the W. M. Keck Observatory, which is operated as a scientific partnership among the California Institute of Technology, the University of California and the National Aeronautics and Space Administration. The Observatory was made possible by the generous financial support of the W. M. Keck Foundation. Gemini spectra were processed using the DRAGONS package \citep{2023RNAAS...7..214L}. LRIS spectra were reduced using the Lpipe pipeline \citep{2019PASP..131h4503P}.

\software{Astropy \citep{2013A&A...558A..33A,2018AJ....156..123A,2022ApJ...935..167A}, emcee \citep{2013PASP..125..306F}.}

\facilities{Gaia (DR2 $\&$ DR3), Gemini-North (GMOS-N), Gemini-South (GMOS-S), Keck:I (LRIS).}

\section*{Data Availability}

The initial WD cluster member candidate sample was developed using the \citet{2023A&A...673A.114H} and \citet{2021MNRAS.508.3877G} catalogs, both of which are publicly available. Reduced Gemini GMOS spectra are available upon request, while raw data can be accessed from the Gemini archive following the designated embargo period.


\bibliographystyle{aasjournal}
\bibliography{main}



\appendix


\section{Cluster-by-Cluster Discussion}
\label{sec:appendix-clusters}

As discussed in Section~\ref{sec:results_wdsample}, this appendix provides a detailed discussion of each cluster in which we identified at least one spectroscopically confirmed DA WD, either from new observations or from the literature. Clusters where no candidates were observed and where no DA WDs are known from the literature are not discussed. We also exclude clusters where the only known WD members are likely products of nonstandard or binary evolution (e.g., helium-core WDs; \citealt{2007ApJ...671..748K}).

For literature WDs, including those identified in pre-Gaia or Gaia DR2-based studies, we examine Gaia DR3 astrometry as a sanity check on cluster associations, provided the source is included in Gaia and has astrometry precise enough for a meaningful membership comparison. While we do not perform a full membership reanalysis, we verify that each WD remains broadly consistent with its host cluster in proper motion and parallax. Deviations exceeding $3\sigma$ in either quantity raise concerns, though we do not enforce this as a strict criterion. In particular, sources previously identified as likely cluster escapees or those that are especially well studied, such as members of the nearby Hyades, are granted greater leeway when assessing astrometric consistency.


\subsection{AB Doradus Moving Group}

While our analysis primarily focuses on open clusters, we make an exception for the AB Doradus Moving Group (ABDMG) due to the presence of GD 50, an ultramassive WD with strong membership support. Although we generally avoid moving groups due to the difficulty in robustly determining their ages, ABDMG is a well-studied nearby association with a well-constrained age. GD 50 was previously proposed as a possible escapee from the Pleiades \citep{2006MNRAS.373L..45D}, but \citet{2018ApJ...861L..13G} used Gaia DR2 kinematics and the BANYAN~$\Sigma$ Bayesian classification algorithm \citep{2018ApJ...856...23G} to show that GD 50 has a $99.7\%$ probability of membership in ABDMG and only negligible probability of association with the Pleiades. \citet{2018ApJ...866...21C} included GD 50 in their IFMR as a Pleiades member, noting the possibility of ABDMG membership and stating that the distinction was unimportant given the similar ages of the two groups and the possibility that they are coeval.

ABDMG was not identified by \hunt, which is unsurprising given its proximity and the limitations of clustering algorithms in resolving nearby, extended associations. We therefore adopt the age of $133^{+15}_{-20}\,\mathrm{Myr}$ from \citet{2018ApJ...861L..13G}. Spectroscopic parameters are taken from \citet{2011ApJ...743..138G}, and we assume solar metallicity for progenitor mass estimation. We find a mass of $1.249\pm0.019\msun$, making GD 50 one of the most massive WDs in our sample. Given the strong Gaia-based membership support, we include GD 50 in our IFMR as a member of the AB Doradus Moving Group.


\subsection{Alpha Persei}

Melotte 20, better known as Alpha Persei, has been the subject of several studies investigating its white dwarf population. \citet{2015MNRAS.451.4259C} conducted a photometric search using the UKIRT Infrared Deep Sky Survey (UKIDSS), with candidate selection further refined using proper motions from the SuperCOSMOS survey, ultimately identifying 14 new WD candidates. Optical spectroscopic follow-up of the brightest 11 sources with the William Herschel Telescope confirmed 7 as DA WDs, 3 as DB WDs, and one as a non-WD object. However, only three WDs were determined to be young enough to have originated from the cluster, and none appear to be genuine members when evaluated against the IFMR. Among these 14 candidates, two were included in the initial kinematic candidate list of \citet{2019A_A...628A..66L}. Neither object lies within the cluster's tidal radius, and further analysis of their ages ruled one out as a candidate and made the other a questionable association. Ultimately, \citet{2019A_A...628A..66L} concluded that neither object is likely kinematically connected to the cluster, a conclusion we adopt in this work.

\citet{2022ApJ...926L..24M} identified three ultramassive WDs whose spectra were consistent with having escaped from the cluster, building directly upon the earlier work of \citet{2021arXiv211004296H}. Each is among the most massive WDs with the most massive progenitors known to have strong cluster associations. Though no new candidates were identified in this work, we reanalyzed their cooling parameters using the spectra from \citet{2022ApJ...926L..24M} and include them in our IFMR.

As discussed in Section~\ref{sub_sec:cluster ages}, we adopt the cluster's age from \citet{2021arXiv211004296H} of $81\pm6\,\mathrm{Myr}$, which was also utilized in \cite{2022ApJ...926L..24M}. This kinematic age is consistent with ages from the lithium depletion boundary (e.g., \citealt{1999ApJ...527..219S}; \citealt{1999ApJ...510..266B}; \citealt{2004ApJ...614..386B}) and Gaia-based cluster isochrones (e.g., \citealt{2018A&A...616A..10G}). Recently, \citet{2025ApJ...988..113D} used the binary HD~21278 to infer a younger cluster age of $49\pm7\,\mathrm{Myr}$ using PARSEC isochrones alongside 2MASS-based CMD of cluster members built from a crossmatch to the Gaia-based membership list of \citet{2023AJ....166...14B}. They use this younger age to rule out membership for the most massive WD from \citet{2022ApJ...926L..24M}. While their techniques are not without merit, their 2MASS CMD exhibits significant scatter compared to their companion Gaia CMD, and the isochrone fit to HD~21278 provides a poor fit to the global 2MASS CMD. Given the significant agreement between Gaia-based isochrone, lithium depletion boundary, and kinematic age estimates, we retain the older age estimates, and thus retain all of the WDs from \citet{2022ApJ...926L..24M}.


\subsection{ASCC 47}

A WD associated with ASCC 47 was identified as a high-fidelity cluster member using Gaia DR2 data by \citet{2021ApJ...912..165R}. This WD is notable for being one of the hottest known, with a temperature of $114{,}000 \pm 3{,}000\,\mathrm{K}$, as analyzed in \citet{2020ApJ...901L..14C}. The WD is magnetic, with a field strength of approximately 1.6~MG, and its spectra and apparent magnetism were further examined in \citet{2020ApJ...901L..14C}. Despite the magnetic nature, the authors determined that the WD's cooling age and and cluster membership are consistent with a single stellar evolution origin, without requiring a cooling delay associated with past mergers.

Due to the degeneracy between magnetic line broadening and surface gravity effects, \citet{2020ApJ...901L..14C} concluded that spectroscopy could not reliably measure $\log g$ for this object. They instead used photometric fitting to estimate the surface gravity, which we adopt in our analysis. It is worth noting that magnetic fits tend to yield slightly lower temperatures than the values used here.

\citet{2020ApJ...901L..14C} derived an age of $90 \pm 20\,\mathrm{Myr}$ using Gaia DR2. Although ASCC 47 is included in the literature crossmatch list of \hunt, it was not matched to any of the clusters identified by their algorithm and therefore does not appear in their final catalogue. This is somewhat surprising given that ASCC 47 is a well-known, well-populated cluster, but even though the \hunt catalogue offers the most complete coverage of any single cluster catalogue to date, some omissions are expected. As a result, we adopt the age estimate of \citet{2020ApJ...901L..14C} in our analysis, since our age-determination techniques cannot be directly applied.


\subsection{ASCC 113}

ASCC 113 has several age estimates in the literature, with those referenced here all derived using isochrone fitting. Our isochrone fit gives an age of $282^{+24}_{-9}\,\mathrm{Myr}$ for a slightly supersolar metallicity and $A_v=0.128$. This is very similar to the 288~Myr age found by \citet{2020A&A...640A...1C} with $A_v=0.13$. \citet{2021ApJ...912..165R} derived a younger age of $240\pm40\,\mathrm{Myr}$, while \citet{2019A&A...623A.108B} reported a significantly older age of 426~Myr for solar metallicity and a lower reddening of $A_v=0.094$.

While we found no WDs associated with ASCC 113 in our crossmatch, one has a credible association in \citet{2021ApJ...912..165R}, based on Gaia DR2 data. The absence of this source in \hunt is likely primarily driven by the WD's updated parallax, which dropped from 1.662~mas in DR2 to 1.035~mas in DR3; however, given its DR3 parallax error of 0.347~mas, it is only just beyond $2\sigma$ from the \hunt cluster center at 1.771~mas. Furthermore, its proper motion remains strongly consistent with cluster membership, and our reanalysis of the WD cooling parameters and inferred progenitor mass further supports membership. Given this combined evidence, we retain the WD as a likely ASCC 113 member. Future Gaia data releases should allow us to better assess whether the distance is in reasonable agreement with the cluster.


\subsection{BH 99}

BH 99, also known as vdB Hagen 99, has an age estimate of 81~Myr derived by \citet{2019A&A...623A.108B} for solar metallicity with $A_v=0.203$. Our isochrone fit finds a similar age of $76^{+10}_{-28}\,\mathrm{Myr}$ but with a lower reddening of $A_v=0.132$ for a slightly supersolar metallicity cluster. Although the cluster's lone giant star suggests a significantly younger age, we prefer the MSTO driven age estimates, as they rely on a well-populated sequence and offer more robust, statistically reliable constraints than those driven heavily by a single evolved star.

We identified one candidate cluster member in BH 99, which we followed up with GMOS. Its Gaia photometry indicates it is potentially among the most massive WDs in our crossmatch sample.
While the GMOS spectra provide results broadly consistent with cluster membership, the atmospheric parameter fitting raises concerns. The derived $\log g$ appears reliable; however, the H$\beta$ spectral line exhibits a significant downward spike at its center (see Fig.~\ref{fig:spectrafits_a}), rendering the effective temperature highly questionable and, consequently, the derived cooling age highly questionable. Additional follow-up spectroscopy is necessary to determine whether the observed shape of H$\beta$ is a genuine feature of the WD or a potential issue with the obtained spectrum. That said, several factors support it being a true cluster member. The derived mass is primarily constrained by the wings of the spectral fit and is therefore less affected by the issues at the center of the H$\beta$ line. Furthermore, the estimated mass corresponds closely to the cluster's turnoff mass. Given the astrometric membership support alongside the mass estimate, we retain the object in the sample despite the uncertainties in its cooling age.


\subsection{HSC 601}

HSC 601 is one of the newly identified clusters from \hunt and, as such, has no prior discussion in the literature. We identified one WD candidate in this cluster. Although the cluster is sparsely populated at the MSTO, making accurate age determination difficult, we obtained follow-up spectroscopy of this WD as its Gaia photometry suggested it was among the most massive candidates in our sample.

Our analysis finds a WD with a mass of $0.89\msun$, which, while reasonably massive, is notably lower than the mass inferred from Gaia photometry. Although the cluster age ($126^{+44}_{-22}\,\mathrm{Myr}$) carries moderate uncertainty, it remains broadly consistent with the WD mass expected from the cluster MSTO. While the WD falls slightly off the typical IFMR sequence, its substantial error bars make it far from a clear outlier. Given these considerations, we retain it in our IFMR as a plausible cluster member. 


\subsection{Hyades}

At approximately 47~pc from the Sun, the Hyades (Melotte 25 in \hunt) is the closest open cluster. As a result, it has been extensively studied in the literature, with many confirmed WD members. Our discussion begins with the sample of 10 confirmed cluster-member WDs identified by \citet{1998AJ....115.1536V}, which includes three binaries and seven single WDs. \citet{2012A&A...537A.129S} referred to these as the classical Hyades WDs. The seven single classical Hyades WDs are WD 0352+096 (EGGR 26; HZ 4), WD 0406+169 (EGGR 29; LB 227), WD 0421+162 (EGGR 36; VR 7; LP 415-46), WD 0425+168 (EGGR 37; VR 16; LP 415-415), WD 0431+126 (EGGR 39; HZ 7), WD 0437+138 (EGGR 316; LP 475-242), and WD 0438+108 (EGGR 42; HZ 14).

\citet{2012A&A...537A.129S} recovered all 10 classical Hyades WDs and identified an additional 17 candidates, mostly expected to be escaping members, using PPMXL position and proper motion data. \citet{2012A&A...547A..99T} relaxed the relative velocity cutoff from \citet{2012A&A...537A.129S}, identifying 14 additional candidates, of which only three were retained after spectroscopic examination. They re-evaluated the 17 candidates from \citet{2012A&A...537A.129S} and their three new ones. Of these, six were ruled out due to spectroscopically derived distances that would require unreasonably large residual velocities to remain connected to the cluster. Of the remaining 14, two were in binaries, six were classified as single WDs young enough to originate from the cluster, five were likely too old, and one was considered questionable. Most of these WDs lay beyond the tidal radius and were thought to be in the process of escaping.

\citet{2012A&A...547A..99T} had assessed membership based on kinematic and spectroscopic distances, tangential velocities, and cluster-centric distances. Building on this, \citet{2013ApJ...770..140Z} expanded the analysis specifically for the seven promising new candidates, namely the six single WDs young enough to originate from the cluster and the one questionable case, using radial velocity measurements from Balmer lines obtained with the HIRES echelle spectrometer on Keck. They confirmed three as likely true escaping members and another as questionable. The newly confirmed members were WD 0348+339 (GD 52), HS 0400+1451 (HG7-85), and WD 0625+415 (GD 74), with WD 0637+477 (GD 77) deemed a questionable member. \citet{2018ApJ...866...21C} included the seven classical single WDs and these four likely or questionable members from \citet{2013ApJ...770..140Z} in their IFMR analysis.

\citet{2018MNRAS.480.3170S} examined the Hyades with Gaia DR2, starting from the cluster members determined by \citet{2018A&A...616A...1G}, which included nine WDs. Seven of the nine were the classical single Hyades WDs, while the remaining two were HS 0400+1451 and WD 0348+339, both confirmed as likely members by \citet{2013ApJ...770..140Z}. They relaxed the membership criteria of \citet{2018A&A...616A...1G} to identify additional WD candidates. Using a relaxed proper motion criterion, they re-evaluated the 25 candidates from \citet{2012A&A...547A..99T}, identifying three additional candidates (GD 38, GD 43, and LP 475-249). Each had been previously rejected as a nonmember by \citet{2012A&A...547A..99T}. Their low parallax values raised suspicions of systematic errors due to unresolved binarity, and while \citet{2018MNRAS.480.3170S} excluded them from their analysis, they suggested that future Gaia releases might clarify their status. We examined these sources in Gaia DR3 and found all three, GD 38 (Gaia DR3 142709256302827776), GD 43 (Gaia DR3 62271429450749824), and LP 475-249 (Gaia DR3 3307915199278868096), to be extreme outliers in both proper motion and parallax space, ruling them out as members.

\citet{2019A&A...623A..35L} analyzed the cluster in 3D space with Gaia DR2, using the kinematic approach of \citet{1998A&A...331...81P}, originally developed by \citet{1971MNRAS.152..231J}, which identifies members based on velocities relative to the cluster barycenter. Restricting candidates to those within 30~pc of the cluster center returned 710 likely members. Cross-matching this sample with the 37 WD candidates from \citet{2012A&A...537A.129S} yielded 13 WDs: the 10 classical WDs along with HS WD 0400+1451, WD 0348+339, and an additional binary. The sample of single WDs was the same as that identified by \citet{2018MNRAS.480.3170S}. They excluded WD 0437+138 from their IFMR analysis due to its spectral type of DBA.

In this work, we identified ten candidate Hyades white dwarfs (WDs 1-10). Because the Hyades is so close, it appears very extended on the sky, and clustering algorithms often struggle to identify candidate members consistently. Among our ten candidates, WDs 1-4 correspond to confirmed cluster members, including classical members WD 0437+138 (WD1) and WD 0352+096 (WD3), as well as later-confirmed members HS 0400+1451 (WD2) and WD 0348+339 (WD4). WD 5 is a new detection but its low mass makes membership unlikely, so we did not follow it up. WD 6 (LB 228) is classified as a DC WD in the literature and remains a candidate, though we excluded it from our DA-focused analysis. Finally, WDs 7--10 were given lower priority because their inferred ages are significantly older than the Hyades.

Since WD1 is a DBA WD, we exclude it from our DA IFMR. We include WDs 2-4 in our IFMR, along with the five classical Hyades WDs we did not independently identify, whose omission is unsurprising given the challenges automated algorithms face with such an extended, nearby cluster. WD 0625+415 and WD 0637+477, both considered candidates by \citet{2012A&A...547A..99T} and included in the IFMR of \citet{2018ApJ...866...21C}, appear to be unlikely members based on Gaia DR3 proper motions and were therefore excluded. Finally, we include the extremely massive WD cluster escapee from \citet{2023ApJ...956L..41M}, which is strongly supported as a true escapee and is the most massive WD in our IFMR sample, with a mass of $1.317^{+0.015}_{-0.018}\msun$.

The age of the Hyades cluster has been widely debated, with estimates varying significantly across different studies. \citet{2019A&A...623A..35L} found an age of $640^{+67}_{-49}\,\mathrm{Myr}$using the WD sequence. \citet{2018ApJ...856...40M} estimated an age of $650\pm75\,\mathrm{Myr}$ from the lithium depletion boundary, using brown dwarf evolutionary models from \citet{2015A&A...577A..42B}. \citet{2018ApJ...863...67G} employed MESA-based stellar evolution models and derived ages ranging from $589$--$776\,\mathrm{Myr}$, depending on photometry and rotational effects, with higher rotation rates corresponding to older ages. A broader compilation by \citet{2023MNRAS.518..662B} placed the cluster's age at $635\pm 135\,\mathrm{Myr}$ based on eight independent literature estimates. We determine an age of $646^{+85}_{-84}\,\mathrm{Myr}$, which aligns well with several previous estimates. The high uncertainty in the cluster age makes the progenitor mass of the extremely massive WD escapee from \citet{2023ApJ...956L..41M} particularly poorly constrained.


\subsection{Melotte 111}

Melotte 111, also known as the Coma Berenices cluster, is one of the nearest open clusters to the solar system but is difficult to study due to its sparse population. \citet{1960PASP...72...42S} identified five WD candidates using photometric plates. \citet{1998MNRAS.295..959B} later used FAUST ultraviolet data to identify additional WD candidates; however, \citet{2009MNRAS.395.1591D} demonstrated that all of them lie outside the cluster's tidal radius and are unlikely members.

Using SDSS photometry, \citet{2009MNRAS.395.1591D} identified four WD candidates, three of which had been previously identified by \citet{1960PASP...72...42S}. SuperCOSMOS proper motions revealed that one of these, SDSS J122420.51+264738.9, was likely a field star. Spectroscopic analysis by \citet{2005ApJS..156...47L} further demonstrated that another candidate, PG 1214+268, was at a distance of $\sim 315\,\mathrm{pc}$ and thus not associated with the cluster. The remaining two candidates were followed up with spectroscopy using the WHT Intermediate Dispersion Spectrograph and Imaging System. From this, \citet{2009MNRAS.395.1591D} found that WD 1216+270 was likely a field star behind the cluster, while WD 1216+260 was a high-probability cluster member, a conclusion supported by radial velocity measurements.

WD 1216+260 is the same source we identify as a cluster member candidate based on the \hunt catalog. Other studies using Gaia, including \citet{2021ApJ...912..162P} and \citet{2021MNRAS.503.3279S}, have also identified it as a candidate member. Spectra of WD 1216+260 have been analyzed several times, including work by \citet{2020ApJ...898...84K}, who reanalyzed the SDSS spectrum. We adopt their atmospheric parameter determinations in this work.

We derive an age of $617^{+11}_{-42}\,\mathrm{Myr}$ for the cluster, assuming no reddening and an approximately solar metallicity. This result is consistent with the 646~Myr age determined by \citet{2020A&A...640A...1C}. Given this age, our cooling model analysis supports the cluster membership of WD 1216+260.


\subsection{NGC 752}

Our crossmatch with the \hunt catalog identified two candidate WDs in NGC 752 as potential cluster members. However, both were considered low priority based on Gaia DR3 photometric masses that fell below our threshold for further consideration. While we are not aware of any spectroscopically confirmed WDs identified as members through Gaia DR3 astrometry, NGC 752 does house at least one spectroscopically confirmed WD member identified through Gaia DR2. \citet{2020NatAs...4.1102M} studied Gaia DR3 342902771504611712, finding it consistent with a cluster WD. Although it was not included as a member by the \hunt clustering algorithm, its Gaia DR3 proper motion and parallax place it within $2\sigma$ of the \hunt cluster mean, and combined with its prior Gaia DR2 support and spectroscopic confirmation, we include this WD as a Gaia-supported member in our IFMR analysis.

We find a best-fit age for NGC 752 of $1549^{+157}_{-156}\,\mathrm{Myr}$ with near-solar metallicity and very low extinction. This aligns with recent isochrone-based age estimates, such as the 1.58~Gyr found by \citet{2024A&A...688A.152J} and the $1.45\pm0.05\,\mathrm{Gyr}$ determined by \citet{2022MNRAS.514.3579B}, who took particular care to remove radial velocity binaries from their fitting.


\subsection{NGC 1039}

\citet{2008AJ....135.2163R} conducted a detailed photometric and spectroscopic study of 44 WD candidates in the field of NGC 1039, also know as M34. They identified five likely single evolution cluster members: LAWDS 9, LAWDS 15, LAWDS 17, LAWDS 20, and LAWDS S2. They also identified one potential binary candidate, LAWDS S1. Later work by \citet{2009MNRAS.395.2248D} found that LAWDS 9, LAWDS 20, and LAWDS S1 were likely field stars due to proper motions that significantly deviate from the cluster mean. The remaining three, LAWDS 15, LAWDS 17, and LAWDS S2, were included in the IFMR study of \citet{2018ApJ...866...21C}, who obtained spectra with Keck to better characterize them.

Despite their inclusion in the \citet{2018ApJ...866...21C} IFMR sample, these three objects are not identified as cluster members in Gaia due to significant astrometric errors. LAWDS 15 and LAWDS S2 lacked complete astrometric data in Gaia DR2, with no parallax or proper motion measurements, while LAWDS 17 had extreme astrometric errors. \citet{2021ApJ...912..165R}, who studied the IFMR using Gaia DR2, noted the possibility of cluster membership for LAWDS 17 but excluded it due to its significant astrometric uncertainties.

In Gaia DR3, all three objects have complete astrometric data, but their parallax over error values remain low. LAWDS 15 has the highest astrometric uncertainty, with a parallax over error of 0.80 and a fractional pmRA error exceeding 1.0. LAWDS 17 has a parallax over error of 2.40 and a fractional pmRA error of 0.56. LAWDS S2 has the highest parallax over error at 6.45, but its fractional pmRA error remains high at 0.89.

While Gaia astrometry does not allow us to reliably support or rule out cluster membership for these sources, we retain them in our IFMR as non-Gaia-supported astrometric sources. Future Gaia data releases may improve the quality of the astrometry enough to allow for a fair assessment of their cluster membership.

We find a cluster age of $313^{+15}_{-12}\,\mathrm{Myr}$ from our heuristic method. Studies based on Gaia DR2 have generally reported younger ages (e.g., 215~Myr \citep{2018A&A...616A..10G}, 266~Myr \citep{2021MNRAS.504..356D}). This discrepancy appears to stem primarily from Gaia DR3 data indicating a lower interstellar reddening.


\subsection{NGC 2099}

NGC 2099, also known as M37, is a rich open cluster located approximately 1.4~kpc away, making it the most distant cluster in our sample with a WD candidate member identified in \hunt. The first identification of WD candidates in NGC 2099 was made by \citet{2001AJ....122.3239K}, who used deep CCD imaging with the Canada-France-Hawaii Telescope (CFHT) to examine the cluster's central field. They identified 67 WD candidates, estimating that 50 were likely cluster members. Using these WDs, they derived a cluster age of $566^{+154}_{-176}\,\mathrm{Myr}$ consistent with their MSTO-derived age of 520~Myr. Later work by \citet{2004MNRAS.351..649K} used synthetic CMDs and Monte Carlo simulations to study the cluster's age. They found that two of their examined models suggested a younger age near 400~Myr, while the third agreed with the MSTO age from \citet{2001AJ....122.3239K}.

\citet{2005ApJ...618L.123K} used Gemini GMOS and Keck LRIS to obtain spectra of 24 WD candidates in the field of NGC 2099, confirming 21 as WDs. The spectroscopic analysis of these objects was presented in a companion paper, \citet{2005ApJ...618L.129K}, where they found that all 21 WDs were consistent with DA spectral types. Of these, 18 could be fit using their spectroscopic models, while three could not, though the authors did not specify the reason.

\citet{2005ApJ...618L.123K} used two multi-object spectral masks with LRIS. The F1 mask provided the spectra used in their analysis, while the F2 mask data were presented but not analyzed due to low SNR. \citet{2015ApJ...807...90C} followed up on the F2 observations with deeper exposures and reexamined the F1 spectra. Using these, they analyzed 20 of the original 24 WDs. One was identified as a main sequence field star, while the remaining 19 were confirmed as genuine WDs. They had sufficient SNR to characterize the masses and cluster membership of 14 objects. They found that 11 were consistent with single stellar evolution cluster membership, while the others were likely field stars.

One WD candidate identified by \citet{2015ApJ...807...90C} appeared to have a high mass but was not presented in their analysis due to insufficient SNR. \citet{2016ApJ...820L..18C} reobserved this candidate using LRIS, along with eight new WD candidates selected based on the parameters of the likely single evolution members from \citet{2015ApJ...807...90C}. Of these nine WDs, three were consistent with single stellar evolution cluster membership, including the previously excluded high-mass WD.

The 11 likely single evolution WDs from \citet{2015ApJ...807...90C}, along with the three additional WDs identified by \citet{2016ApJ...820L..18C}, were included in the IFMR work of \citet{2018ApJ...866...21C}. In their study, \citet{2018ApJ...866...21C} reobserved eight WDs with limited SNR in earlier studies. Four of these were found to be inconsistent with cluster membership or had errors too large for analysis. The remaining four were coadded with the original spectra and included in their final IFMR analysis.

More recent work by \citet{2022MNRAS.515.1841G} used wide-field $ugi$ imaging, supported by Gaia data, to identify seven WD candidates in the cluster (WD1--WD7). Their selection was based on Gaia colors, BP magnitudes, and astrometric cuts within $3\sigma$ of the cluster's proper motion and parallax. They noted that all previously known WD members, including those from the IFMR work of \citet{2018ApJ...866...21C}, were too faint to meet their selection criteria. None of the previously known cluster member WDs are in Gaia DR2 or Gaia DR3. We identify one WD candidate in NGC 2099, which corresponds to WD1 in \citet{2022MNRAS.515.1841G}. Spectroscopic follow-up by \citet{2022MNRAS.515.1841G} with LRIS revealed it to be a very hot, hydrogen-deficient PG1159 spectral type WD. This object had previously been identified as the central star of a planetary nebula \citep{d23d02b8b8d54f9dba277fbfad0093df,2021A&A...656A.110C}. Although astrometry supports cluster membership, its unique atmosphere prevents reliable parameter estimation, and we exclude it from our IFMR analysis.

The remaining six WDs identified by \citet{2022MNRAS.515.1841G} were further examined with spectroscopy by \citet{2023A_A...678A..89W}. They found that only WD2 (Gaia DR3 3451182182857026048) is consistent with cluster membership through a single stellar evolution channel. WD3 was determined to be a background star, while WD4, WD5, and WD7 were found to be foreground objects. WD6, despite a spectroscopic distance consistent with the cluster, has a mass too low to have formed through single stellar evolution, suggesting it may be a product of binary evolution.

Based on the consistent Gaia astrometry and spectroscopic analysis from \citet{2023A_A...678A..89W}, we include their WD2 in our analysis. We also include 13 of the 14 WDs included in \citet{2018ApJ...866...21C} as non-Gaia-supported astrometric cluster WDs. We exclude NGC 2099-WD 33, an ultramassive WD whose position on the IFMR is highly discrepant and suggests a significant cooling delay, potentially due to a past merger event. As all of the \citet{2018ApJ...866...21C} sources fall below the Gaia magnitude limit, it is not expected that any of them will be examinable in future Gaia data releases.

NGC 2099 has a significant extended MSTO. \citet{2023MNRAS.524..108G} obtained data with CFHT, supplemented with archival data from an earlier study by \citet{2001AJ....122..257K}, to determine proper motions and separate field stars from cluster members. By comparing the WD luminosity function to theoretical predictions, they ruled out age spread as the source of the eMSTO. This is important for progenitor lifetime determination, as it relies on a single age to describe the cluster population. We adopt the blue edge of the eMSTO for fitting because it reflects the true turnoff mass of the cluster, providing the most reliable age estimate under the assumption of a single stellar population. In contrast, the red edge of the eMSTO is broadened by factors like stellar rotation, binarity, or differential reddening, which can make stars appear redder and thus the cluster appear older than its true age. Using this approach, we find an age of $535^{+27}_{-43}\,\mathrm{Myr}$, consistent with earlier work by \citet{2001AJ....122.3239K} and more recent Gaia DR2 results from \citet{2021MNRAS.504..356D}, who found an age of $608^{+63}_{-58}\,\mathrm{Myr}$.


\subsection{NGC 2168}
NGC 2168, also known as M35, is a young open cluster with an estimated age of approximately 150~Myr at a distance of around 800~pc. \citet{1980ApJ...235..992R} first identified four faint blue objects as potential WD members, designating them as NGC 2168-1 through NGC 2168-4. \citet{1988A&A...202...77R} later obtained spectra for three of these objects: NGC 2168-1, 3, and 4. They found NGC 2168-3 and 4 to be very hot WDs, with cooling ages and distances consistent with cluster membership, while NGC 2168-1 was deemed too cool and faint to be a member. The fourth WD candidate was too dim to be observed spectroscopically and was also considered unlikely to be a cluster member.

\citet{2002AJ....124.1555V} conducted a photometric study of the cluster and identified one additional dim WD candidate. Subsequent deep imaging with CFHT by \citet{2003AJ....126.1402K} revealed at least a dozen more WD candidates. Using $UBV$ imaging, \citet{2004ApJ...615L..49W} selected eight candidate WDs in the field of NGC 2168, including all four candidates from \citet{1980ApJ...235..992R} and the candidate from \citet{2002AJ....124.1555V}. They followed up with LRIS and found seven of the eight candidates to be likely cluster members or escapees, with only LAWDS 11 identified as a likely field star.

Expanding on this work, \citet{2009ApJ...693..355W} identified 41 candidate WDs in the field of NGC 2168. They obtained new spectra for six WDs, re-reduced spectra for three WDs previously observed (LAWDS 5, 6, and 22), and reanalyzed spectra for five additional WDs from earlier observations. Their analysis covered 14 WDs in total, with 12 classified as DA WDs, one as a DB WD, and one as a hot DQ WD. These 12 DA WDs were reanalyzed by \citet{2018ApJ...866...21C} and included in their IFMR. Of these, four appear in the Gaia catalog (LAWDS 5, 6, 22, and 29), though LAWDS 29 lacks both parallax and proper motion data, while the remaining eight are too faint to be detected by Gaia. We find that LAWDS 22 is a proper motion outlier and therefore an unlikely cluster member. Additionally, LAWDS 5 and LAWDS 6 exhibit substantial Gaia astrometric errors.

Five additional WD candidates from \citet{2009ApJ...693..355W} that were not followed up with spectroscopy are present in Gaia DR3: LAWDS 10 (Gaia DR3 3426266145561778048), LAWDS 41 (Gaia DR3 3426288002653120512), LAWDS 45 (Gaia DR3 3426286211647529088), LAWDS 49 (Gaia DR3 3426280164337959296), and LAWDS 52 (Gaia DR3 3426335174276034560), though LAWDS 52 lacks both parallax and proper motion data. Three candidates, LAWDS 10, LAWDS 49, and LAWDS 6, were not identified as cluster members in our work due to failing the astrometric fidelity requirement used in the \hunt catalog. However, their updated astrometric fidelity values, calculated using the refined criteria from \citet{2022MNRAS.510.2597R}, would meet the \hunt cutoff, suggesting they warrant further consideration as potential cluster members. LAWDS 10 and LAWDS 49 both show significant parallax and proper motion inconsistencies with the cluster, making membership improbable. LAWDS 41 and LAWDS 45 have even larger deviations in both parallax and proper motion, definitively ruling them out as cluster members. Collectively, we exclude all four objects as candidate cluster members.
Despite its well-studied WD population, NGC 2168 has no WDs that are strongly supported as cluster members based on Gaia DR3 astrometry. That said, of the 12 DA WDs included in the IFMR of \citet{2018ApJ...866...21C}, only one (LAWDS 22) can be ruled out as a cluster member based on Gaia astrometry. Therefore, we include the remaining 11 WDs as non-Gaia-based astrometric cluster members in our IFMR. Many of these sources are in Gaia, and we expect future data releases will allow for a more reliable assessment of their cluster membership.

\citet{2024ApJ...962...41C} found an age of 213~Myr from Bayesian analysis of available Gaia DR3, Pan-STARRS, and 2MASS photometry using the BASE-9 software suite. We find a similar age of $208^{+34}_{-21}\,\mathrm{Myr}$ from our heuristic isochrone fitting method.


\subsection{NGC 2287}

NGC 2287, also known as M41 and sometimes the Little Beehive Cluster, was first studied for potential WD members by \citet{1980ApJ...235..992R}, who used photographic plates to identify five faint blue objects in the cluster field, designated as NGC 2287-1 through NGC 2287-5. \citet{1981A&A....99L...8K} attempted to follow up on each of these five candidates. However, two objects (NGC 2287-1 and NGC 2287-4) were too dim to be discerned from the background during their observations and were discarded. NGC 2287-2 and NGC 2287-5 were identified as typical DA WDs consistent with cluster membership, while NGC 2287-3 was a likely foreground WD, though the authors did not entirely rule out its potential membership.

\citet{2009MNRAS.395.2248D} revisited these candidates using higher-quality spectroscopy obtained with the ESO VLT to better assess their atmospheric parameters and potential cluster membership. Their findings aligned with those of \citet{1981A&A....99L...8K}, confirming that NGC 2287-2 and NGC 2287-5 were likely cluster members, while NGC 2287-3 was a likely foreground WD. \citet{2012MNRAS.423.2815D} conducted a deeper CCD-based survey of the cluster, identifying four additional candidate cluster member WDs that had not been examined by \citet{2009MNRAS.395.2248D}. Follow-up VLT spectroscopy was obtained for two of these candidates, confirming NGC 2287-4 as a likely cluster member. This WD had initially been identified as a candidate by \citet{1980ApJ...235..992R} but was discarded by \citet{1981A&A....99L...8K} due to its faintness, which had made it difficult to separate from the background.

Based on these membership determinations, \citet{2018ApJ...866...21C} included NGC 2287-2, NGC 2287-4, and NGC 2287-5 in their IFMR analysis. While all three WDs are present in the Gaia catalog, NGC 2287-4 (Gaia DR3 2927020766282196992) lacks both parallax and proper motion data. The proper motions of NGC 2287-2 (Gaia DR3 2927203353930175232) and NGC 2287-5 (Gaia DR3 2926996577021773696) align with the cluster mean, but their parallax over error values below 1.0 make their distance measurements unreliable, leading us to classify both as questionable members. Future Gaia releases warrant further examination of the known candidate WDs from the cluster. We include these three sources as non-Gaia-supported astrometric sources in our IFMR.

We find an age of $206^{+14}_{-16}\,\mathrm{Myr}$ for NGC 2287. Though a rich young cluster, it is not well studied in the literature, with minimal recent age estimates for comparison. \citet{2021MNRAS.504..356D} used an automated isochrone fitting technique with Gaia DR2 and found an older cluster age of 302~Myr. We attribute part of the discrepancy to our finding a lower cluster reddening.


\subsection{NGC 2323}

Also known as M50, \citet{2016ApJ...818...84C} observed ten candidate WDs in the field of NGC 2323. Of these, they identified two as likely cluster members with single-stellar evolution origins. \citet{2018ApJ...866...21C} included both as cluster members in their IFMR analysis. NGC 2323-WD10 (Gaia DR3 3051559974457298304) has incomplete Gaia data, with no available parallax or proper motion. The other candidate, NGC 2323-WD11 (Gaia DR3 3051568873629418624), has a parallax over error of 0.48, making any assessment of its cluster membership with Gaia unreliable. We include both WDs as non-Gaia-supported astrometric sources in our IFMR. Both are high-mass WDs and among the hottest in our sample, with temperatures exceeding $50{,}000\,\mathrm{K}$.

We find a cluster age of $200^{+33}_{-38}\,\mathrm{Myr}$. This is consistent with work by \citet{2019JApA...40...33O}, who found an isochrone age of $200\pm50\,\mathrm{Myr}$ from Sierra San Pedro Mártir National Astronomical Observatory open cluster survey photometry.


\subsection{NGC 2516}

Several white dwarf candidates in NGC 2516 were first identified by \citet{1982A&A...116..341R}, who examined ten objects and found that three candidates, NGC 2516-1, NGC 2516-2, and NGC 2516-5, were consistent with cluster membership. \citet{1996A&A...313..810K} reobserved the cluster using the ESO Faint Object Spectrograph, confirming these three members and adding NGC 2516-3 as an additional candidate. These four objects were later included in the IFMR study of \citet{2018ApJ...866...21C}.

We match NGC 2516-1, NGC 2516-2, NGC 2516-WD3, and NGC 2516-5 to this works NGC 2516 WD1, WD3, WD6, and WD2, respectively. \citet{2021ApJ...912..165R} identified NGC 2516-1, NGC 2516-2, and NGC 2516-5 in their work, but only considered NGC 2516-5 to be a cluster member. NGC 2516 WD6 was suggested to be a background object by \citet{2021A&A...645A..13P} based on Gaia DR2, but was considered a cluster member in follow-up work by \citet{2023A&A...678A..20P} using Gaia DR3. This object was initially a lower-priority target due to its total age estimate from Gaia photometry being well older than the cluster mean, but given available spectroscopy supporting membership, we include it in our IFMR.

Two additional candidates were identified in Gaia DR2 by \citet{2019RNAAS...3...49H}. These were also flagged by \citet{2021A&A...645A..13P} and noted as potential cluster escapees by \citet{2021ApJ...912..165R}. These objects are NGC 2516 WD4 and WD5 in our work. We followed up both objects with spectroscopy and find both to be likely cluster members.  

NGC 2516 age estimates are generally consistent across the literature. \citet{2019A&A...623A.108B} reported an age of 251~Myr with $A_v=0.22$, while \citet{2020A&A...640A...1C} found 240~Myr with $A_v=0.11$. Our isochrone fit suggests a slightly younger age of $206^{+37}_{-12}\,\mathrm{Myr}$ with $A_v=0.185$ and a metallicity of $Z=0.017$. Although the cluster's giants suggest an age closer to 150~Myr, we adopt the MSTO-driven estimate because it provides a more self-consistent fit to the CMD.


\subsection{NGC 2682}

NGC 2682, commonly referred to as M67, is one of the oldest open clusters in the solar neighbourhood, located at a distance of approximately 900~pc with an age of a few Gyr \citep{2015MNRAS.450.2500B}.\citet{2018ApJ...867...62W} conducted an LRIS spectroscopic study of WD candidates in the cluster field, confirming 24 as DA WDs. Of these, 18 had spectra of sufficient quality for membership analysis, finding 13 likely cluster members WDs, two of which are potential binary systems. The authors noted that none of these objects had sufficiently precise Gaia DR2 parallaxes to better constrain cluster membership.

\citet{2021AJ....161..169C} later expanded the sample to 22 WDs, including some non-DA types, nine of which have Gaia DR3 data. We identified four candidate cluster member WDs in our search (NGC 2682 WD1 through WD4), two of which overlap with their sample. These include NGC 2682 WD1, which they classify as a photometric nonmember, and NGC 2682 WD3, found to be a DB WD. The candidates not shared with their sample are NGC 2682 WD2 and WD4: NGC 2682 WD2 was on our follow-up list but was given low priority due to its low cooling age, which would require significant merger history to be consistent with cluster membership; we consider this a questionable member. NGC 2682 WD4 was given low priority due to its very low mass from Gaia photometry. NGC 2682 WD3 was initially considered low-priority due to its low expected mass based on Gaia photometry, which assumed a DA composition. While literature spectra suggest WD3 may be a member, it is not a single DA WD, so we exclude it from our IFMR.

After detailed analysis, \citet{2021AJ....161..169C} included seven well-measured DA WDs as likely cluster members for use in their IFMR. Of these, two objects, M67:WD14 (Gaia DR3 604922164839680384) and M67:WD25 (Gaia DR3 604916353749379712), have Gaia DR3 data, although M67:WD14 lacks parallax and proper motion information. None of the seven DA WDs included by \citet{2021AJ....161..169C} have Gaia astrometric support for cluster membership, but we retain them in our IFMR as non-Gaia-supported astrometric members.

\citet{2024ApJ...976...87N} examined a large sample of subgiant branch stars and statistically estimated cluster ages to be between about 3.8 and 4.3~Gyr. Isochrone fitting using Gaia DR3 by \citet{2024MNRAS.532.2860R} found a similarly advanced cluster of $3.95^{+0.16}_{-0.15}\,\mathrm{Gyr}$. Our analysis yielded a comparable age of $3{,}819^{+292}_{-86}\,\mathrm{Myr}$ for a subsolar metallicity cluster. We note that the \hunt catalog estimated a significantly younger cluster age of $\sim 1.7\,\mathrm{Gyr}$, though this did not affect our WD candidate selection since no objects were discarded based on presumed advanced ages.


\subsection{NGC 3532}

WDs in NGC 3532 were first identified by \citet{1989A&A...218..118R}, who found seven candidates based on their colors. Follow-up spectroscopy confirmed three as DA WDs, designated NGC 3532-1, 5, and 6. The remaining four were found not to be WDs. \citet{1993A&A...275..479K} expanded the search to a wider field around the cluster, identifying seven additional WD candidates. Of these, three were confirmed as DA WDs (NGC 3532-8, 9, and 10). Together, these six spectroscopically confirmed WDs were considered strong cluster member candidates. However, the available spectra had significant uncertainties, which led to uncertain cooling parameters, making it difficult to associate them with the cluster confidently.

\citet{2009MNRAS.395.2248D} re-examined these six WDs using the ESO VLT and FORS1, finding that four were consistent with cluster membership based on their derived distance moduli: NGC 3532-1, 5, 9, and 10. The other two, NGC 3532-6 and 8, were excluded due to inconsistent distances. Later, deep, wide-field CCD photometry was performed by \citet{2011AJ....141..115C} using the Cerro Tololo Inter-American Observatory (CTIO) 0.9~m telescope. They recovered eight of the nine previously confirmed spectroscopic WDs from earlier studies, which included the six mentioned above along with three additional candidates obtained via private communication. Additionally, they identified approximately 30 more WD candidates with high probabilities of being cluster members.

\citet{2012MNRAS.423.2815D} followed up on seven new WD candidates identified through CCD photometry, observing four of them with the VLT and FORS2. Three of these were confirmed as likely cluster member DA WDs (J1106-590, J1106-584, and J1107-584), while the fourth (J1105-585) was deemed a non-member due to a proper motion greater than $3\sigma$ from the cluster mean. The three unobserved candidates were brighter and less likely to be higher-mass WDs, which was the primary focus of their study. All seven candidates had been previously identified by \citet{2011AJ....141..115C}.

\citet{2016ApJ...818...84C} included the four WDs found to be members in \citet{2009MNRAS.395.2248D} and the three from \citet{2012MNRAS.423.2815D} in their IFMR analysis. \citet{2016MNRAS.457.1988R} later used VPHAS+ DR2 photometry to search for additional WD candidates in 11 clusters; in NGC 3532, they identified three new candidates (VPHAS J1103-5837, J1104-5830, and J1105-5842). Follow-up spectroscopy found J1104-5830 to be a background DC WD. J1105-5842 was determined to be a foreground object, while J1103-5837 had a distance consistent with possible cluster membership. \citet{2018ApJ...866...21C} included the seven WDs from \citet{2016ApJ...818...84C} and VPHAS J1103-5837 in their IFMR analysis.

We identified 12 candidate WDs in NGC 3532 as potential cluster members. Among these, NGC 3532 WD7, WD6, and WD8 correspond to the literature WDs NGC 3532-1, -9, and -10, respectively, while the remaining nine lack available spectra. We followed up four of these with GMOS (NGC 3532 WD2, WD4, WD9, and WD10). Four others were considered cluster candidates but were not followed up due to their lower-priority status based on inferred mass, while the final one had an expected mass from Gaia photometry below $0.51\msun$, for which we do not compute total ages. Our analysis finds that NGC 3532 WD10 has a very low WD mass given its inferred progenitor mass. Due to its small final mass uncertainty, this places it as a strong outlier well below the IFMR trend. As such, we consider it almost certainly a nonmember of NGC 3532, and exclude it from our IFMR. NGC 3532 WD2 is also somewhat of an outlier, showing a lower than expected final mass compared to its initial mass. However, given its substantially larger uncertainties, we do not rule out cluster membership and retain it in our IFMR. We also retain WD4 and WD9, both of which are consistent with the IFMR and considered likely cluster members. As our IFMR fitting routine is robust to outliers, the inclusion of WD2 does not significantly impact our results. 

We failed to recover the other five WDs (NGC 3532-5, J1106-590, J1106-584, J1107-584, and VPHAS J1103-5837) included in the IFMR of \citet{2018ApJ...866...21C} in our crossmatch. NGC 3532-5 is likely not in \hunt due to its parallax being more than $2.5\sigma$ from the cluster mean. However, \citet{2022MNRAS.509.1664J}, using Gaia data along with Gaia-ESO Survey spectroscopy, assigned it a $96\%$ probability of cluster membership. Given this strong Gaia-based evidence, we will include it in our IFMR. J1106-584 is also included in the Gaia-based IFMR of \citet{2021ApJ...912..165R} and is retained in our sample. They additional include J1106-590, but we find its Gaia association uncertain given its extremely low astrometric fidelity of just 0.033. Nevertheless, we include it as a non-Gaia-supported astrometric source based on non-Gaia literature membership support. J1107-584 has parallax over error in Gaia DR3 of less than 1; as such, we exclude it from our analysis, considering it a questionable member. Similarly, as VPHAS J1103-5837 has substantial Gaia astrometric error and very low astrometric fidelity, we also exclude it as a questionable association.

The eMSTO observed in NGC 3532 complicates precise age determination. Several studies have suggested ages close to 400~Myr (e.g., \citealt{2019A&A...623A.108B}; \citealt{2020A&A...640A...1C}; \citealt{2022ApJ...931..156P}), while others prefer an age closer to 300~Myr (e.g., \citealt{2019A&A...622A.110F}). More recently, \citet{2024arXiv241209217H} derived an age of 340~Myr, where they found that differential reddening, which has been found to significantly impact the eMSTO in some clusters, has little impact on the observed broadening of the NGC 3532 MSTO. Our isochrone fitting results are largely consistent with these studies. We find that an age of $385^{+43}_{-33}\,\mathrm{Myr}$ with $A_V=0.060$ and a solar metallicity provides a robust fit to both the blue end of the eMSTO and the majority of the cluster's giant stars.


\subsection{NGC 6121, NGC 6819, and NGC 7789}

We include low-mass WDs from NGC 6121 (M4), NGC 6819, and NGC 7789\footnote{Grouped here because their only spectroscopically confirmed DA WDs are too faint for Gaia; other clusters with faint literature WDs have at least one WD assessable with Gaia and are discussed individually (see Table~\ref{tab:wd_results_nongaia}).}, in our IFMR analysis, using data from \citet{2018ApJ...866...21C}. While these clusters are all in the \hunt catalog, their WDs are too faint for Gaia. The cluster sequences are also poorly represented in the catalog, making accurate age determination difficult. Therefore, we adopt the PARSEC-based isochrone ages from \citet{2018ApJ...866...21C} for our analysis.

NGC 6121, better known as M4, is a nearby globular cluster located at a distance of approximately 1.7~kpc. With an age of $10.2\pm1.0\,\mathrm{Gyr}$, it is the oldest cluster included in our analysis and provides critical insight into the low-mass end of the IFMR. NGC 6819 and NGC 7789 are both moderately distant open clusters. NGC 6819, located roughly 2.4~kpc away, has an age of $2.43\pm0.15\,\mathrm{Gyr}$, while NGC 7789, at a similar distance of about 2.3~kpc, is somewhat younger, with an age of $1.56\pm0.10\,\mathrm{Gyr}$.


\subsection{Pleiades}

The Pleiades, or Melotte 22 in \hunt, is one of the nearest and most well-studied open clusters. Despite this, due to its young age, confirmed WD identifications remain limited. The only currently accepted cluster member WD is EGGR 25 (Gaia DR3 66697547870378368), originally identified by \citet{1965ApJ...141...83E}. This source has been extensively studied in both pre-Gaia and Gaia-era works, with its cluster membership confirmed by numerous studies. We also identified EGGR 25 as a candidate in this work, with no other candidates.

\citet{2022ApJ...926..132H} included EGGR 25 in their Gaia DR2-based study and also identified two likely cluster escapees, Lan 532 (Gaia DR3 228579606900931968) and WD 0518-105 (Gaia DR3 3014049448078210304). Both were classified as single-stellar evolution DA WDs consistent with former cluster membership based on analysis of existing literature spectra. We incorporate these three WDs in our IFMR, using the spectral fits provided by \citet{2022ApJ...926..132H}.

The extremely high-mass WD GD 50 was previously proposed as a possible escapee from the Pleiades \citep{2006MNRAS.373L..45D}, but as discussed above, more recent Gaia DR2-based analyses \citep{2018ApJ...861L..13G} strongly support its membership in the AB Doradus Moving Group. An additional ultramassive WD, PG 0136+251, was also considered a Pleiades candidate by \citet{2006MNRAS.373L..45D} and included in the IFMR of \citet{2018ApJ...866...21C}. \citet{2020ApJ...901L..14C} examined this source with Gaia DR2, applying an analysis method similar to that used by \citet{2018ApJ...861L..13G} in their study associating GD 50 with the AB Doradus moving group. They concluded that it was farther from both the Pleiades and AB Doradus moving groups in the past and is thus a nonmember. We follow their conclusion and do not include PG 0136+251 in our IFMR.

As with Alpha Persei, here we adopt the kinematically determined cluster age from \citet{2021arXiv211004296H} of $128\pm7\,\mathrm{Myr}$, in agreement with other literature age estimates from lithium depletion (e.g., \citealt{1998ApJ...499L.199S}), and isochrone fitting (e.g., \citealt{2022ApJ...926..132H}; \citealt{2023A&A...677A.162B}).


\subsection{Praesepe}

Also known as NGC 2632, the Beehive Cluster, and M44, Praesepe is a heavily populated open cluster and one of the closest to the solar system, making it one of the most well-studied clusters. White dwarfs from this cluster have been identified in numerous studies. The first candidate was discovered through a proper motion study by \citet{1962LB....C31....1L}. \citet{1965ApJ...141...83E} later used spectroscopy to confirm the WD nature of that object, as well as a second candidate cluster member. \citet{1982ApJ...255..245A} reconfirmed the second candidate through photometry and proper motion analysis, and identified two additional candidates. Follow-up work by \citet{1984AJ.....89..267A} added two more photometric candidates.
\citet{2001ApJ...563..987C} obtained spectra for all six candidate WDs identified in these earlier works, confirming five as likely cluster members: EGGR 59, EGGR 60, EGGR 61, LB 5893, and LB 1876. However, \citet{2009MNRAS.395.1795C} identified a naming error in \citet{2001ApJ...563..987C} involving EGGR 59 and EGGR 61, based on differences in the presence of a magnetic field between the two objects. This error propagated through subsequent studies. Specifically, WD 0836+201 should be labeled as EGGR 59 or LB 309, while WD 0837+199 should be identified as LB 393 or EGGR 61.

\citet{2004MNRAS.355L..39D} recovered four of these five previously confirmed WD members, missing LB 5893 due to a proper motion criterion. The missing detection was possibly caused by astrometric contamination from a nearby source. Notably, LB 5893 (Gaia DR3 661270898815358720) has since been reconfirmed as a likely cluster member by several studies using Gaia data. \citet{2004MNRAS.355L..39D} also identified six new WD candidates, two of which, WD 0837+185 and WD 0837+218, were spectroscopically confirmed as cluster members. Follow-up work by \citet{2006MNRAS.369..383D} added four more spectroscopically confirmed cluster members: WD 0833+194, WD 0840+190, WD 0840+205, and WD 0843+184, bringing the total sample to 11 probable members.

\citet{2009MNRAS.395.1795C} obtained high-resolution spectra for nine of these 11 WDs, excluding WD 0836+197 and WD 0840+205. Their analysis revealed that WD 0836+201 is magnetic, leading them to exclude it from their IFMR work. They also excluded WD 0837+218, concluding it is a field star based on its radial velocity, proximity to the cluster, and deviation from the IFMR. This interpretation was challenged by \citet{2018A&A...616A..10G}, who considered WD 0837+218 a likely cluster member. \citet{2009MNRAS.395.1795C} also argued that WD 0837+185 is a double degenerate system, a claim that \citet{2012ApJ...759L..34C} expanded on by suggesting a potential brown dwarf companion. However, this claim was refuted by \citet{2019MNRAS.483.3098S}, who found no evidence for a double degenerate system in Gaia DR2. Our analysis of Gaia DR3 data supports this finding.

Using Gaia DR2, \citet{2019MNRAS.483.3098S} recovered 11 known Praesepe WDs and identified one new WD (SDSS J082839.87+194343.1, Gaia DR3 662998983199228032). In their Gaia DR2 analysis, \citet{2021A&A...645A..13P} recovered the same 12 WDs and did not identify any additional candidates. In a study focused on Praesepe's tidal tails, \citet{2019A&A...627A...4R} identified these 12 WDs, along with three new WDs in the cluster's tidal tails, which they considered highly probable members. These new identifications are US 2088 (Gaia DR3 1008837274655408128), SDSS J092713.51+475659.6 (Gaia DR3 825640117568825472), and LAMOST J081656.19+204946.4 (Gaia DR3 675767959626673152).

\citet{2019MNRAS.486.5405G} also analyzed Praesepe with Gaia DR2 and identified 13 WD candidates. Of these, 11 were previously known, one matched the new candidate from \citet{2019MNRAS.483.3098S}, and one was a new candidate labeled G19 (Gaia DR3 659335994570713600), which remains unconfirmed as a cluster member due to a lack of spectroscopic follow-up.

In our \hunt crossmatch, we identified 15 candidate WDs in Praesepe. Two of these, Praesepe WD14 and WD15, were very low priority due to advanced age estimates inconsistent with cluster membership. One of these is the previously identified WD named G19 from \citet{2019MNRAS.486.5405G}. The other, Praesepe WD15, is included in our IFMR based on Gaia astrometry and existing spectral data that support its cluster membership.

Of the remaining 13 WDs, all but one (Praesepe WD11) have literature spectra available. This corresponds to the new source identified by \citet{2019MNRAS.483.3098S} and several later studies. Although it remained on our follow-up list, we did not observe it due to constraints in available telescope time. Praesepe WD12 (Gaia DR3 662152977721471488), which has not previously been associated with the cluster, has literature spectra classifying it as a DC WD. Due to this classification, we consider it a questionable member and exclude it from our IFMR analysis. Praesepe WD13 corresponds to the previously discussed DB WD LAMOST J081656.19+204946.4 and is also excluded from our IFMR. We include the remaining 10 WDs in our IFMR using literature spectra, along with US 2088 and SDSS J092713.51+475659.6, which were confirmed as DA WDs by \citet{2019MNRAS.482.5222T} and considered high-probability members from Gaia data by \citet{2019A&A...627A...4R}. Praesepe WD9 (J083936.48+193043.6) has asymmetric $T_\textrm{eff}$ error bars in our spectroscopic source \citep{2019A_A...628A..66L}. Our fitting routine assumes symmetric errors, which is the case for all other WDs in this work. Rather than modifying our routine for this single case, we use the average error in out fit.

The age of Praesepe has been extensively studied, with a broad range of estimates largely due to variations in methods, metallicities, and stellar rotation effects. Recent studies include \citet{2023A&A...675A.167P}, who derived an age of $580\pm230\,\mathrm{Myr}$ from an analysis of delta Scuti stars, and \citet{2022AJ....164...34M}, who estimated an age of $636\pm19\,\mathrm{Myr}$ based on the binary system $\epsilon$~Cnc. Isochrone fitting using Gaia DR2 data by \citet{2019MNRAS.483.3098S} yielded an age of 708~Myr, while \citet{2023A&A...677A.163A} found a slightly older age of $760$~Myr using Gaia DR3. Our isochrone fitting yields a comparable result of $741^{+72}_{-30}\,\mathrm{Myr}$ for a slight supersolar metallicity. Giant stars in the cluster suggest a much younger cluster. Each of the twelve Praesepe WDs included in our IFMR remains a probable cluster member when considering our derived cluster age. 


\subsection{Ruprecht 147}

Ruprecht 147, also known as NGC 6774, is one of the oldest open clusters in the solar neighbourhood. Despite its proximity, no WD members were reliably identified in Ruprecht 147 until the release of Gaia DR2. \citet{2018A&A...616A..10G} used Gaia DR2 astrometry to find 10 probable WD members, and \citet{2019A&A...625A.115O} expanded this list to 15 candidates by applying simple photometric criteria based on known Gaia WDs. However, none of these candidates had been spectroscopically confirmed at the time.

\citet{2020NatAs...4.1102M} also examined the cluster with Gaia DR2, identifying 10 candidate WDs whose astrometry suggested potential cluster membership. Many of these overlap with the candidates from earlier studies. They followed up all 10 candidates with LRIS spectroscopy, confirming eight DA WDs, one DB WD (Gaia DR3 4088108859141437056), and one background star. Of the DA WDs, three had distances inconsistent with cluster membership, while the remaining five DA WDs and the DB WD were considered likely members.

In this work, we identified 10 candidate WDs (designated Ruprecht 147 WD1 to WD10). Initially, we deprioritize WD7 to WD10 due to their low expected masses or apparent advanced ages inconsistent with cluster membership. However, we reconsidered the candidacy of any of these with available literature spectra. Comparing our sample to the spectroscopically examined WDs from \citet{2020NatAs...4.1102M}, we find that WD2, WD6, and WD8 correspond to their likely member DA WDs (R147-WD10, R147-WD01, and R147-WD07, respectively). Meanwhile, WD5 (their R147-WD09), WD7 (their R147-WD03), and WD9 (their R147-WD11) were deemed unlikely members based on spectroscopic distances and were excluded from our analysis. We also exclude WD3, which corresponds to the DB WD R147-WD02 in their work.

Additionally, we include two WDs from \citet{2020NatAs...4.1102M} that were not identified in our initial crossmatch: R147-WD04 and R147-WD08. Both were found to be likely members, and we include them in our IFMR. WD1, WD4, and WD10 do not have available literature spectra. While WD10 was discarded based on its low Gaia photometric mass, WD1 and WD4 remain follow-up candidates. Both were previously identified as likely cluster members in multiple studies, including \citet{2018A&A...616A..10G} and \citet{2019A&A...625A.115O}. However, we did not observe these WDs due to telescope time constraints.

Several WDs in Ruprecht 147, including some from our sample, were considered for inclusion in Gaia-based IFMR studies by \citet{2021ApJ...912..165R}, \citet{2021A&A...645A..13P}, and \citet{2023A&A...678A..20P}. To our knowledge, no candidate WDs from this cluster have been spectroscopically followed up outside of the work by \citet{2020NatAs...4.1102M}.

Age estimates for Ruprecht 147 are consistent across several studies. \citet{2020ApJ...896..162T} inferred an age of $2.67^{+0.39}_{-0.55}\,\mathrm{Gyr}$ from the triple system EPIC 219552514. Our analysis finds an age of $2{,}754^{+408}_{-124}\,\mathrm{Myr}$, in agreement with the aforementioned estimate and consistent with other recent determinations. Given our age determination, the five candidates we reanalyze for our IFMR remain likely members.


\subsection{Stock 1}

We identified one candidate WD in Stock 1, which we followed up with LRIS spectroscopy. Our isochrone-derived age of $447^{+19}_{-40}\,\mathrm{Myr}$ closely matches recent estimates by \citet{2019A&A...623A.108B} (474~Myr) and \citet{2020A&A...640A...1C} (417~Myr). Spectroscopic analysis confirms the WD as a DA type with no signs of a magnetic field, and its parameters are consistent with single-star evolution and cluster membership, making it the first known WD member of Stock 1.


\subsection{Stock 2}

Stock 2 is a heavily populated open cluster with several Gaia-based works identifying candidate WDs (e.g., \citealt{2018A&A...616A..10G}; \citealt{2021A&A...645A..13P}; \citealt{2021MNRAS.503.3279S}; \citealt{2021ApJ...912..165R}). Among these candidates, only one has been followed up with spectroscopy: Gaia DR3 506862078583709056, which was confirmed as a likely member DA WD by \citet{2021ApJ...912..165R}.

Given the cluster's moderate age, close distance, and rich membership, Stock 2 likely hosts many additional WDs beyond the one which has been spectroscopically confirmed. We identified 20 candidates (WD1-WD20), the most of any cluster we studied. Seven were made very low priority based on Gaia photometric cooling parameters that indicated either low masses or advanced ages. Of the remaining 13 candidates, WD1 corresponds to the previously confirmed WD from \citet{2021ApJ...912..165R}, while the other 12 were considered for follow-up spectroscopy to confirm membership.

We followed up WD2 and WD3 with LRIS and WD4 through WD10 with GMOS. WD11 through WD13 were considered lower priority due to photometric properties and were not observed. As previously discussed, we find that WD6 is a DZA WD which is a nonmember, and thus do not include it in IFMR analysis. The remaining eight WDs were all found to be DA WDs, one of which (WD2) exhibited notable Zeeman splitting dictating the use of photometrically determined surface gravity.

Age determination for Stock 2 presents challenges due to a substantial extended MSTO. \citet{2021A&A...656A.149A} analyzed the cluster and concluded that the eMSTO likely results from significant differential reddening. In their study of 46 cluster stars, including giants, they estimated an age of $450\pm150\,\mathrm{Myr}$ and solar-like metallicity.

To address these challenges, we estimated the cluster's age using only sources with complete Gaia DR3 reddening data (both $A_g$ and $E(\mathrm{Bp} - \mathrm{Rp})$ with associated uncertainties). This allowed us to deredden individual stars rather than applying a uniform cluster-wide reddening correction. From this subset of stars, we applied our standard isochrone fitting technique without adjusting the reddening, deriving an age of $437^{+22}_{-16}\,\mathrm{Myr}$ with solar metallicity. While this approach excludes giant stars, none of which have complete Gaia reddening data, we consider this limitation minor, especially given the agreement with the age derived by \citet{2021A&A...656A.149A}, who incorporated giant stars in their analysis. Furthermore, our focus remains on the MSTO region, which we find more reliable for cluster age estimation.

Based on our spectroscopic analysis and derived ages, the seven DA white dwarfs without notable Zeeman splitting are consistent with single-star cluster evolution and are therefore included in our IFMR. Although the presence of a magnetic field in WD2 could suggest a merger origin, our analysis indicates that its properties are also consistent with a single-stellar evolution origin, and we therefore retain it in our IFMR sample. This interpretation aligns with previous findings suggesting that magnetic white dwarfs, particularly those born from intermediate mass stars, can arise from single-stellar evolution without requiring a merger history \citep{2020ApJ...901L..14C}. Conversely, we find the WD member spectroscopically confirmed by \citet{2021ApJ...912..165R} appears to be an questionable member. This alternate conclusion stems from our substantially older cluster age estimate of $437^{+22}_{-16}\,\mathrm{Myr}$, compared to their younger estimate of $225\pm50\,\mathrm{Myr}$. They acknowledged difficulties in fitting the MSTO due to high scatter and instead derived the cluster age from giant stars, noting that the scatter could result from differential reddening or other factors, which added uncertainty to their estimate. We include WD1 in our IFMR for completeness, given the uncertainties surrounding the cluster's age and its classification as a likely member in previous Gaia-based studies.


\subsection{Stock 12}

We identified one candidate WD in Stock 12 but did not follow it up due to its apparent age being significantly older than the cluster. A different candidate was identified with Gaia DR2 by \citet{2021ApJ...912..165R}. Their spectroscopic study using GMOS revealed a $\log g$ significantly higher than suggested by photometry and larger than values obtained from an SDSS spectrum of the same object. They found no clear evidence of a magnetic field but noted that its presence could explain the unexpectedly high $\log g$. \citet{2021A&A...645A..13P} also independently recovered this candidate.

Age estimates for Stock 12 vary widely, with automated methods yielding younger values (e.g., 154~Myr in \citealt{2019A&A...623A.108B}; 112~Myr in \citealt{2020A&A...640A...1C}) compared to manual isochrone fitting (e.g., $300\pm50\,\mathrm{Myr}$ in \citet{2021ApJ...912..165R}). Our fit suggests $234^{+49}_{-11}\,\mathrm{Myr}$ with $A_v=0.292$, better aligning with the MSTO, whereas younger estimates fit the upper main sequence better.


\subsection{Theia 248}

Theia 248 is a young, moderately populated cluster located approximately 500~pc away. To date, no studies have specifically examined this cluster. We identified one candidate WD, which we followed up with LRIS spectroscopy, confirming it as a young, moderate-mass DA WD. Our age estimate for the cluster is $490^{+18}_{-14}\,\mathrm{Myr}$, significantly higher than the $112^{+77}_{-45}\,\mathrm{Myr}$ age estimate by \hunt. Given this age, the WD is consistent with a single-stellar evolution origin and is included in our IFMR as a likely cluster member.


\subsection{Theia 517}

We identified one candidate cluster member WD in Theia 517, also known as NGC 7092 or M39. This source was previously recognized as a cluster member candidate with Gaia DR2 by \citet{2021ApJ...912..165R}, who followed up with GMOS spectroscopy, confirming its likely association with the cluster. \citet{2020ApJ...901L..14C} examined its magnetic field, finding a 1.4~MG field that impacted the $\log g$ values derived from spectroscopy. To account for this, they obtained an alternative $\log g$ estimate from photometry, which we adopt in our reanalysis. Despite these complications, the WD remains consistent with a single-stellar evolution origin. The object was also identified as a likely cluster member by \citet{2021A&A...645A..13P}. 

Age estimates for Theia 517 show some variation. \citet{2021ApJ...912..165R} derived an age of $280\pm20\,\mathrm{Myr}$ from manual isochrone fitting, closely matching the 310~Myr estimate from \citet{2019A&A...623A.108B}. In contrast, \citet{2020A&A...640A...1C} found an older age of 398~Myr. Our isochrone fitting aligns well with this older estimate, yielding an age of $407^{+32}_{-20}\,\mathrm{Myr}$ for solar metallicity and low reddening.


\subsection{Theia 817}

Theia 817 is an understudied cluster located approximately 360~pc away. We identified two candidate WDs in this cluster, one of which was deprioritized due to an advanced age estimate inconsistent with cluster membership. The other, Theia 817 WD1, was followed up with LRIS spectroscopy.

The MSTO and upper main sequence are notably sparse, with only two high-probability members spanning nearly two magnitudes at the bright end of the main sequence. This sparsity complicated our heuristic isochrone fitting, as minor deviations from the best fit significantly inflated the weighted chi-squared value. To address this issue, we increased the chi-squared threshold and derived an age of $302^{+16}_{-25}\,\mathrm{Myr}$ for a super-solar metallicity of $Z=0.020$ with a reddening of $A_v=0.152$. However, given the limited upper main sequence data, our age uncertainty may be underestimated. Our fit strongly supports cluster membership for Theia 817 WD1, but we caution that this assessment should be revisited as more cluster member data becomes available.


\subsection{UPK 303}

UPK 303 is a moderately populated cluster located approximately 200~pc away. Despite its proximity, the cluster remains largely understudied. In this work, we identified one candidate WD member, which we followed up with LRIS spectroscopy. Combined with an age estimate of $263^{+31}_{-14}\,\mathrm{Myr}$ for solar metallicity and moderate reddening, the spectroscopic analysis strongly supports its cluster membership, making it the first probable WD member of UPK 303.


\section{Additional Figures and Tables}
\label{sec:appendix-figsandtables}

\begin{figure*}[ht]
\centering
  \setlength{\unitlength}{1cm}
  \begin{picture}(17,14)(-0.3,0)
\put(0.58,7.12){\includegraphics[width=0.1808\textwidth,clip,trim=0.900in 0.75in 0 0]{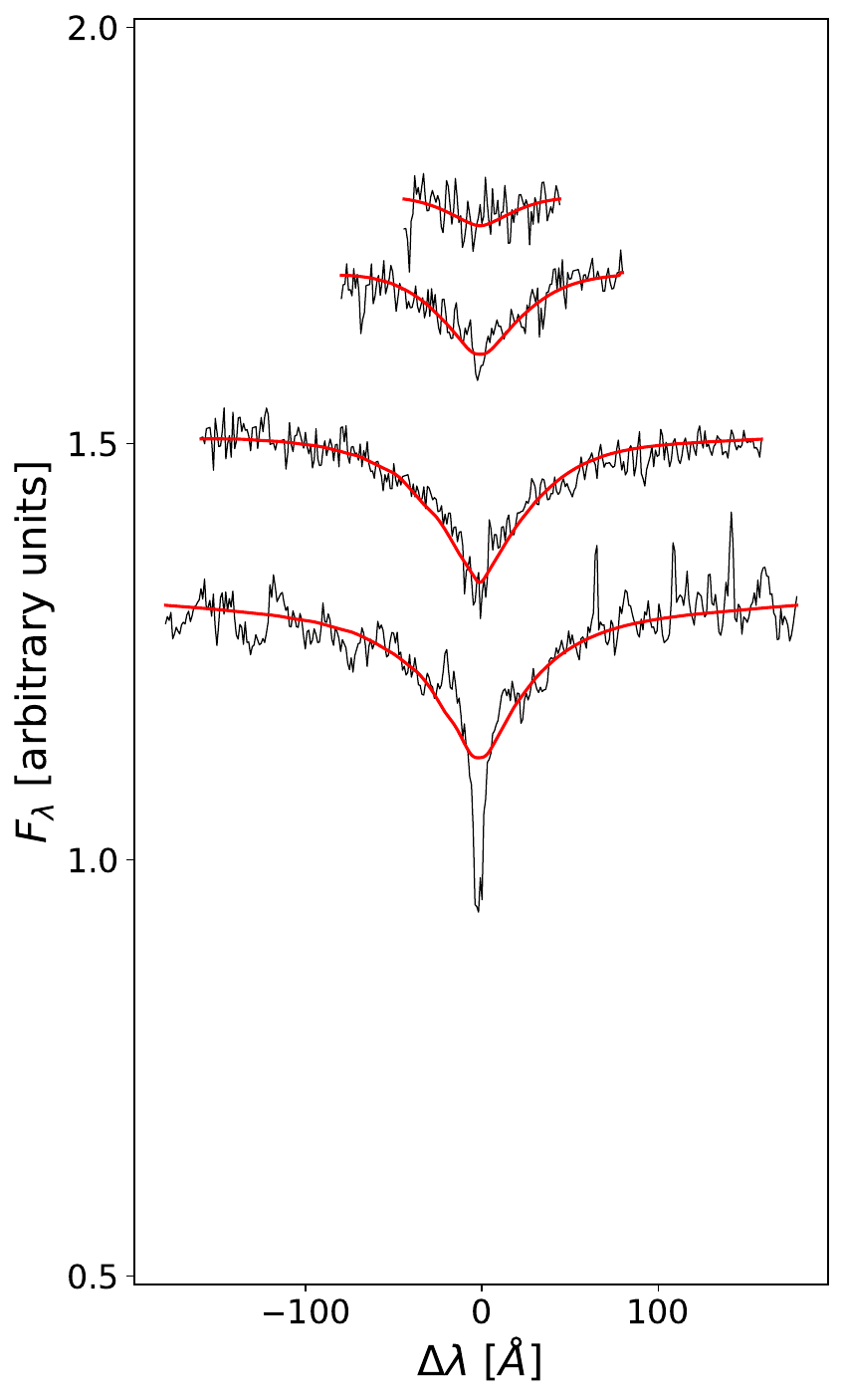}}
\put(3.78,7.12){\includegraphics[width=0.18\textwidth,clip,trim=0.92in 0.75in 0 0]{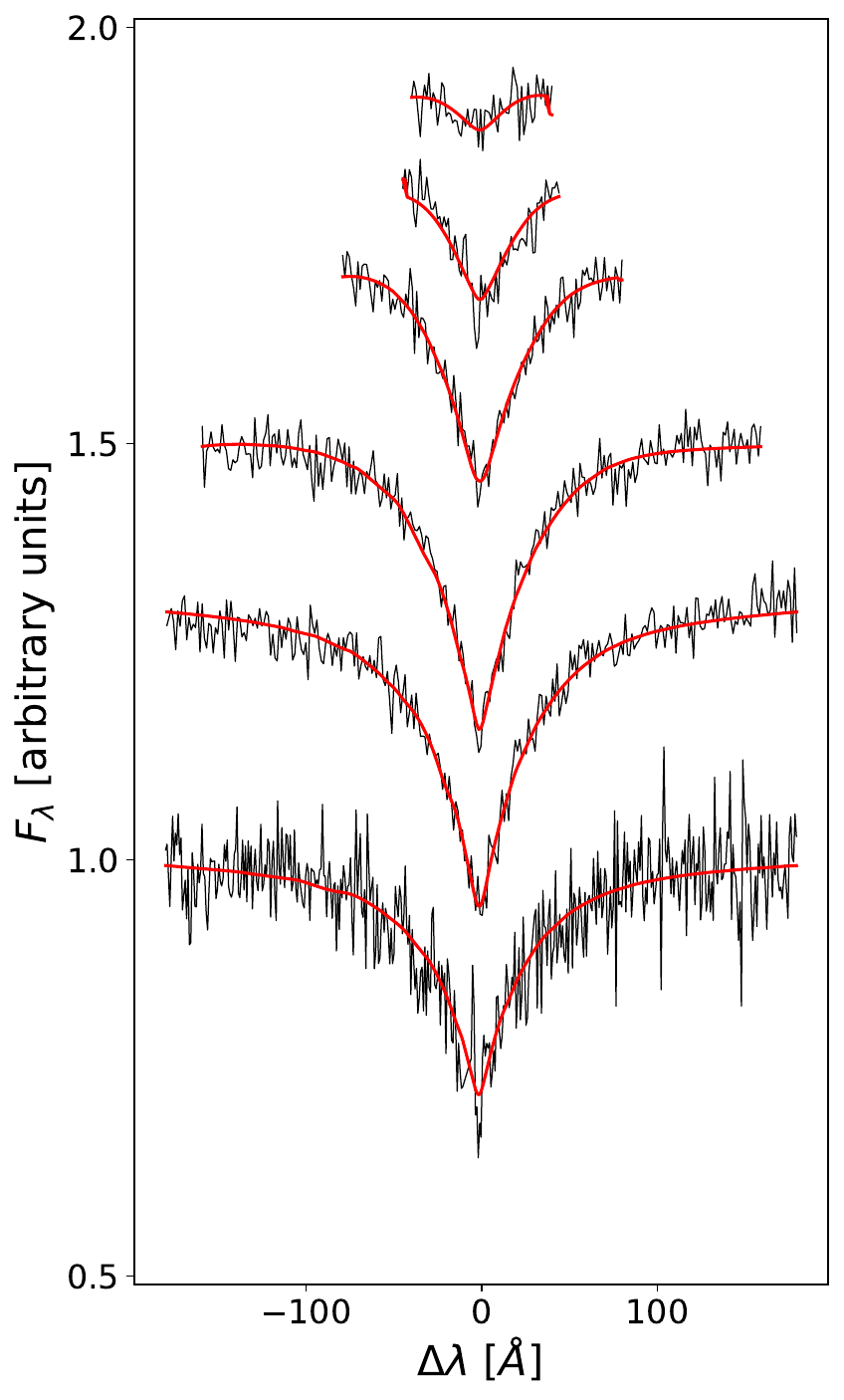}}
\put(6.965,7.12){\includegraphics[width=0.18\textwidth,clip,trim=0.92in 0.75in 0 0]{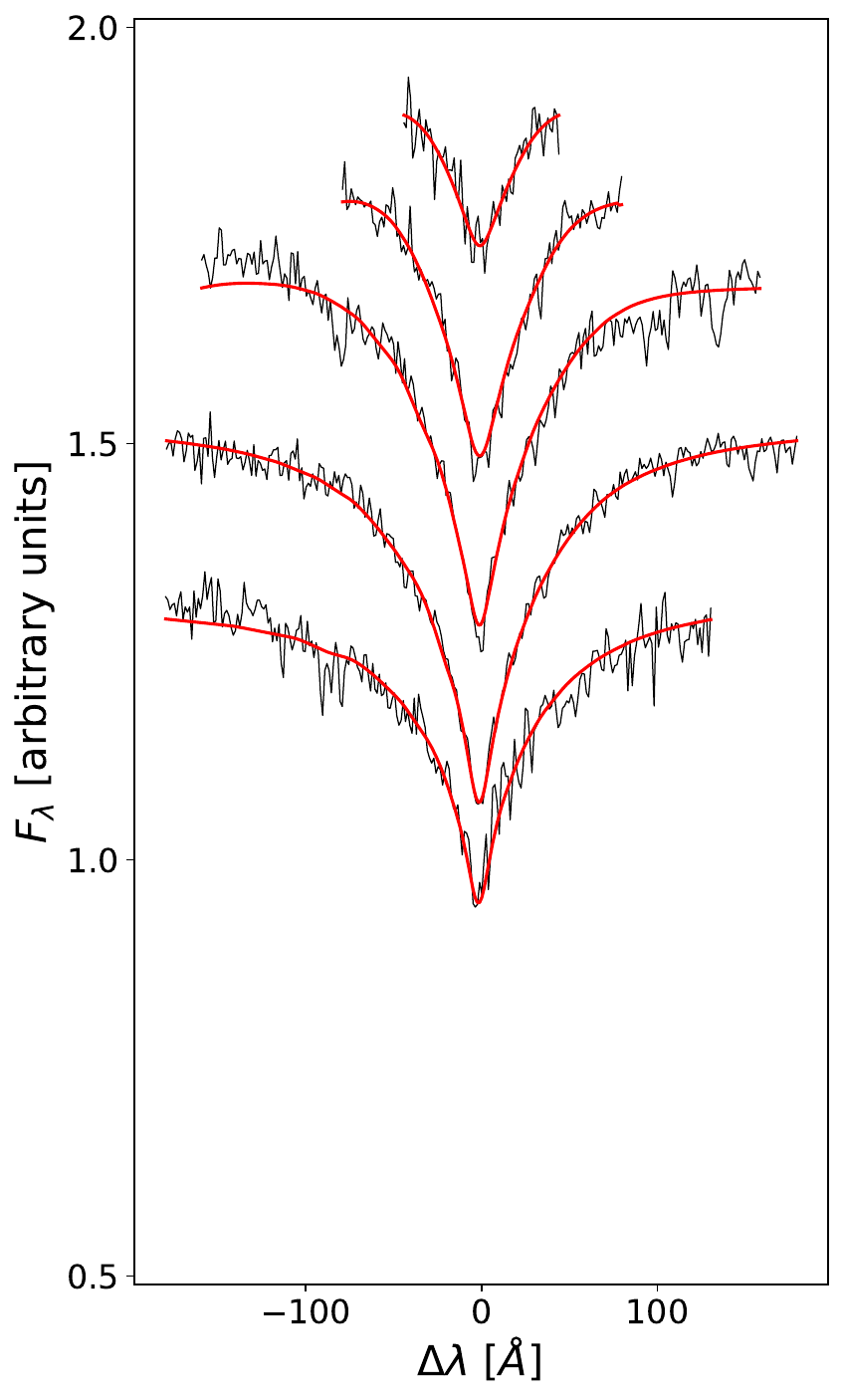}}
\put(10.15,7.12){\includegraphics[width=0.18\textwidth,clip,trim=0.92in 0.75in 0 0]{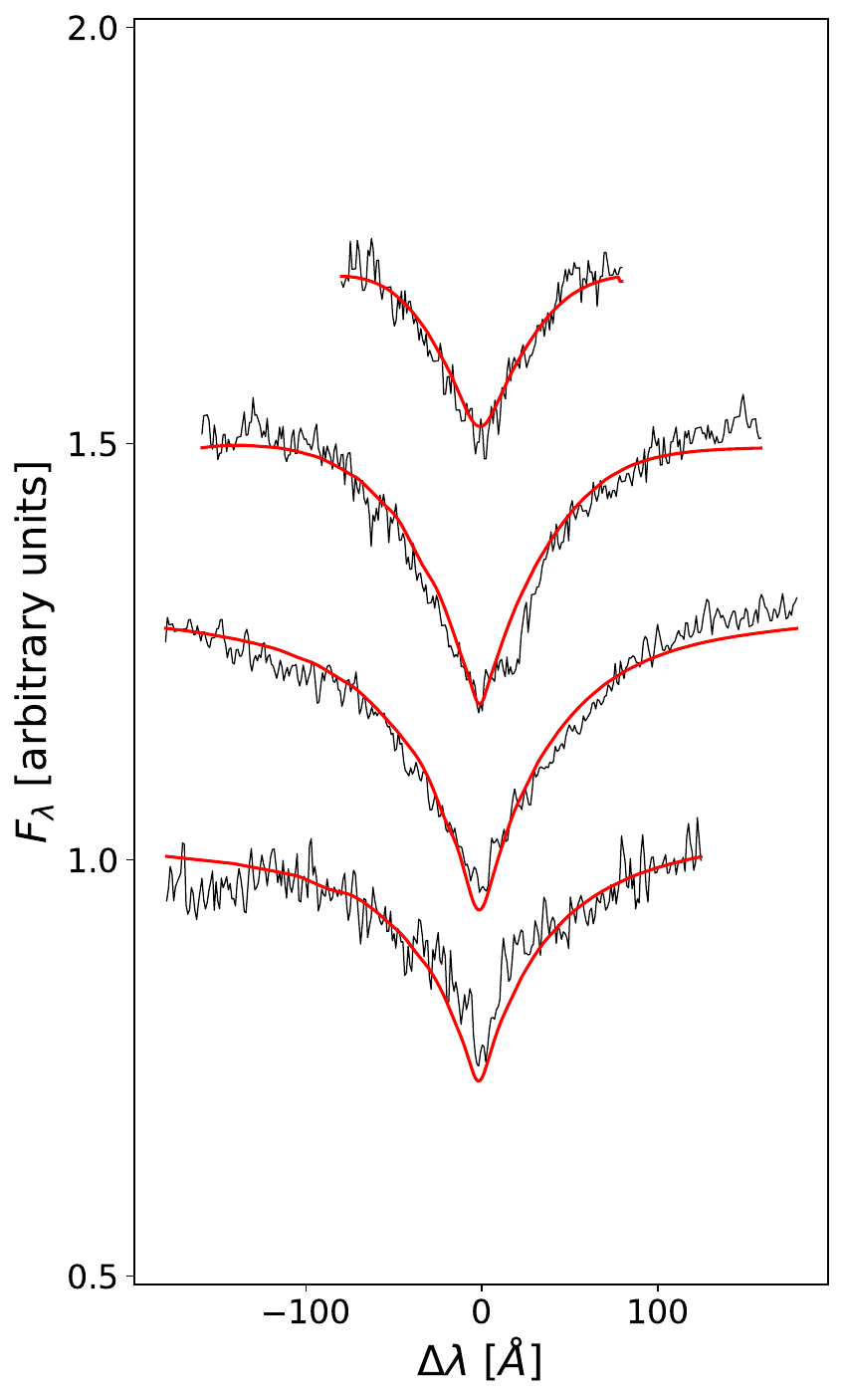}}
\put(0.58,0.72){\includegraphics[width=0.1808\textwidth,clip,trim=0.900in 0.75in 0 0]{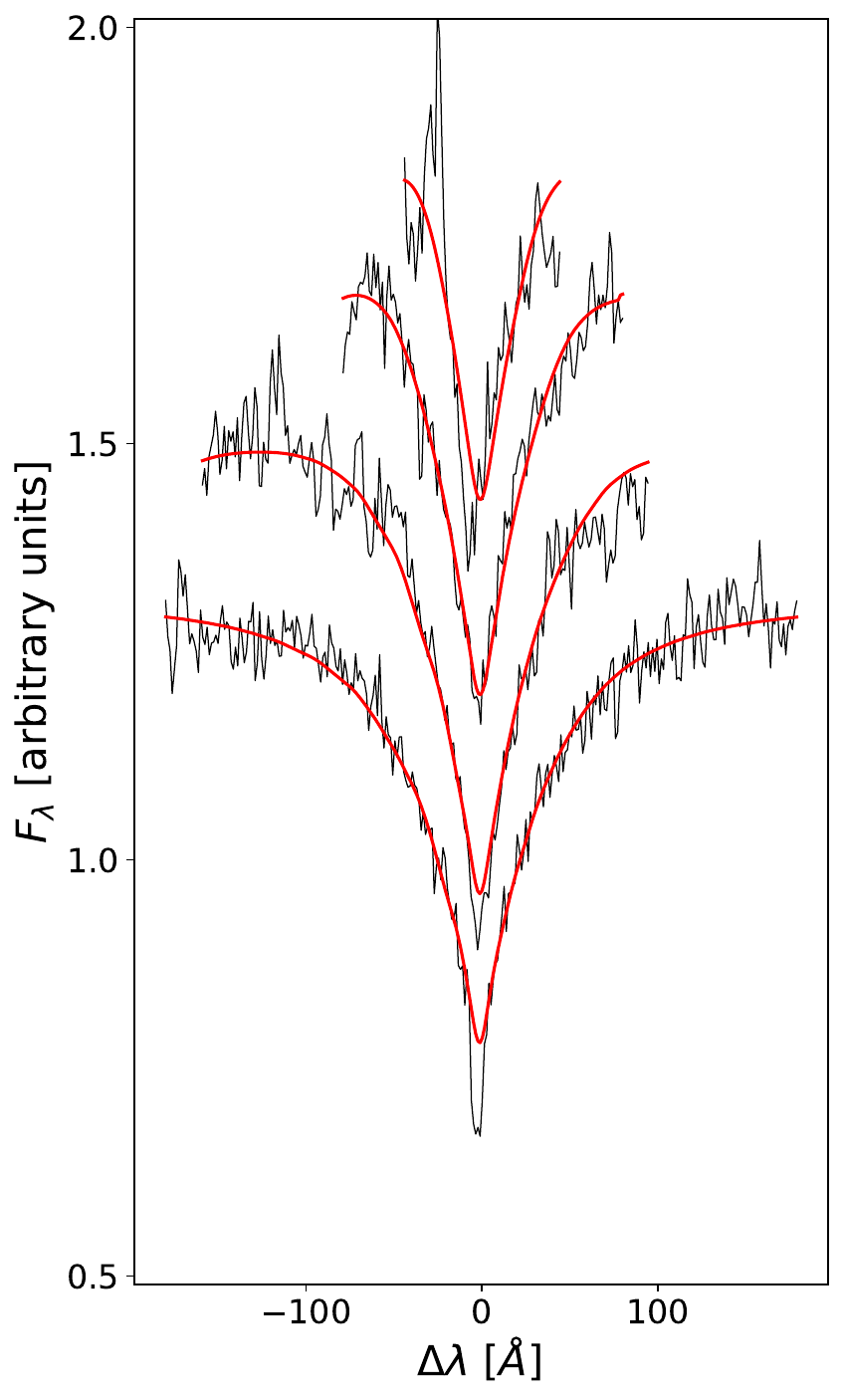}}
\put(3.78,0.72){\includegraphics[width=0.18\textwidth,clip,trim=0.92in 0.75in 0 0]{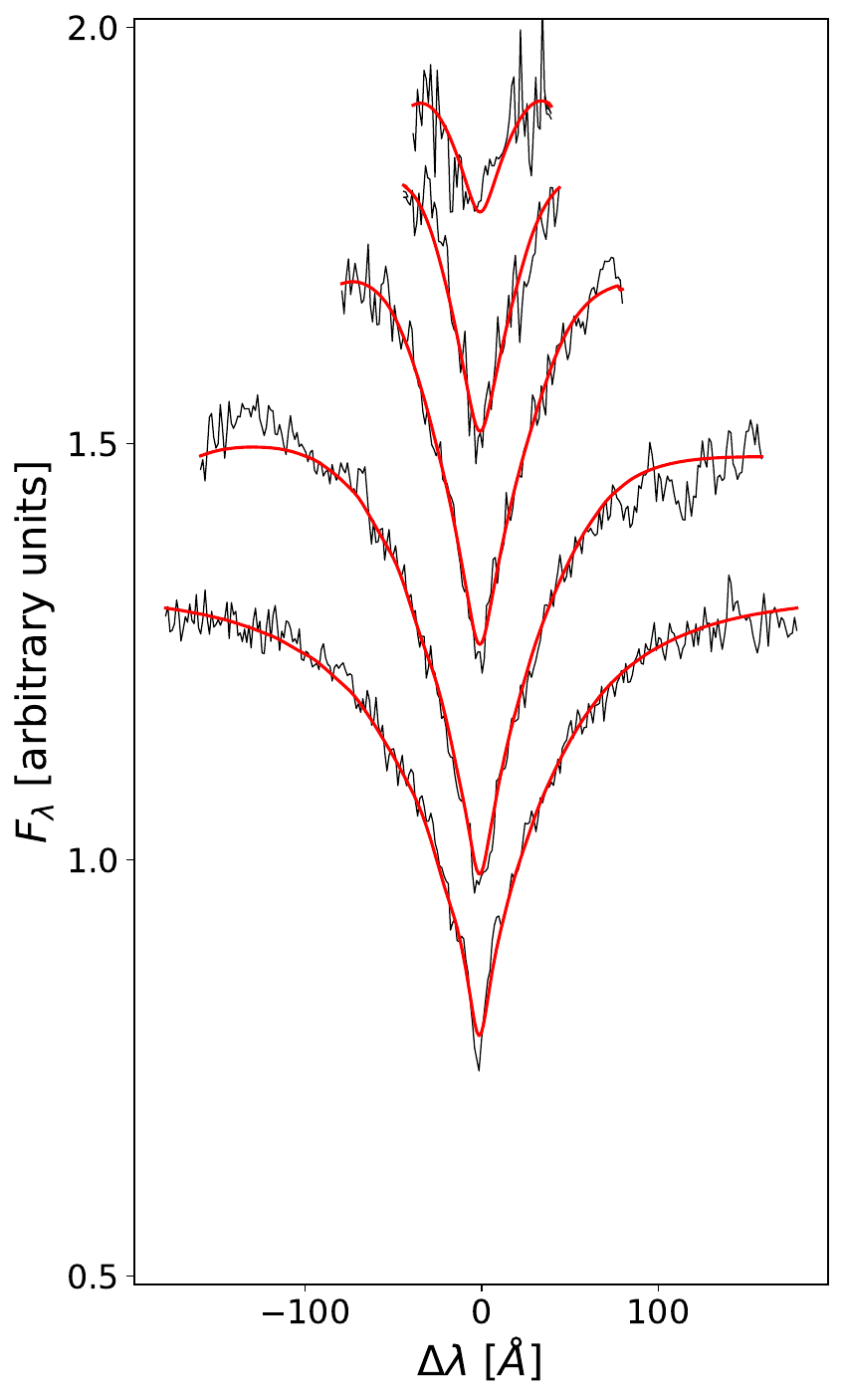}}
\put(6.965,0.72){\includegraphics[width=0.18\textwidth,clip,trim=0.92in 0.75in 0 0]{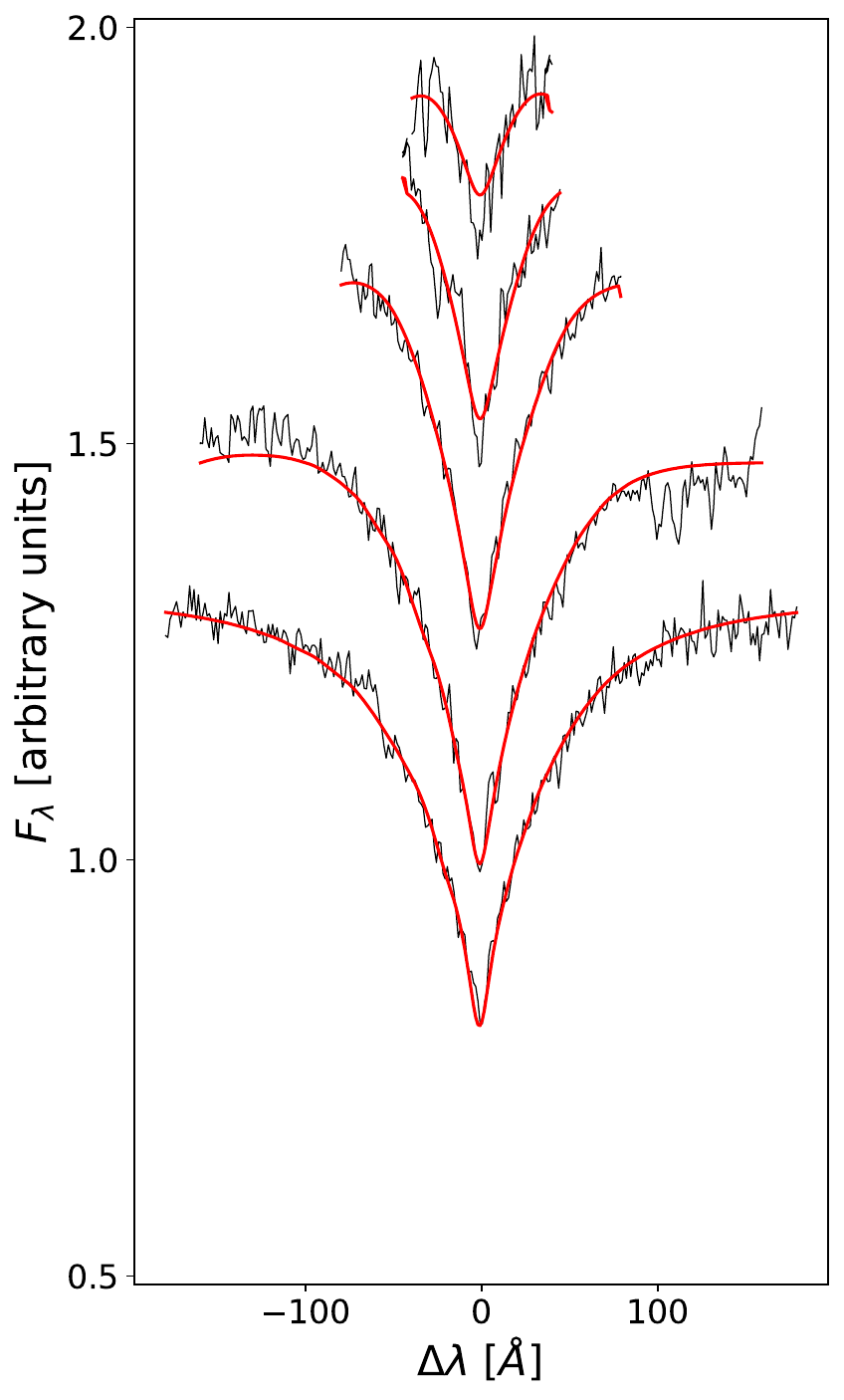}}
\put(10.15,0.72){\includegraphics[width=0.18\textwidth,clip,trim=0.92in 0.75in 0 0]{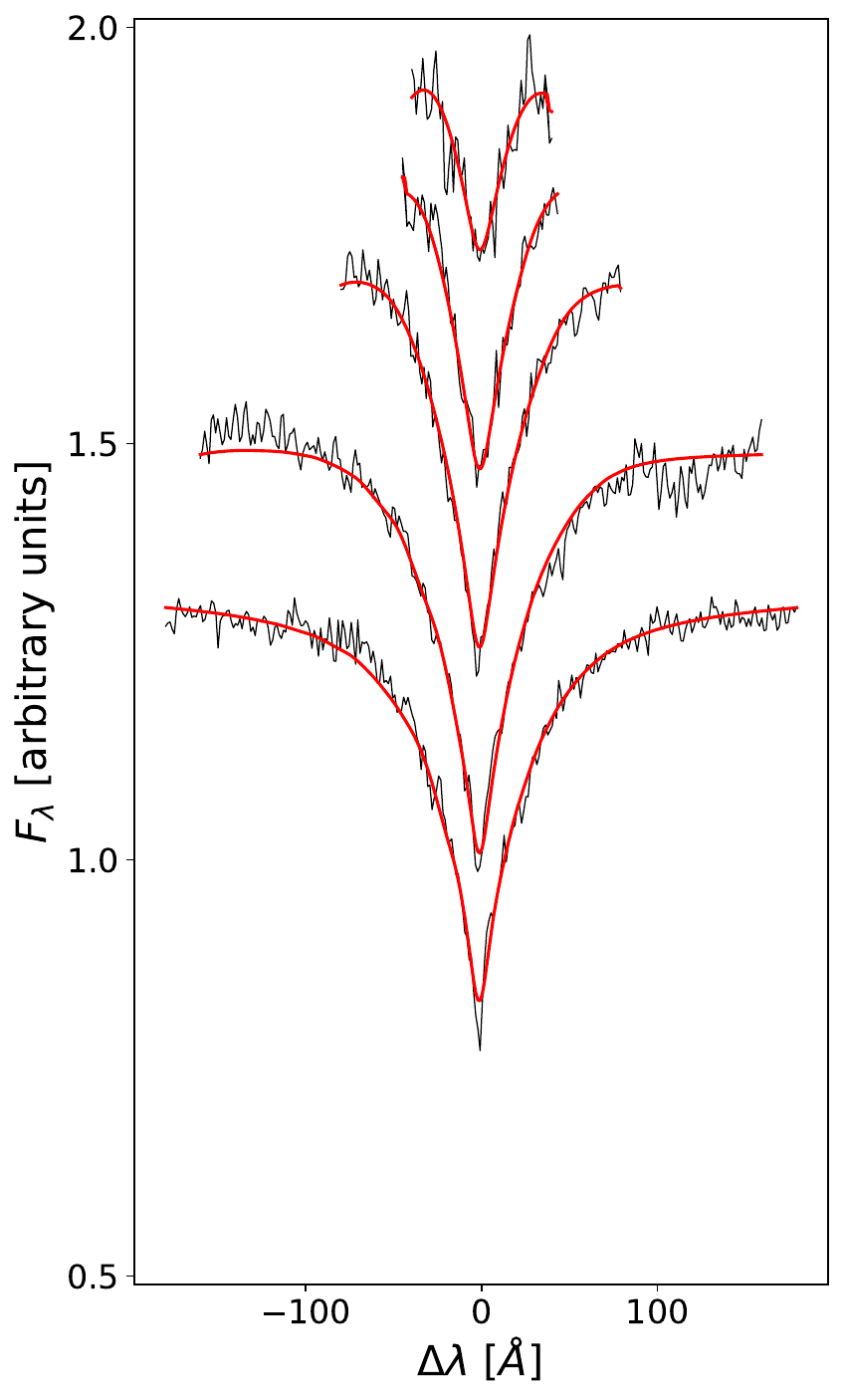}}
\put(13.335,0.72){\includegraphics[width=0.18\textwidth,clip,trim=0.92in 0.75in 0 0]{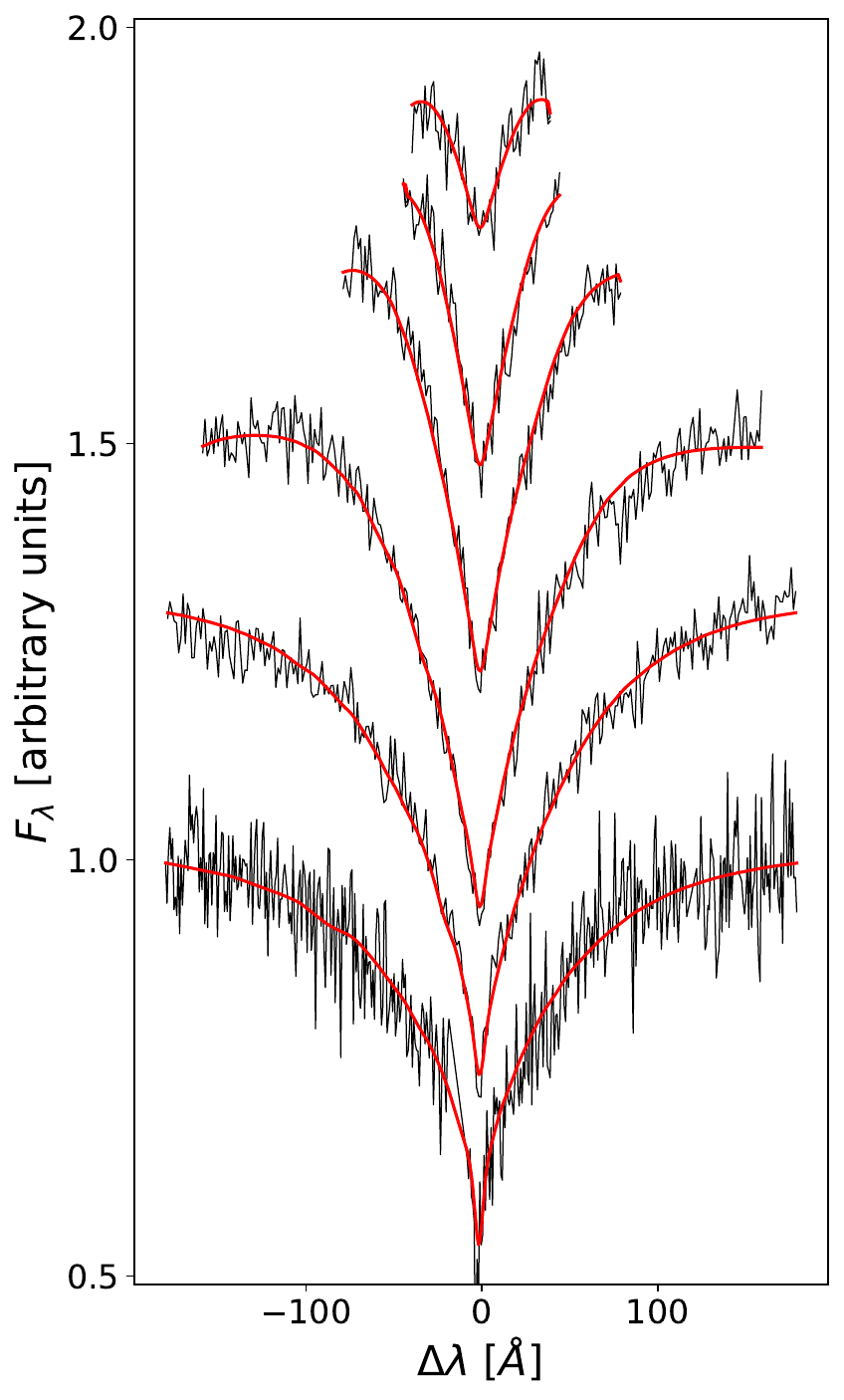}} 
\put(1.65,13.0){BH 99}
\put(4.65,13.0){HSC 601}
\put(7.35,13.0){NGC 2516 WD4}
\put(10.55,13.0){NGC 2516 WD5}
\put(1.00,6.60){NGC 3532 WD2}
\put(4.15,6.60){NGC 3532 WD4}
\put(7.30,6.60){NGC 3532 WD9}
\put(10.40,6.60){NGC 3532 WD10}
\put(14.25,6.60){Stock 1}      
\put(2.08,0.4){$0$}
\put(0.86,0.4){$-100$}
\put(2.72,0.4){$100$}
\put(1.68,0){$\Delta \lambda$ [\AA]}
\put(5.27,0.4){$0$}
\put(4.03,0.4){$-100$}
\put(5.90,0.4){$100$}
\put(4.85,0){$\Delta \lambda$ [\AA]}
\put(8.45,0.4){$0$}
\put(7.22,0.4){$-100$}
\put(9.08,0.4){$100$}
\put(8.05,0){$\Delta \lambda$ [\AA]}
\put(11.63,0.4){$0$}
\put(10.40,0.4){$-100$}
\put(12.25,0.4){$100$}
\put(11.18,0){$\Delta \lambda$ [\AA]}
\put(14.80,0.4){$0$}
\put(13.5861,0.4){$-100$}
\put(15.447,0.4){$100$}
\put(14.40,0){$\Delta \lambda$ [\AA]}
\put(-0.3,8){\rotatebox{90}{$F_\lambda$ [arbitrary units]}}
\put(0.1,7.09){0.5}
\put(0.1,9.00){1.0}
\put(0.1,10.9){1.5}
\put(0.1,12.82){2.0}
\put(-0.3,2){\rotatebox{90}{$F_\lambda$ [arbitrary units]}}
\put(0.1,0.70){0.5}
\put(0.1,2.60){1.0}
\put(0.1,4.55){1.5}
\put(0.1,6.42){2.0}
\end{picture}
    \caption{Balmer series from H$\alpha$ to H$\zeta$ (where useable) for followed-up WDs in various clusters. Spectra for BH 99, NGC 2516, and NGC 3532 WDs from GMOS on Gemini-South, with HSC 601 and Stock 1 obtained with Keck LRIS. Best fitting hydrogen model superimposed in red.}
    \label{fig:spectrafits_a}
\end{figure*}
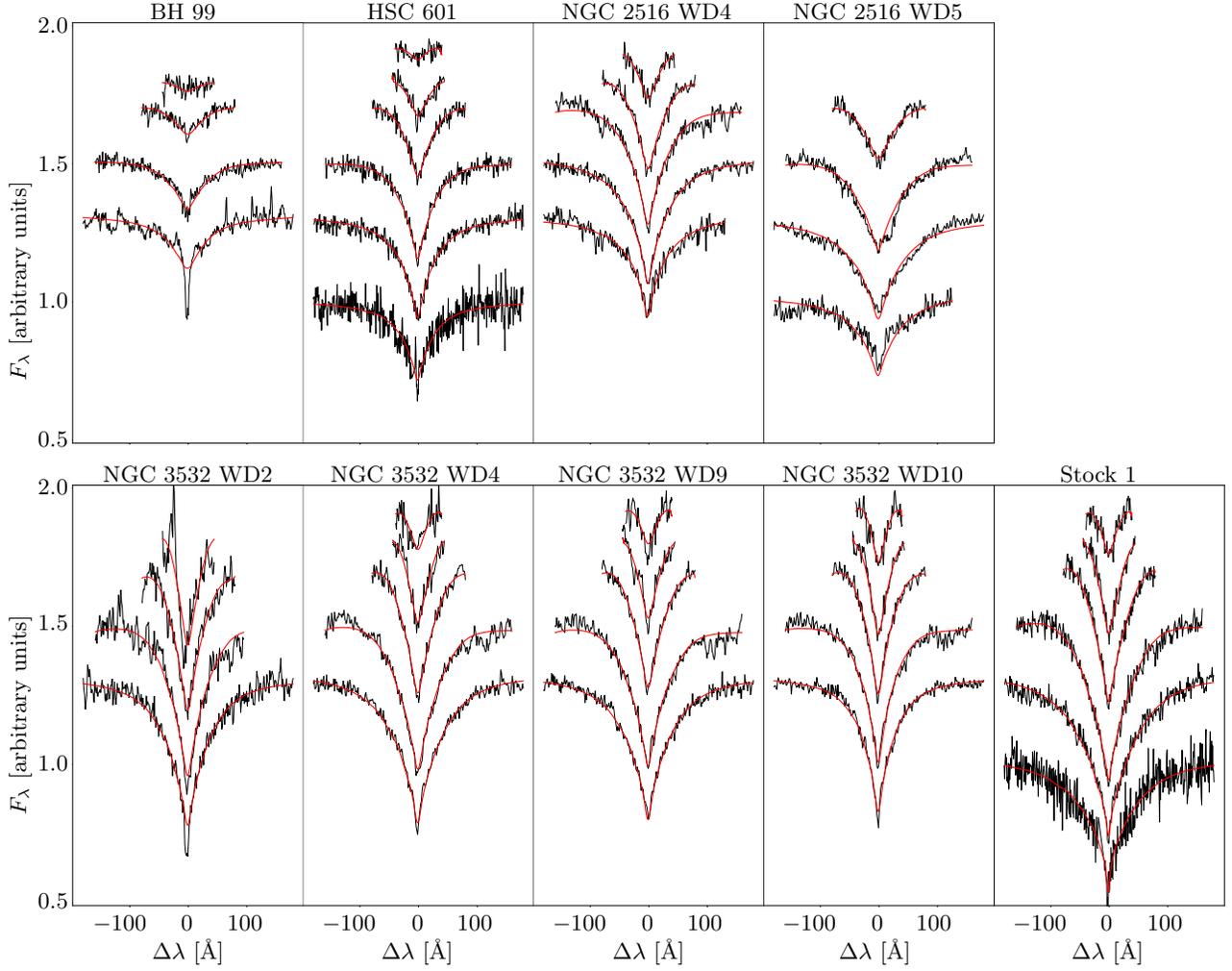

\begin{figure*}[ht]
\centering
  \setlength{\unitlength}{1cm}
  \begin{picture}(17,14)(-0.3,0)
\put(0.58,7.12){\includegraphics[width=0.1808\textwidth,clip,trim=0.90in 0.75in 0 0]{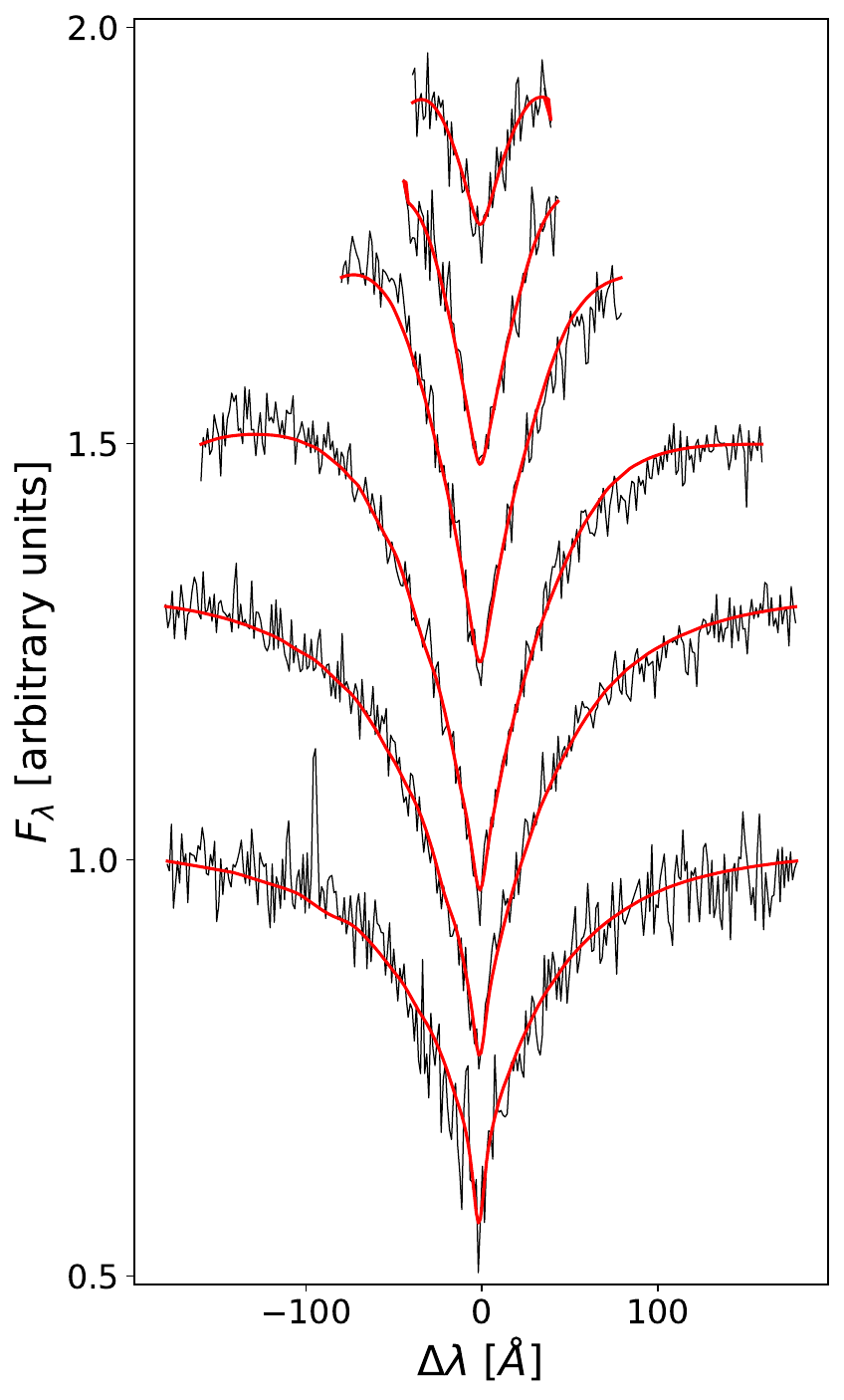}}
\put(3.78,7.12){\includegraphics[width=0.18\textwidth,clip,trim=0.92in 0.75in 0 0]{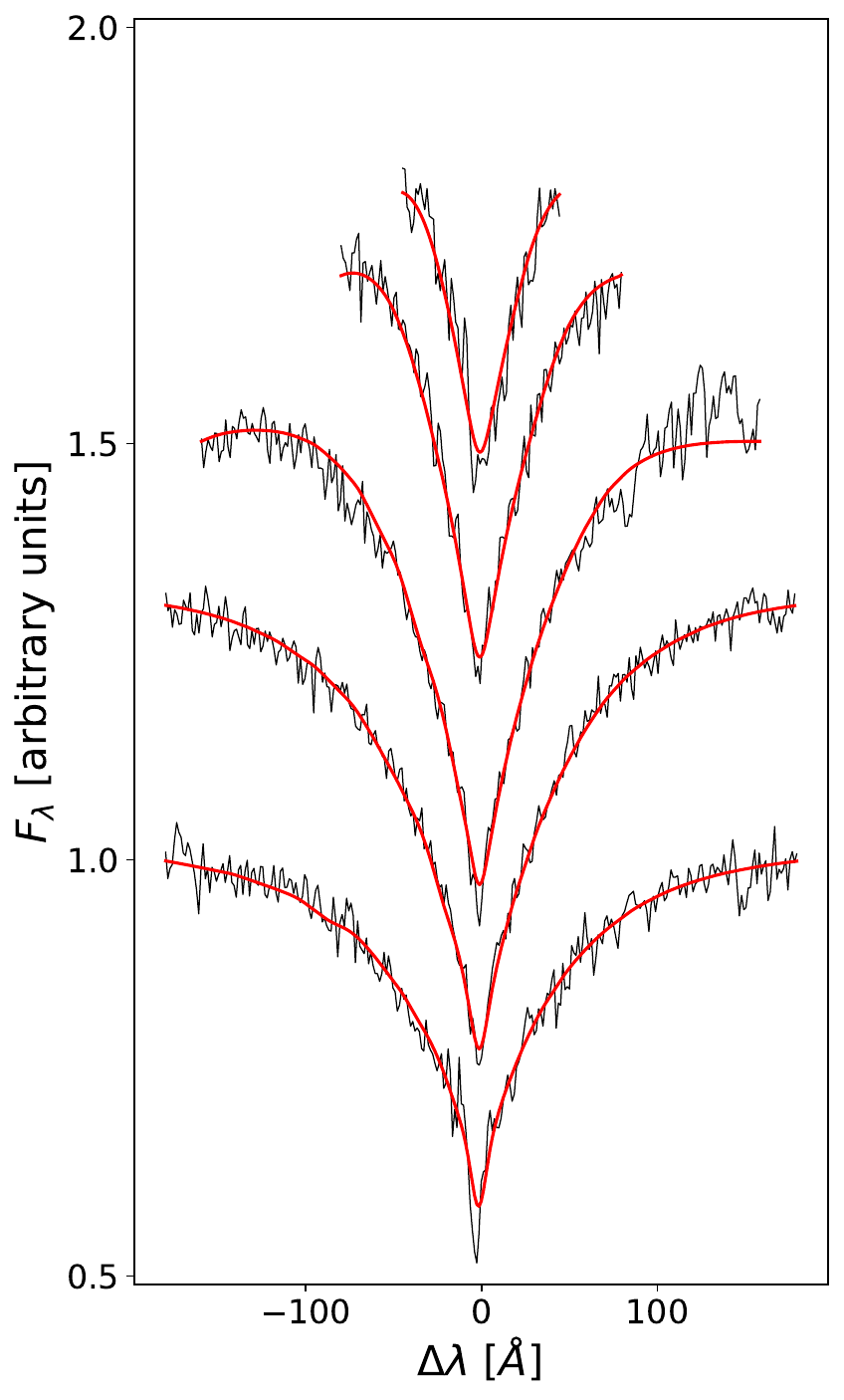}}
\put(6.965,7.12){\includegraphics[width=0.18\textwidth,clip,trim=0.92in 0.75in 0 0]{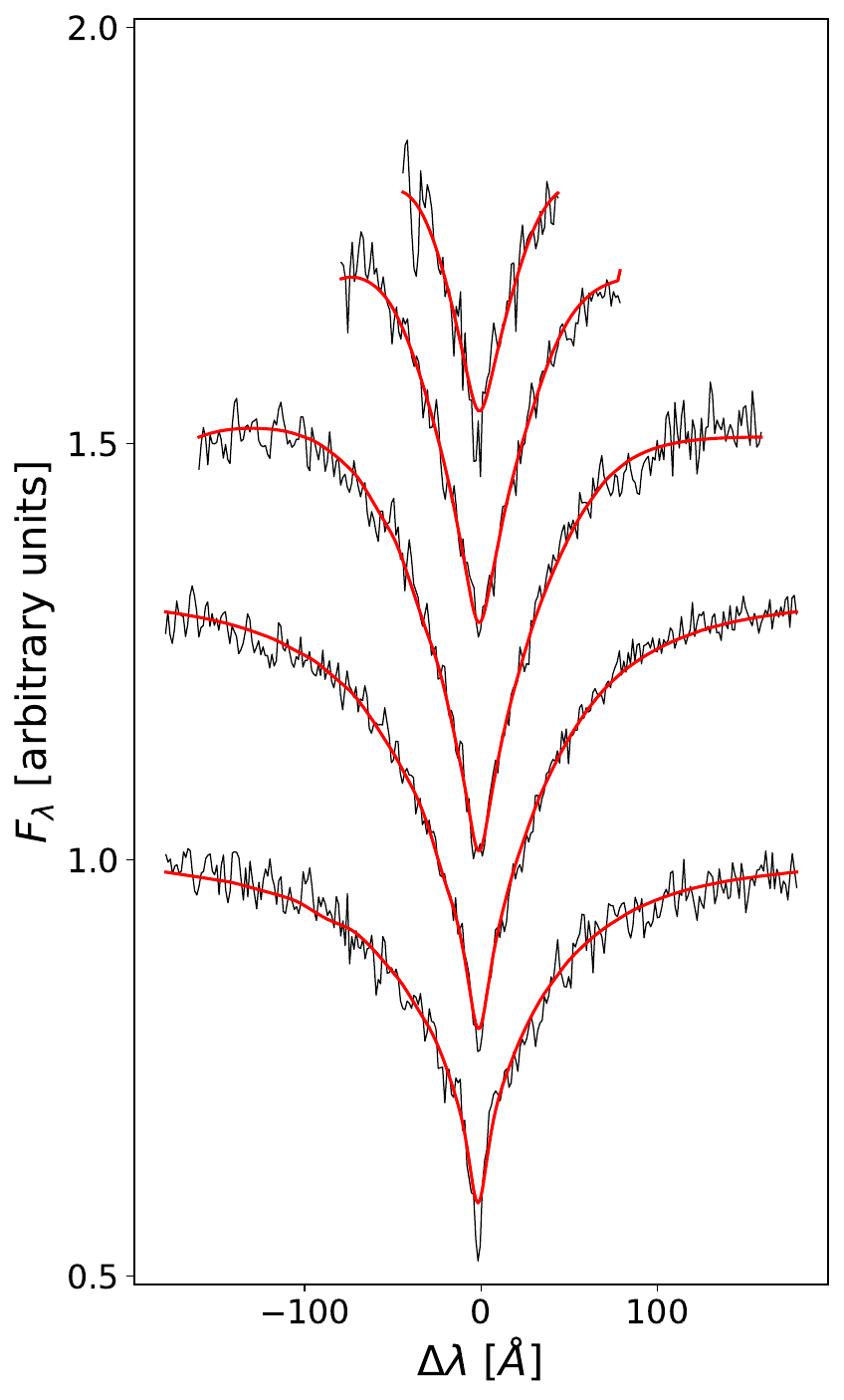}}
\put(10.15,7.12){\includegraphics[width=0.18\textwidth,clip,trim=0.92in 0.75in 0 0]{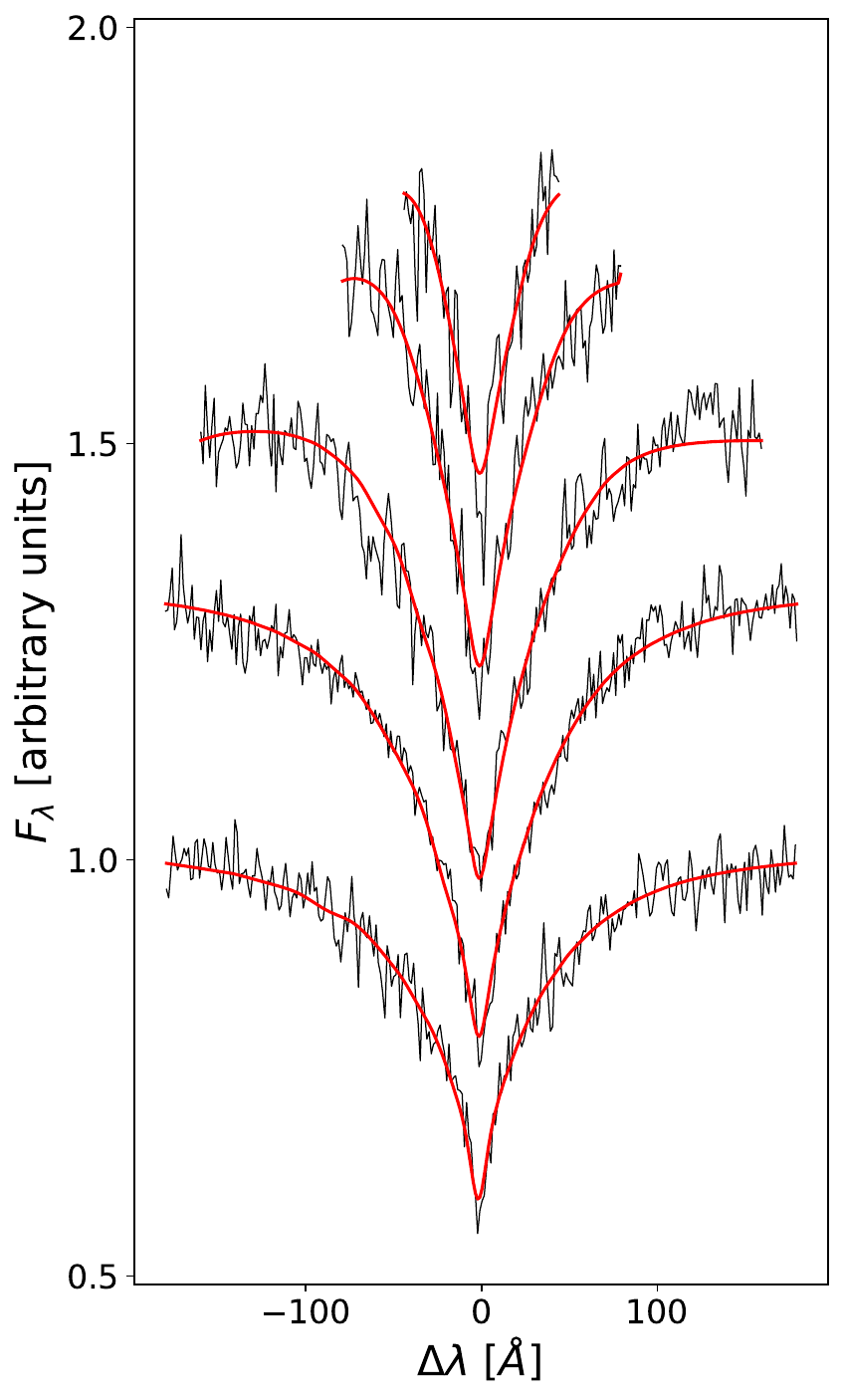}}
\put(13.335,7.12){\includegraphics[width=0.18\textwidth,clip,trim=0.92in 0.75in 0 0]{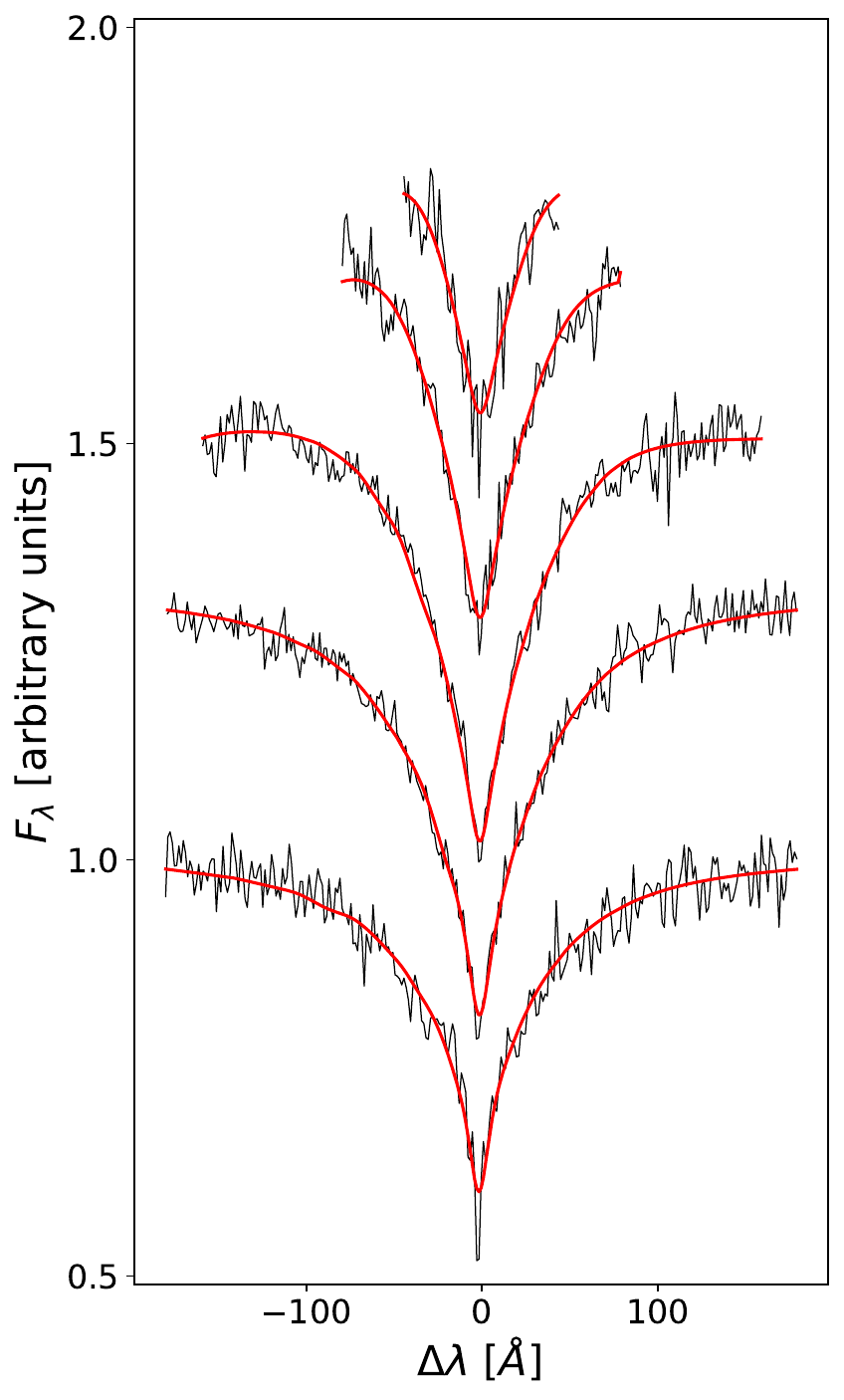}} 
\put(0.58,0.72){\includegraphics[width=0.1808\textwidth,clip,trim=0.90in 0.75in 0 0]{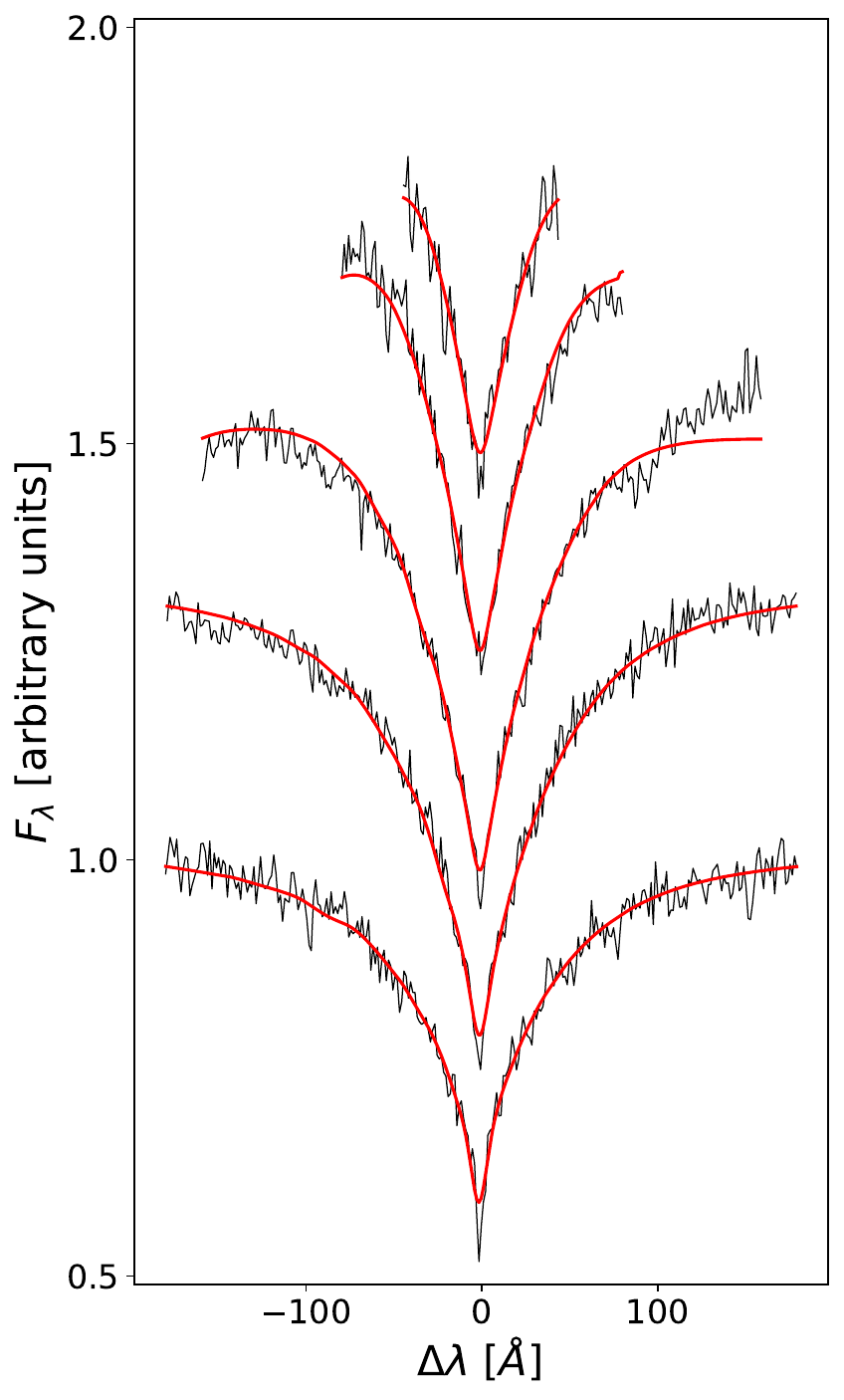}}
\put(3.78,0.72){\includegraphics[width=0.18\textwidth,clip,trim=0.92in 0.75in 0 0]{Stock2_WD10_fit.pdf}}
\put(6.965,0.72){\includegraphics[width=0.18\textwidth,clip,trim=0.92in 0.75in 0 0]{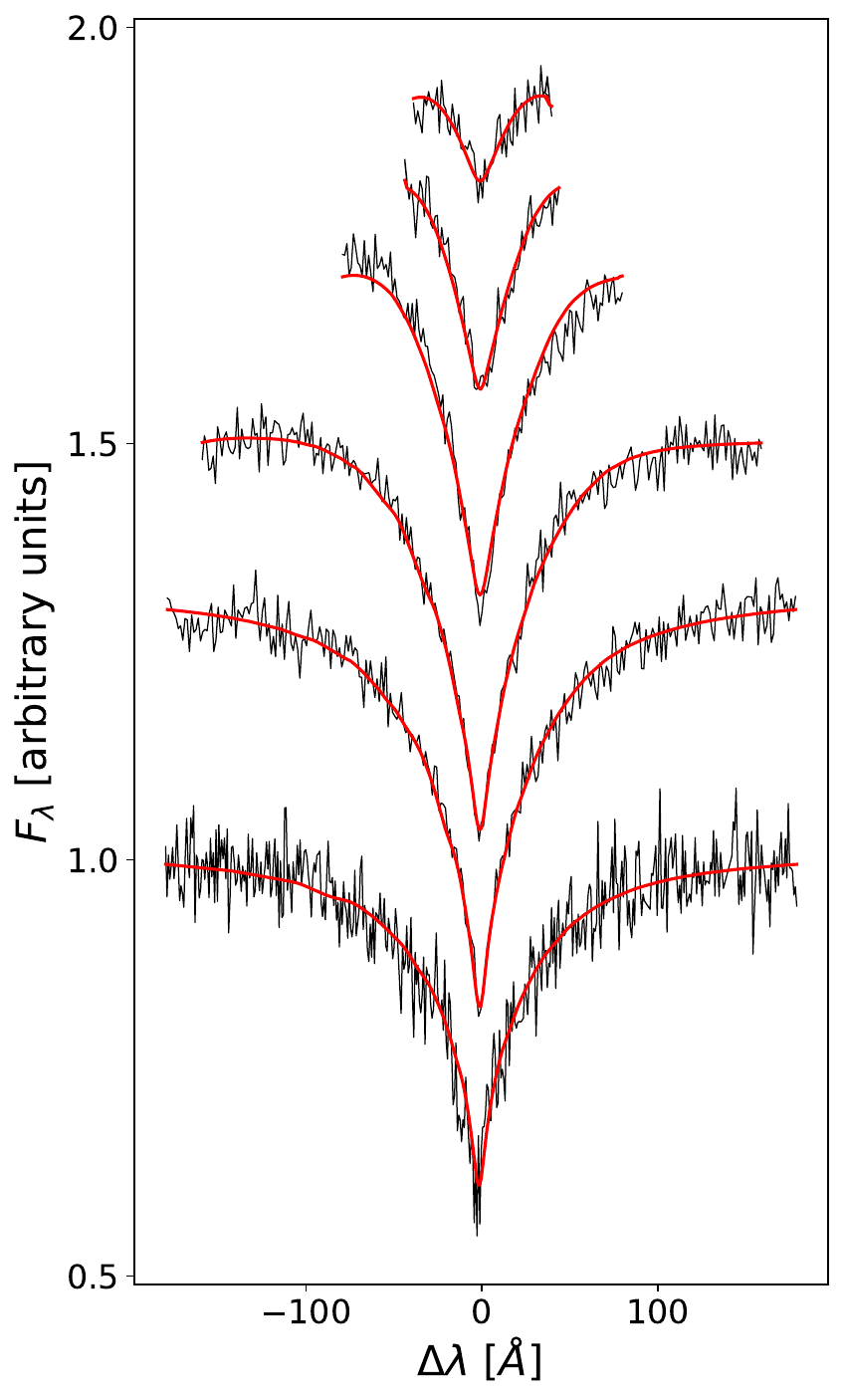}}
\put(10.15,0.72){\includegraphics[width=0.18\textwidth,clip,trim=0.92in 0.75in 0 0]{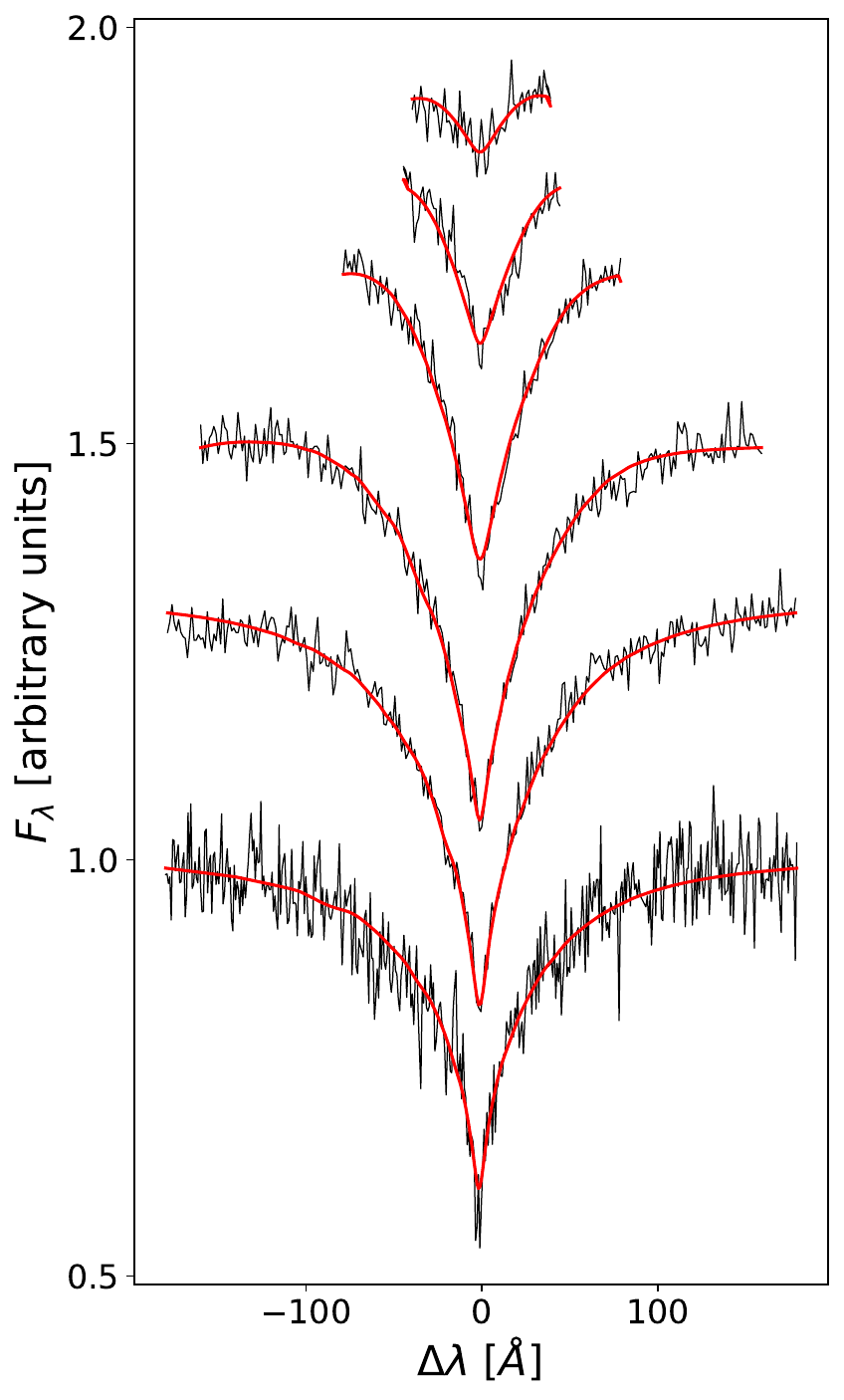}}
\put(13.335,0.72){\includegraphics[width=0.18\textwidth,clip,trim=0.92in 0.75in 0 0]{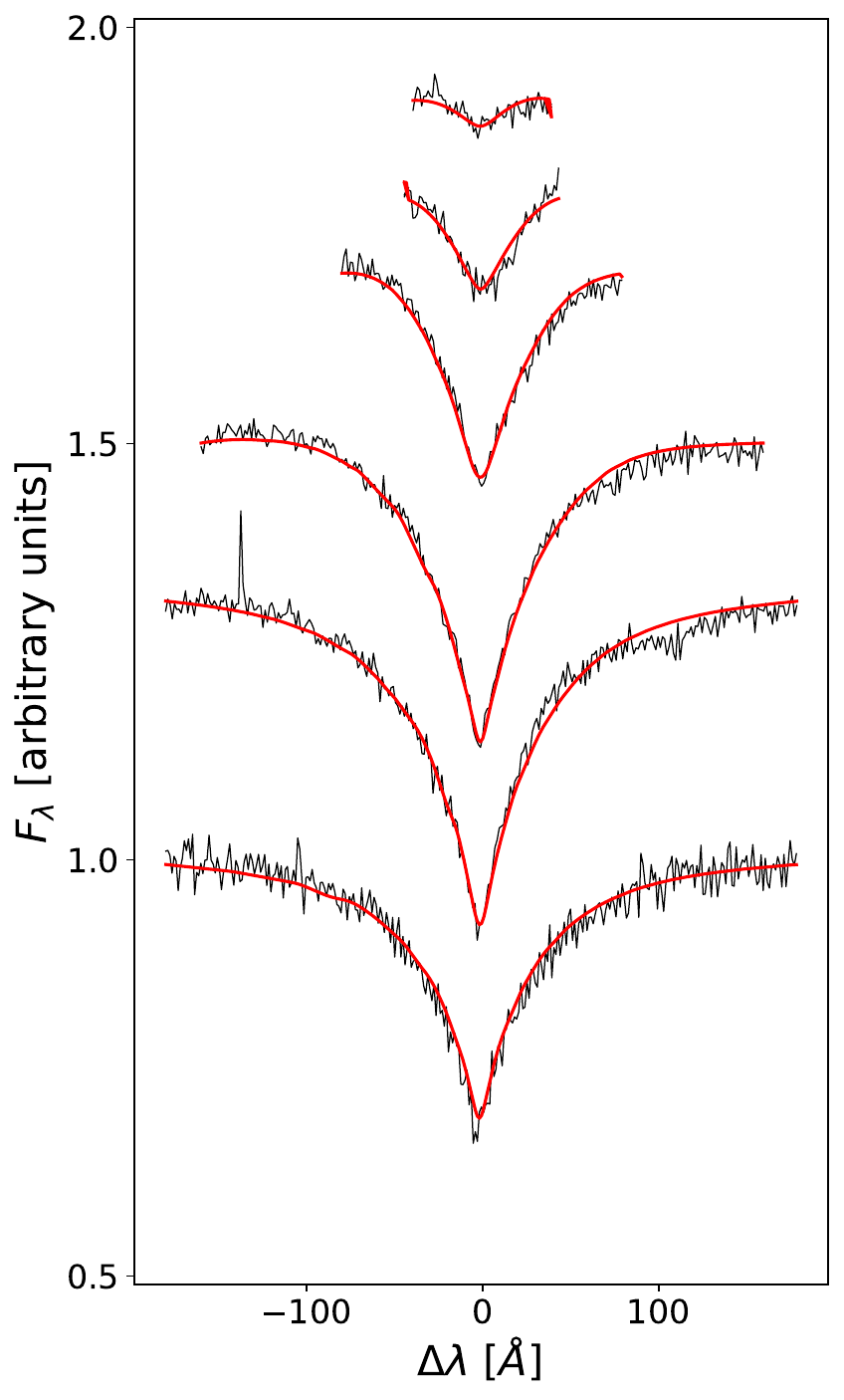}} 
\put(1.19,13.0){Stock 2 WD3}
\put(4.35,13.0){Stock 2 WD4}
\put(7.55,13.0){Stock 2 WD5}
\put(10.72,13.0){Stock 2 WD7}
\put(13.88,13.0){Stock 2 WD8}      
\put(1.19,6.60){Stock 2 WD9}
\put(4.25,6.60){Stock 2 WD10}
\put(7.70,6.60){Theia 248}
\put(10.90,6.60){Theia 817}
\put(14.15,6.60){UPK 303}      
\put(2.08,0.4){$0$}
\put(0.86,0.4){$-100$}
\put(2.72,0.4){$100$}
\put(1.68,0){$\Delta \lambda$ [\AA]}
\put(5.27,0.4){$0$}
\put(4.03,0.4){$-100$}
\put(5.90,0.4){$100$}
\put(4.85,0){$\Delta \lambda$ [\AA]}
\put(8.45,0.4){$0$}
\put(7.22,0.4){$-100$}
\put(9.08,0.4){$100$}
\put(8.05,0){$\Delta \lambda$ [\AA]}
\put(11.63,0.4){$0$}
\put(10.40,0.4){$-100$}
\put(12.25,0.4){$100$}
\put(11.18,0){$\Delta \lambda$ [\AA]}
\put(14.80,0.4){$0$}
\put(13.5861,0.4){$-100$}
\put(15.447,0.4){$100$}
\put(14.40,0){$\Delta \lambda$ [\AA]}
\put(-0.3,8){\rotatebox{90}{$F_\lambda$ [arbitrary units]}}
\put(0.1,7.09){0.5}
\put(0.1,9.00){1.0}
\put(0.1,10.9){1.5}
\put(0.1,12.82){2.0}
\put(-0.3,2){\rotatebox{90}{$F_\lambda$ [arbitrary units]}}
\put(0.1,0.70){0.5}
\put(0.1,2.60){1.0}
\put(0.1,4.55){1.5}
\put(0.1,6.42){2.0}
\end{picture}
    \caption{As in Fig.~\ref{fig:spectrafits_a}, but for different clusters. Stock 2 WDs 4, 5, 7, 8, 9, 10 from GMOS on Gemini-North, rest from Keck LRIS. Stock 2 WD10 is repeated from the main text.}
    \label{fig:spectrafits_b}
\end{figure*}

\begin{figure*}[ht]
    \centering
    \setlength{\unitlength}{1cm}
    \begin{picture}(17, 21)(0, 0)
        \put(0, 13.11){\includegraphics[width=0.260\textwidth]{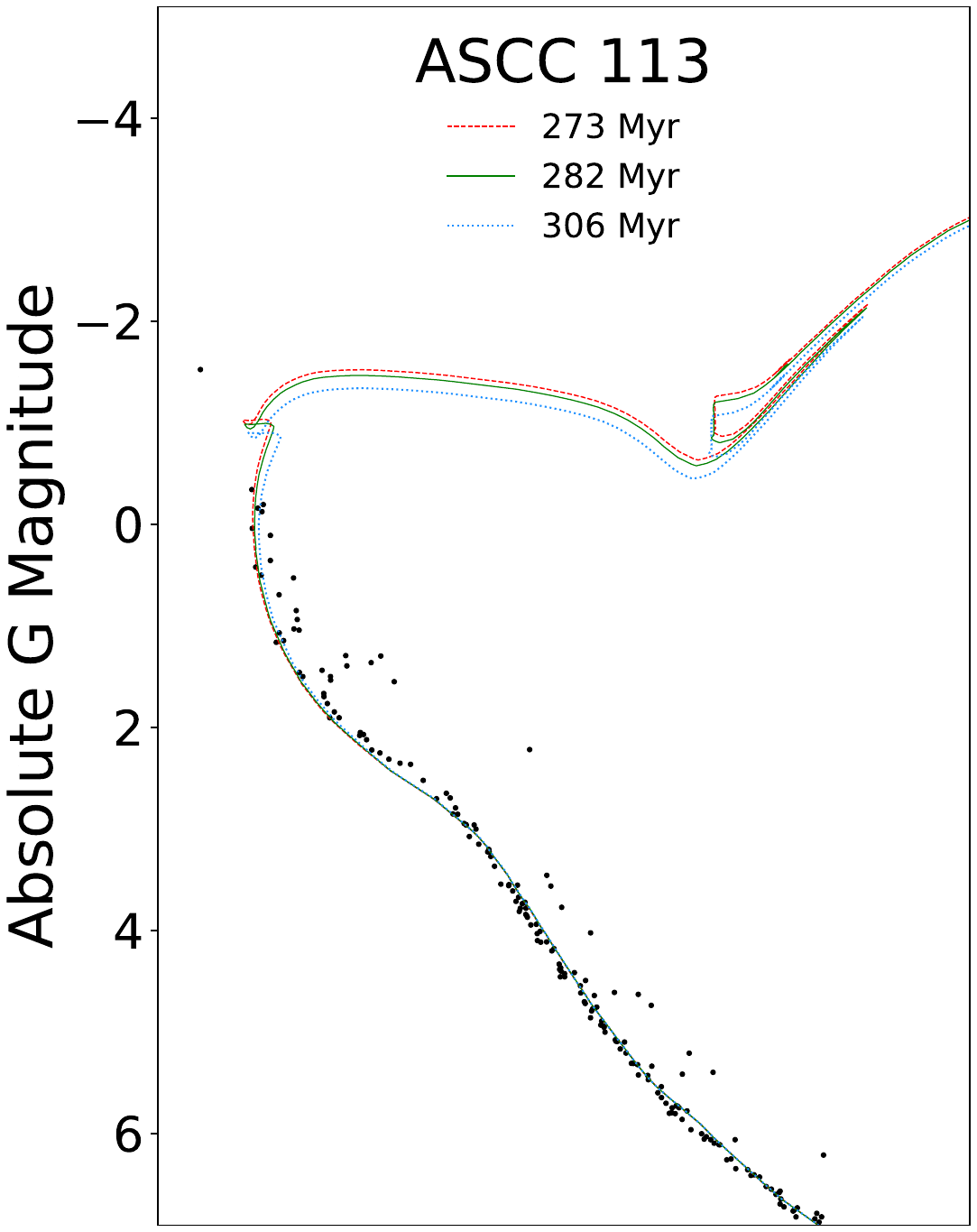}}
        \put(4.66, 13.11){\includegraphics[width=0.22\textwidth]{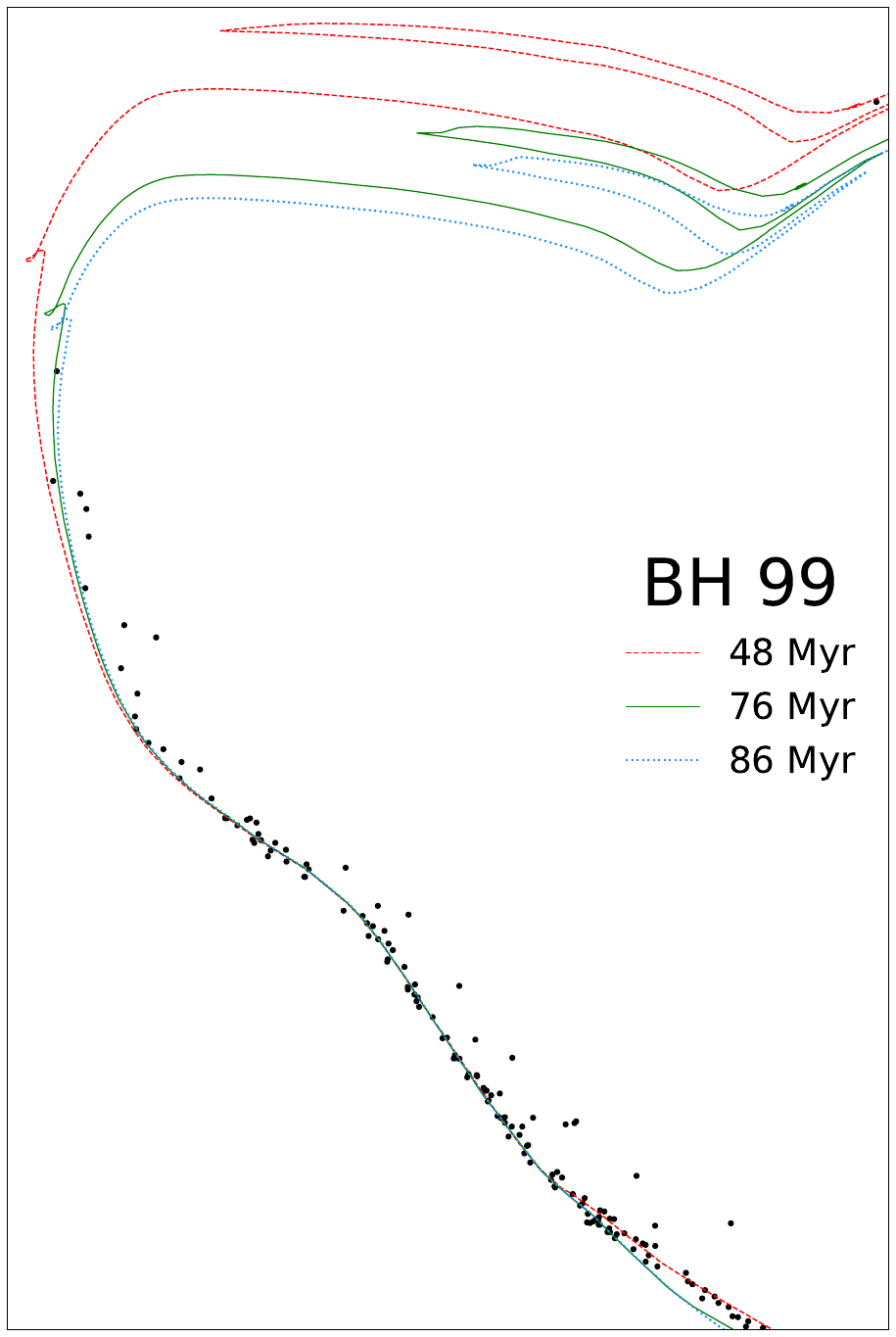}}
        \put(8.597, 13.11){\includegraphics[width=0.22\textwidth]{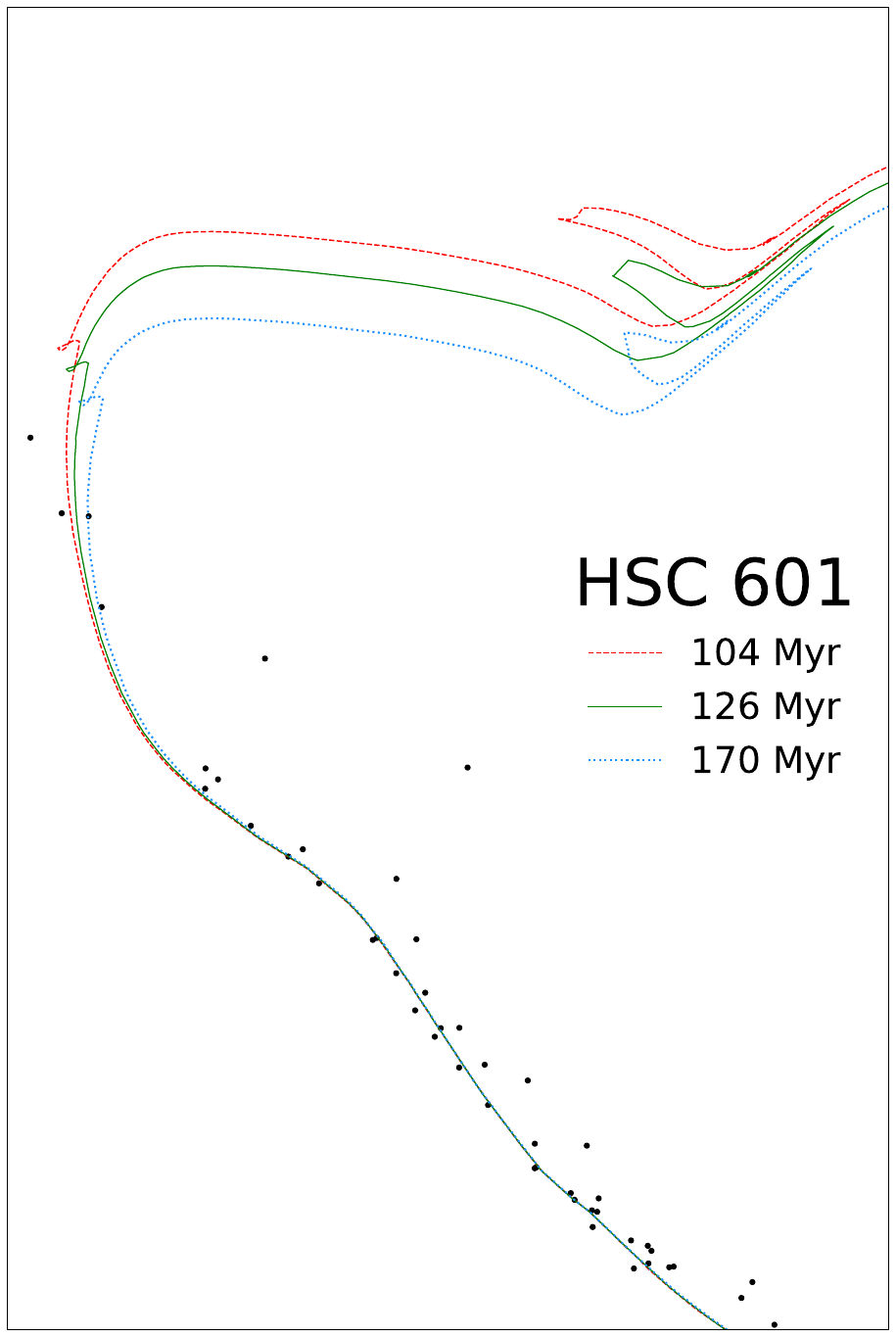}}
        \put(12.533, 13.11){\includegraphics[width=0.22\textwidth]{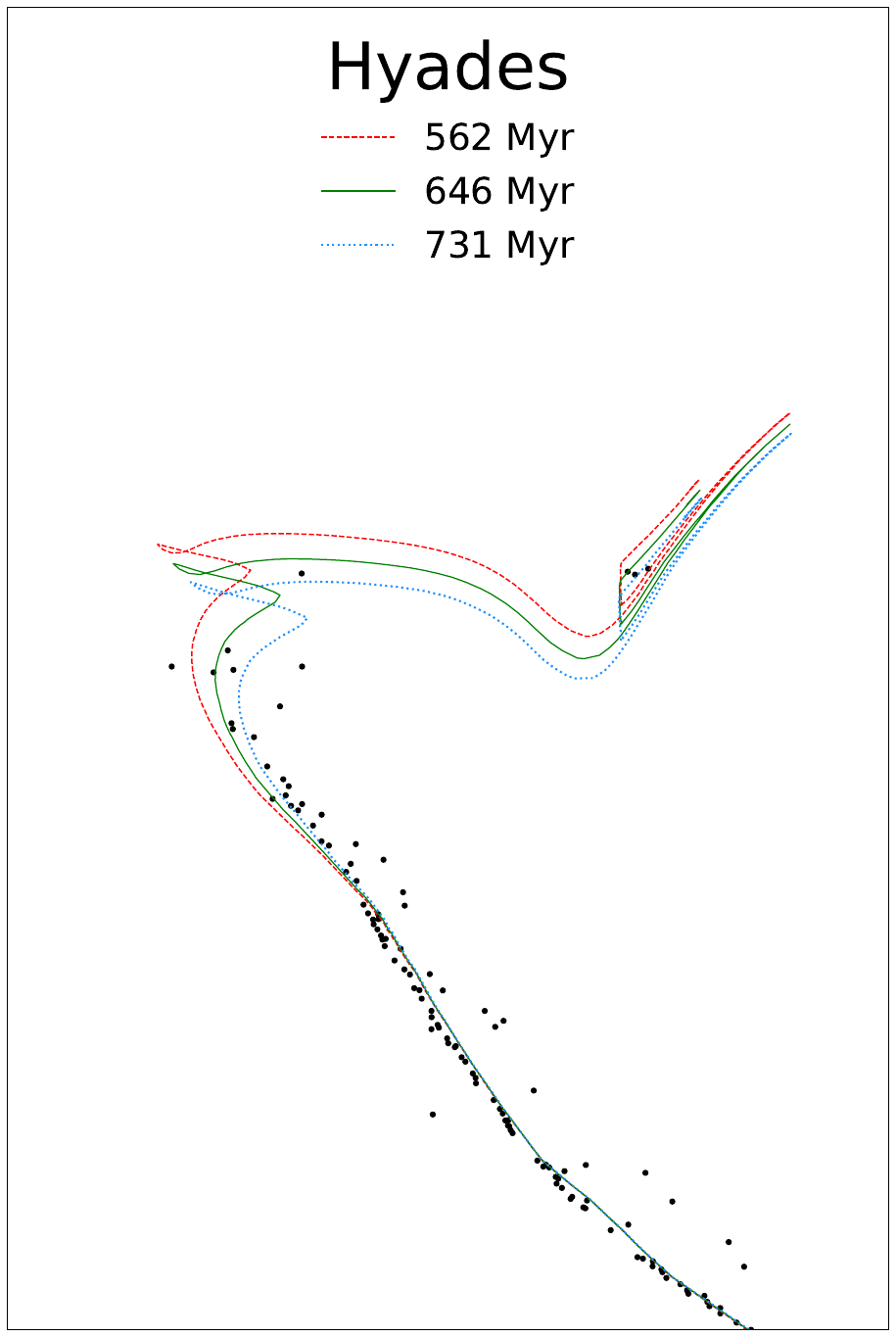}}
        \put(0, 7.22){\includegraphics[width=0.260\textwidth]{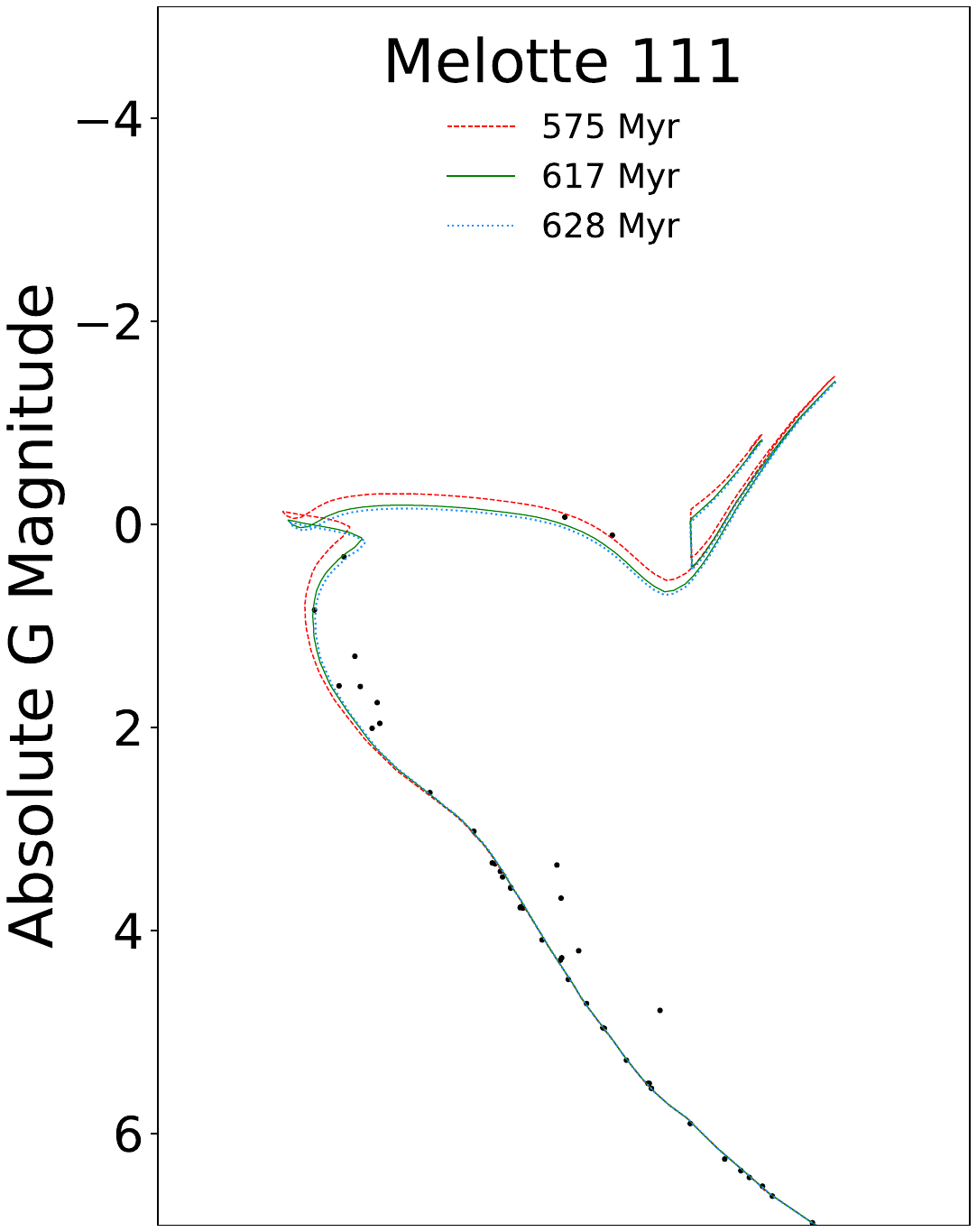}}
        \put(4.66, 7.22){\includegraphics[width=0.22\textwidth]{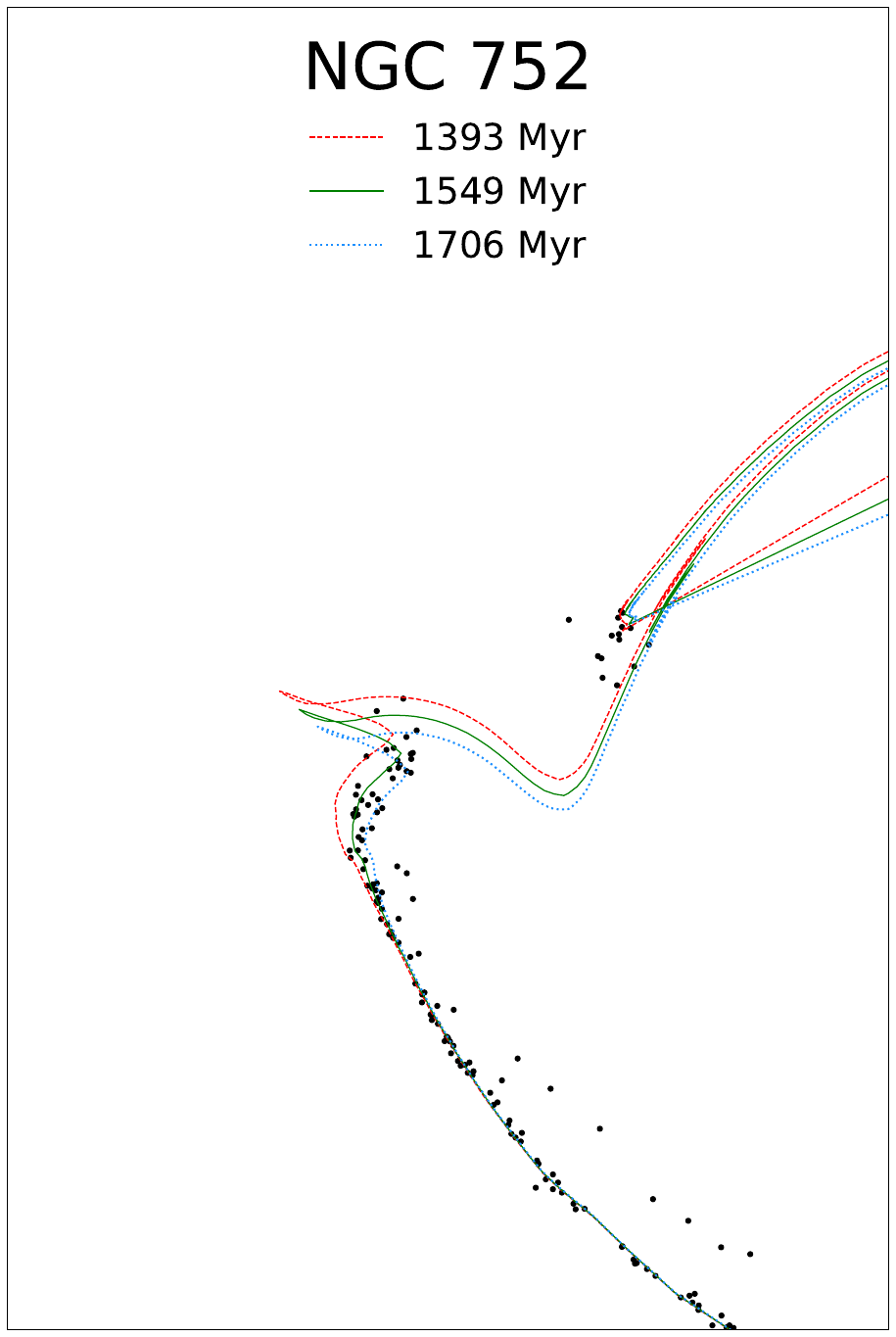}}
        \put(8.597, 7.22){\includegraphics[width=0.22\textwidth]{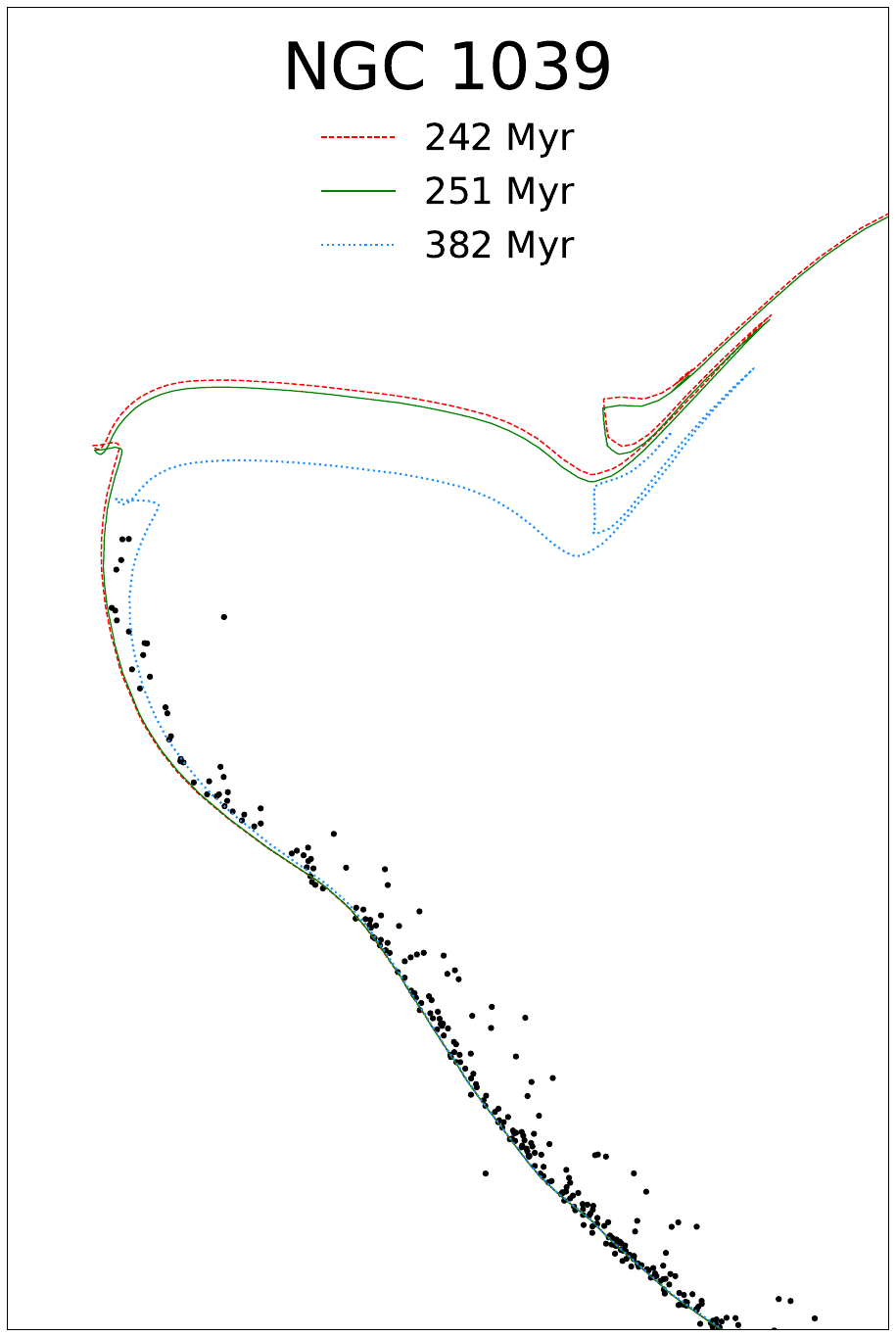}}
        \put(12.533, 7.22){\includegraphics[width=0.22\textwidth]{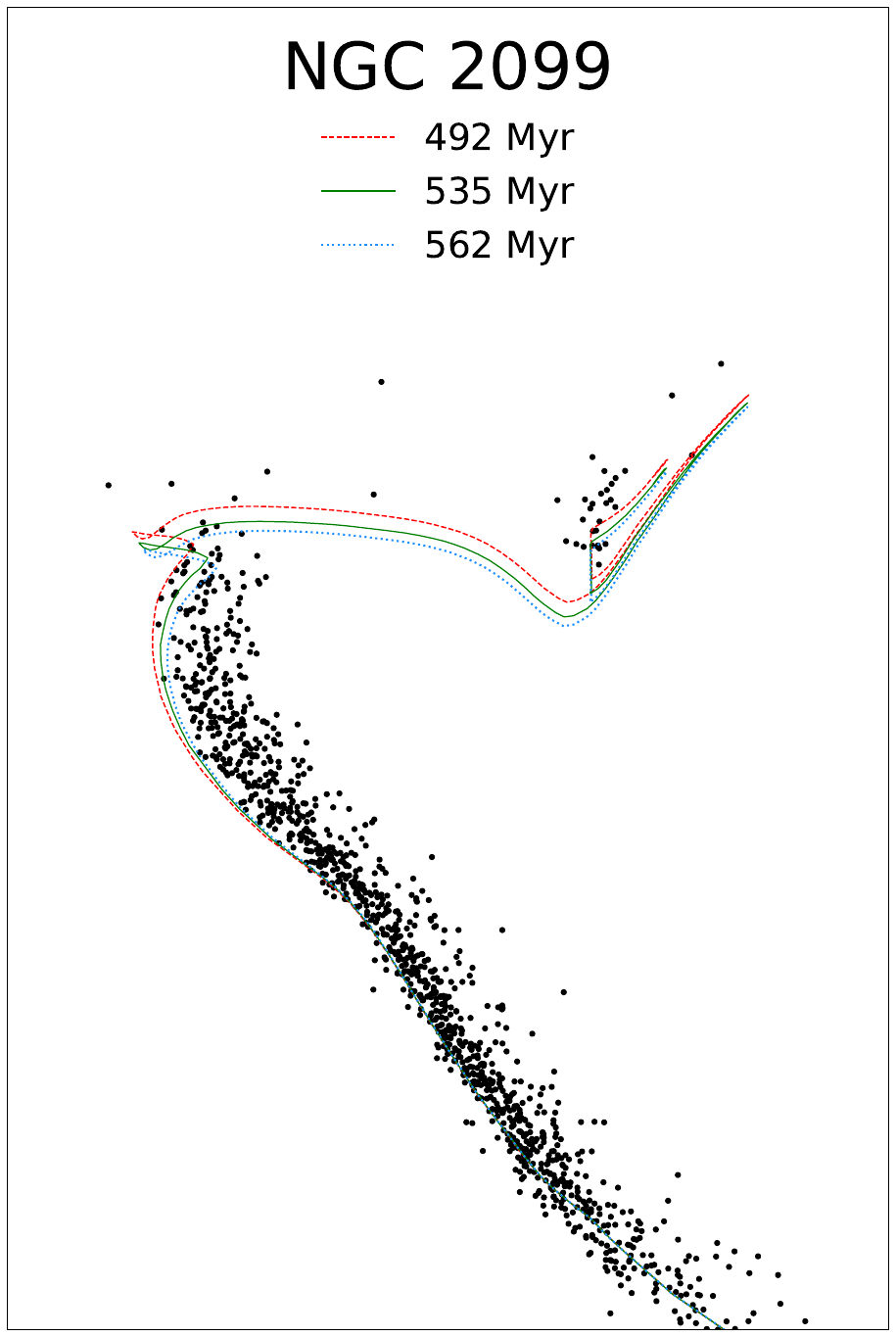}}
        \put(0, 0.72){\includegraphics[width=0.260\textwidth]{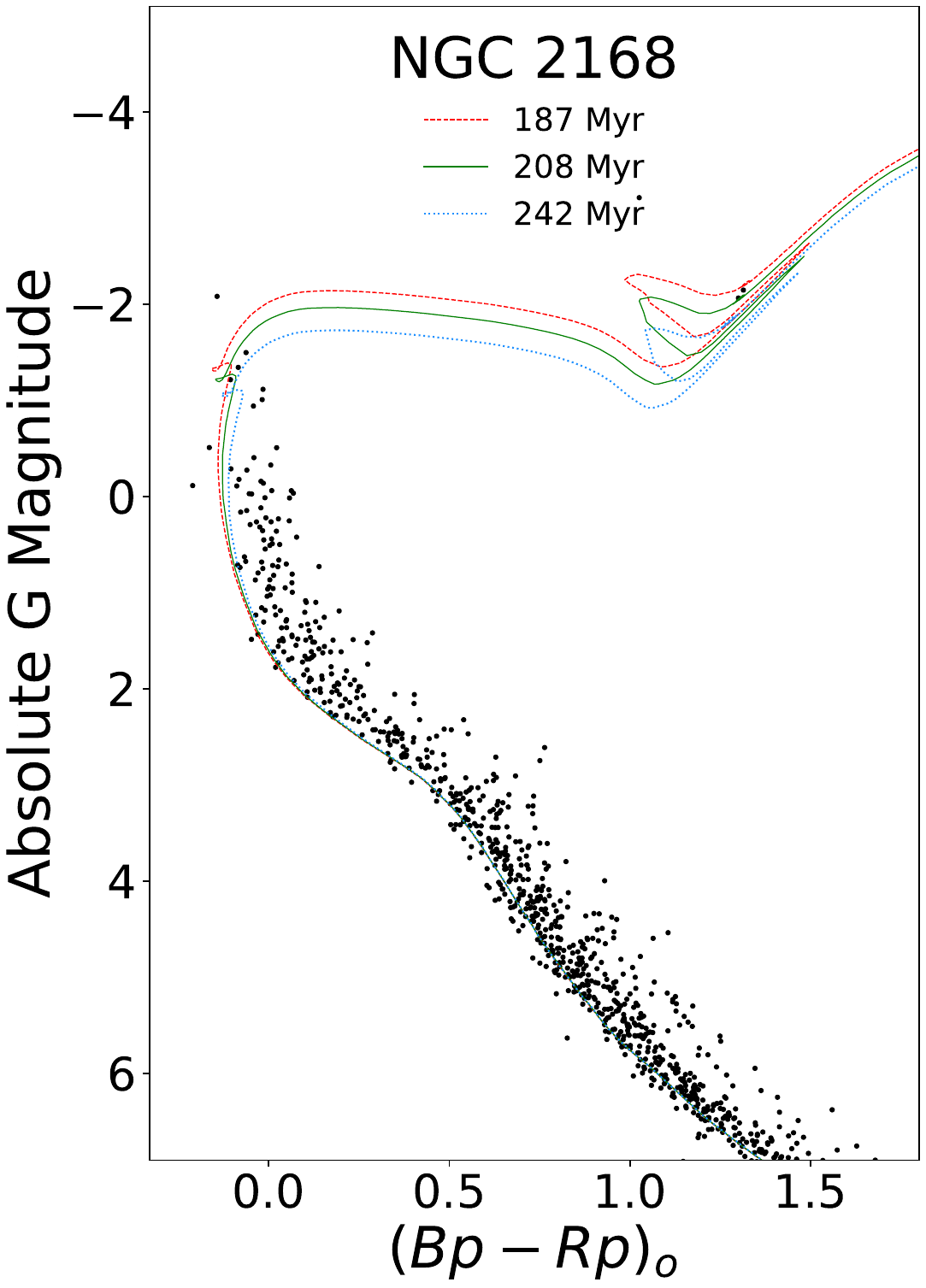}}
        \put(4.66, 0.72){\includegraphics[width=0.22\textwidth]{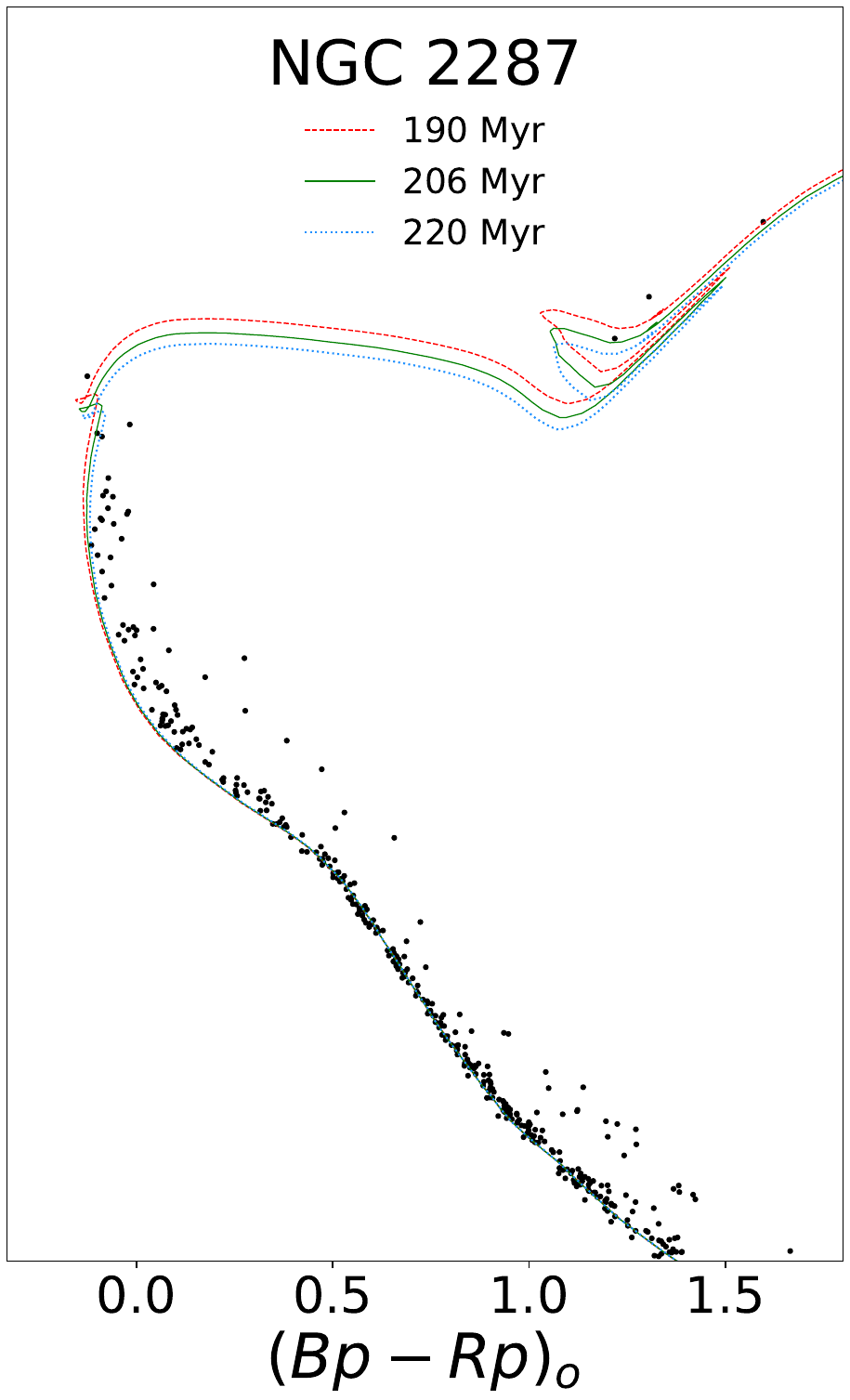}}
        \put(8.597, 0.72){\includegraphics[width=0.22\textwidth]{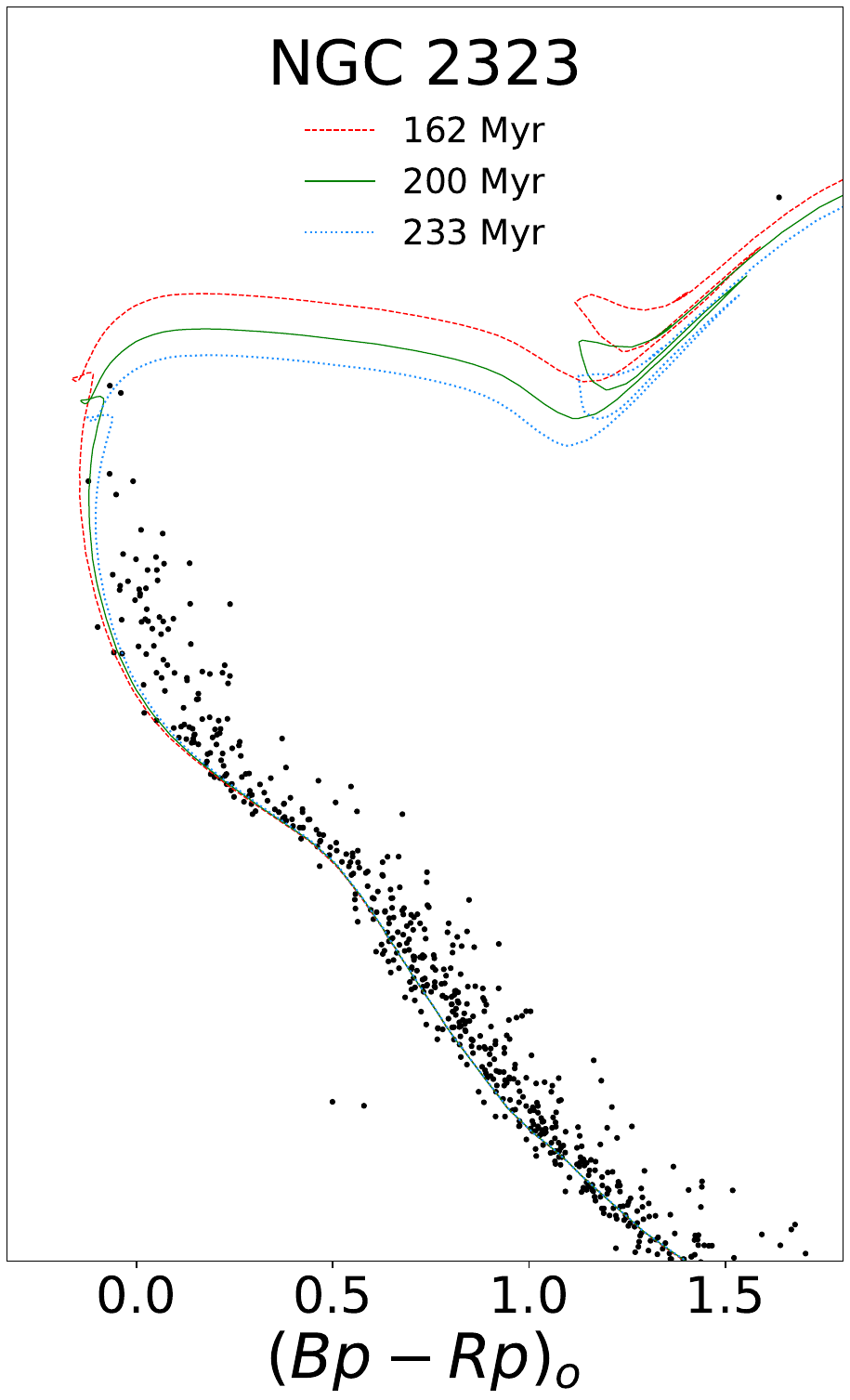}}
        \put(12.533, 0.72){\includegraphics[width=0.22\textwidth]{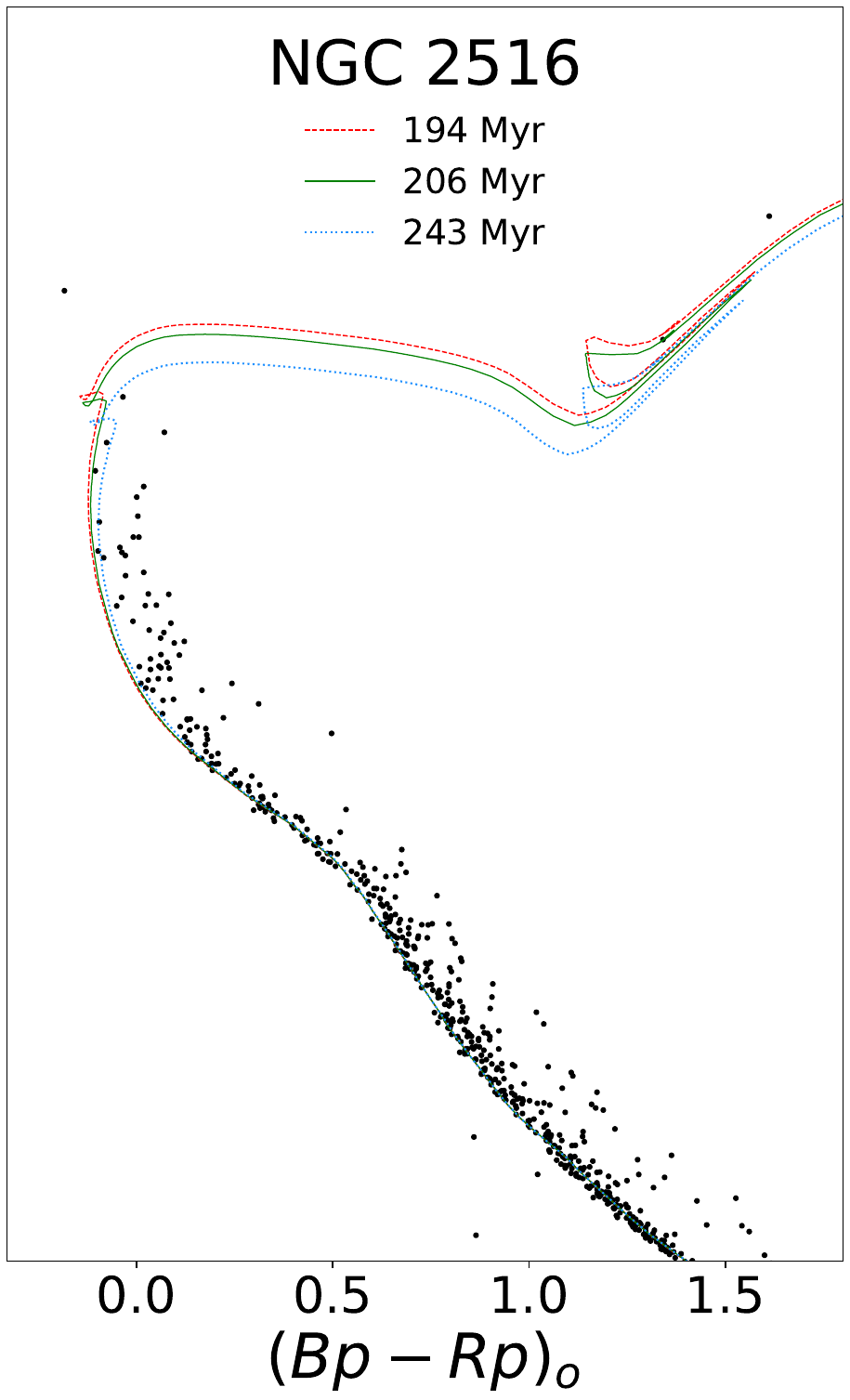}}
    \end{picture}
    \caption{Cluster CMDs using only members with $>50\%$ membership probability in \hunt for first twelve clusters with member WDs. Best-fit (green solid) and $1\sigma$ error (red dashed and blue dotted) PARSEC isochrones from heuristic fitting method are overlaid. ASCC 113 is repeated from the main text.}
    \label{fig:cluster_ages_1}
\end{figure*}

\begin{figure*}[ht]
    \centering
    \setlength{\unitlength}{1cm}
    \begin{picture}(17, 21)(0, 0)
        \put(0, 13.11){\includegraphics[width=0.260\textwidth]{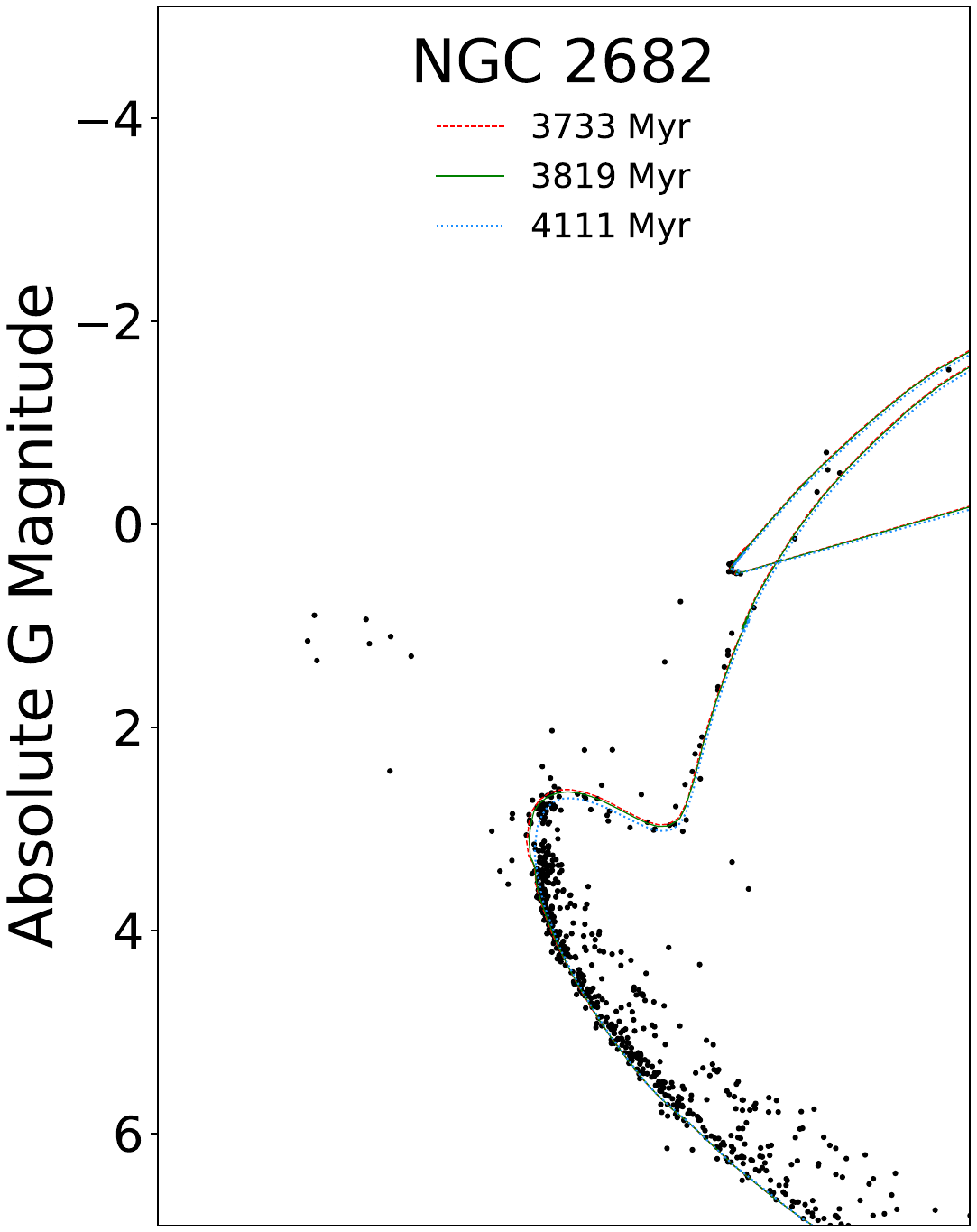}}
        \put(4.66, 13.11){\includegraphics[width=0.22\textwidth]{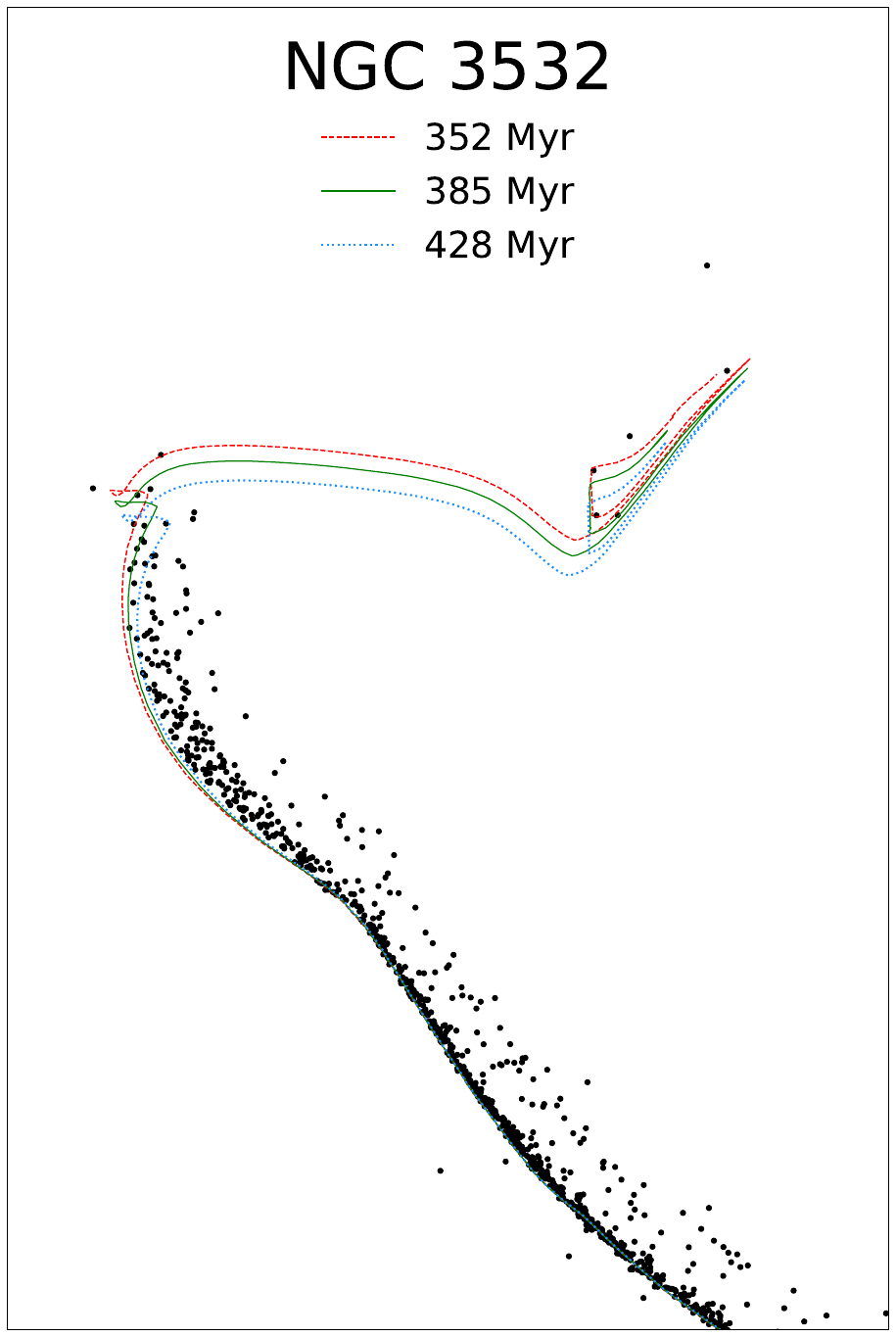}}
        \put(8.597, 13.11){\includegraphics[width=0.22\textwidth]{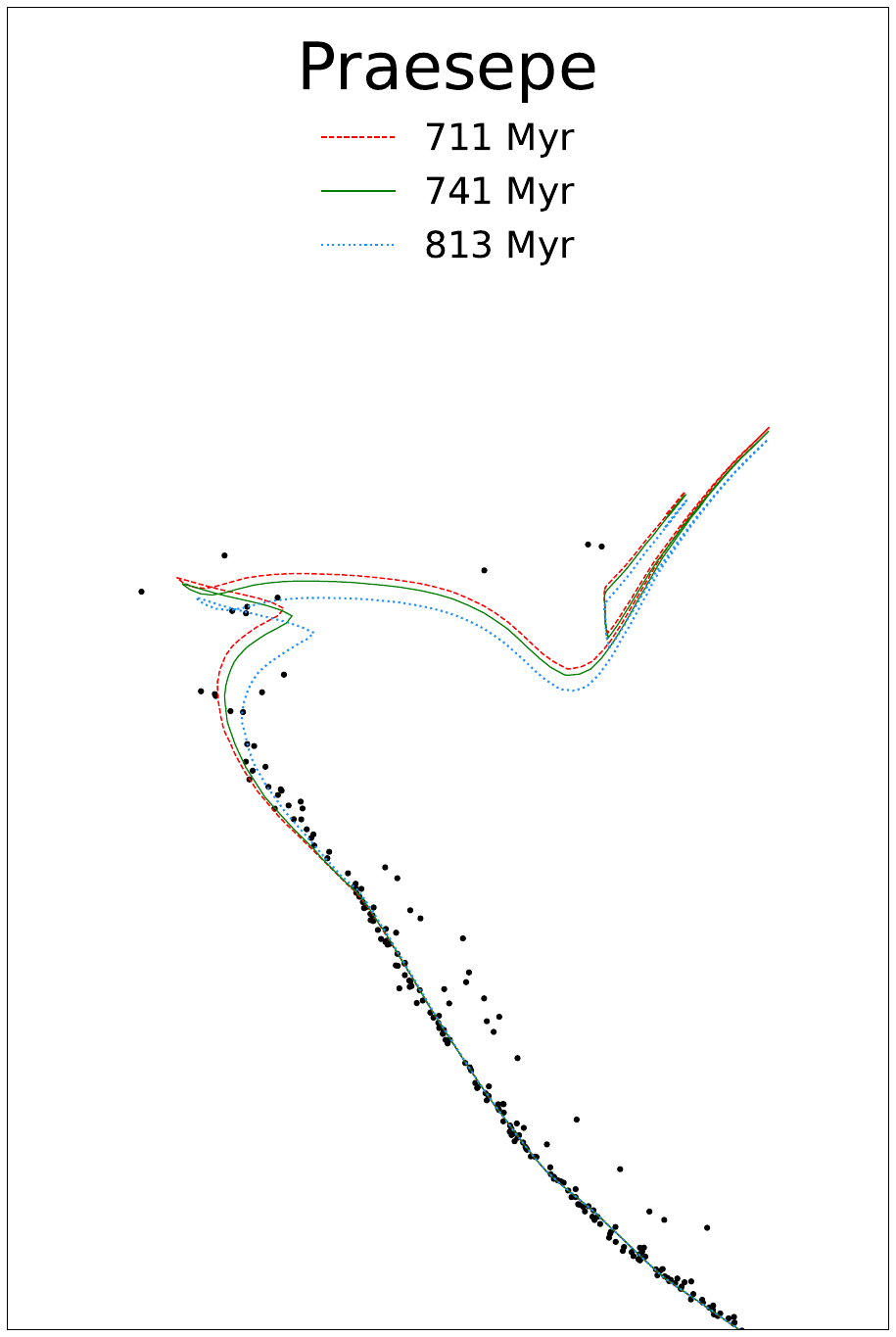}}
        \put(0, 7.22){\includegraphics[width=0.260\textwidth]{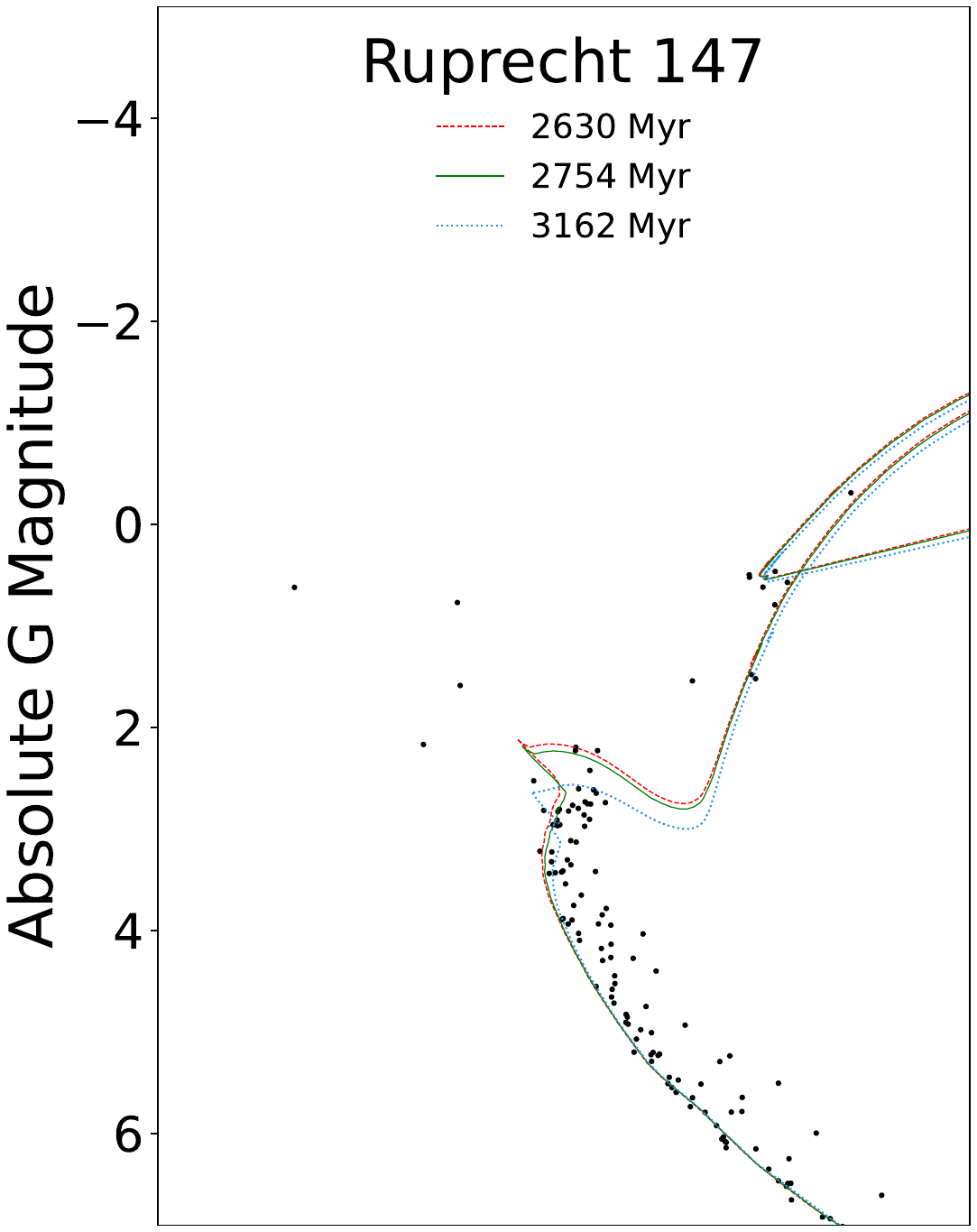}}
        \put(4.66, 7.22){\includegraphics[width=0.22\textwidth]{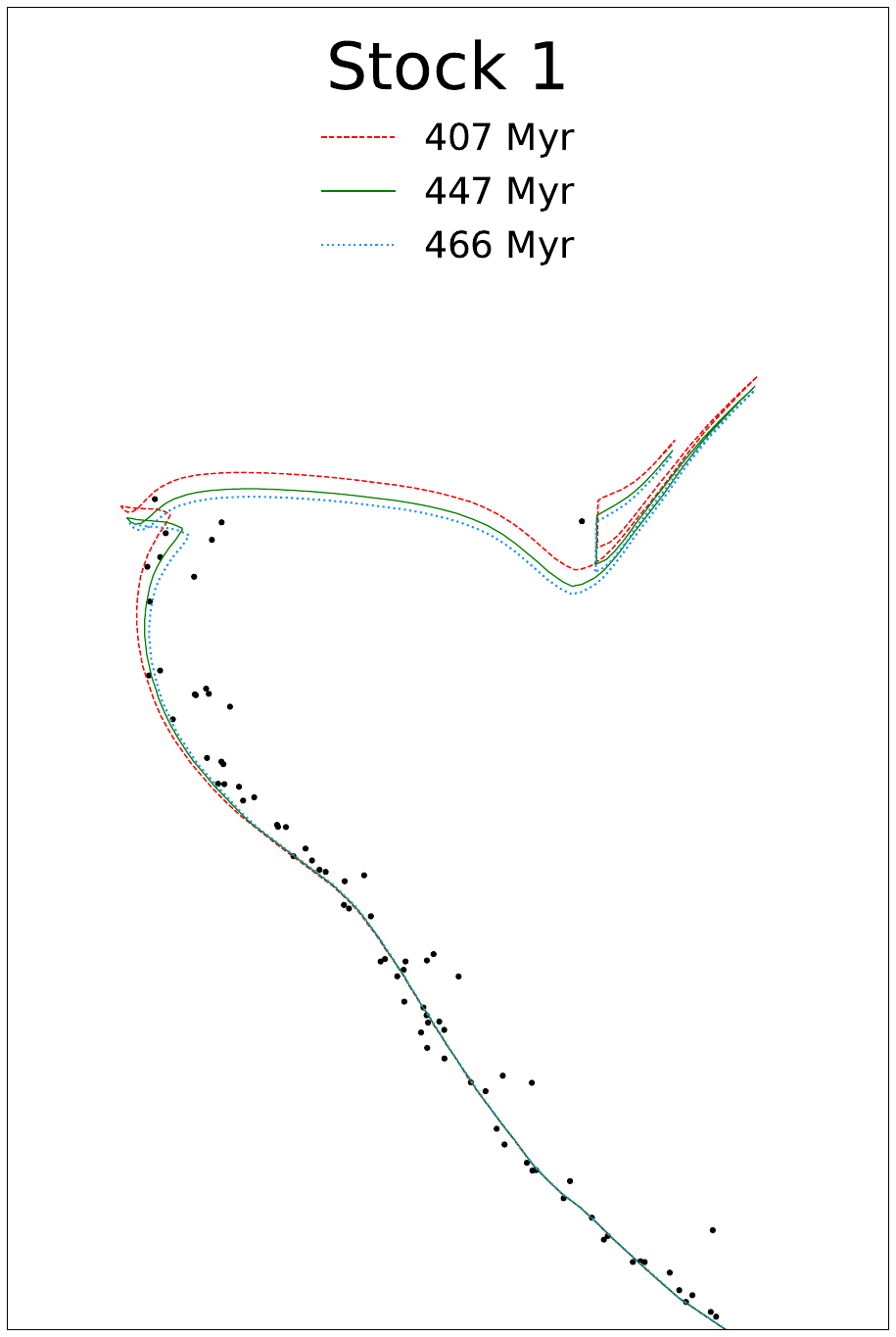}}
        \put(8.597, 7.22){\includegraphics[width=0.22\textwidth]{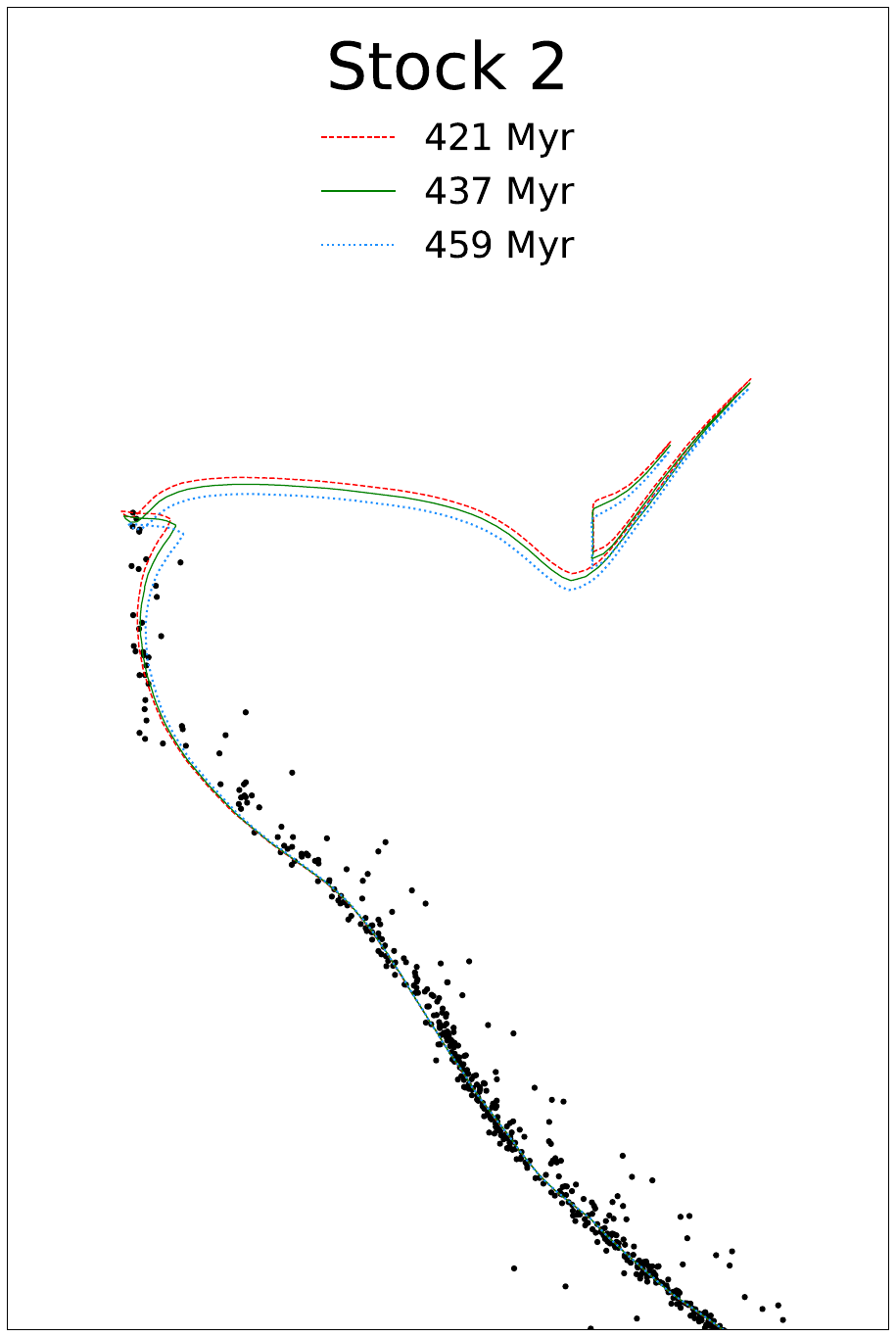}}
        \put(12.533, 7.22){\includegraphics[width=0.22\textwidth]{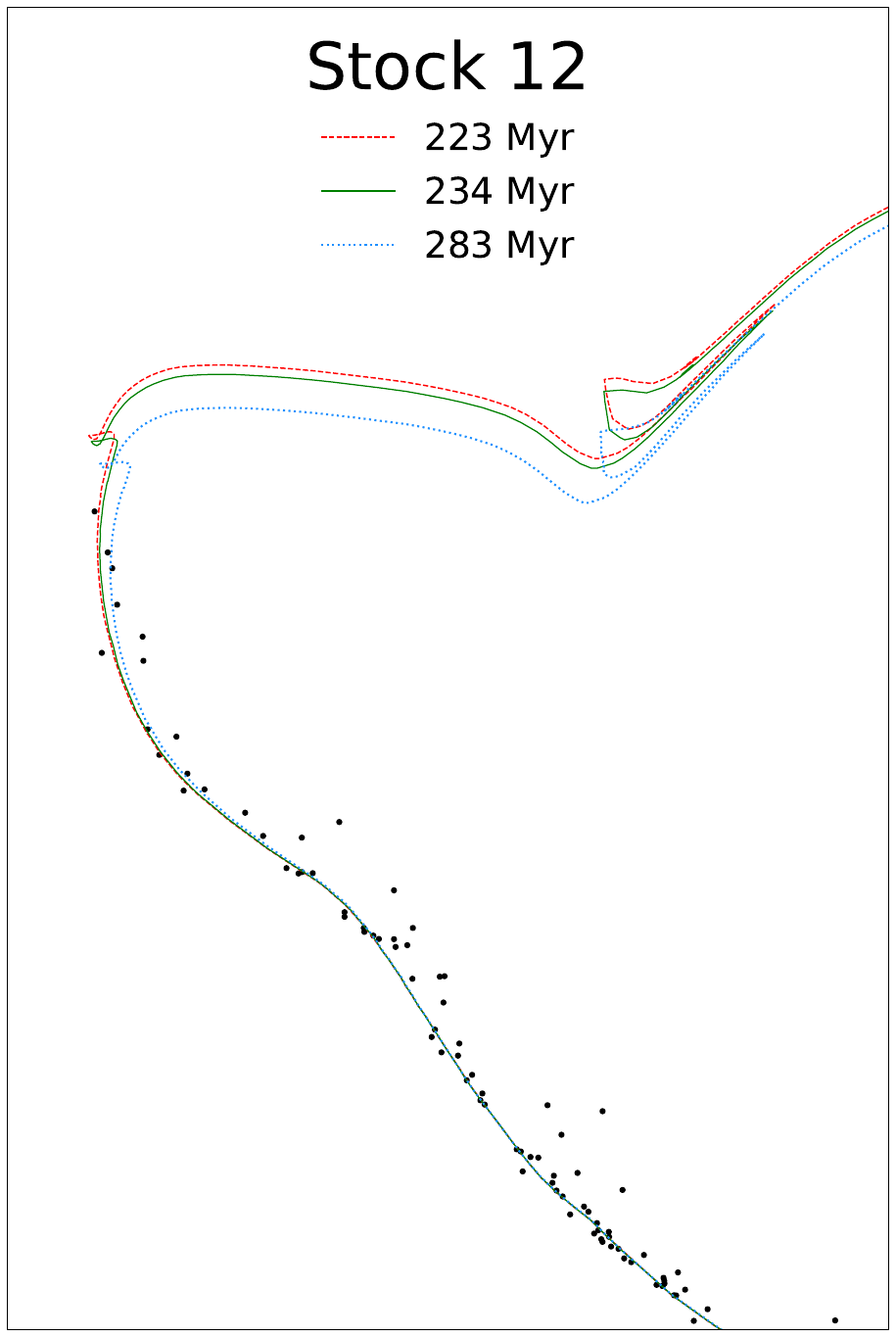}}
        \put(0, 0.72){\includegraphics[width=0.260\textwidth]{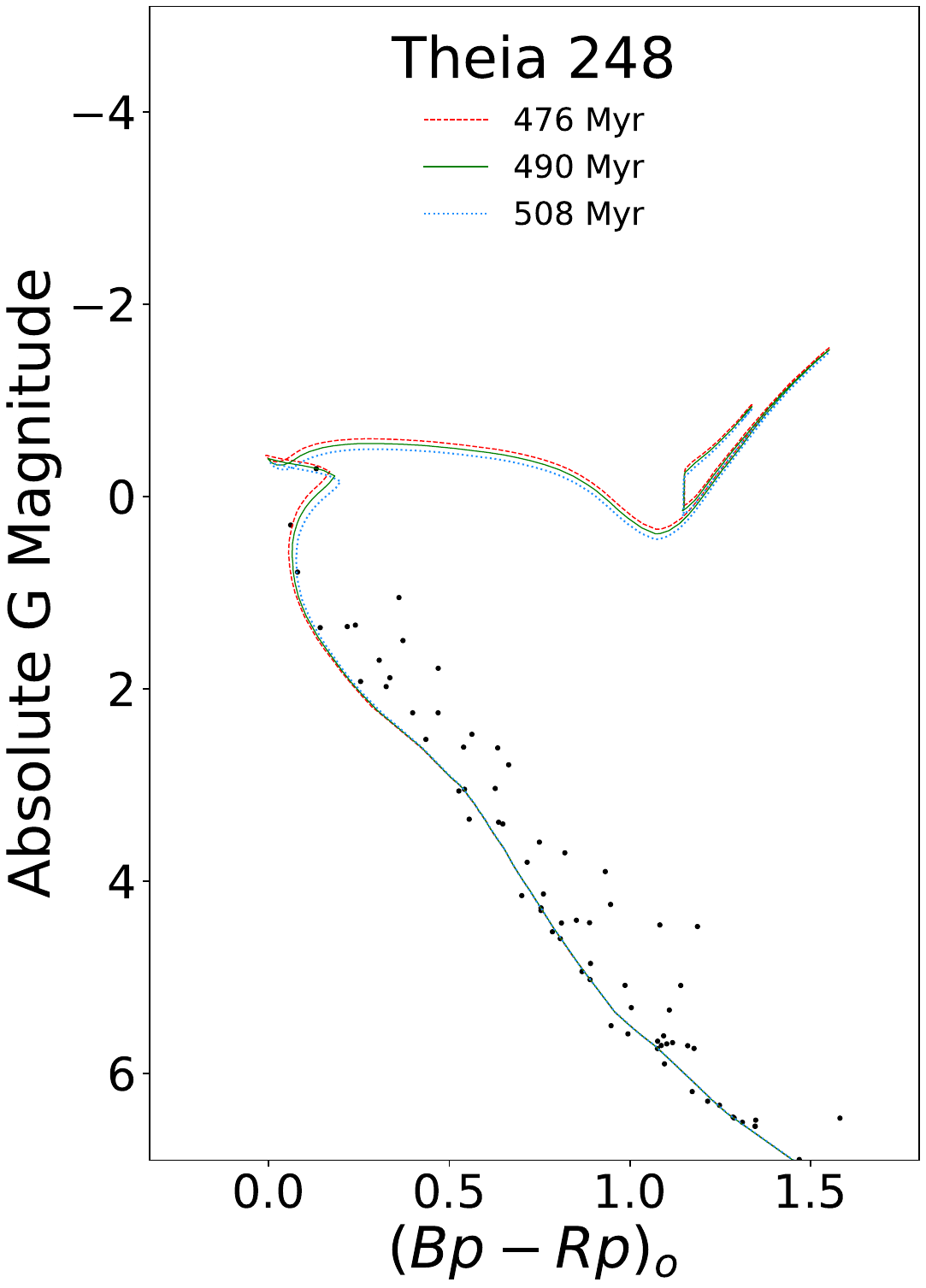}}
        \put(4.66, 0.72){\includegraphics[width=0.22\textwidth]{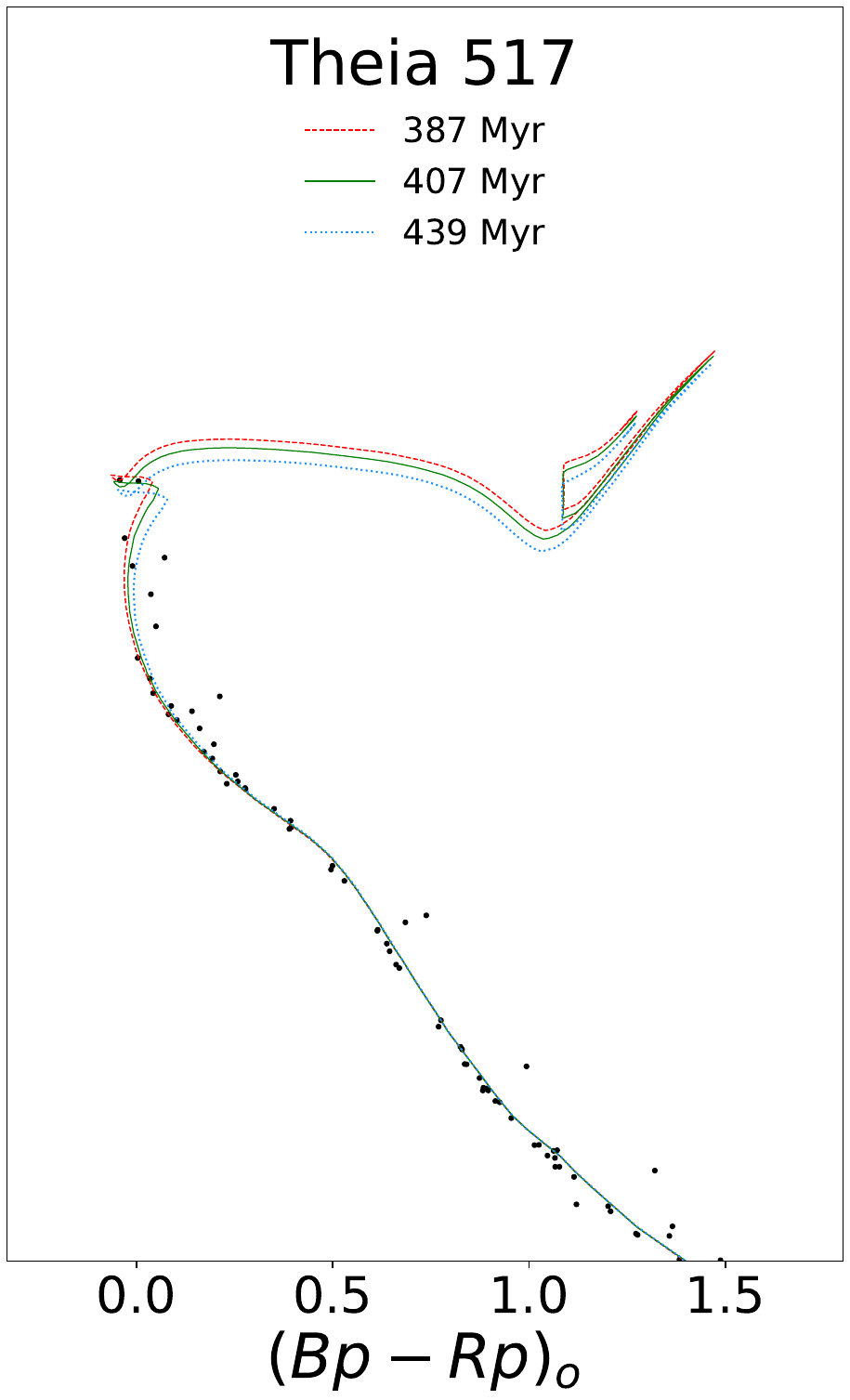}}
        \put(8.597, 0.72){\includegraphics[width=0.22\textwidth]{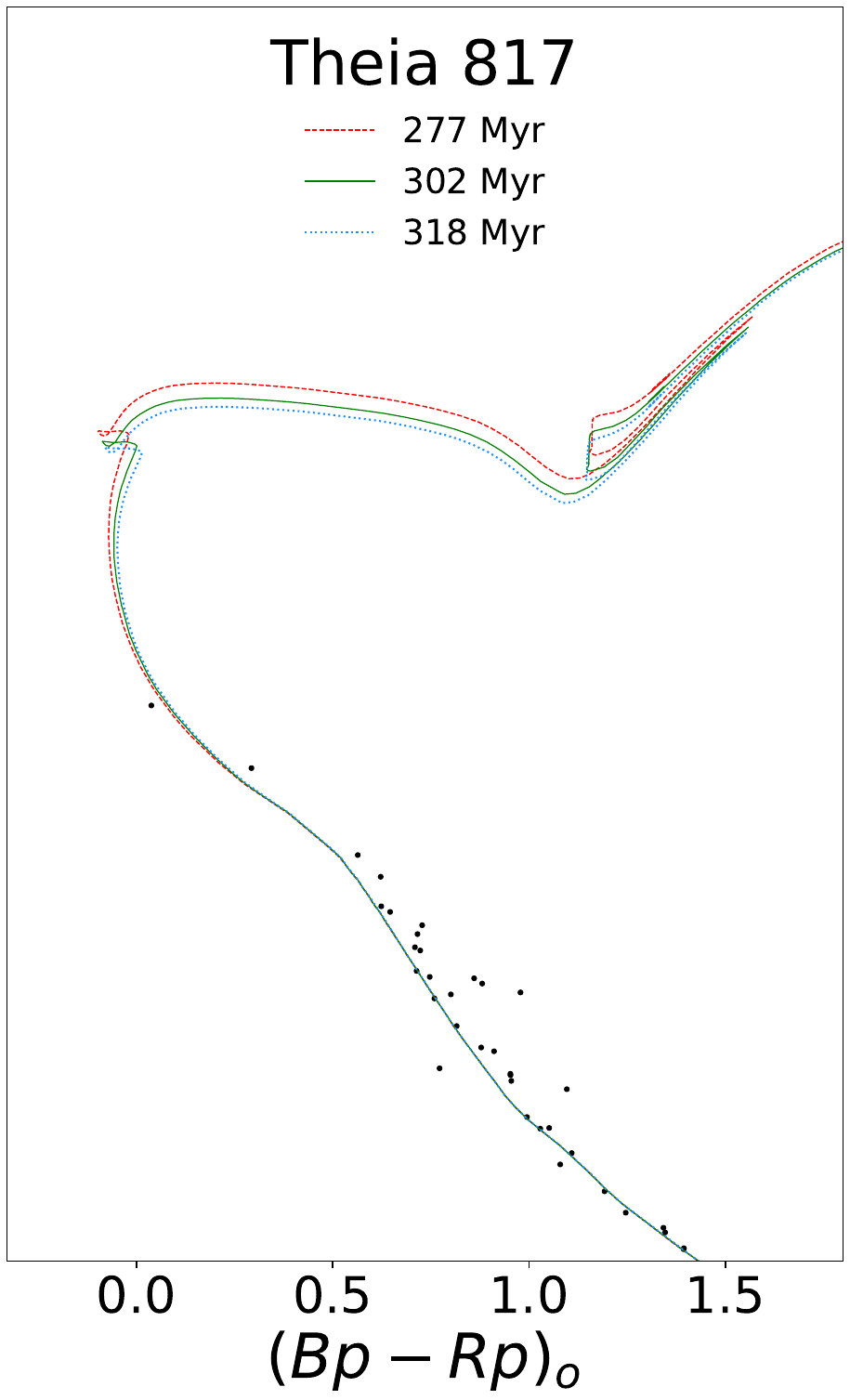}}
        \put(12.533, 0.72){\includegraphics[width=0.22\textwidth]{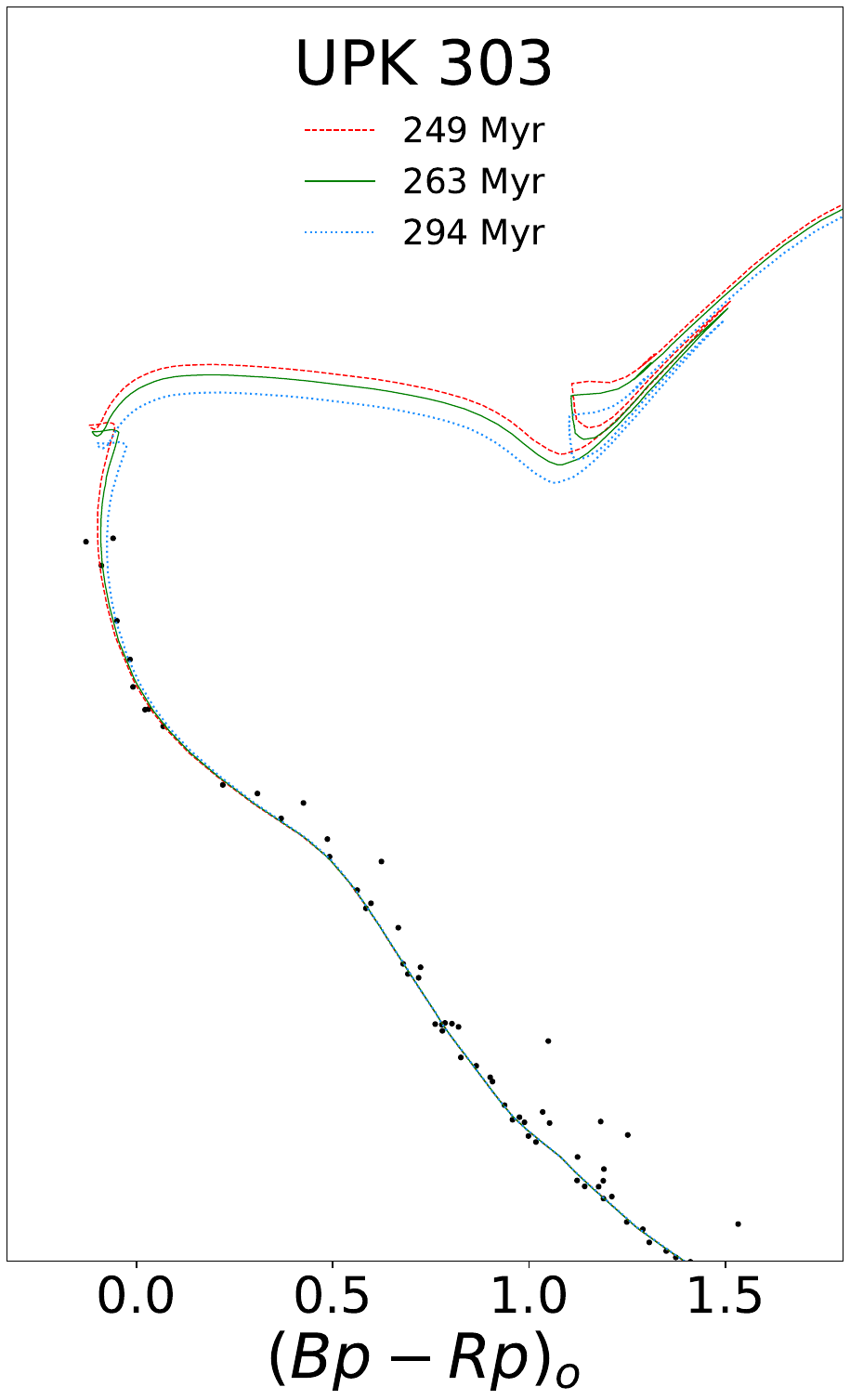}}
    \end{picture}
    \caption{As in Fig.~\ref{fig:cluster_ages_1}, but for the remaining eleven clusters.}
    \label{fig:cluster_ages_2}
\end{figure*}

\begin{figure*}[ht]
    \centering
    \includegraphics[width=1.0\textwidth]{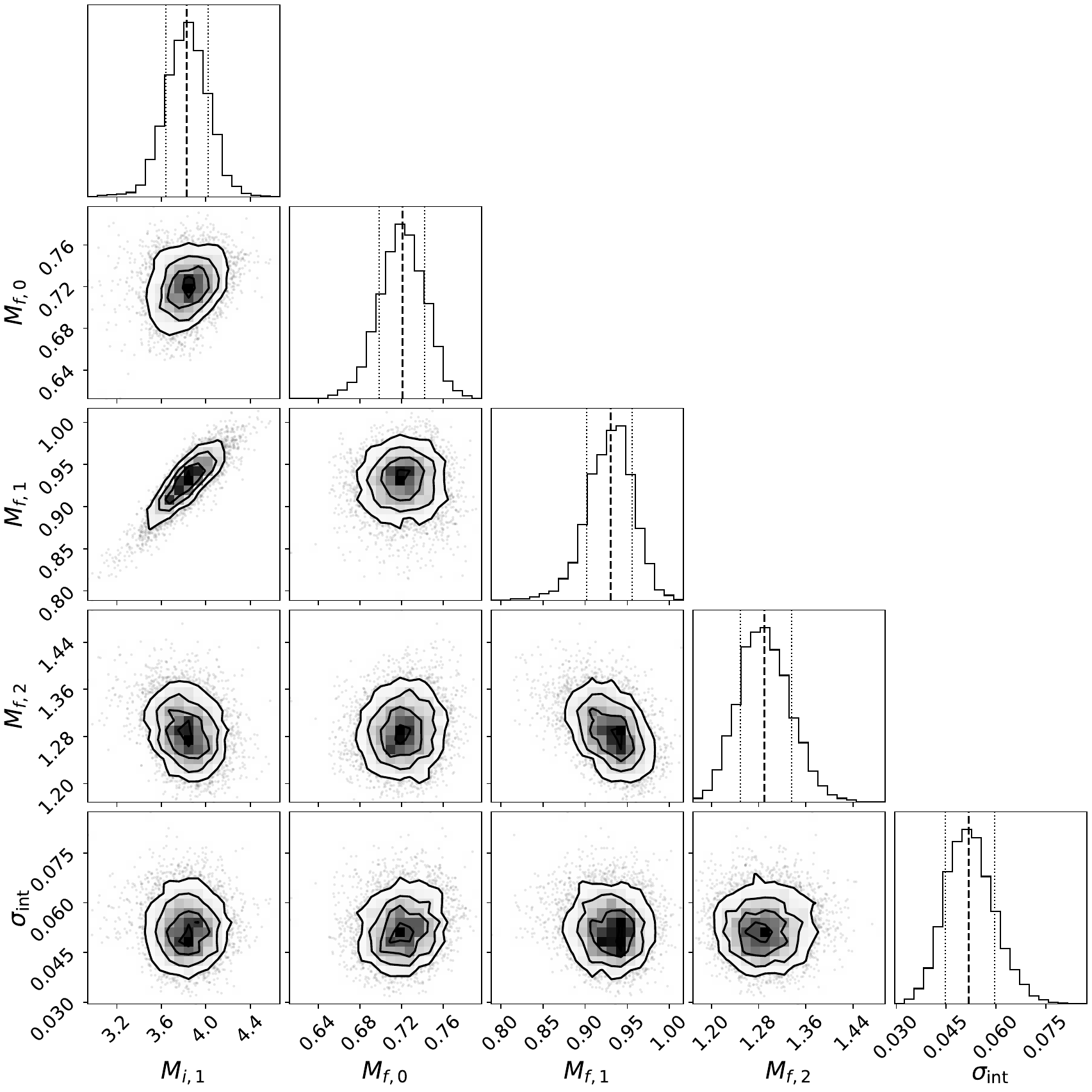}
    \caption{Posterior distributions for Gaia-based IFMR fit, shown in the top panel of Fig.~\ref{fig:IFMR}. Plotted are the initial mass at the interior breakpoint ($M_{i,1}$), the final mass at the interior breakpoint ($M_{f,1}$) and the endpoints ($M_{f,0}$, $M_{f,2}$), and the intrinsic scatter $\sigma_{\mathrm{int}}$. Vertical dashed lines indicate the median and $1\sigma$ credible intervals.}
    \label{fig:corner_gaia_ifmr}
\end{figure*}

\begin{figure*}[ht]
    \centering
    \includegraphics[width=1.0\textwidth]{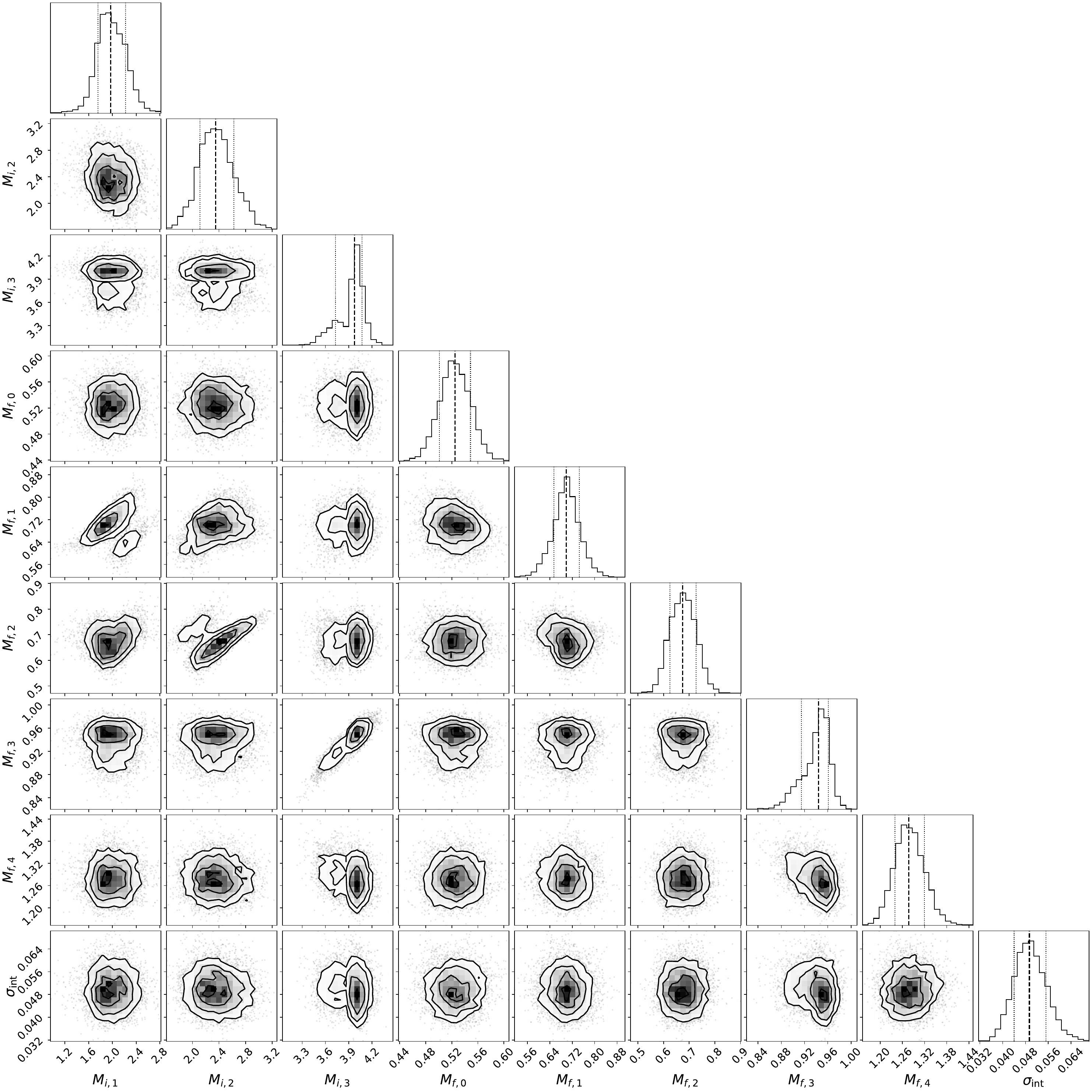}
    \caption{Posterior distributions for Full Sample Fit 1, shown in the middle panel of Fig.~\ref{fig:IFMR}. As in Fig.~\ref{fig:corner_gaia_ifmr}, this figure shows the initial masses at the interior breakpoints ($M_{i,1}$, $M_{i,2}$, $M_{i,3}$), the final masses at the interior breakpoints ($M_{f,1}$, $M_{f,2}$, $M_{f,3}$) and the endpoints ($M_{f,0}$, $M_{f,4}$), and the intrinsic scatter $\sigma_{\mathrm{int}}$.}
    \label{fig:corner_full_fit1_ifmr}
\end{figure*}

\begin{figure*}[ht]
    \centering
    \includegraphics[width=1.0\textwidth]{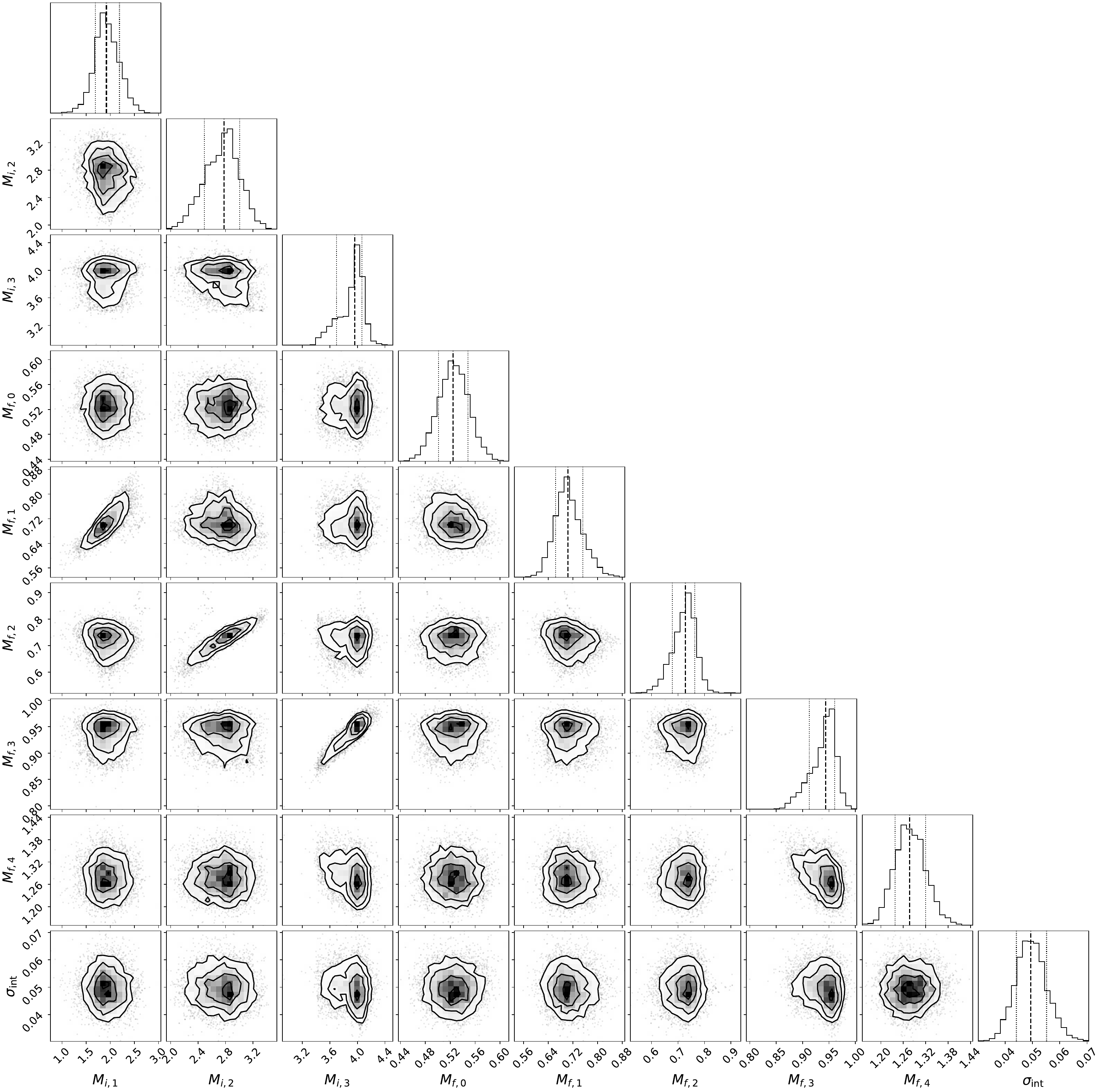}
    \caption{As in Fig.~\ref{fig:corner_full_fit1_ifmr}, but for Full Sample Fit 2, shown in the bottom panel of Fig.~\ref{fig:IFMR}.}
    \label{fig:corner_full_fit2_ifmr}
\end{figure*}

\begin{figure*}[ht]
    \centering
    \includegraphics[width=0.75\textwidth]{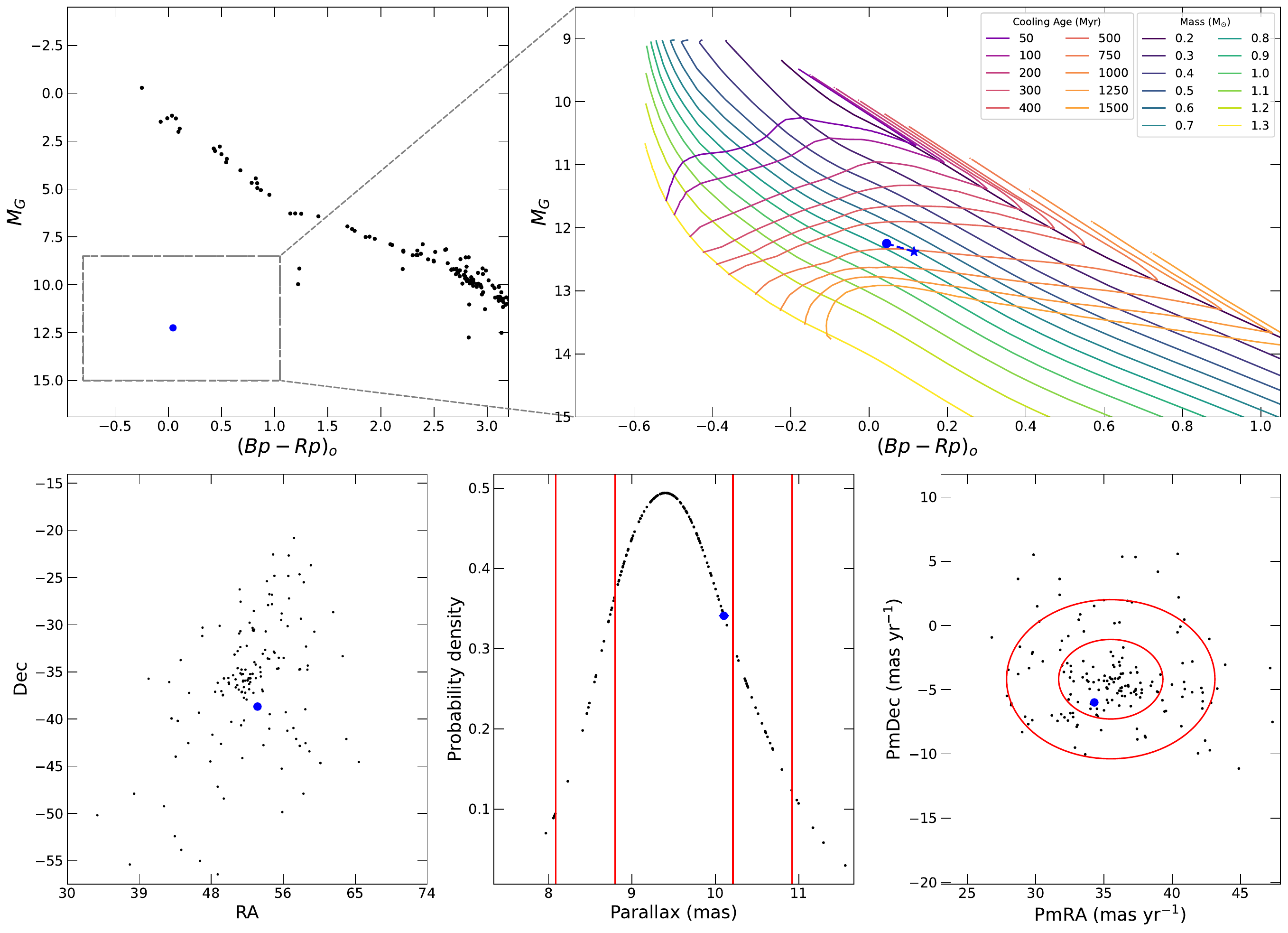}
    \caption{\textit{Top left}: Dereddened cluster CMD for Alessi 13 cluster using \hunt catalog cluster member candidates and the catalog's mean $A_v$. WD member candidates are shown as colored circles in all panels; colors are consistent across panels. \textit{Top right}: Zoom in on the WD sequence, with WD cooling models overlaid. The star symbols show WD candidates without any applied dereddening. The dashed lines connect the \hunt catalog mean reddening and no reddening cases. \textit{Bottom}: Distribution for astrometric data of cluster member candidates in Alessi 13, showing the position (left), kernel density estimation of parallax (center), and proper motion (right). Solid red lines are 1 and $2\sigma$ regions relative to cluster mean values, as calculated in \hunt.}
    \label{fig:CMD_alessi_13}
\end{figure*}

\begin{figure*}[!tp]
    \centering
    \includegraphics[width=0.75\textwidth]{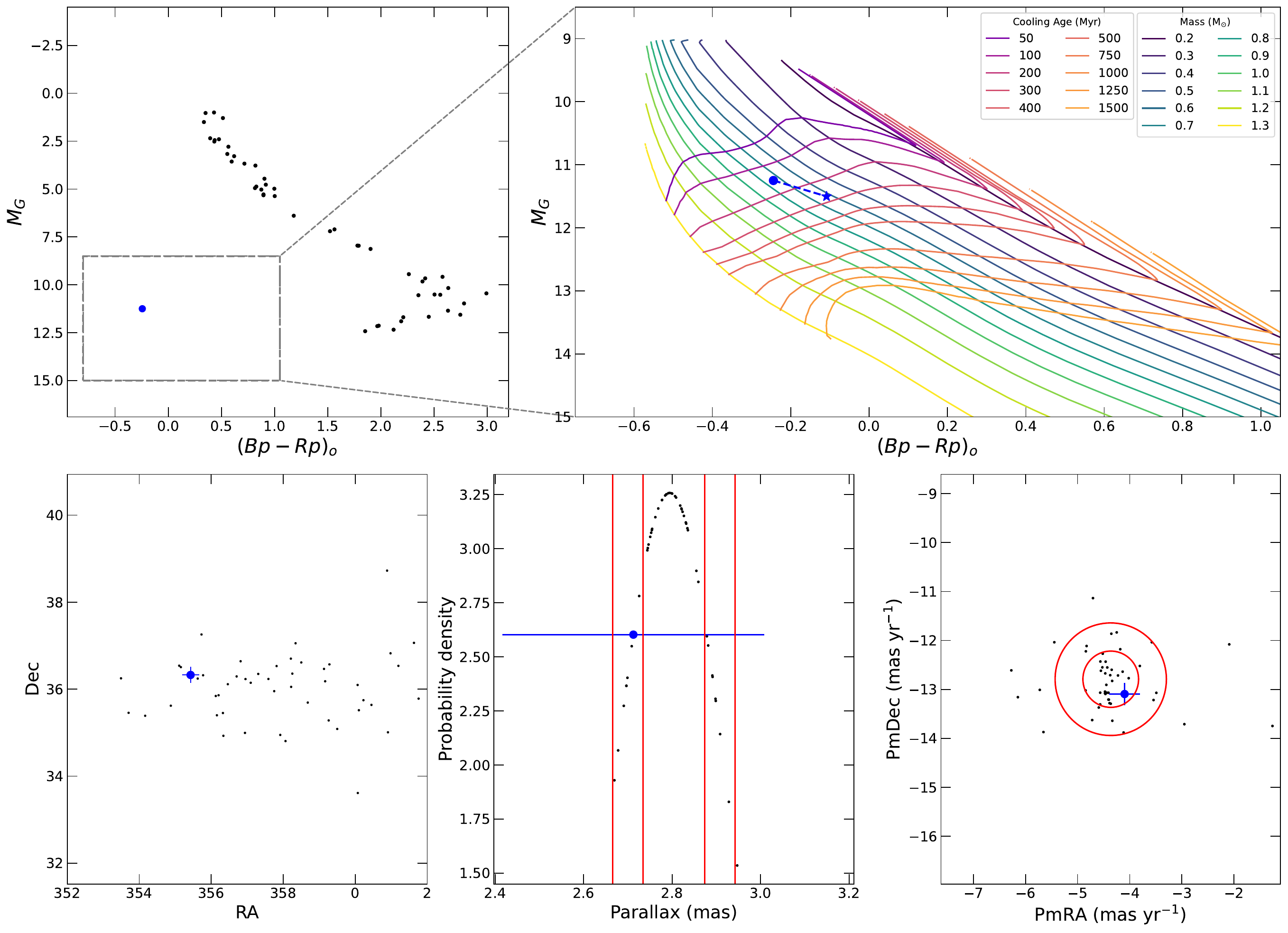}
    \caption{Same as Fig.~\ref{fig:CMD_alessi_13}, but for the Alessi 22 cluster.}
    \label{fig:CMD_alessi_22}
\end{figure*}

\begin{figure*}[!tp]
    \centering
    \includegraphics[width=0.75\textwidth]{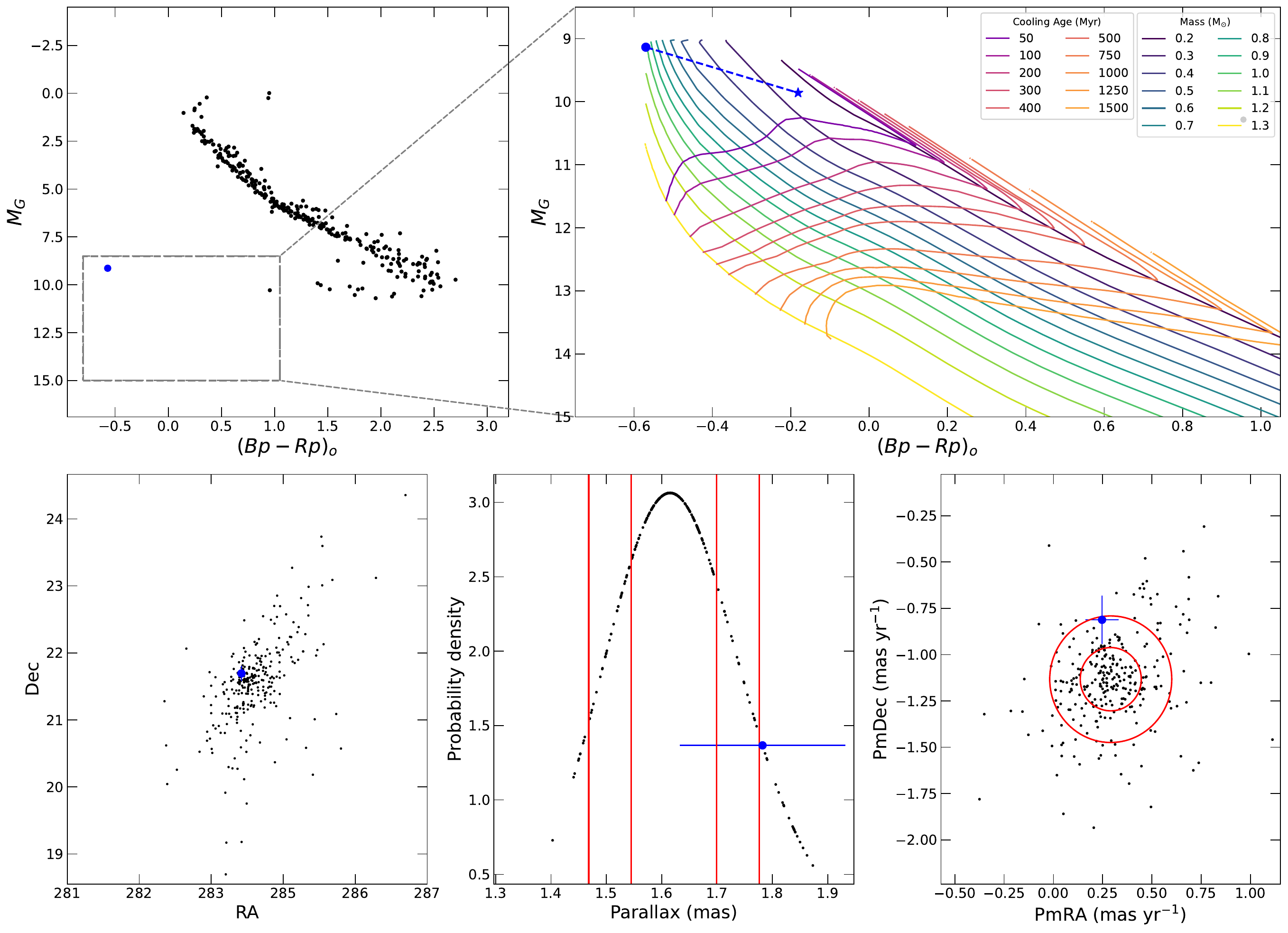}
    \caption{Same as Fig.~\ref{fig:CMD_alessi_13}, but for the Alessi 62 cluster.}
    \label{fig:CMD_alessi_62}
\end{figure*}

\begin{figure*}[!tp]
    \centering
    \includegraphics[width=0.75\textwidth]{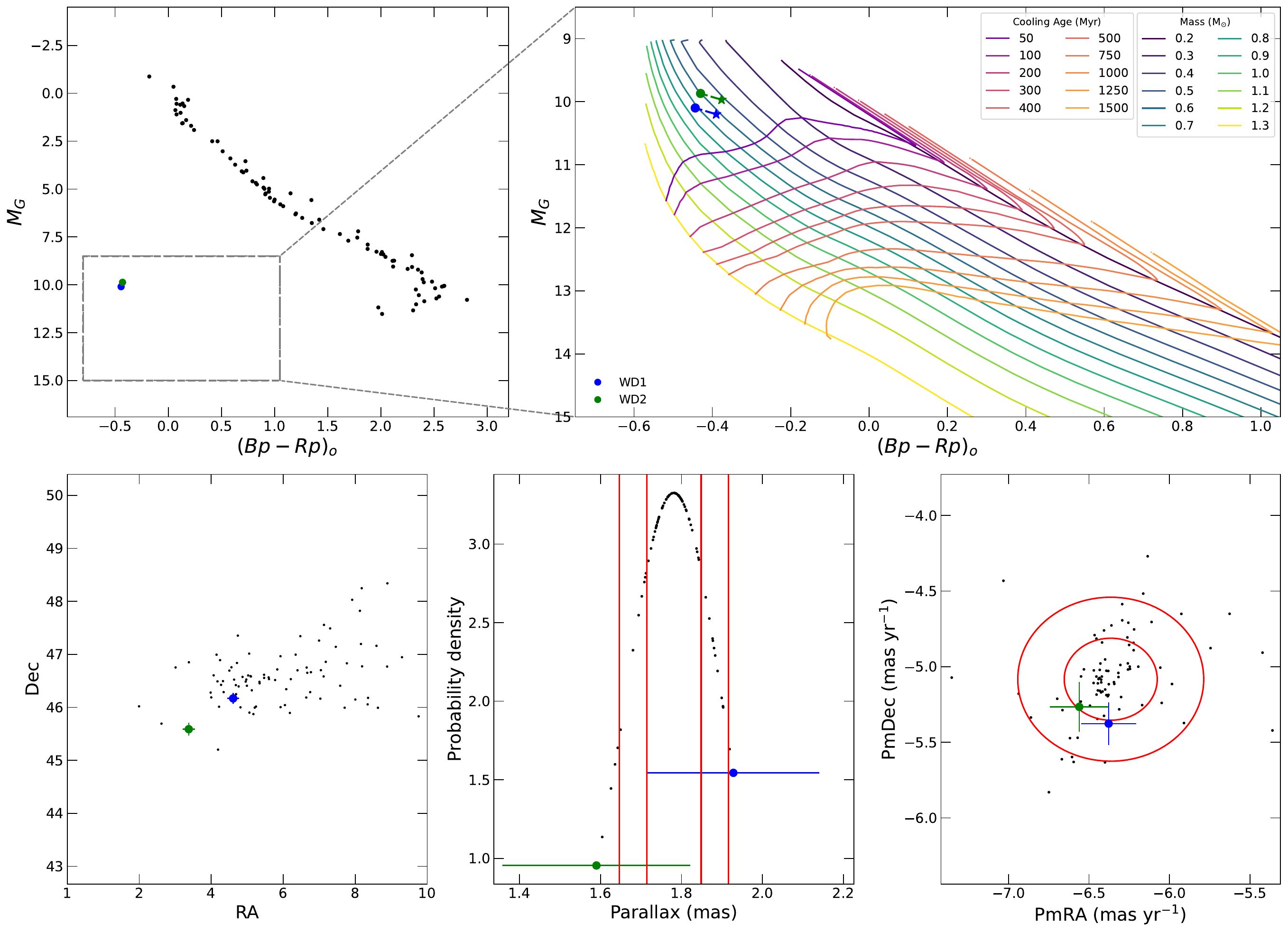}
    \caption{Same as Fig.~\ref{fig:CMD_alessi_13}, but for the Alessi 94 cluster.}
    \label{fig:CMD_alessi_94}
\end{figure*}

\begin{figure*}[!tp]
    \centering
    \includegraphics[width=0.75\textwidth]{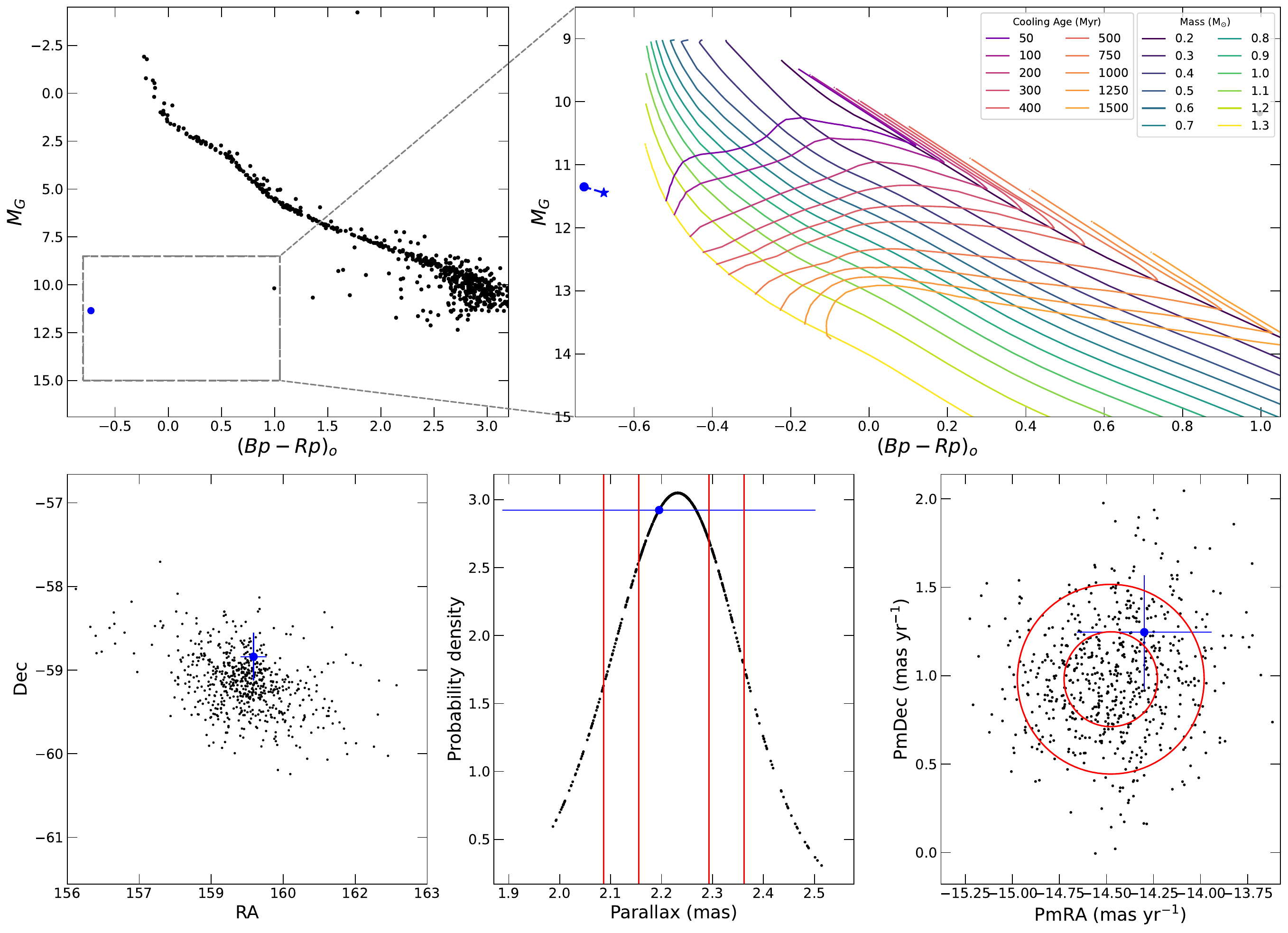}
    \caption{Same as Fig.~\ref{fig:CMD_alessi_13}, but for the BH 99 cluster.}
    \label{fig:CMD_bh_99}
\end{figure*}

\begin{figure*}[!tp]
    \centering
    \includegraphics[width=0.75\textwidth]{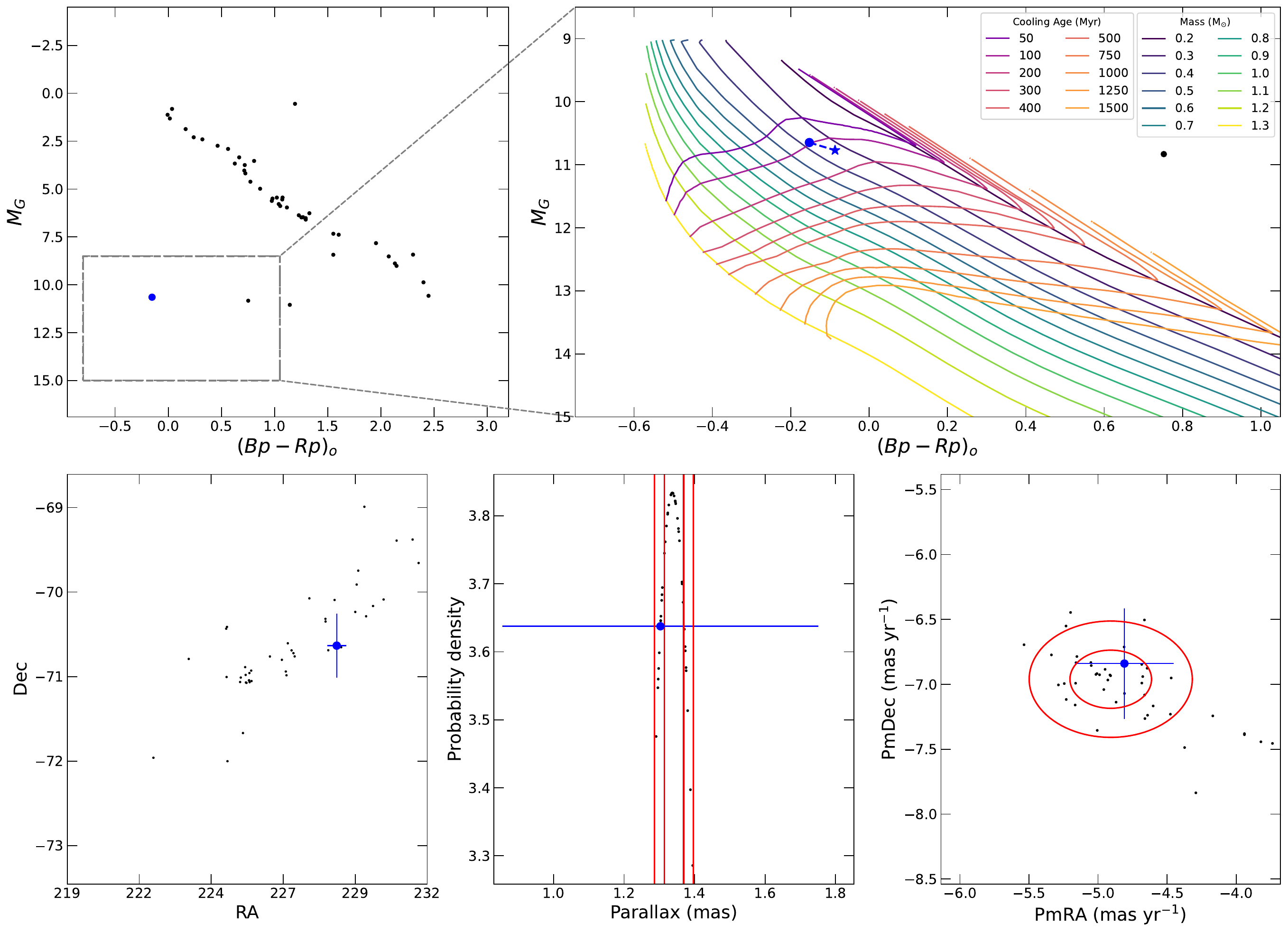}
    \caption{Same as Fig.~\ref{fig:CMD_alessi_13}, but for the CWNU 41 cluster.}
    \label{fig:CMD_cwnu_41}
\end{figure*}

\begin{figure*}[!tp]
    \centering
    \includegraphics[width=0.75\textwidth]{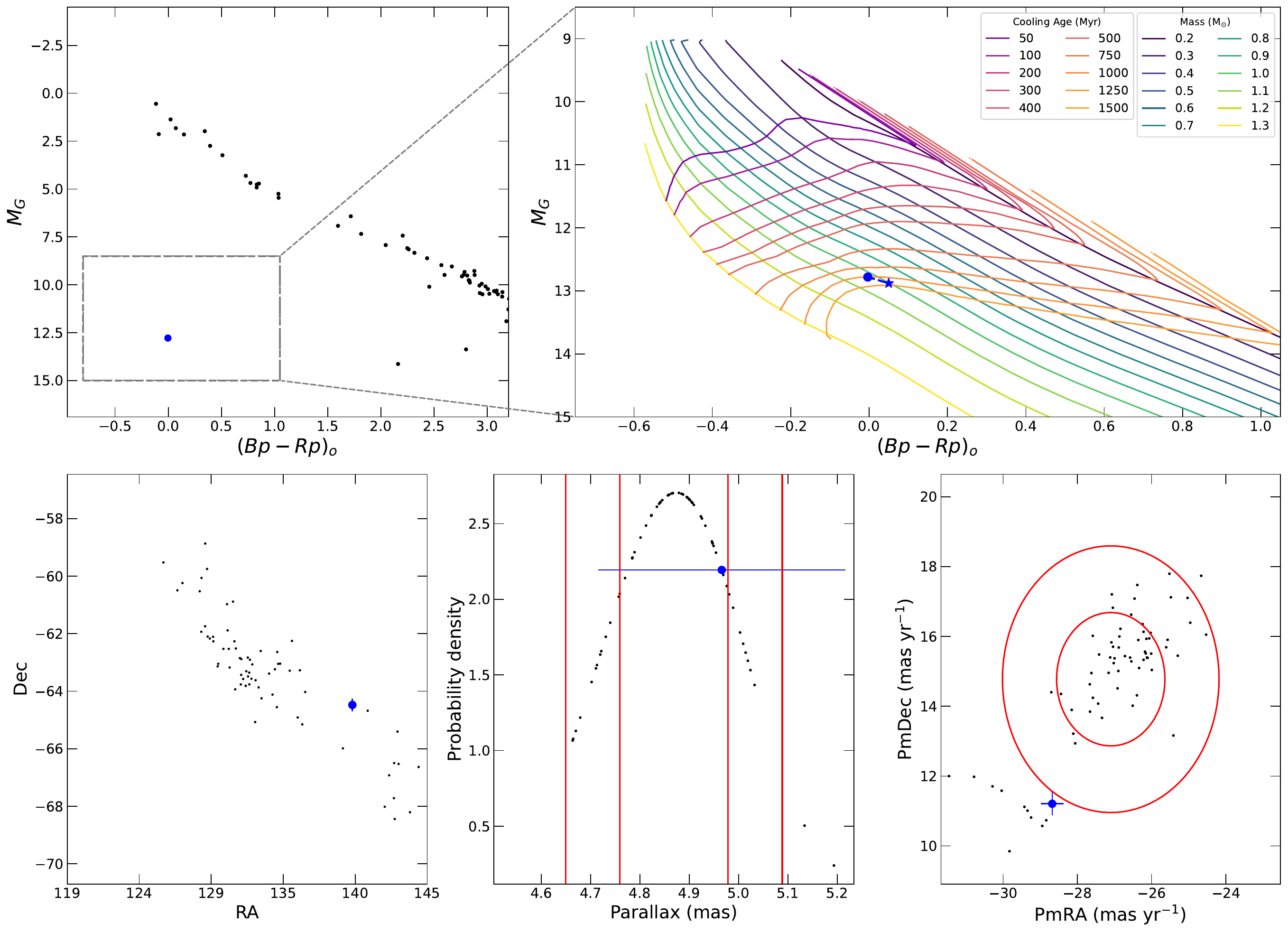}
    \caption{Same as Fig.~\ref{fig:CMD_alessi_13}, but for the CWNU 515 cluster.}
    \label{fig:CMD_cwnu_515}
\end{figure*}

\begin{figure*}[!tp]
    \centering
    \includegraphics[width=0.75\textwidth]{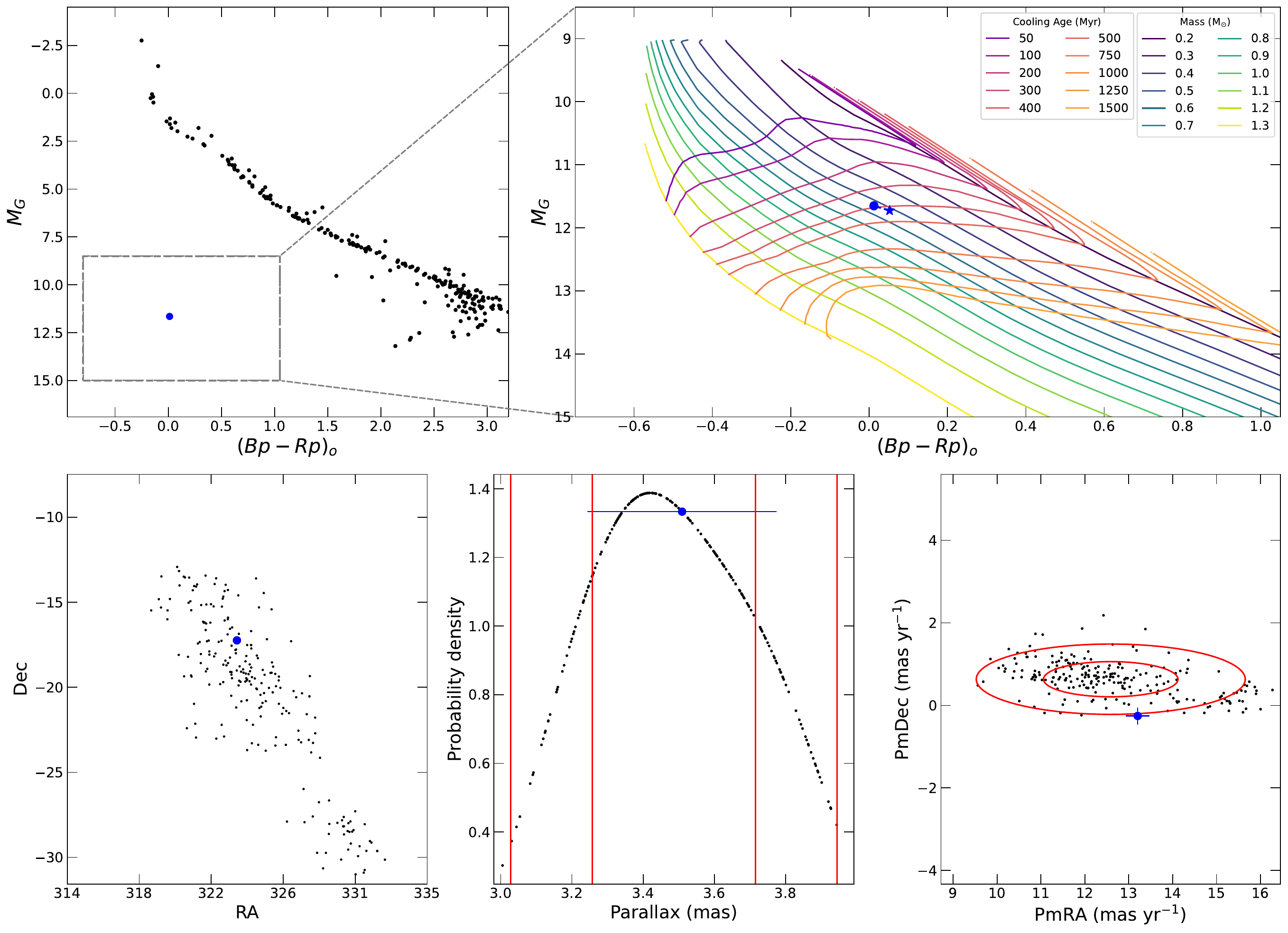}
    \caption{Same as Fig.~\ref{fig:CMD_alessi_13}, but for the CWNU 1012 cluster.}
    \label{fig:CMD_cwnu_1012}
\end{figure*}

\begin{figure*}[!tp]
    \centering
    \includegraphics[width=0.75\textwidth]{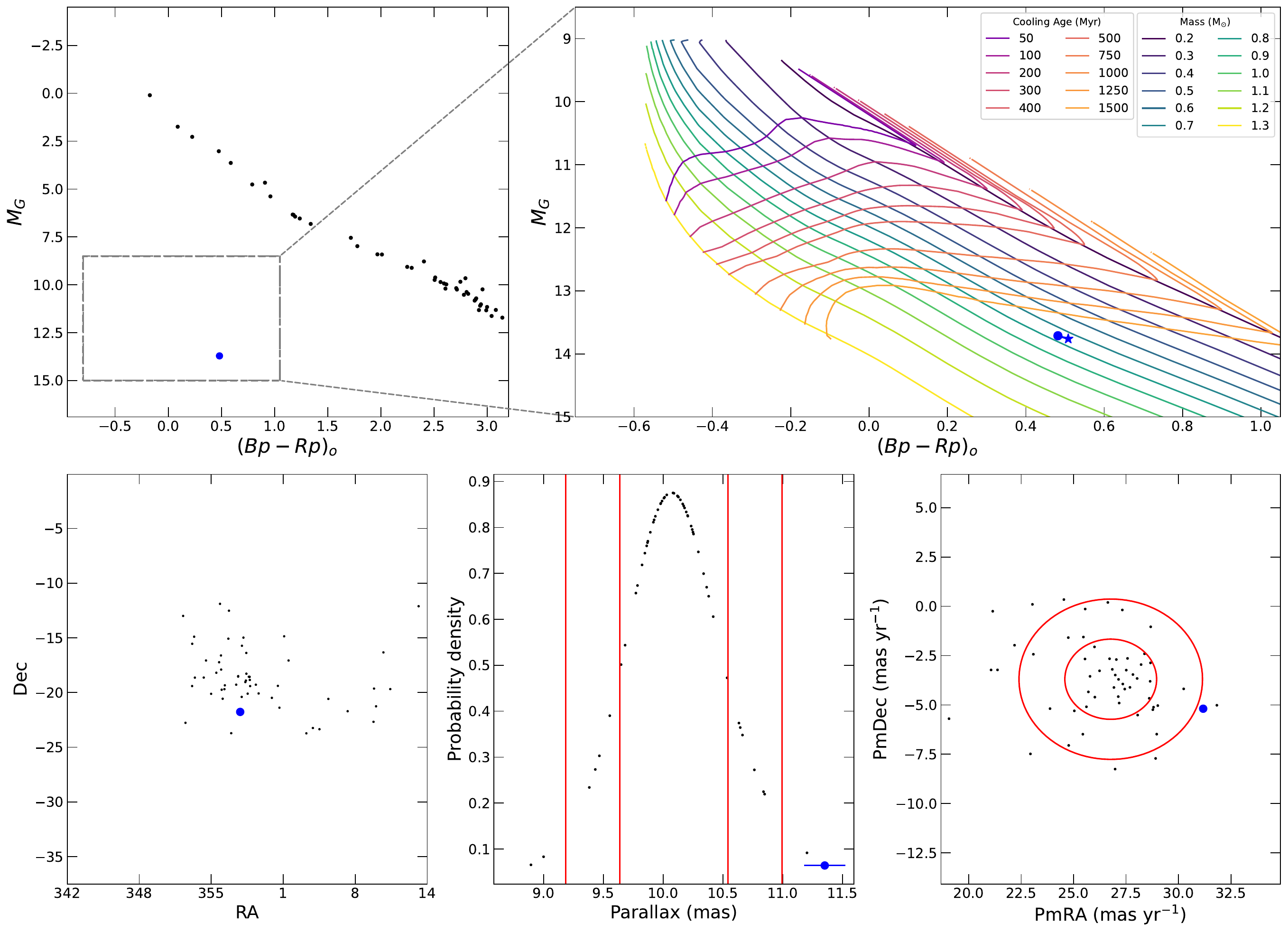}
    \caption{Same as Fig.~\ref{fig:CMD_alessi_13}, but for the CWNU 1066 cluster.}
    \label{fig:CMD_cwnu_1066}
\end{figure*}

\begin{figure*}[!tp]
    \centering
    \includegraphics[width=0.75\textwidth]{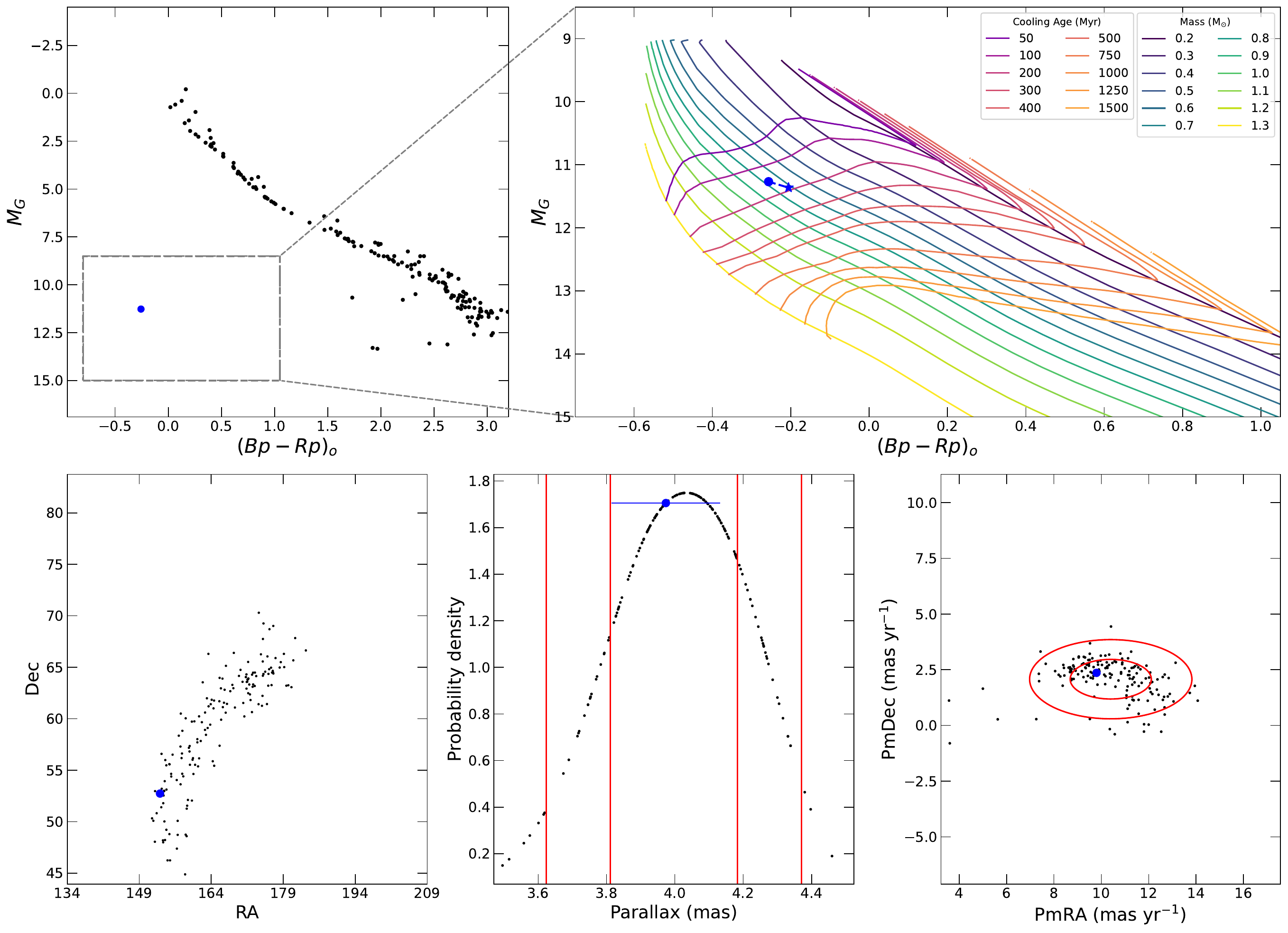}
    \caption{Same as Fig.~\ref{fig:CMD_alessi_13}, but for the CWNU 1095 cluster.}
    \label{fig:CMD_cwnu_1095}
\end{figure*}

\begin{figure*}[!tp]
    \centering
    \includegraphics[width=0.75\textwidth]{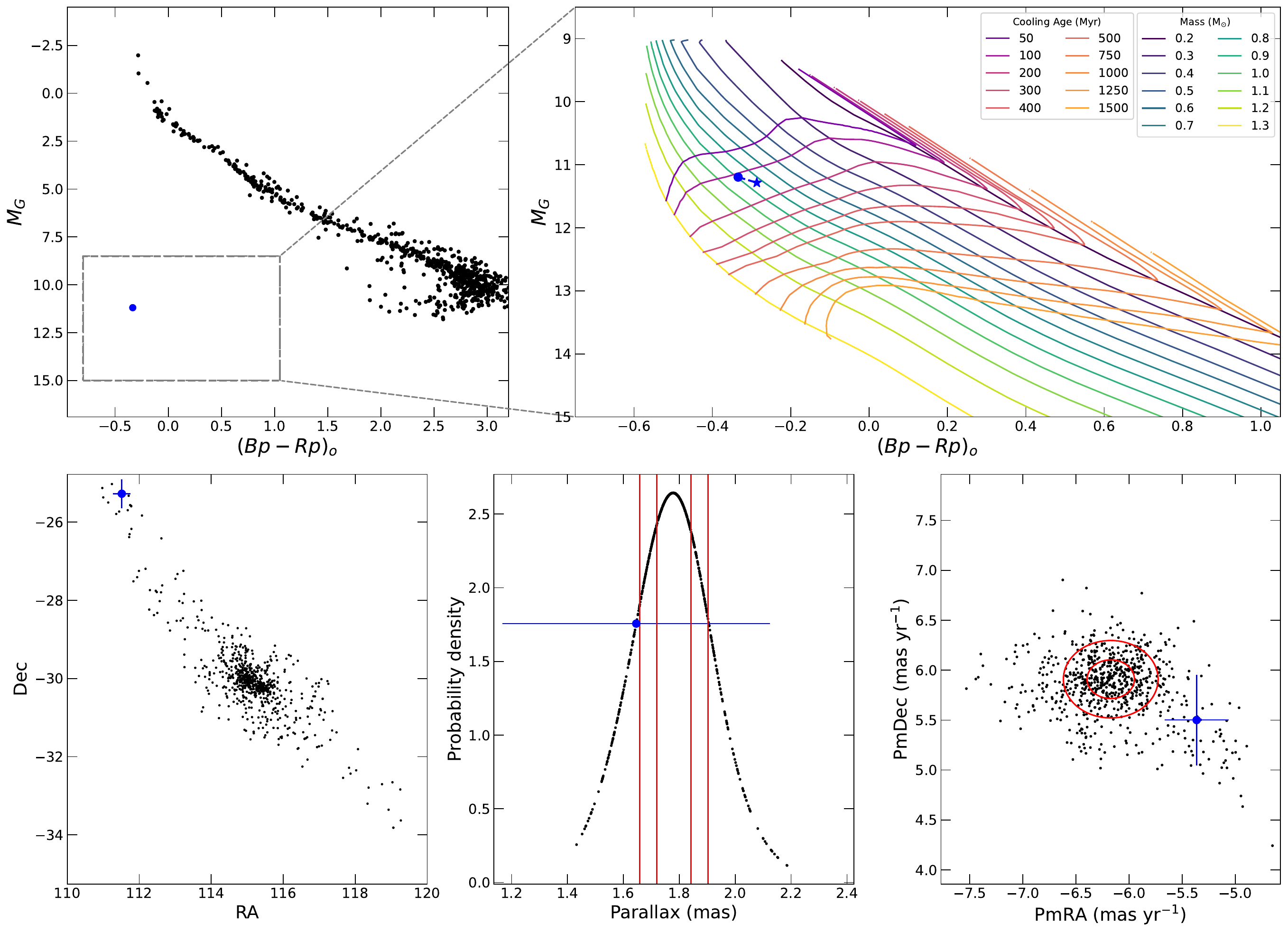}
    \caption{Same as Fig.~\ref{fig:CMD_alessi_13}, but for the Haffner 13 cluster.}
    \label{fig:CMD_haffner_13}
\end{figure*}

\begin{figure*}[!tp]
    \centering
    \includegraphics[width=0.75\textwidth]{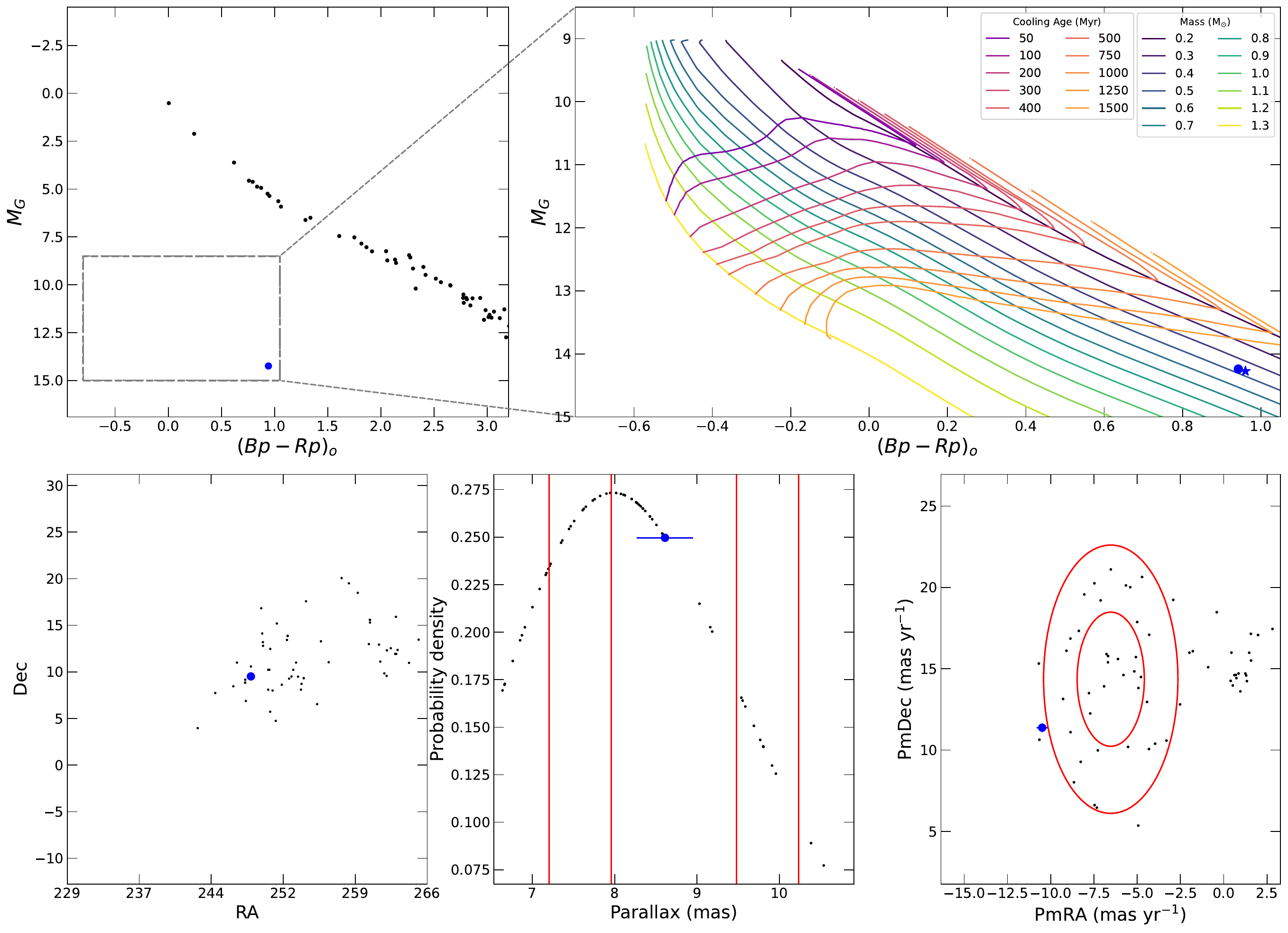}
    \caption{Same as Fig.~\ref{fig:CMD_alessi_13}, but for the HSC 242 cluster.}
    \label{fig:CMD_hsc_242}
\end{figure*}

\begin{figure*}[!tp]
    \centering
    \includegraphics[width=0.75\textwidth]{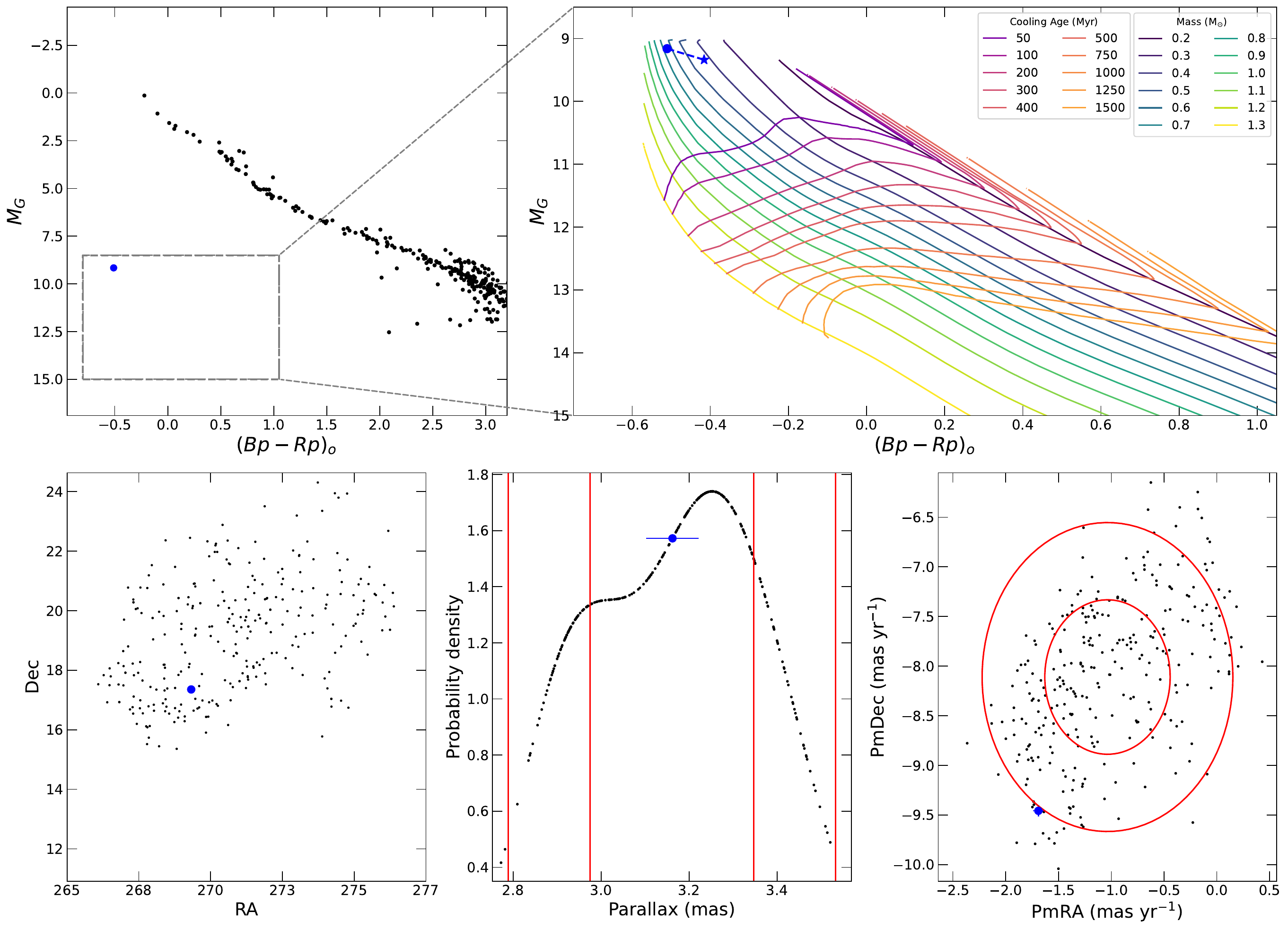}
    \caption{Same as Fig.~\ref{fig:CMD_alessi_13}, but for the HSC 381 cluster.}
    \label{fig:CMD_hsc_381}
\end{figure*}

\begin{figure*}[!tp]
    \centering
    \includegraphics[width=0.75\textwidth]{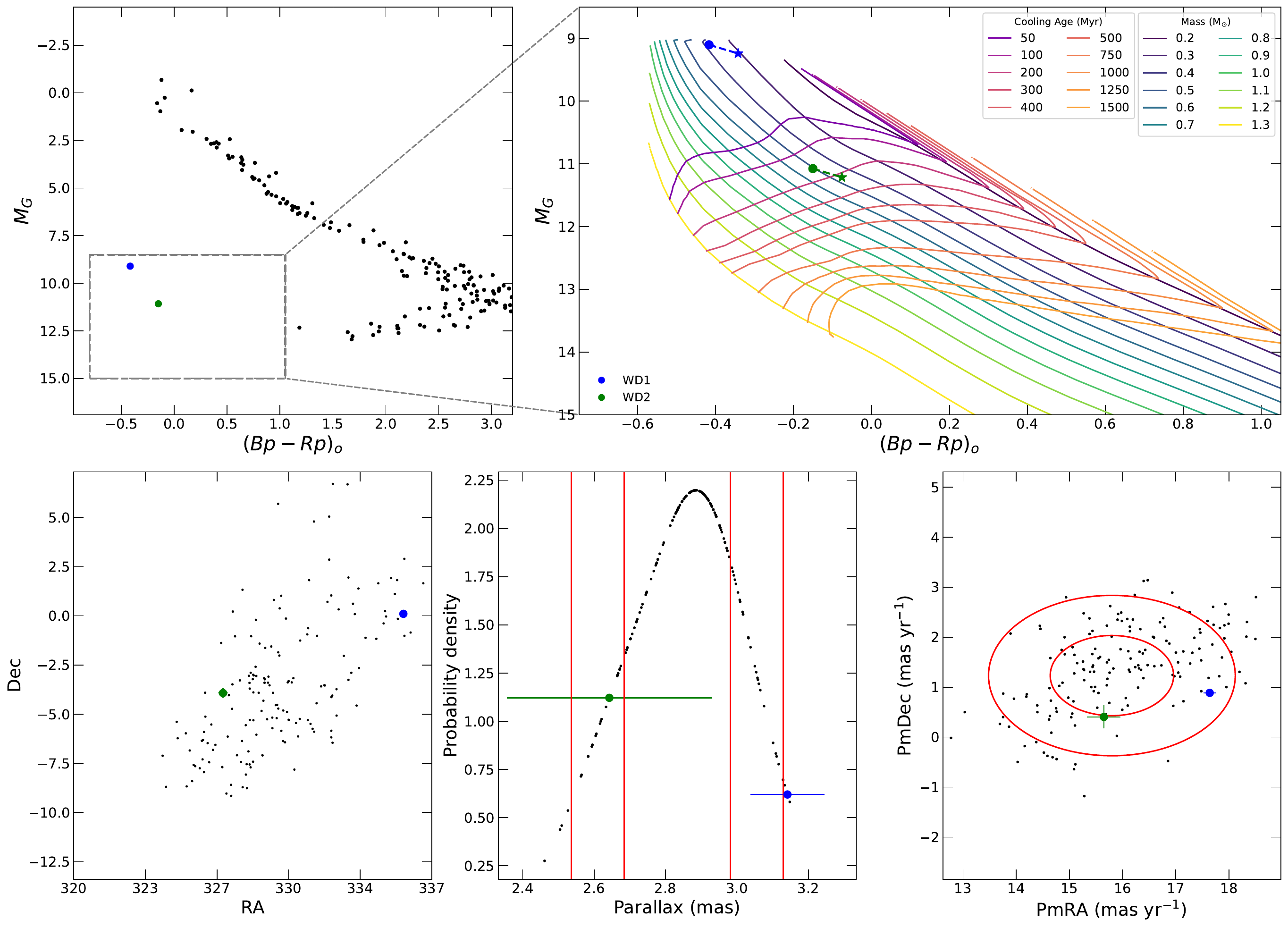}
    \caption{Same as Fig.~\ref{fig:CMD_alessi_13}, but for the HSC 448 cluster.}
    \label{fig:CMD_hsc_448}
\end{figure*}

\begin{figure*}[!tp]
    \centering
    \includegraphics[width=0.75\textwidth]{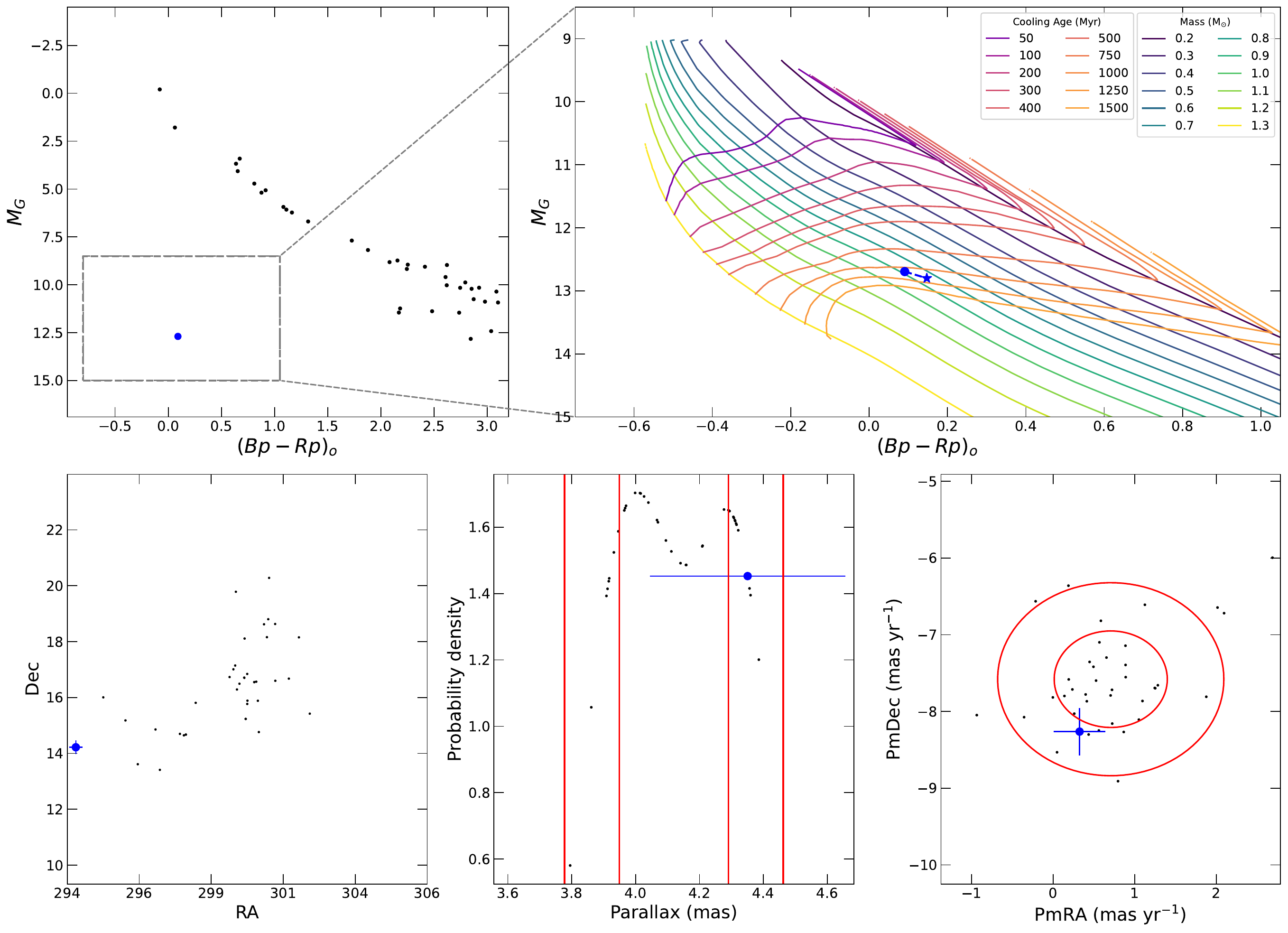}
    \caption{Same as Fig.~\ref{fig:CMD_alessi_13}, but for the HSC 452 cluster.}
    \label{fig:CMD_hsc_452}
\end{figure*}

\begin{figure*}[!tp]
    \centering
    \includegraphics[width=0.75\textwidth]{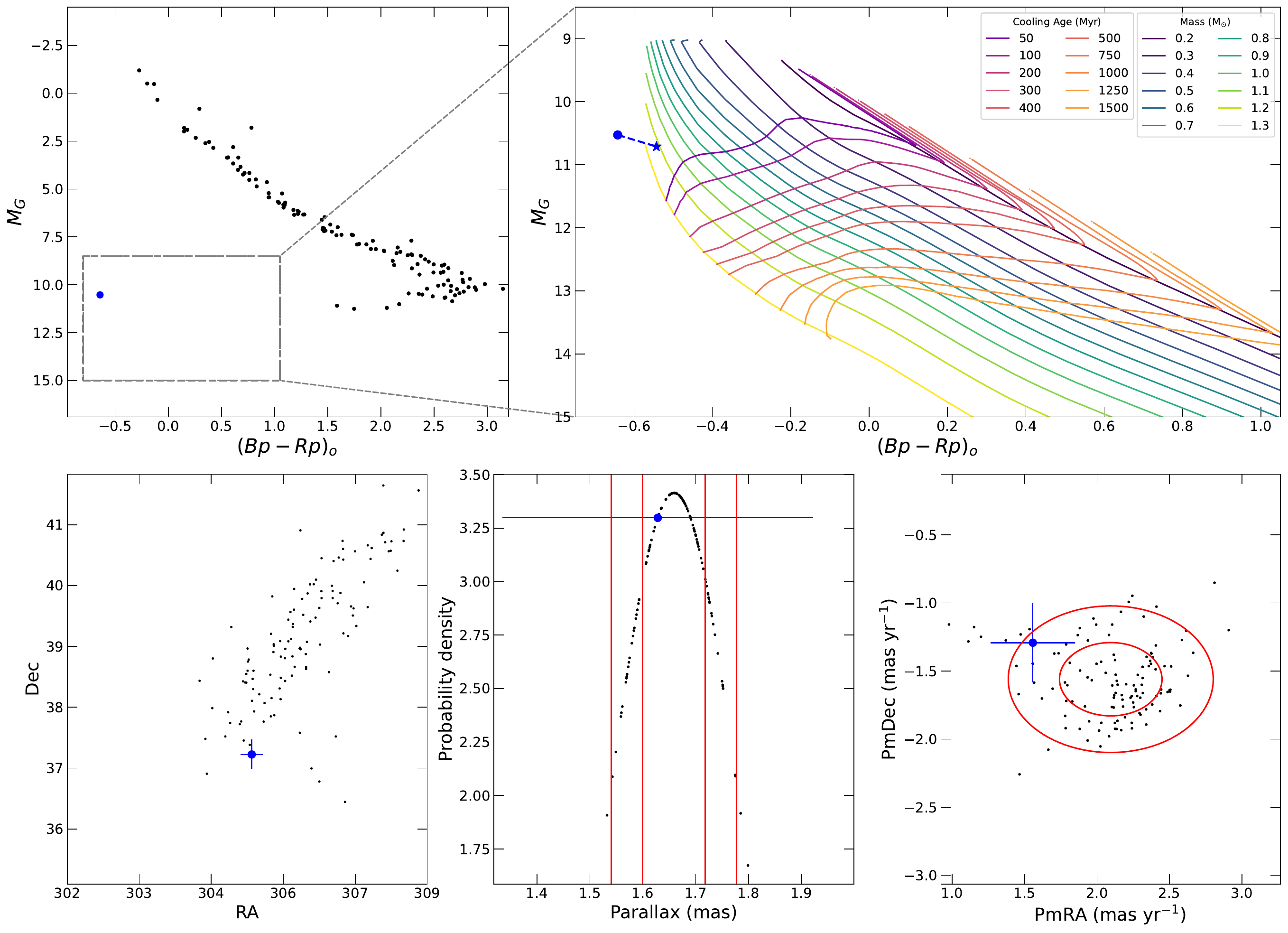}
    \caption{Same as Fig.~\ref{fig:CMD_alessi_13}, but for the HSC 601 cluster.}
    \label{fig:CMD_hsc_601}
\end{figure*}

\begin{figure*}[!tp]
    \centering
    \includegraphics[width=0.75\textwidth]{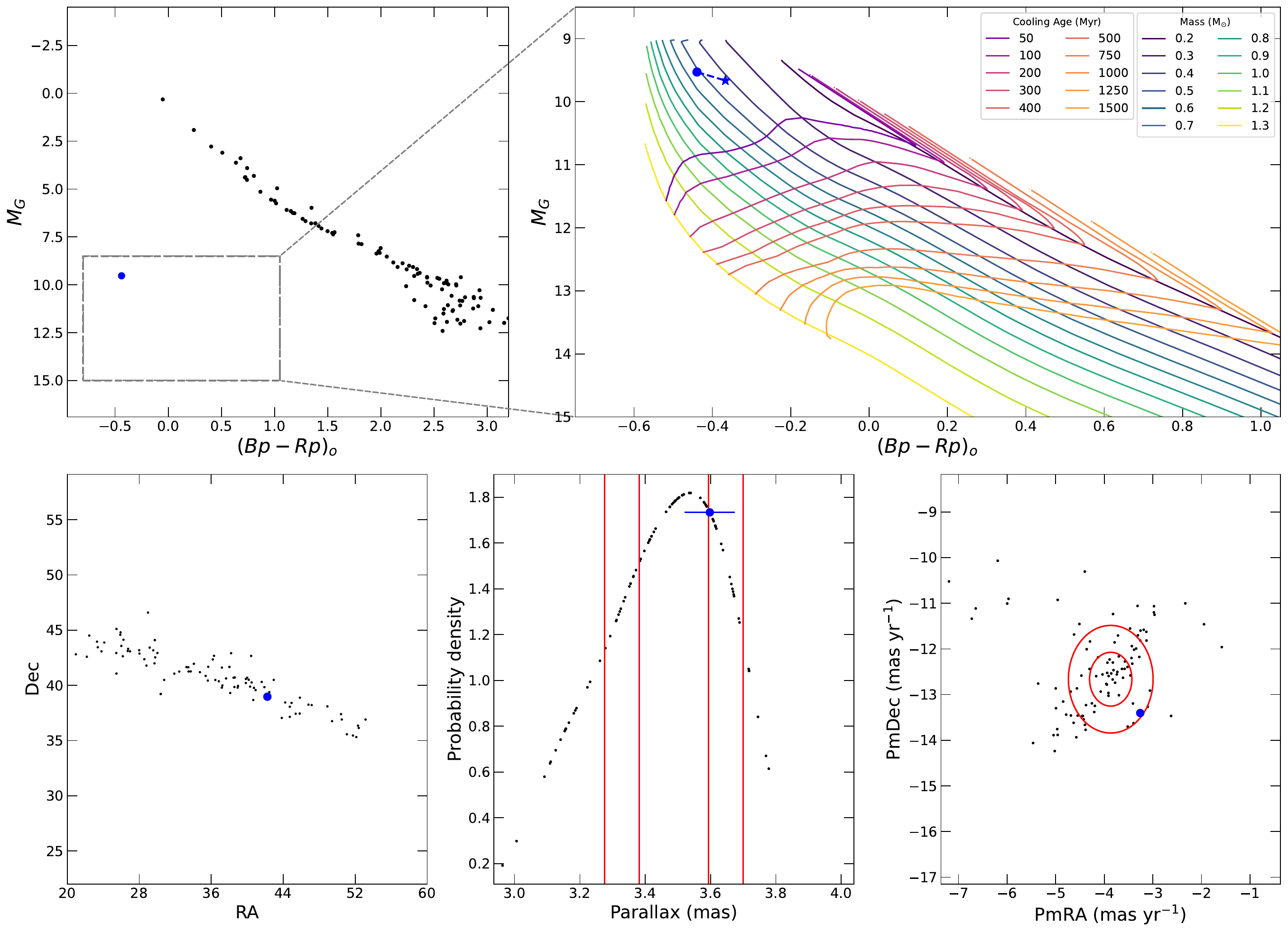}
    \caption{Same as Fig.~\ref{fig:CMD_alessi_13}, but for the HSC 1155 cluster.}
    \label{fig:CMD_hsc_1155}
\end{figure*}

\begin{figure*}[!tp]
    \centering
    \includegraphics[width=0.75\textwidth]{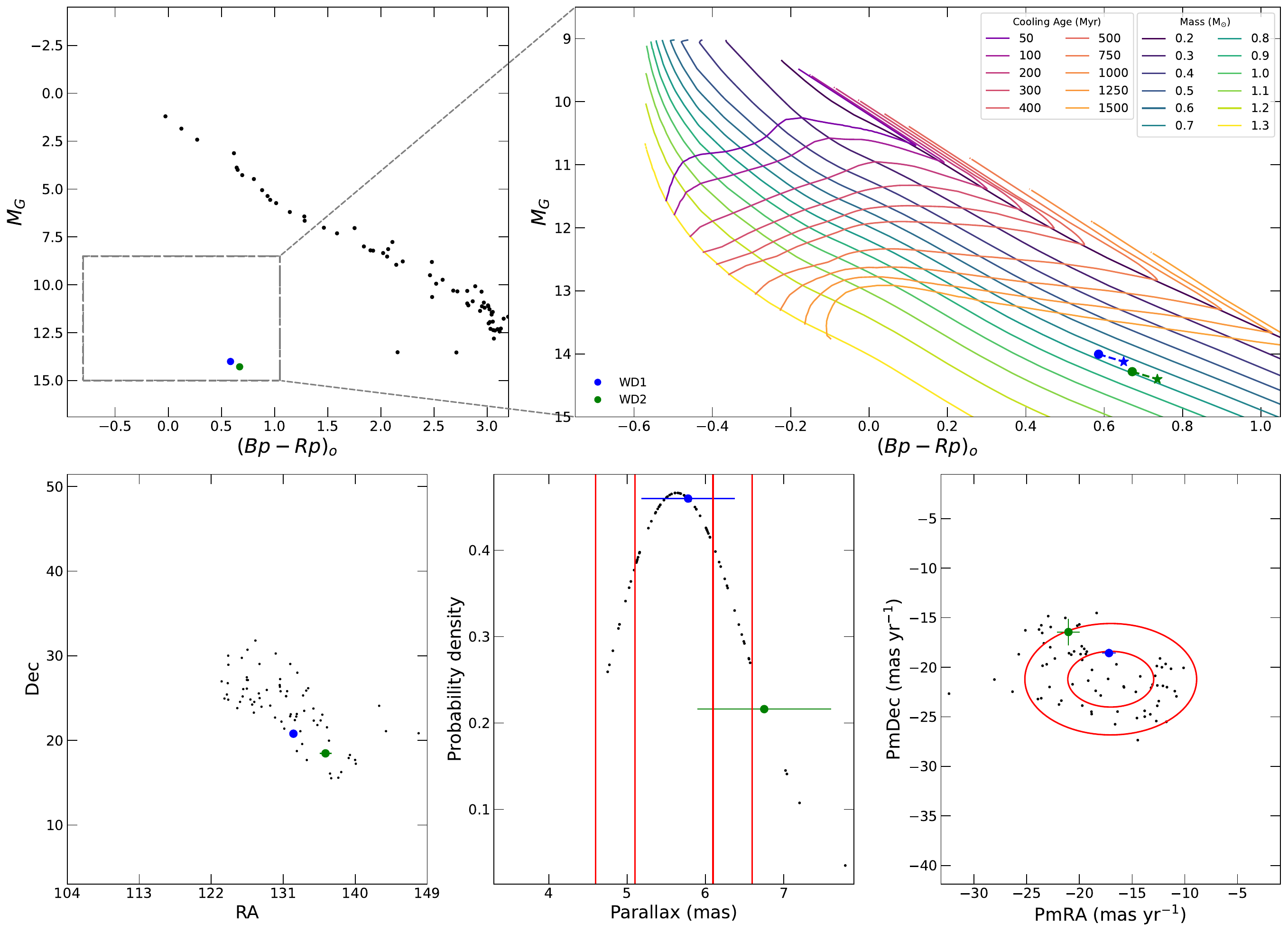}
    \caption{Same as Fig.~\ref{fig:CMD_alessi_13}, but for the HSC 1555 cluster.}
    \label{fig:CMD_hsc_1555}
\end{figure*}

\begin{figure*}[!tp]
    \centering
    \includegraphics[width=0.75\textwidth]{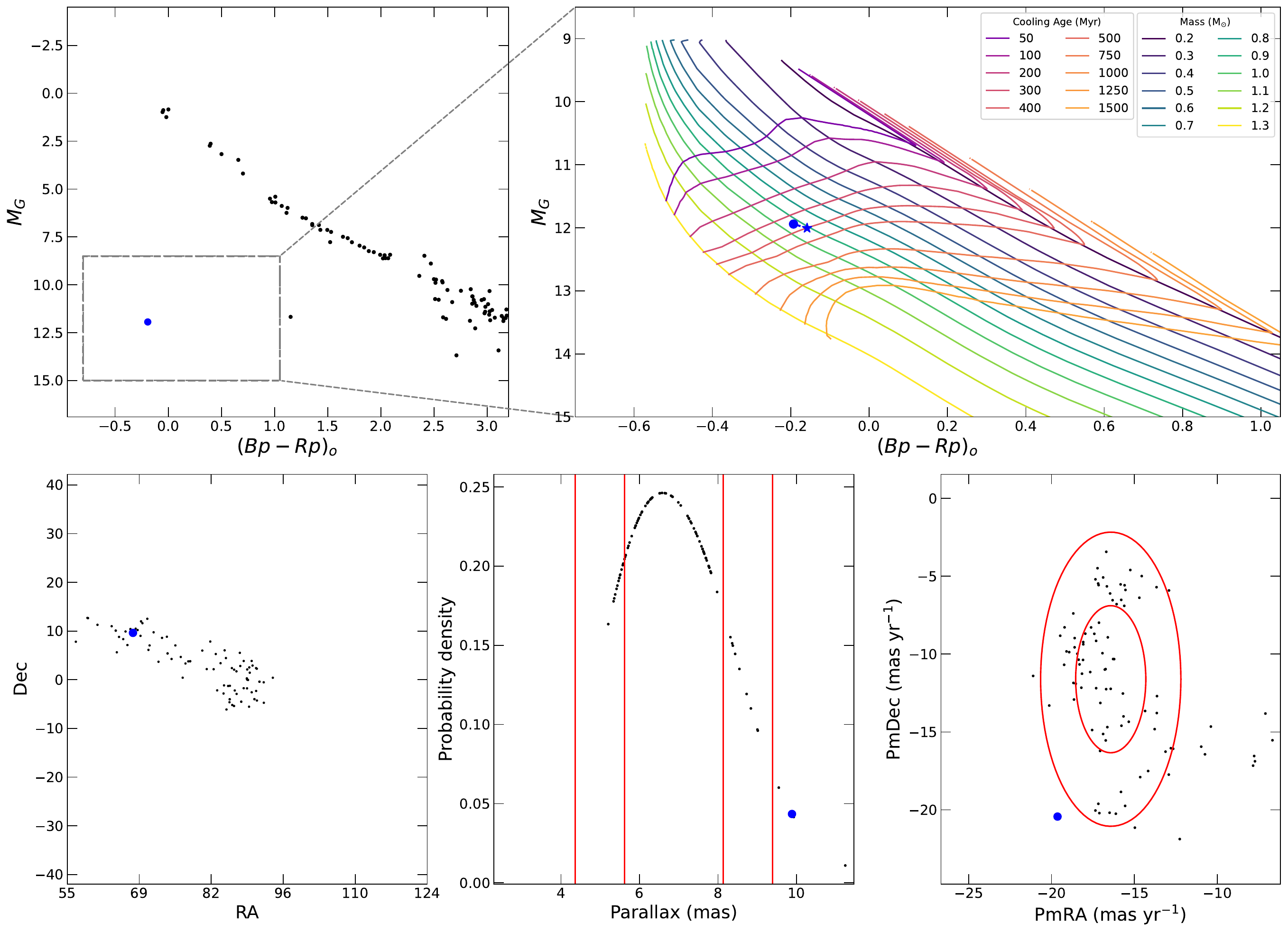}
    \caption{Same as Fig.~\ref{fig:CMD_alessi_13}, but for the HSC 1630 cluster.}
    \label{fig:CMD_hsc_1630}
\end{figure*}

\begin{figure*}[!tp]
    \centering
    \includegraphics[width=0.75\textwidth]{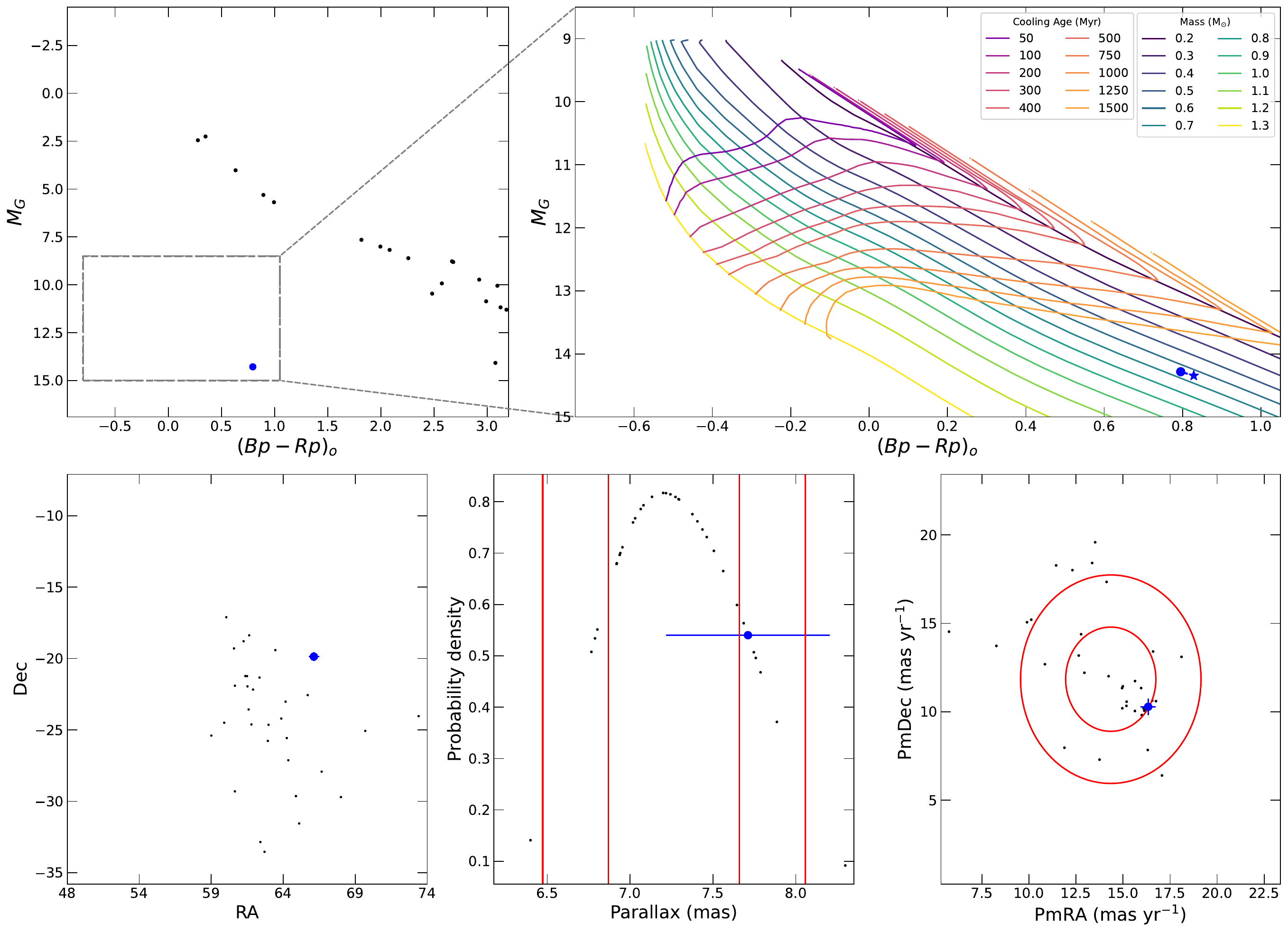}
    \caption{Same as Fig.~\ref{fig:CMD_alessi_13}, but for the HSC 1710 cluster.}
    \label{fig:CMD_hsc_1710}
\end{figure*}

\begin{figure*}[!tp]
    \centering
    \includegraphics[width=0.75\textwidth]{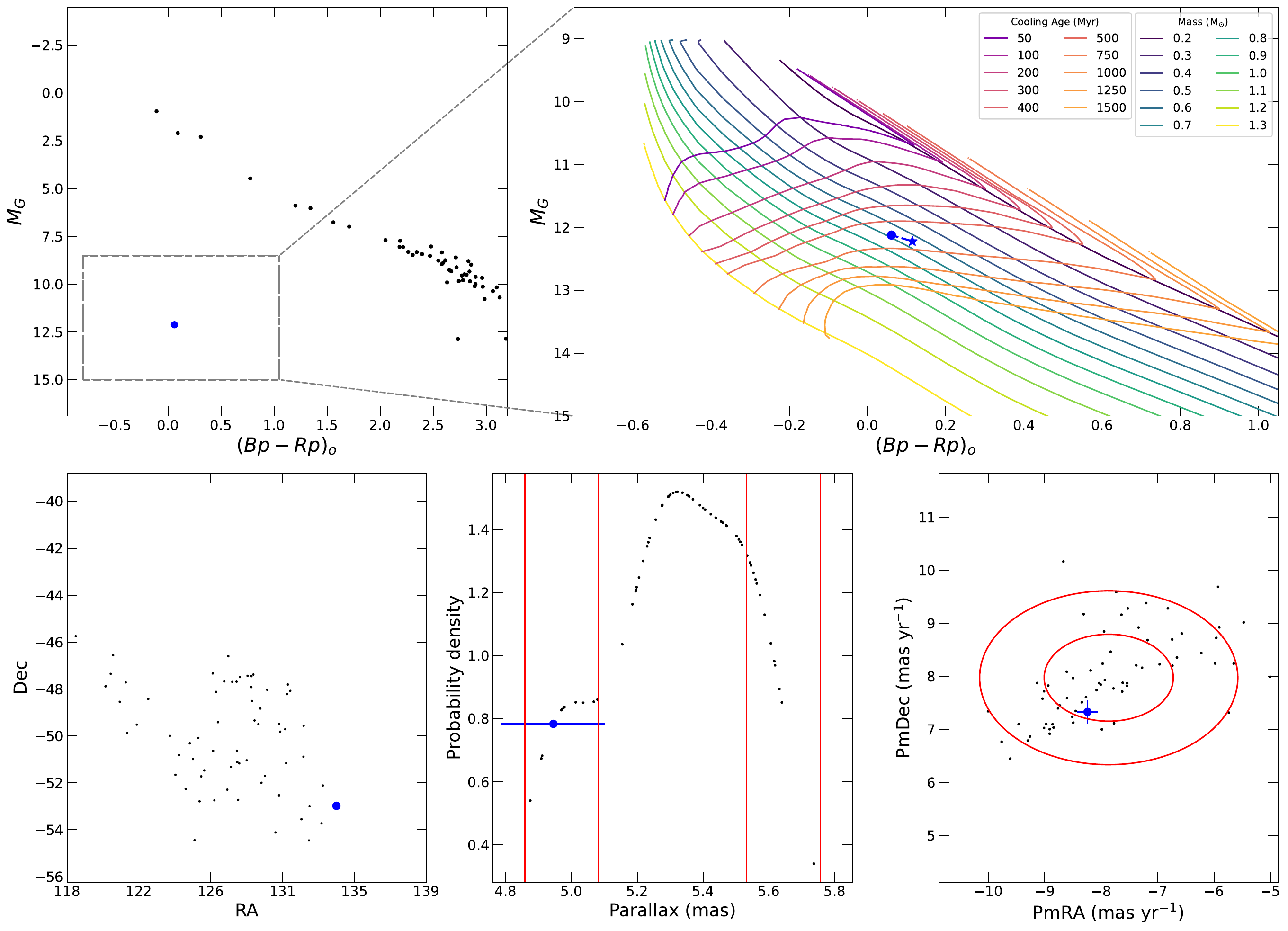}
    \caption{Same as Fig.~\ref{fig:CMD_alessi_13}, but for the HSC 2139 cluster.}
    \label{fig:CMD_hsc_2139}
\end{figure*}

\begin{figure*}[!tp]
    \centering
    \includegraphics[width=0.75\textwidth]{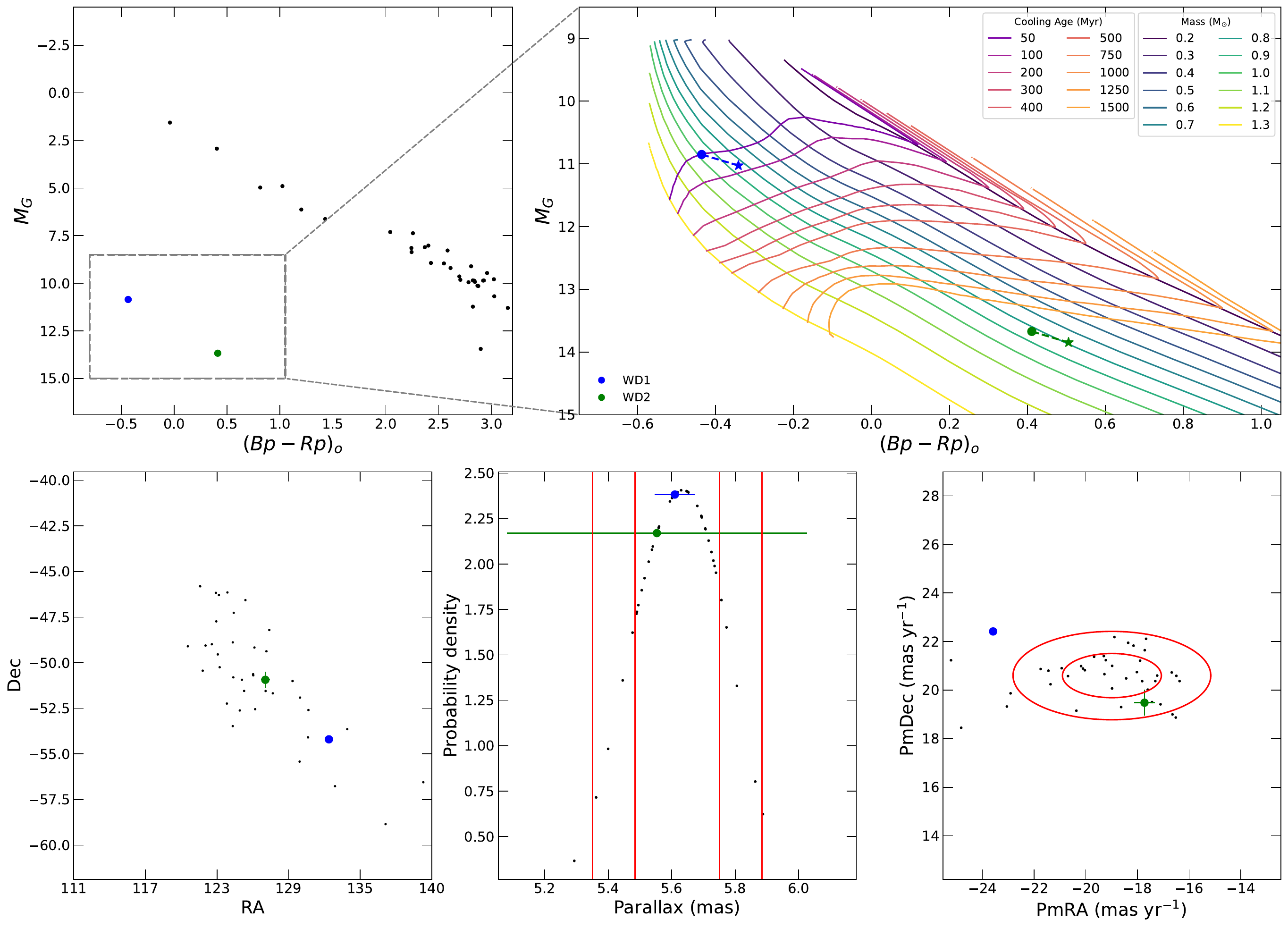}
    \caption{Same as Fig.~\ref{fig:CMD_alessi_13}, but for the HSC 2156 cluster.}
    \label{fig:CMD_hsc_2156}
\end{figure*}

\begin{figure*}[!tp]
    \centering
    \includegraphics[width=0.75\textwidth]{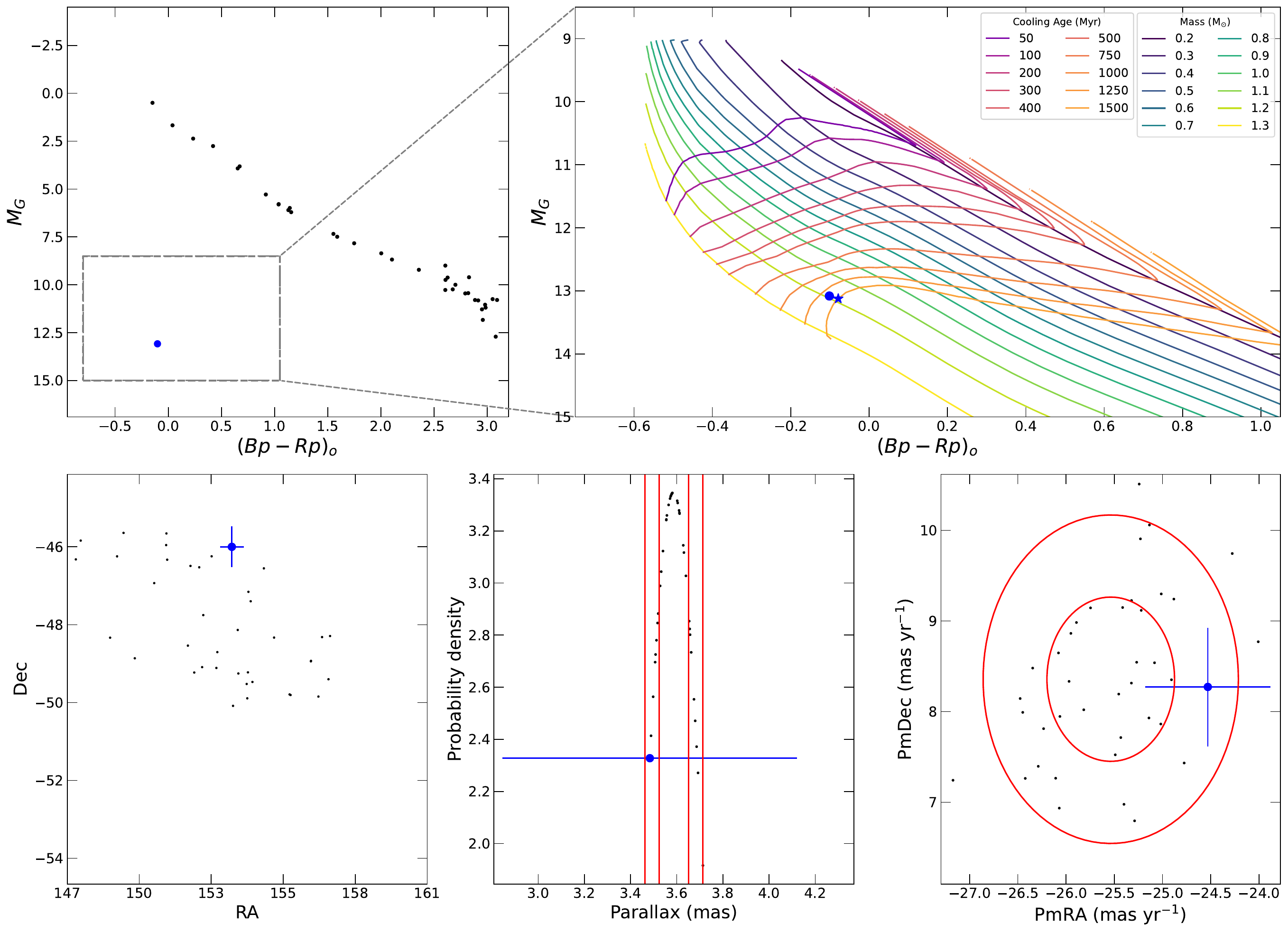}
    \caption{Same as Fig.~\ref{fig:CMD_alessi_13}, but for the HSC 2263 cluster.}
    \label{fig:CMD_hsc_2263}
\end{figure*}

\begin{figure*}[!tp]
    \centering
    \includegraphics[width=0.75\textwidth]{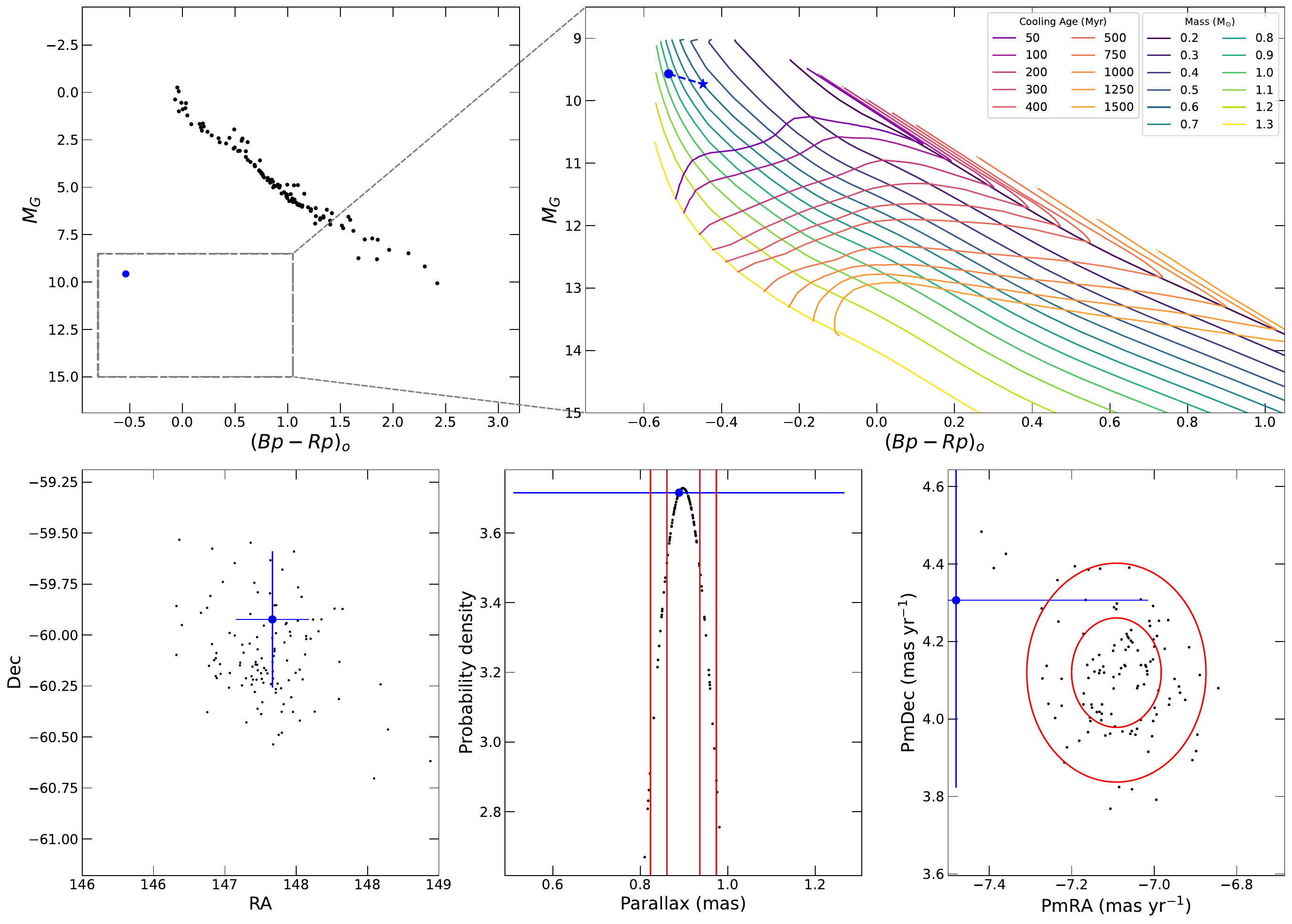}
    \caption{Same as Fig.~\ref{fig:CMD_alessi_13}, but for the HSC 2304 cluster.}
    \label{fig:CMD_hsc_2304}
\end{figure*}

\begin{figure*}[!tp]
    \centering
    \includegraphics[width=0.75\textwidth]{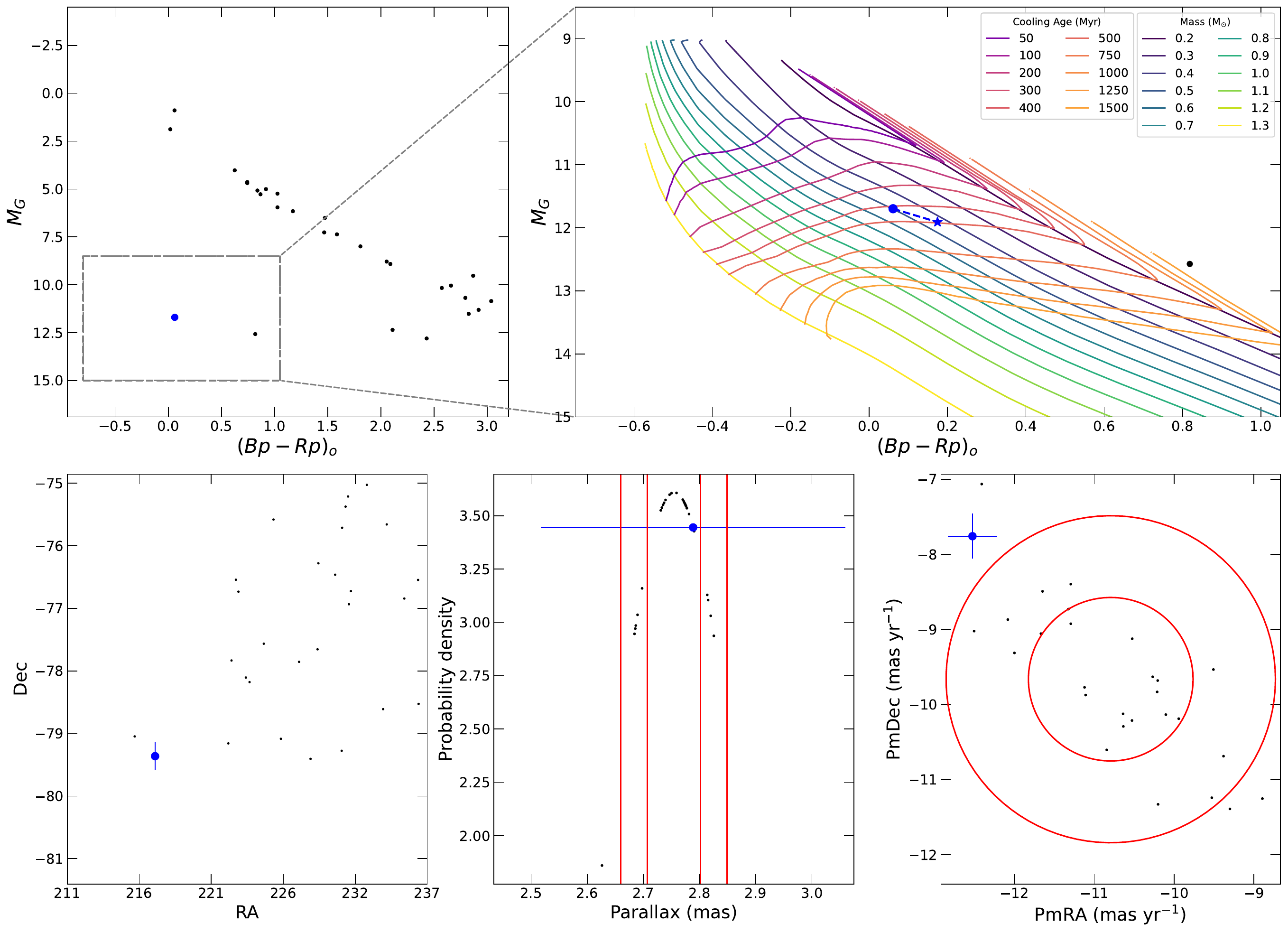}
    \caption{Same as Fig.~\ref{fig:CMD_alessi_13}, but for the HSC 2609 cluster.}
    \label{fig:CMD_hsc_2609}
\end{figure*}

\begin{figure*}[!tp]
    \centering
    \includegraphics[width=0.75\textwidth]{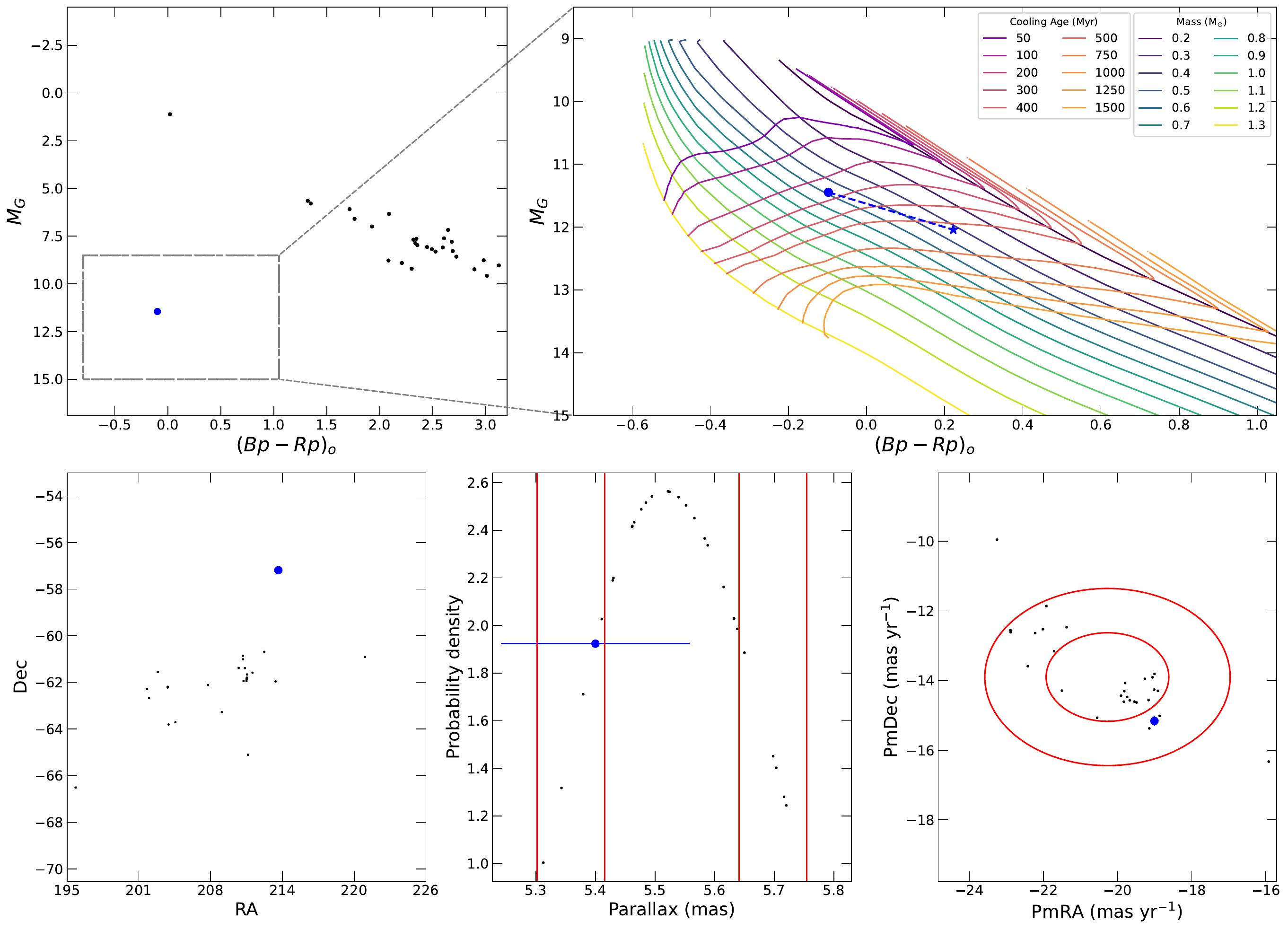}
    \caption{Same as Fig.~\ref{fig:CMD_alessi_13}, but for the HSC 2630 cluster.}
    \label{fig:CMD_hsc_2630}
\end{figure*}

\begin{figure*}[!tp]
    \centering
    \includegraphics[width=0.75\textwidth]{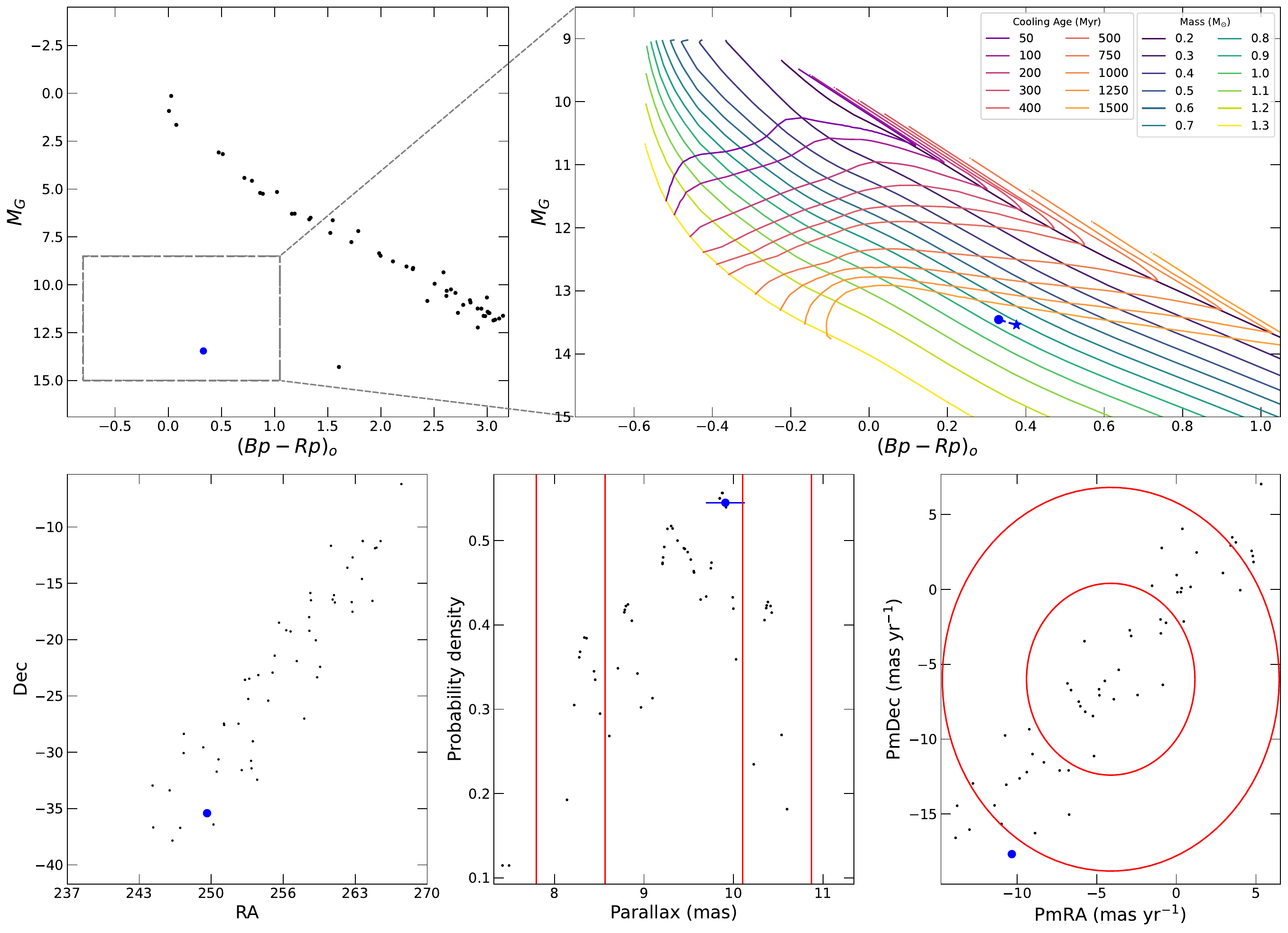}
    \caption{Same as Fig.~\ref{fig:CMD_alessi_13}, but for the HSC 2971 cluster.}
    \label{fig:CMD_hsc_2971}
\end{figure*}

\begin{figure*}[!tp]
    \centering
    \includegraphics[width=0.75\textwidth]{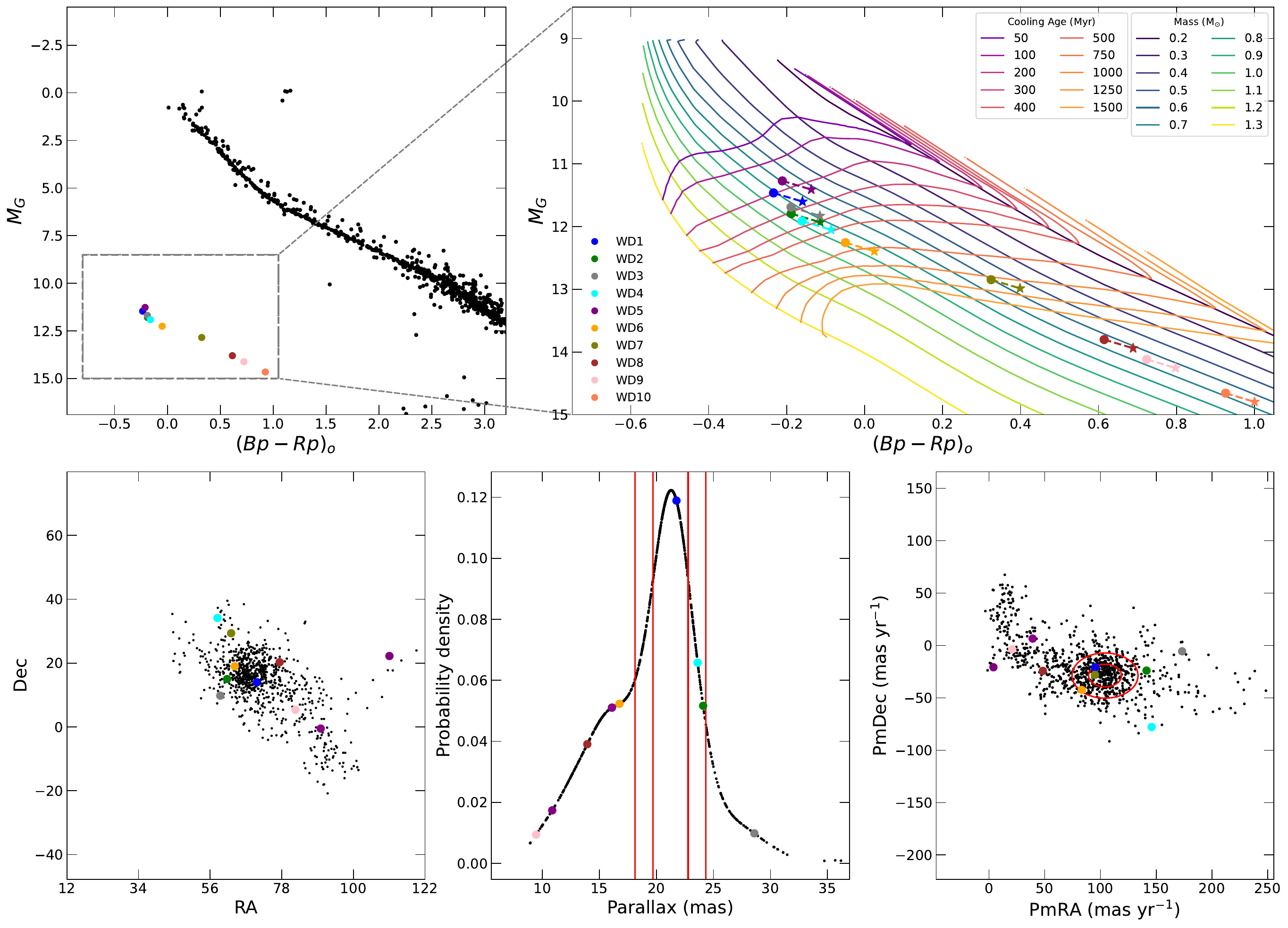}
    \caption{Same as Fig.~\ref{fig:CMD_alessi_13}, but for the Hyades cluster.}
    \label{fig:CMD_hyades}
\end{figure*}

\begin{figure*}[!tp]
    \centering
    \includegraphics[width=0.75\textwidth]{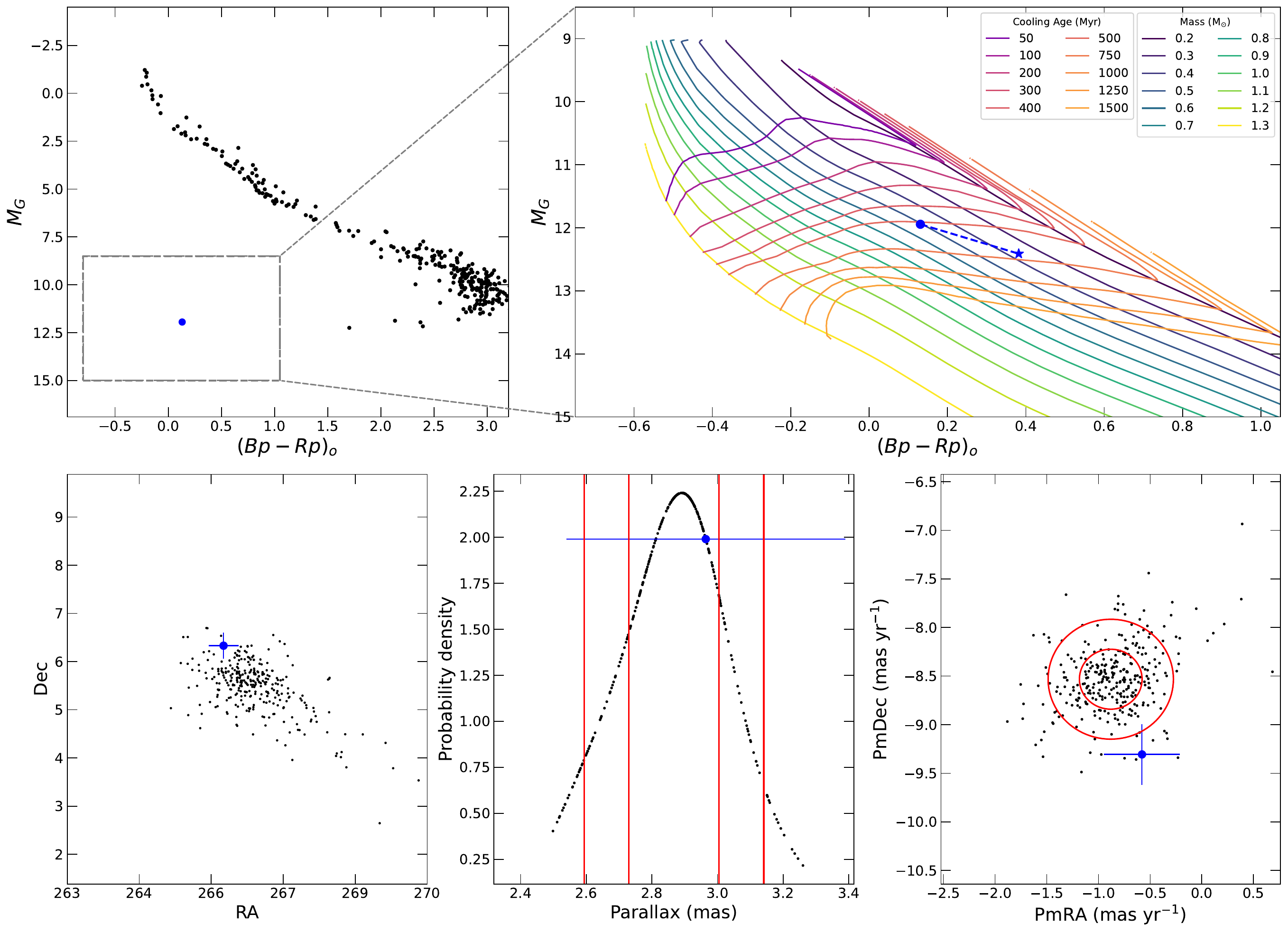}
    \caption{Same as Fig.~\ref{fig:CMD_alessi_13}, but for the IC 4665 cluster.}
    \label{fig:CMD_ic_4665}
\end{figure*}

\begin{figure*}[!tp]
    \centering
    \includegraphics[width=0.75\textwidth]{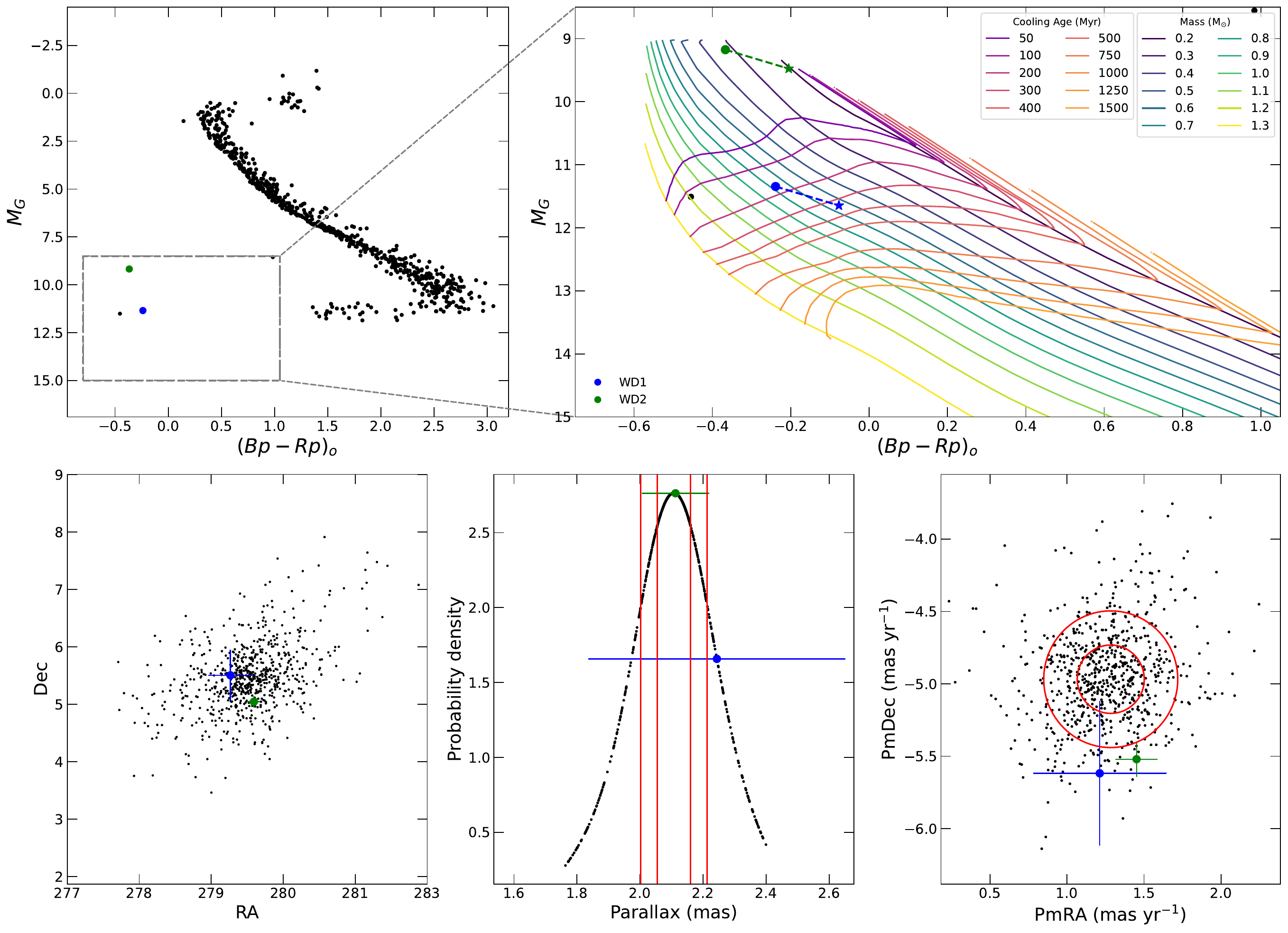}
    \caption{Same as Fig.~\ref{fig:CMD_alessi_13}, but for the IC 4756 cluster.}
    \label{fig:CMD_ic_4756}
\end{figure*}

\begin{figure*}[!tp]
    \centering
    \includegraphics[width=0.75\textwidth]{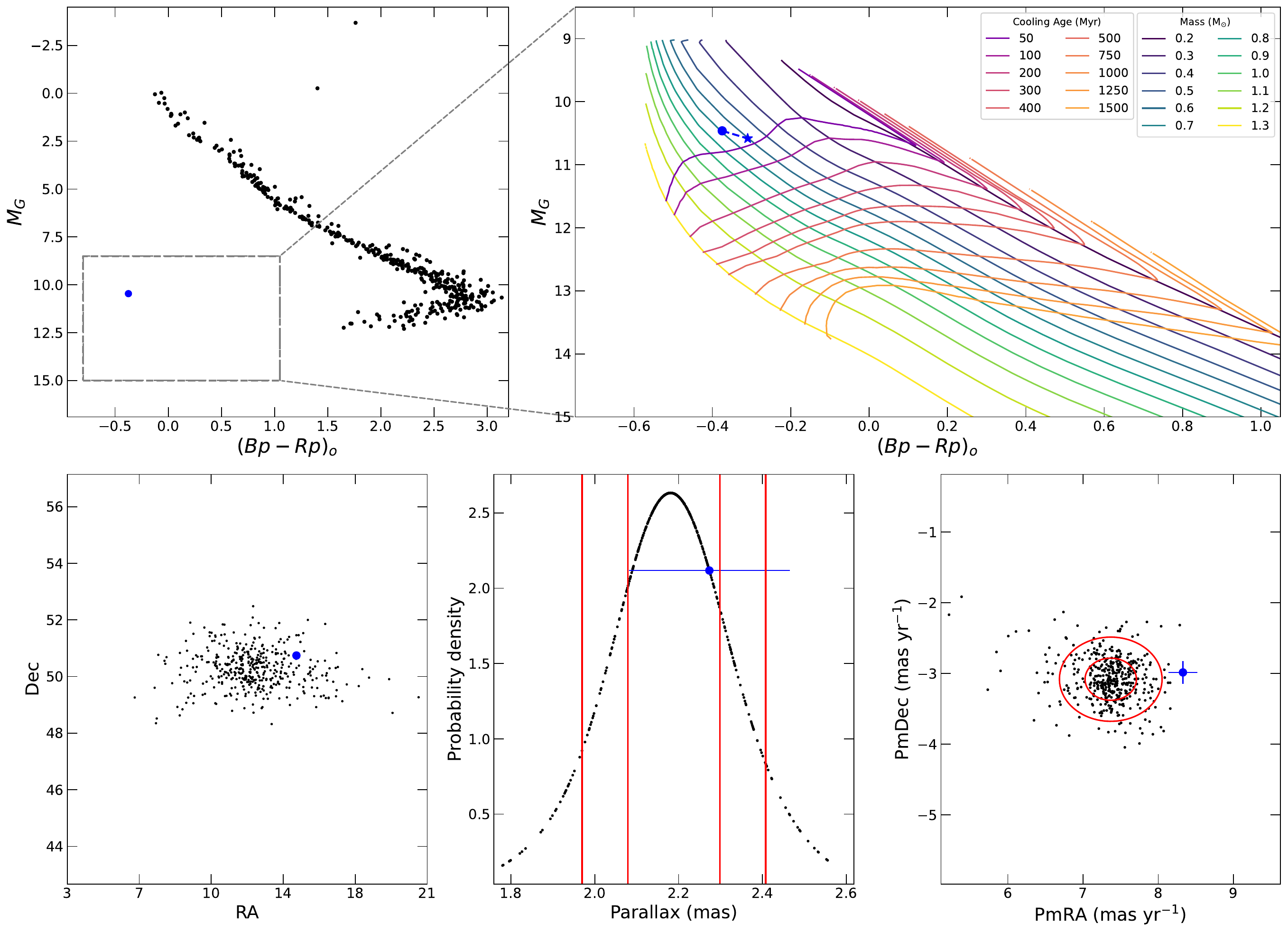}
    \caption{Same as Fig.~\ref{fig:CMD_alessi_13}, but for the LISC 3534 cluster.}
    \label{fig:CMD_lisc_3534}
\end{figure*}

\begin{figure*}[!tp]
    \centering
    \includegraphics[width=0.75\textwidth]{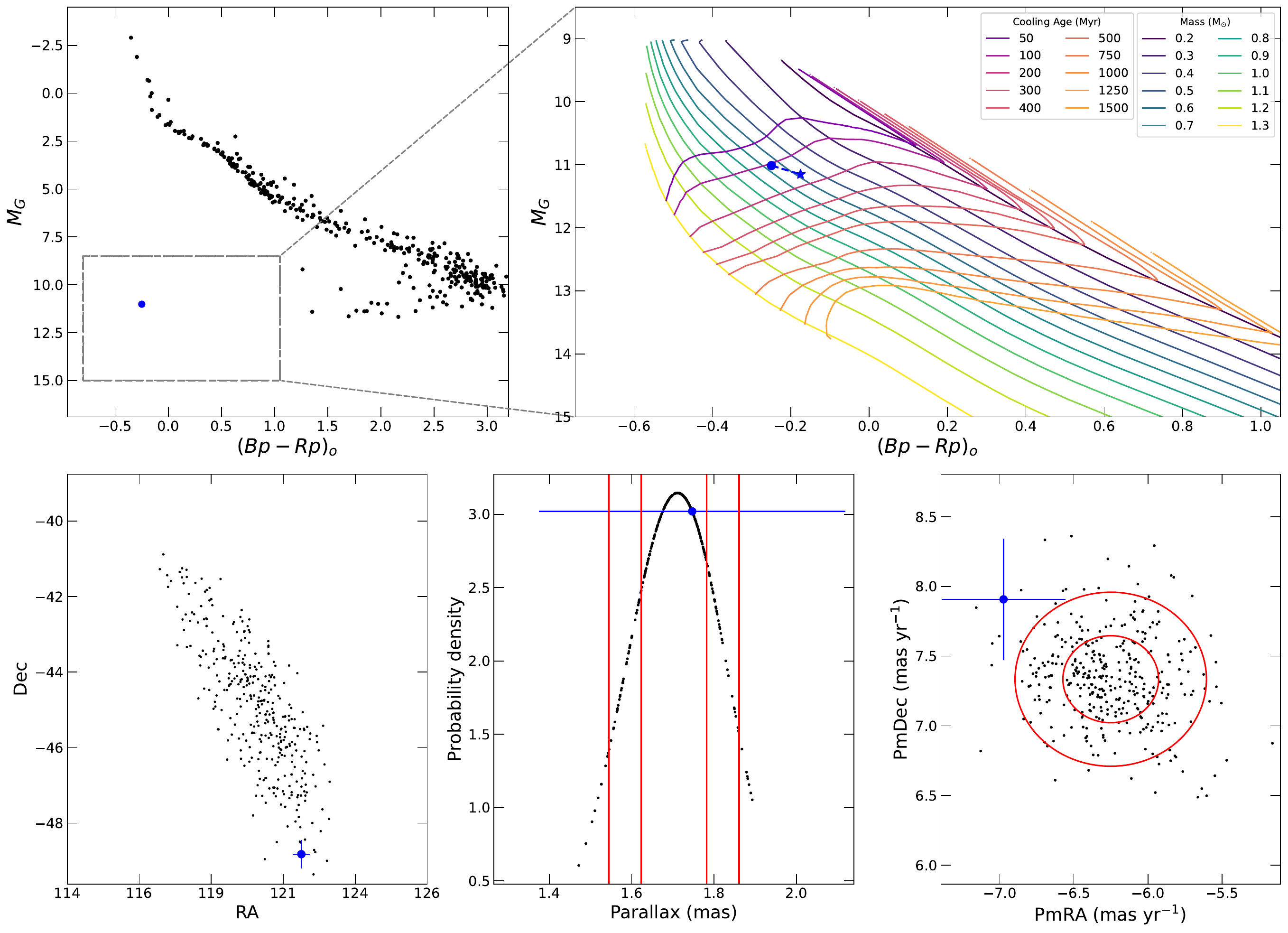}
    \caption{Same as Fig.~\ref{fig:CMD_alessi_13}, but for the LISC-III 3668 cluster.}
    \label{fig:CMD_lisc_iii_3668}
\end{figure*}

\begin{figure*}[!tp]
    \centering
    \includegraphics[width=0.75\textwidth]{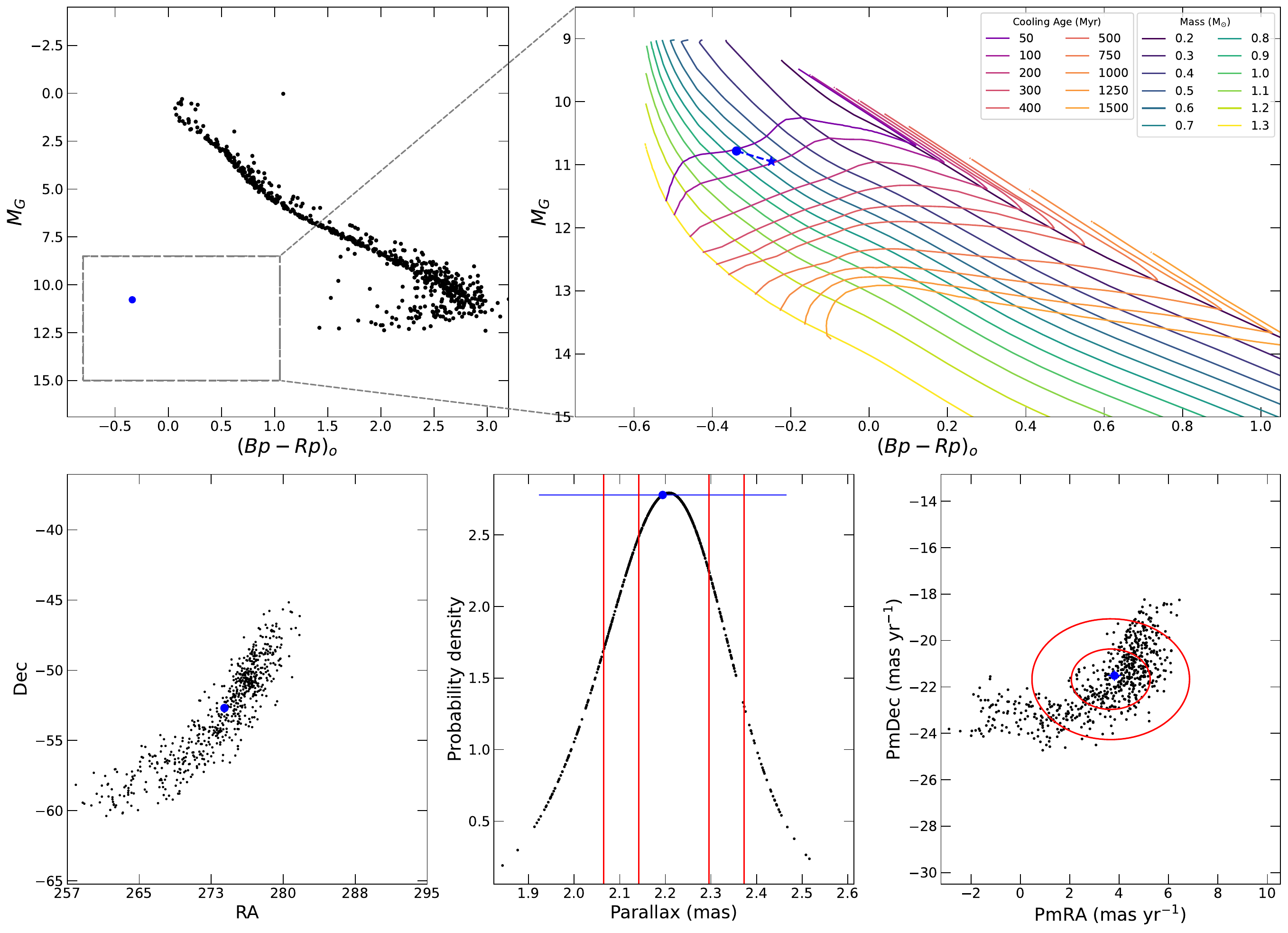}
    \caption{Same as Fig.~\ref{fig:CMD_alessi_13}, but for the Mamajek 4 cluster.}
    \label{fig:CMD_mamajek_4}
\end{figure*}

\begin{figure*}[!tp]
    \centering
    \includegraphics[width=0.75\textwidth]{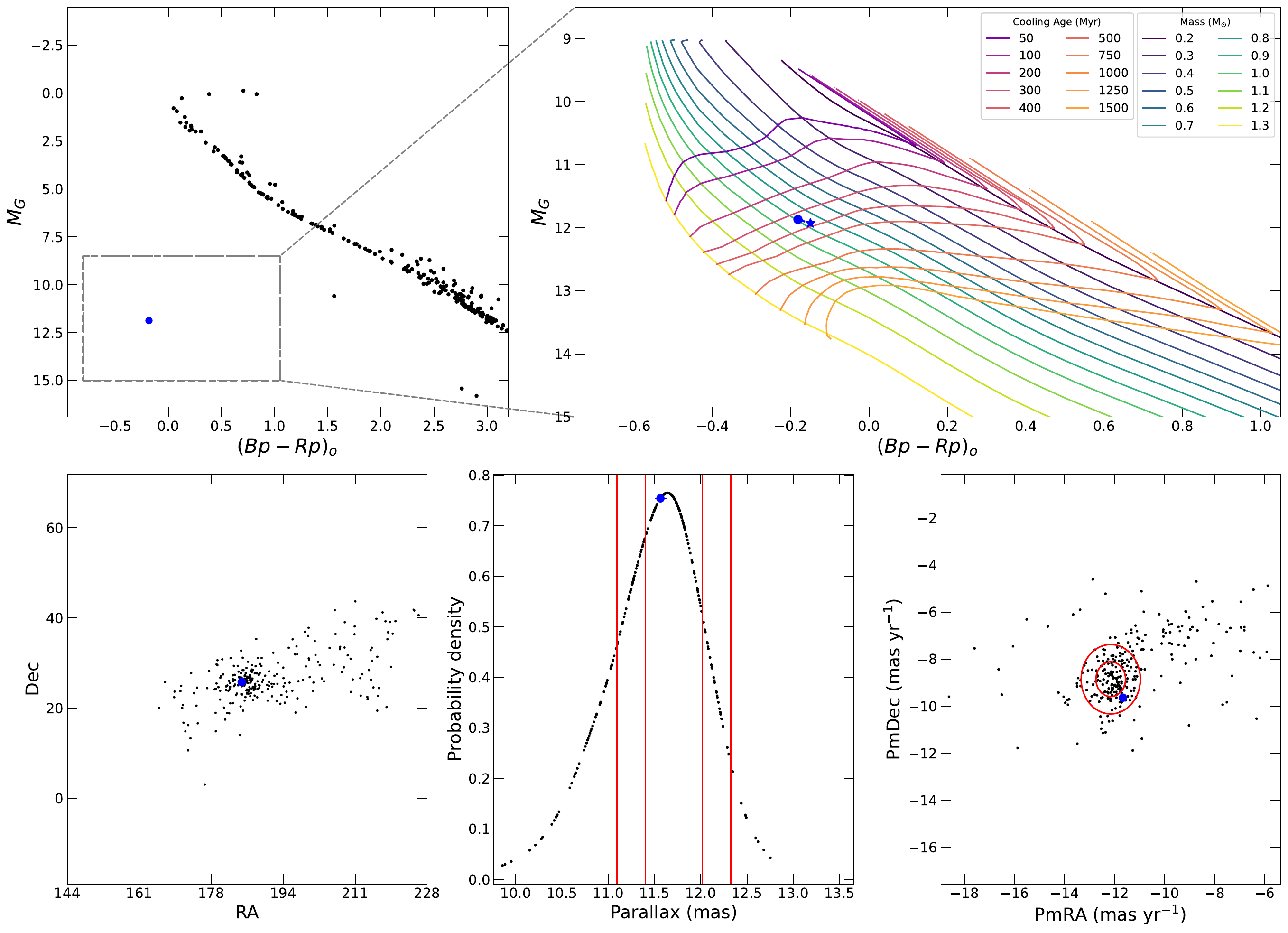}
    \caption{Same as Fig.~\ref{fig:CMD_alessi_13}, but for the Melotte 111 cluster.}
    \label{fig:CMD_melotte_111}
\end{figure*}

\begin{figure*}[!tp]
    \centering
    \includegraphics[width=0.75\textwidth]{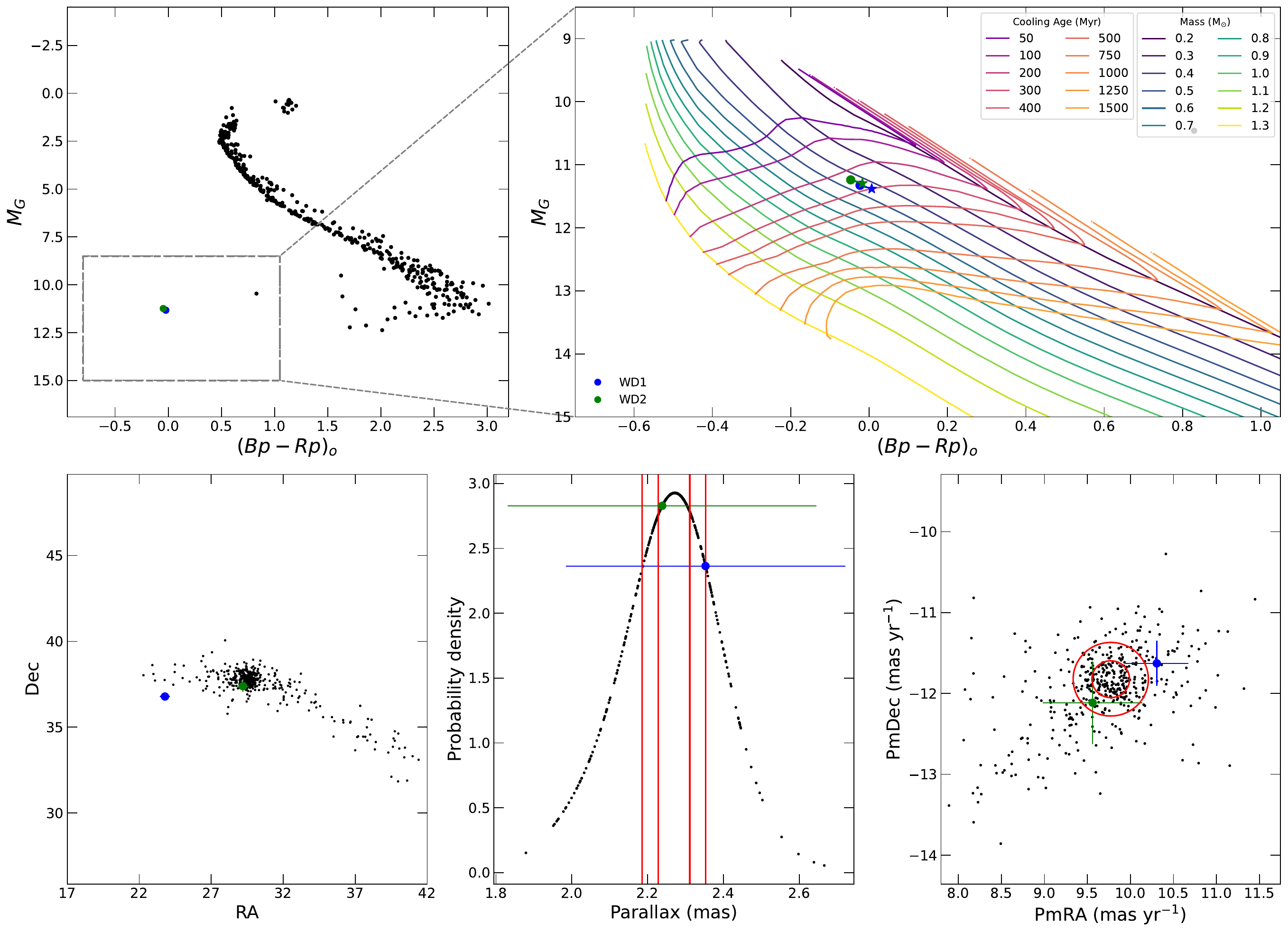}
    \caption{Same as Fig.~\ref{fig:CMD_alessi_13}, but for the NGC 752 cluster.}
    \label{fig:CMD_ngc_752}
\end{figure*}

\begin{figure*}[!tp]
    \centering
    \includegraphics[width=0.75\textwidth]{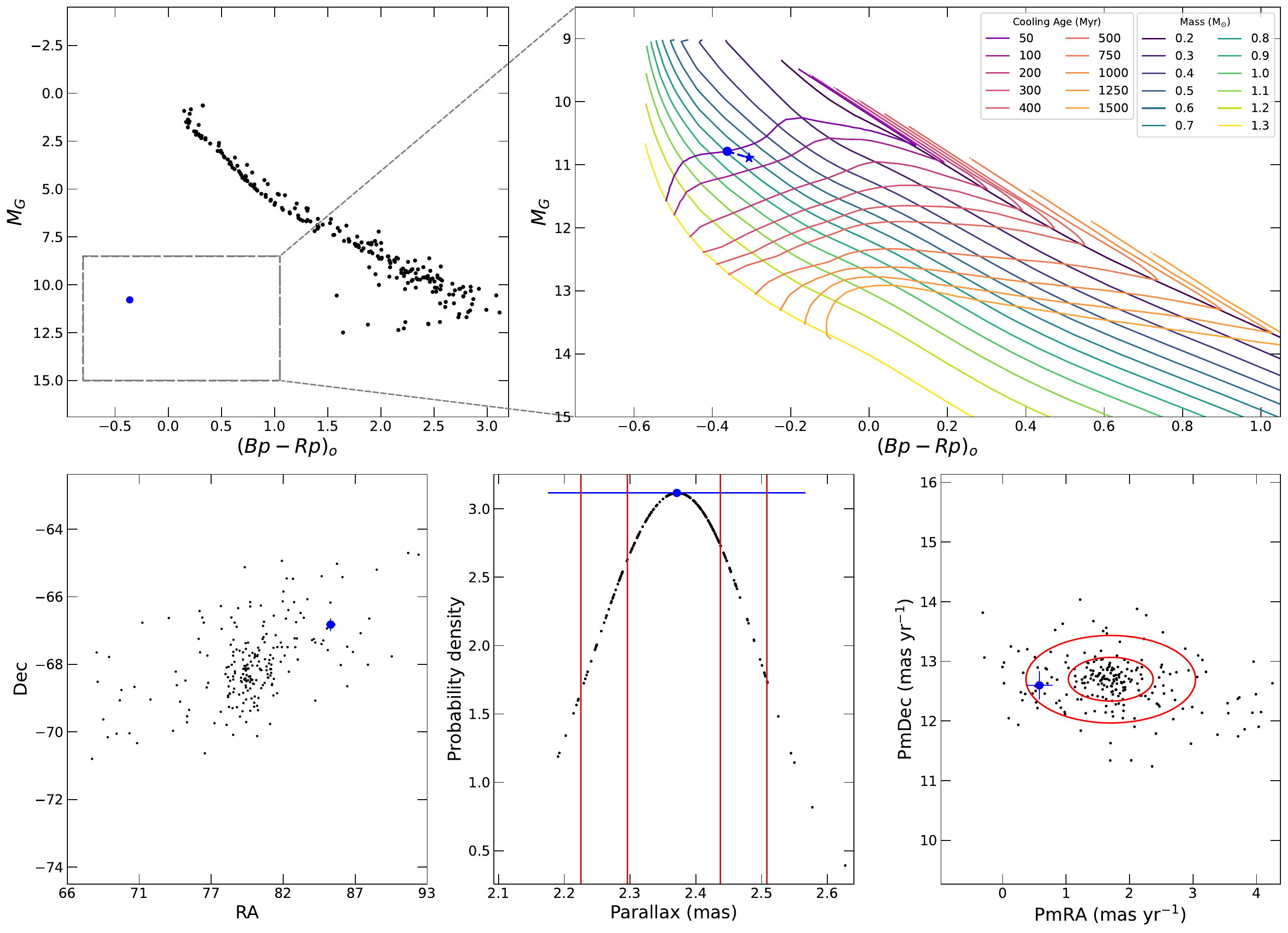}
    \caption{Same as Fig.~\ref{fig:CMD_alessi_13}, but for the NGC 1901 cluster.}
    \label{fig:CMD_ngc_1901}
\end{figure*}

\begin{figure*}[!tp]
    \centering
    \includegraphics[width=0.75\textwidth]{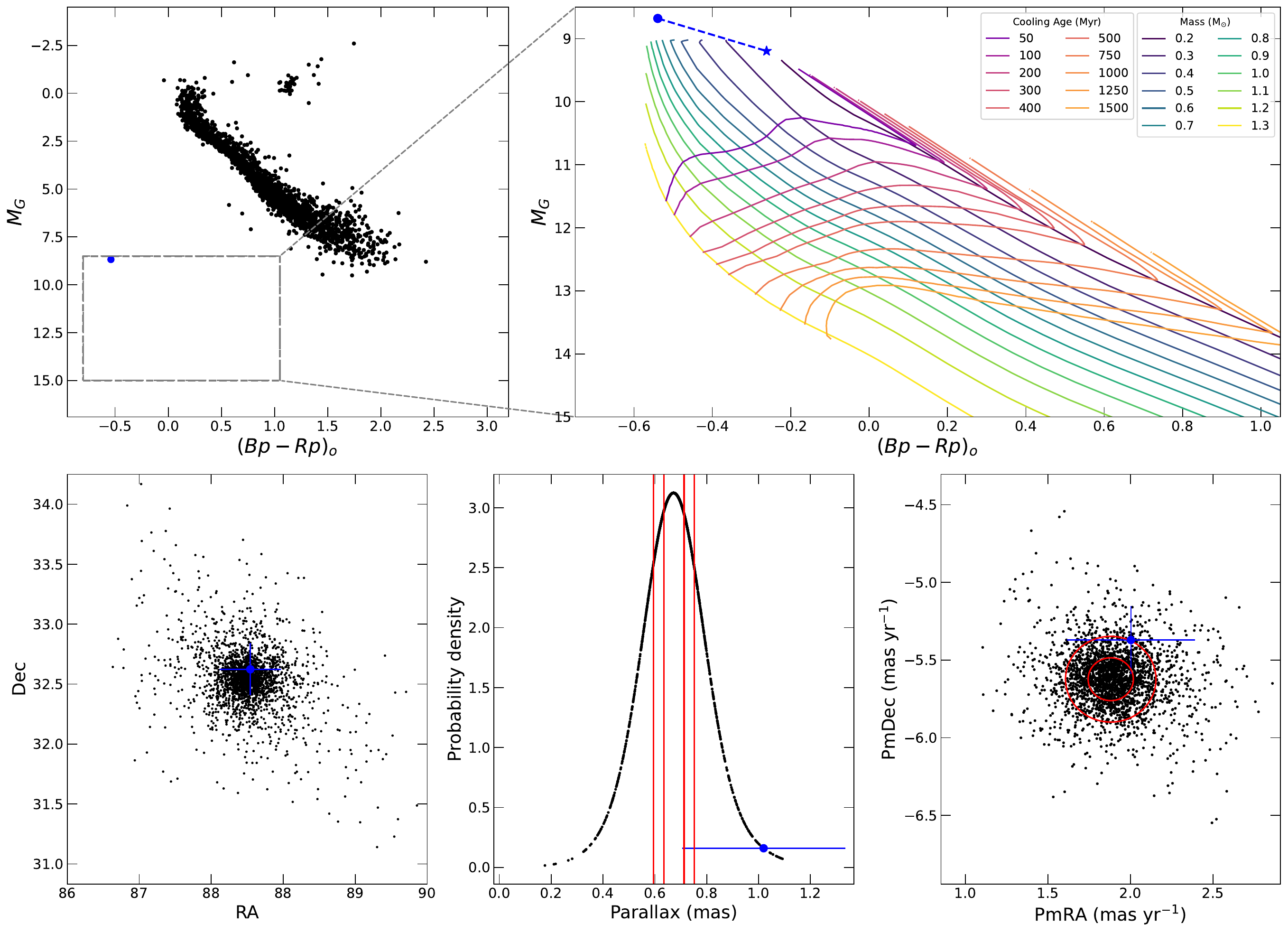}
    \caption{Same as Fig.~\ref{fig:CMD_alessi_13}, but for the NGC 2099 cluster.}
    \label{fig:CMD_ngc_2099}
\end{figure*}

\begin{figure*}[!tp]
    \centering
    \includegraphics[width=0.75\textwidth]{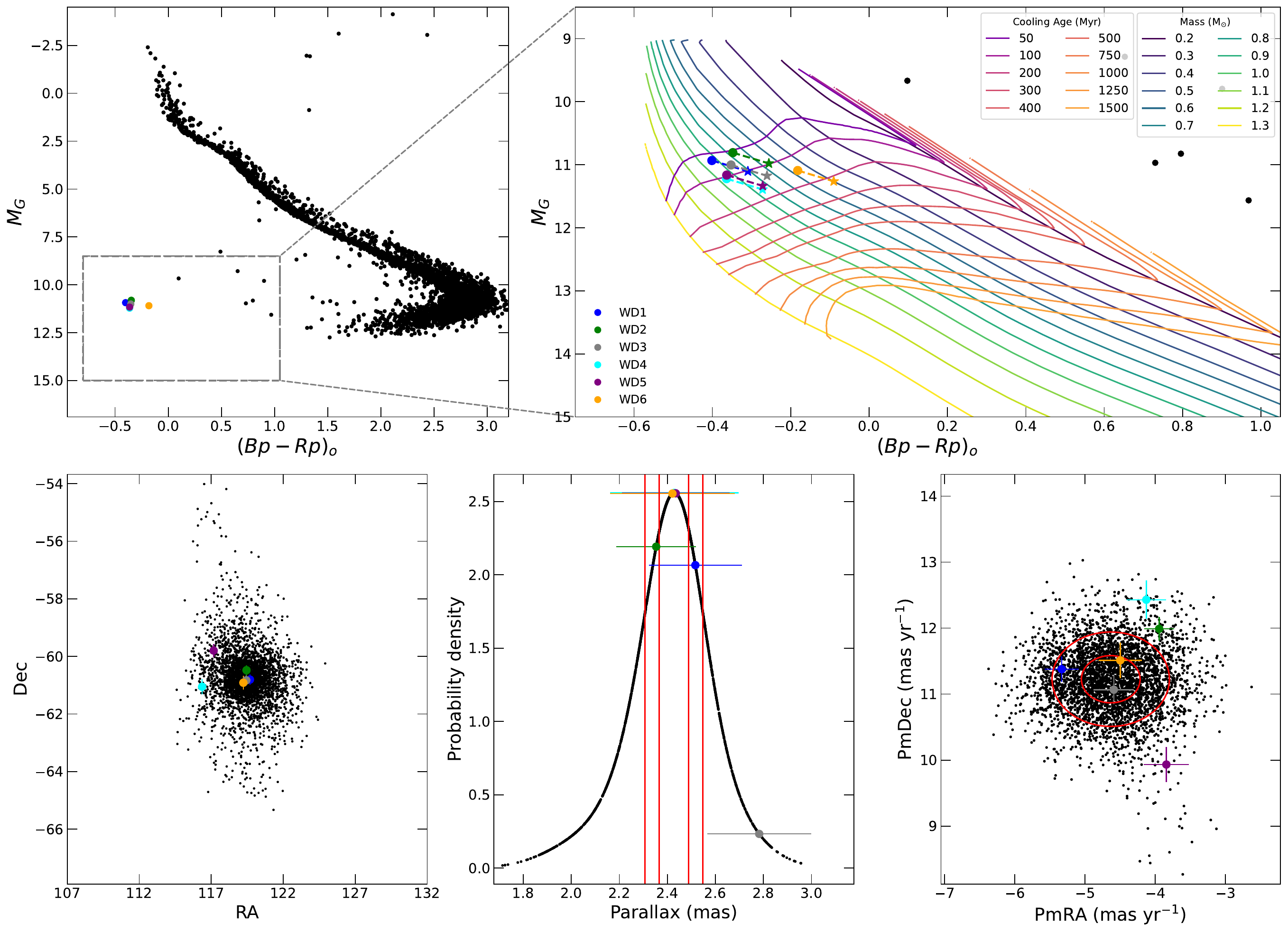}
    \caption{Same as Fig.~\ref{fig:CMD_alessi_13}, but for the NGC 2516 cluster. This figure is repeated from the main text (see Fig.~\ref{fig:CMD_NGC2516}) for completeness.}
    \label{fig:CMD_ngc_2516}
\end{figure*}

\begin{figure*}[!tp]
    \centering
    \includegraphics[width=0.75\textwidth]{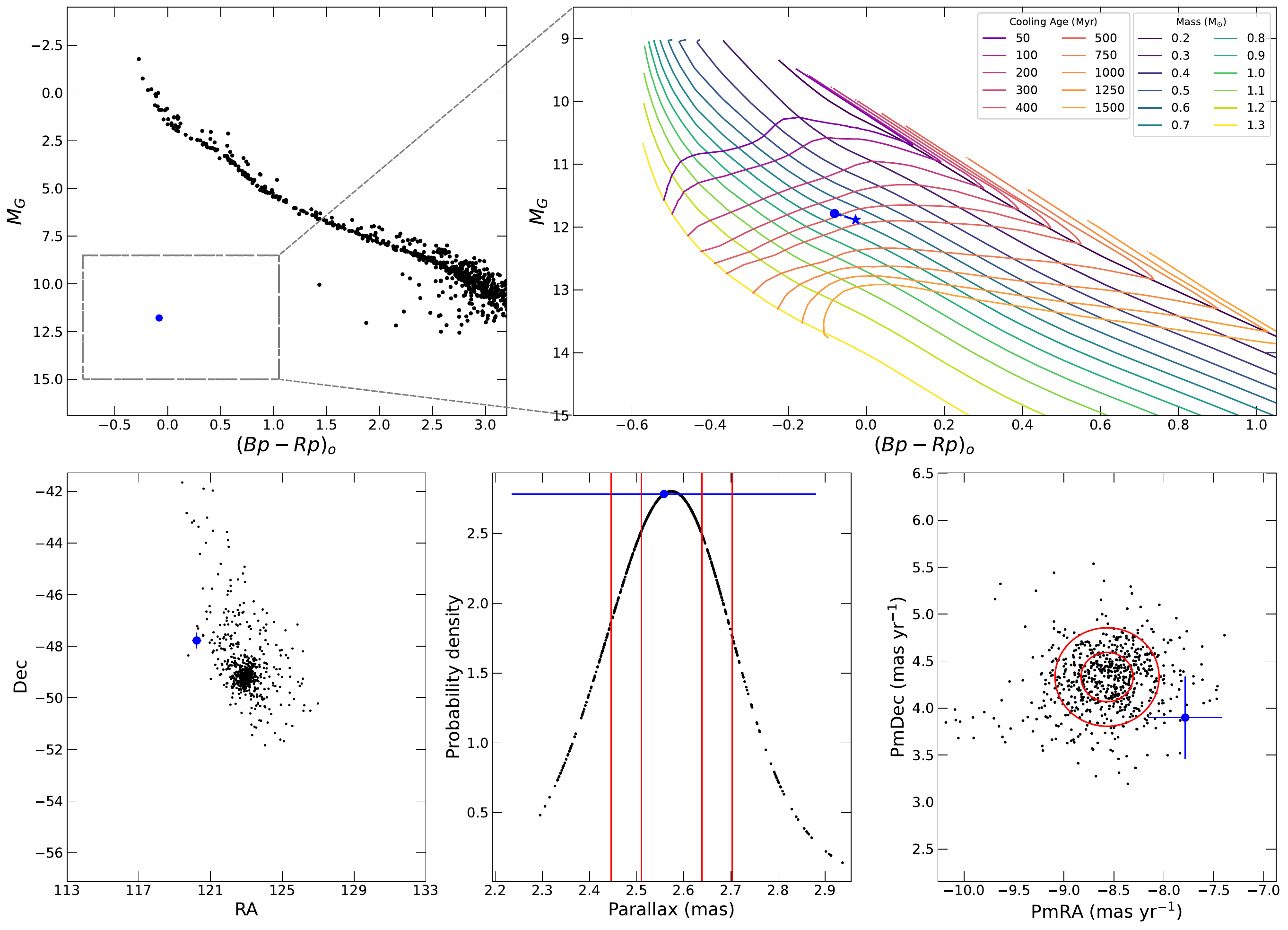}
    \caption{Same as Fig.~\ref{fig:CMD_alessi_13}, but for the NGC 2547 cluster.}
    \label{fig:CMD_ngc_2547}
\end{figure*}

\begin{figure*}[!tp]
    \centering
    \includegraphics[width=0.75\textwidth]{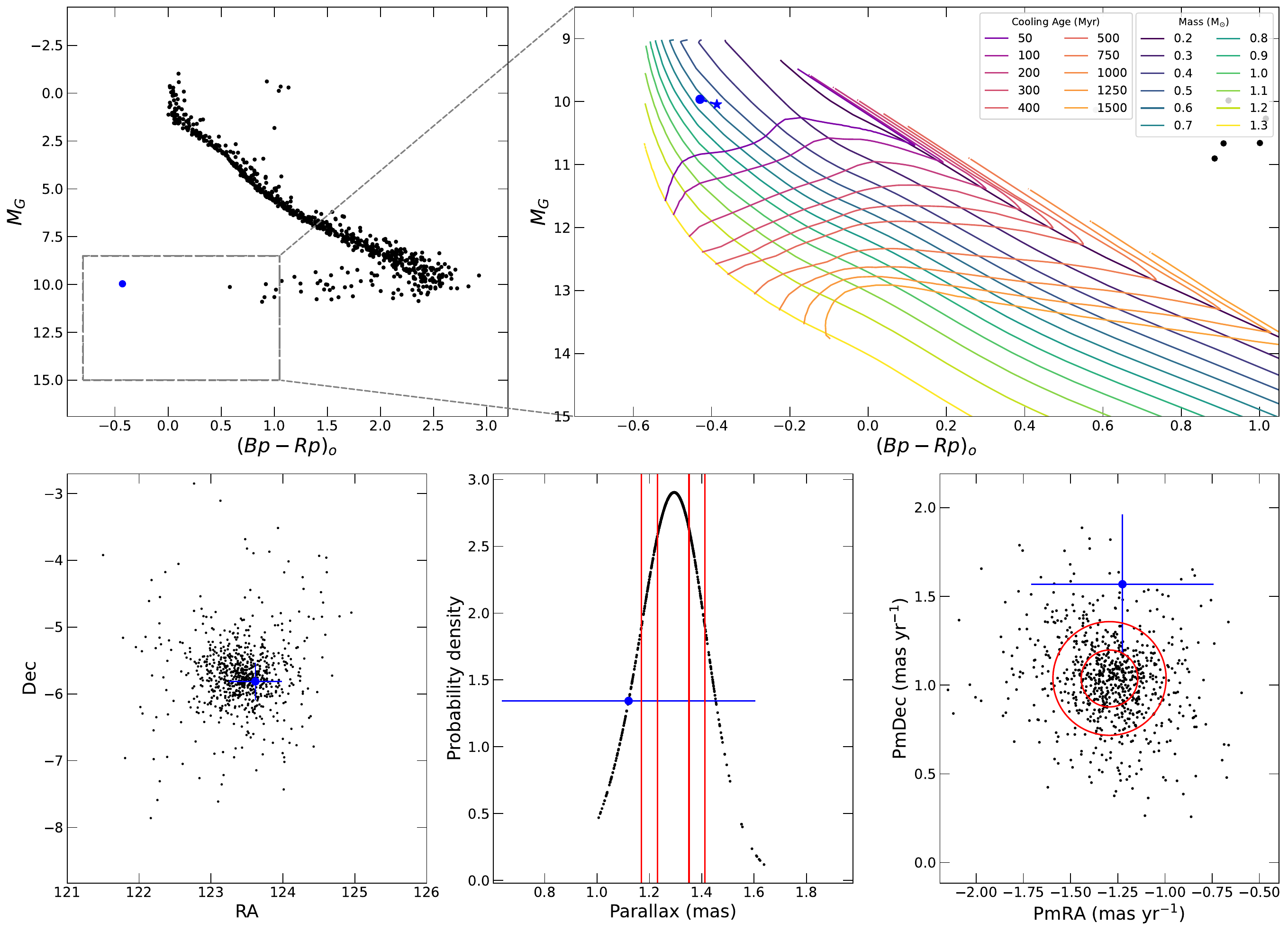}
    \caption{Same as Fig.~\ref{fig:CMD_alessi_13}, but for the NGC 2548 cluster.}
    \label{fig:CMD_ngc_2548}
\end{figure*}

\begin{figure*}[!tp]
    \centering
    \includegraphics[width=0.75\textwidth]{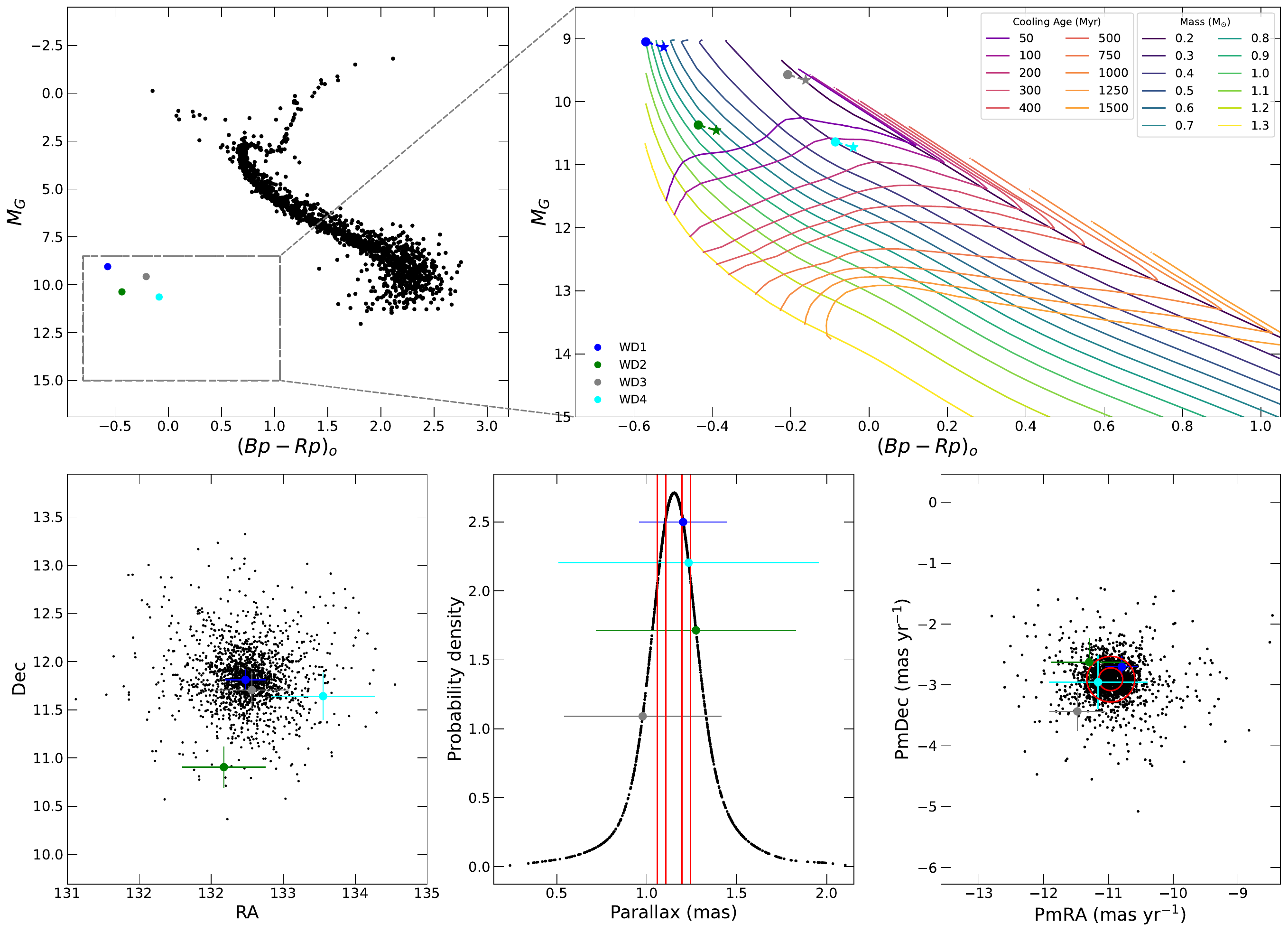}
    \caption{Same as Fig.~\ref{fig:CMD_alessi_13}, but for the NGC 2682 cluster.}
    \label{fig:CMD_ngc_2682}
\end{figure*}

\begin{figure*}[!tp]
    \centering
    \includegraphics[width=0.75\textwidth]{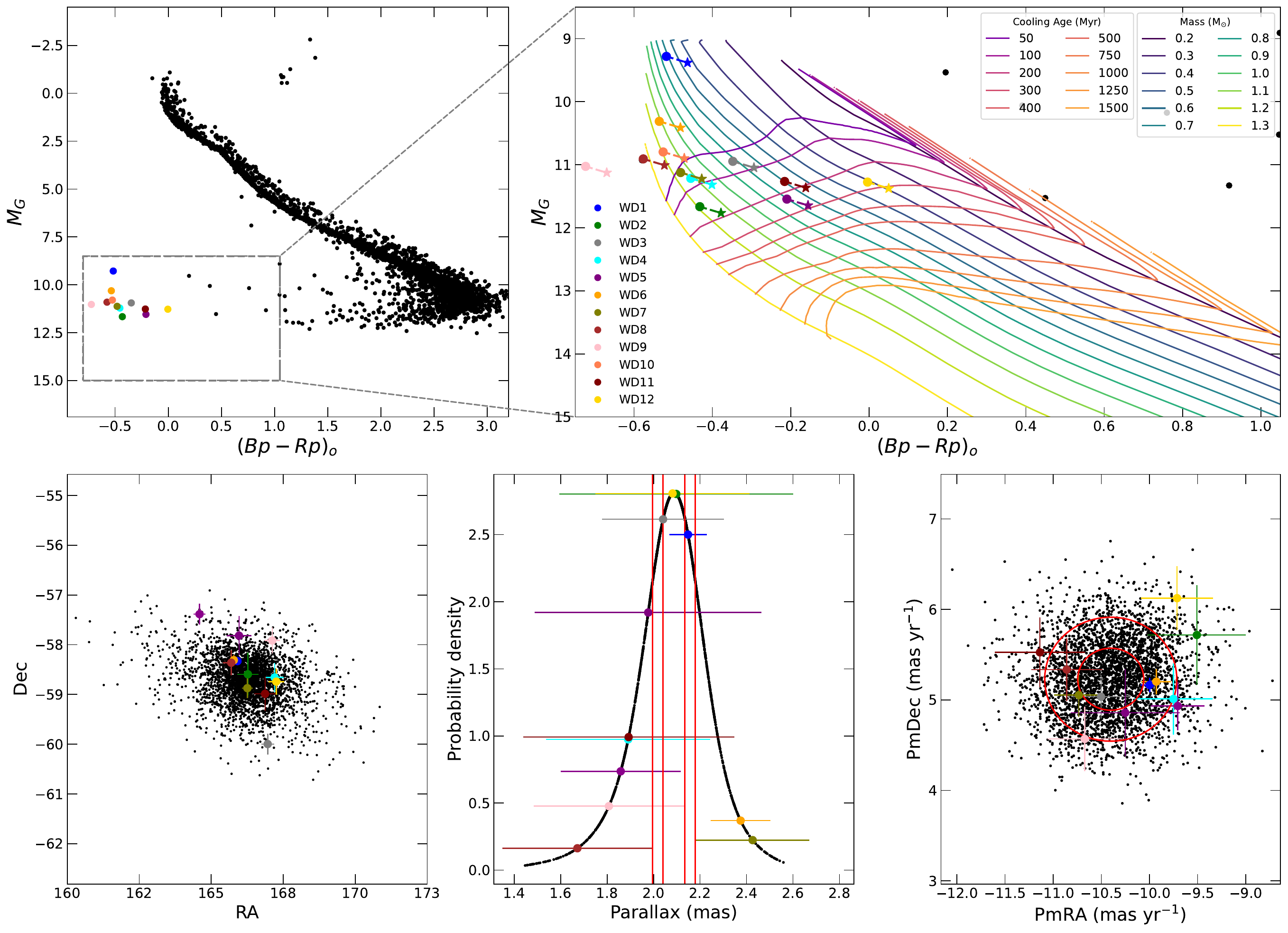}
    \caption{Same as Fig.~\ref{fig:CMD_alessi_13}, but for the NGC 3532 cluster.}
    \label{fig:CMD_ngc_3532}
\end{figure*}

\begin{figure*}[!tp]
    \centering
    \includegraphics[width=0.75\textwidth]{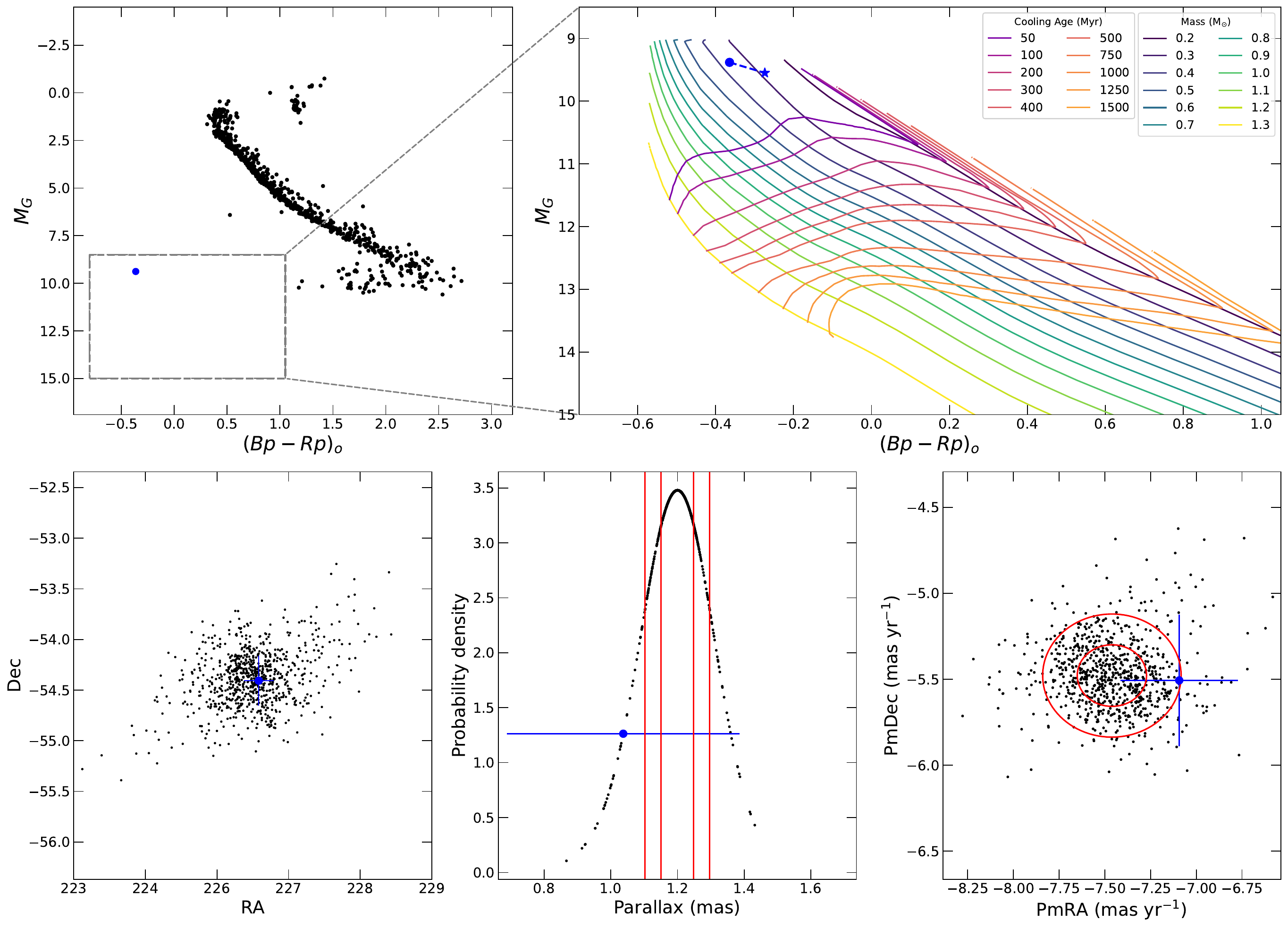}
    \caption{Same as Fig.~\ref{fig:CMD_alessi_13}, but for the NGC 5822 cluster.}
    \label{fig:CMD_ngc_5822}
\end{figure*}

\begin{figure*}[!tp]
    \centering
    \includegraphics[width=0.75\textwidth]{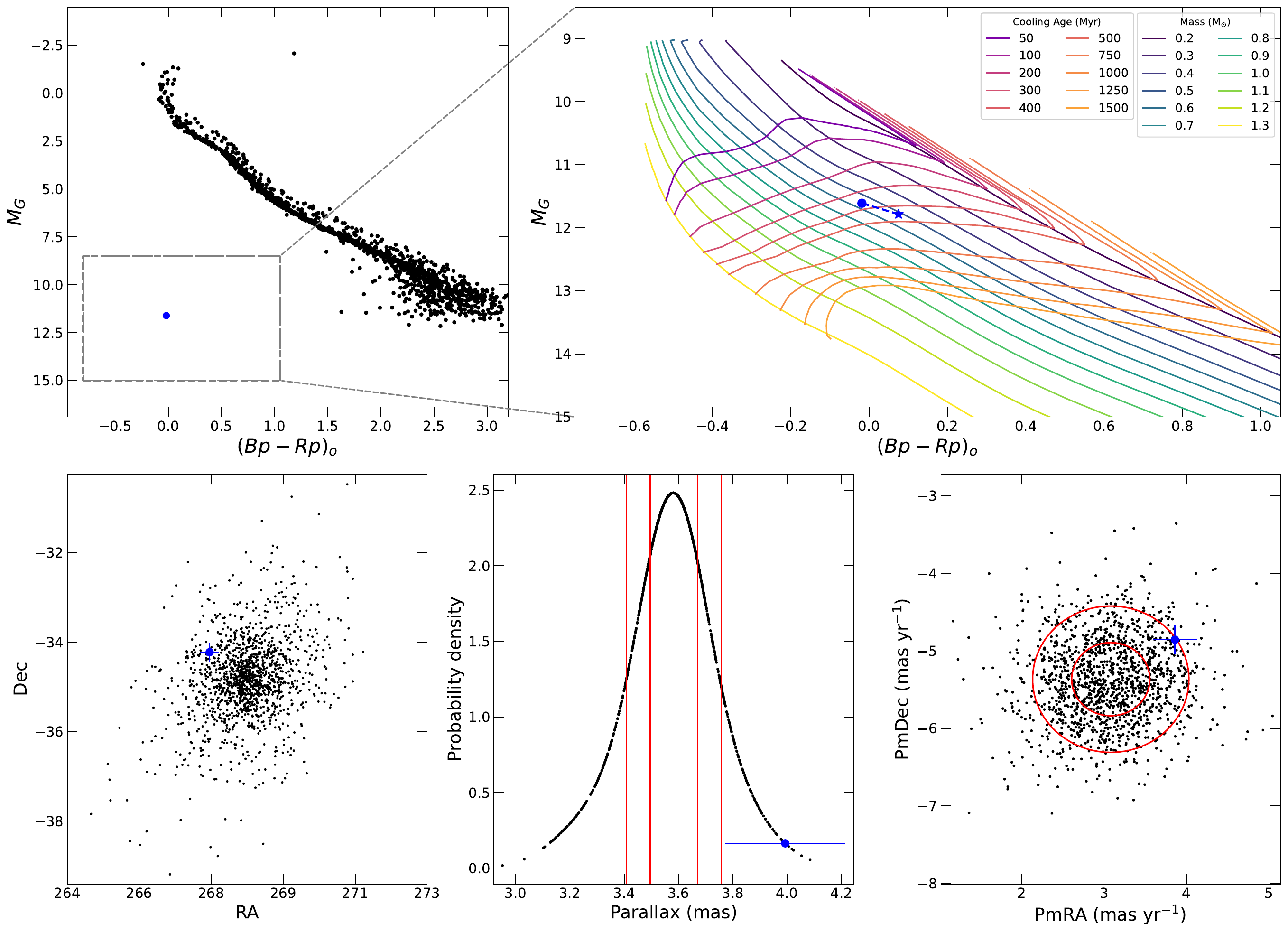}
    \caption{Same as Fig.~\ref{fig:CMD_alessi_13}, but for the NGC 6475 cluster.}
    \label{fig:CMD_ngc_6475}
\end{figure*}

\begin{figure*}[!tp]
    \centering
    \includegraphics[width=0.75\textwidth]{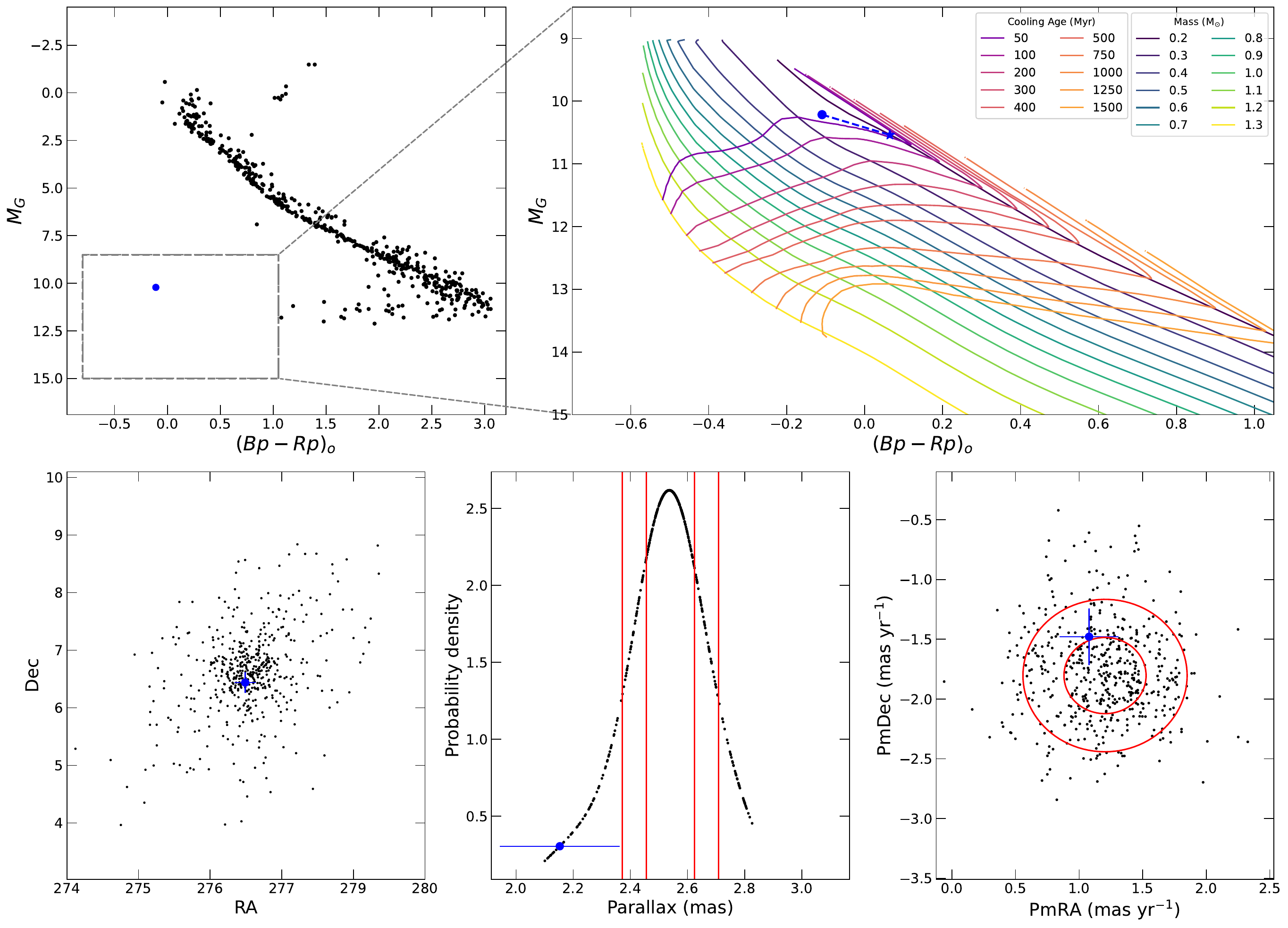}
    \caption{Same as Fig.~\ref{fig:CMD_alessi_13}, but for the NGC 6633 cluster.}
    \label{fig:CMD_ngc_6633}
\end{figure*}

\begin{figure*}[!tp]
    \centering
    \includegraphics[width=0.75\textwidth]{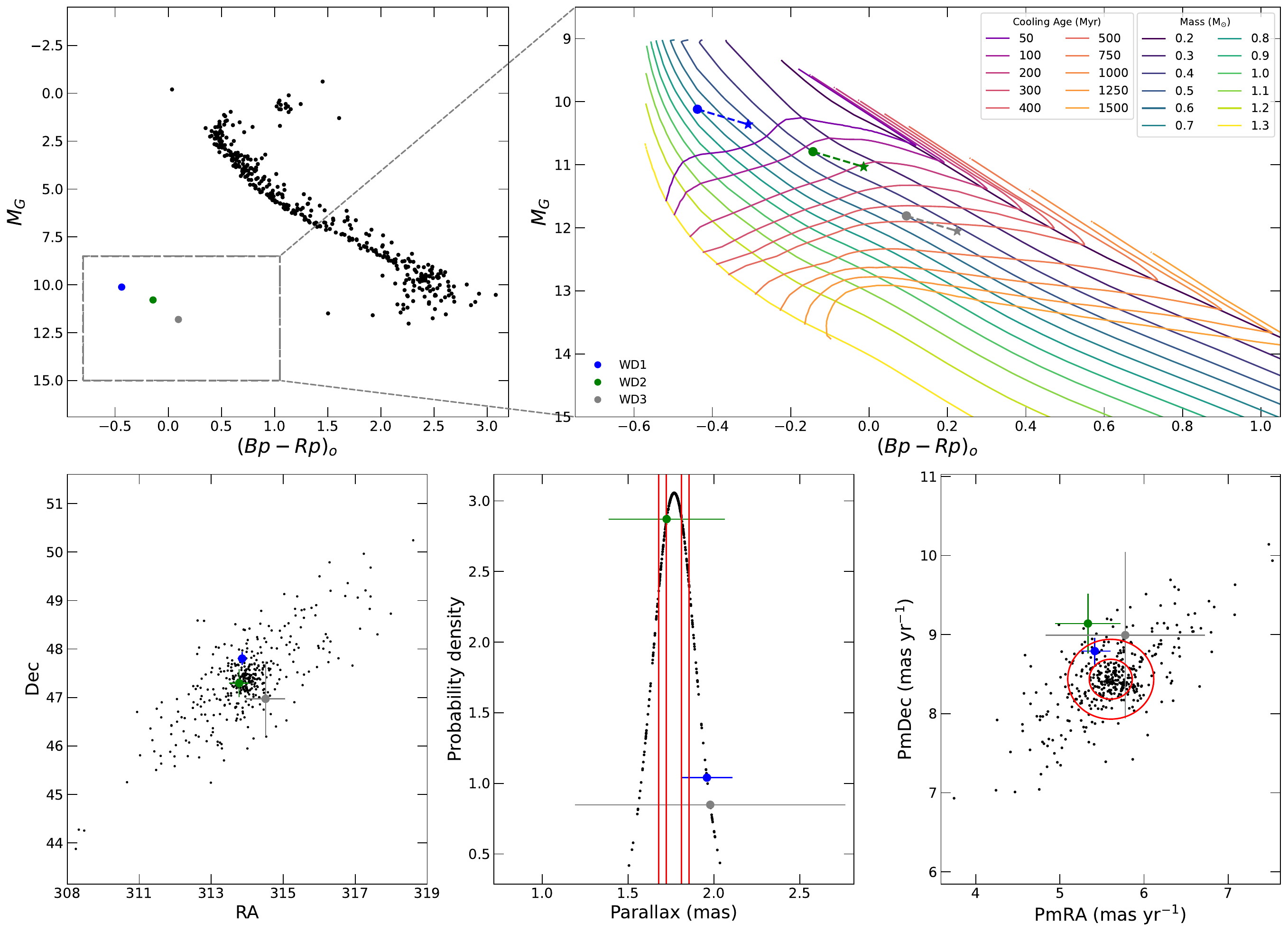}
    \caption{Same as Fig.~\ref{fig:CMD_alessi_13}, but for the NGC 6991 cluster.}
    \label{fig:CMD_ngc_6991}
\end{figure*}

\begin{figure*}[!tp]
    \centering
    \includegraphics[width=0.75\textwidth]{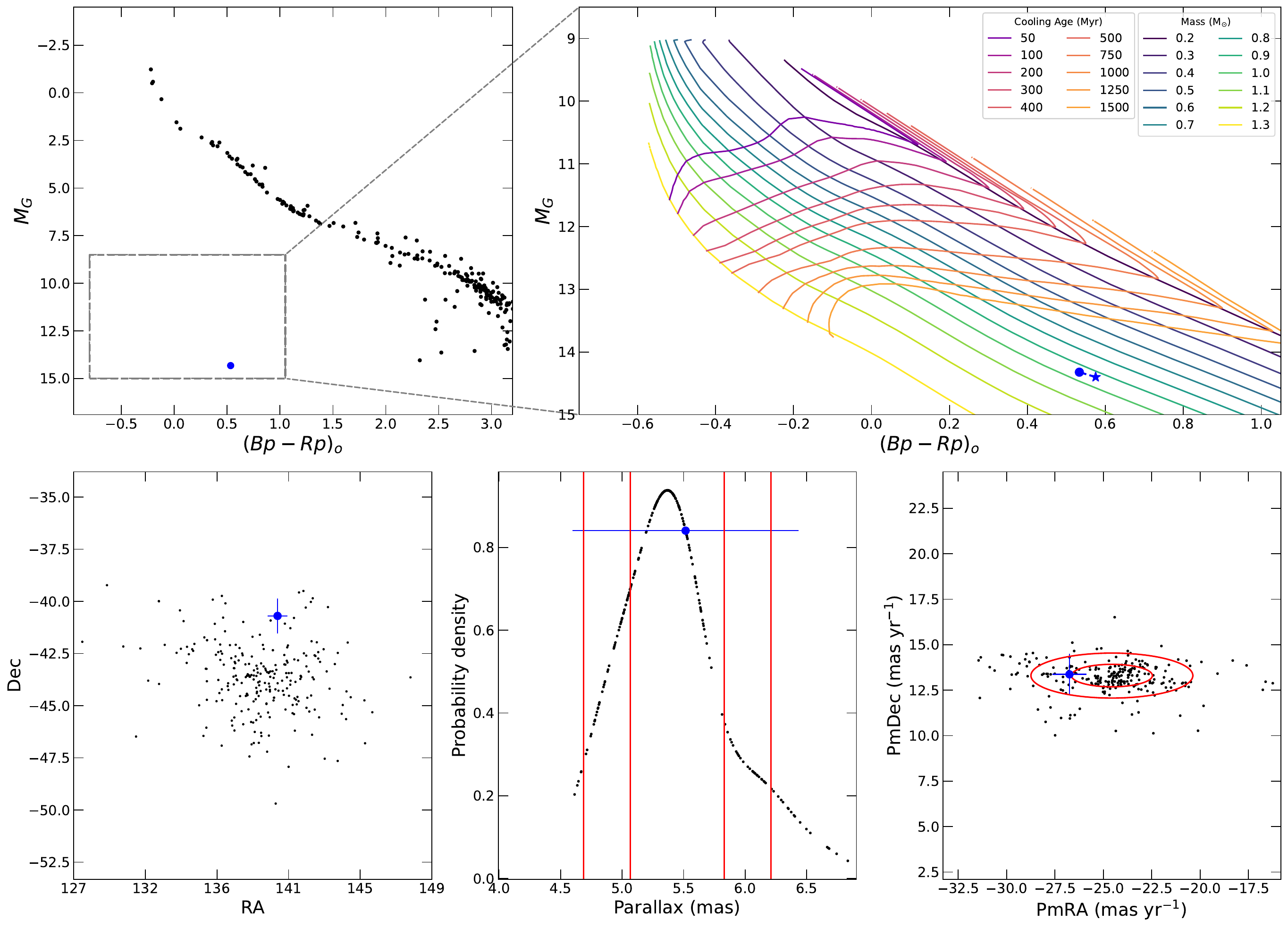}
    \caption{Same as Fig.~\ref{fig:CMD_alessi_13}, but for the Platais 9 cluster.}
    \label{fig:CMD_platais_9}
\end{figure*}

\begin{figure*}[!tp]
    \centering
    \includegraphics[width=0.75\textwidth]{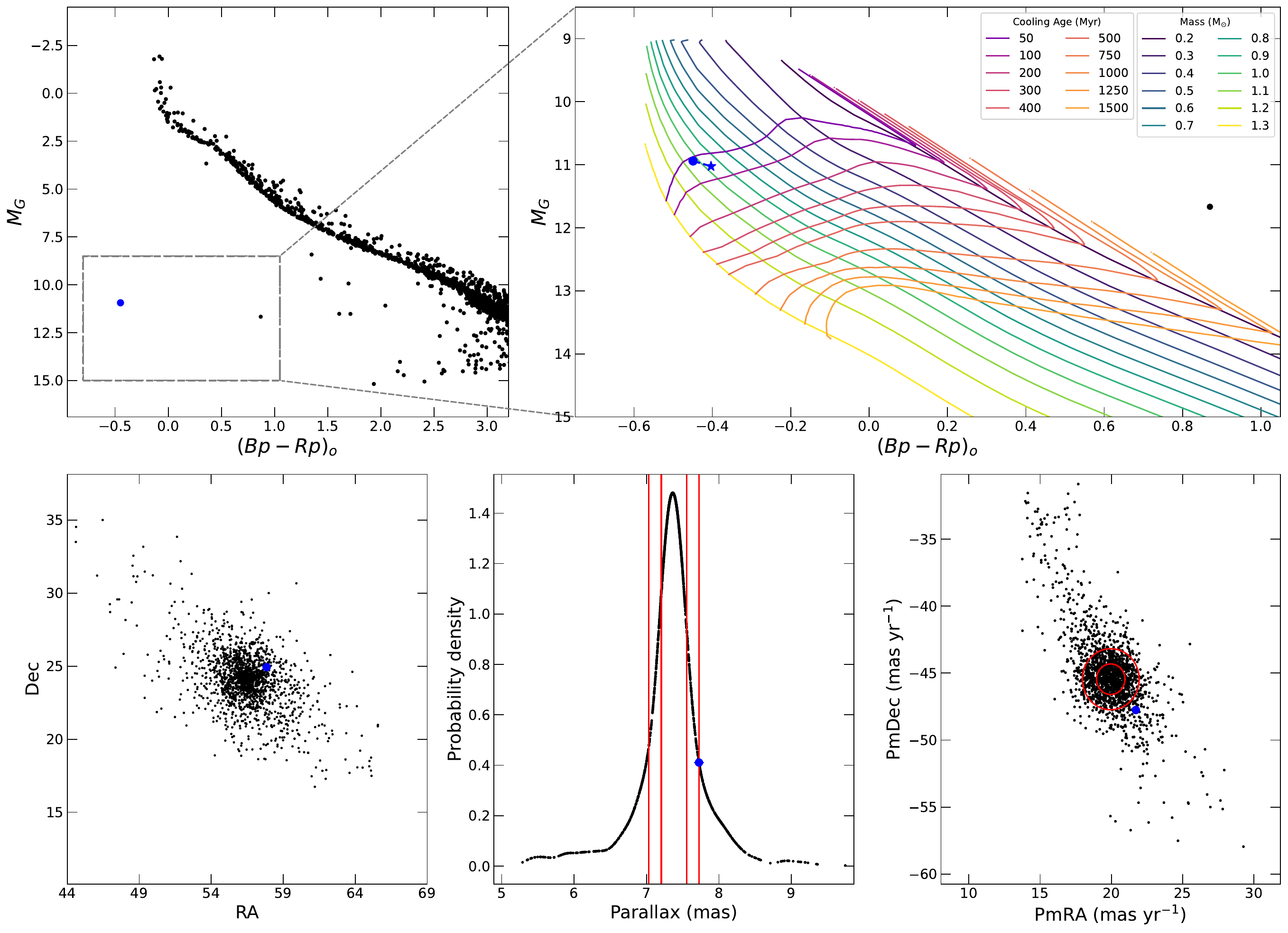}
    \caption{Same as Fig.~\ref{fig:CMD_alessi_13}, but for the Pleiades cluster.}
    \label{fig:CMD_pleiades}
\end{figure*}

\begin{figure*}[!tp]
    \centering
    \includegraphics[width=0.75\textwidth]{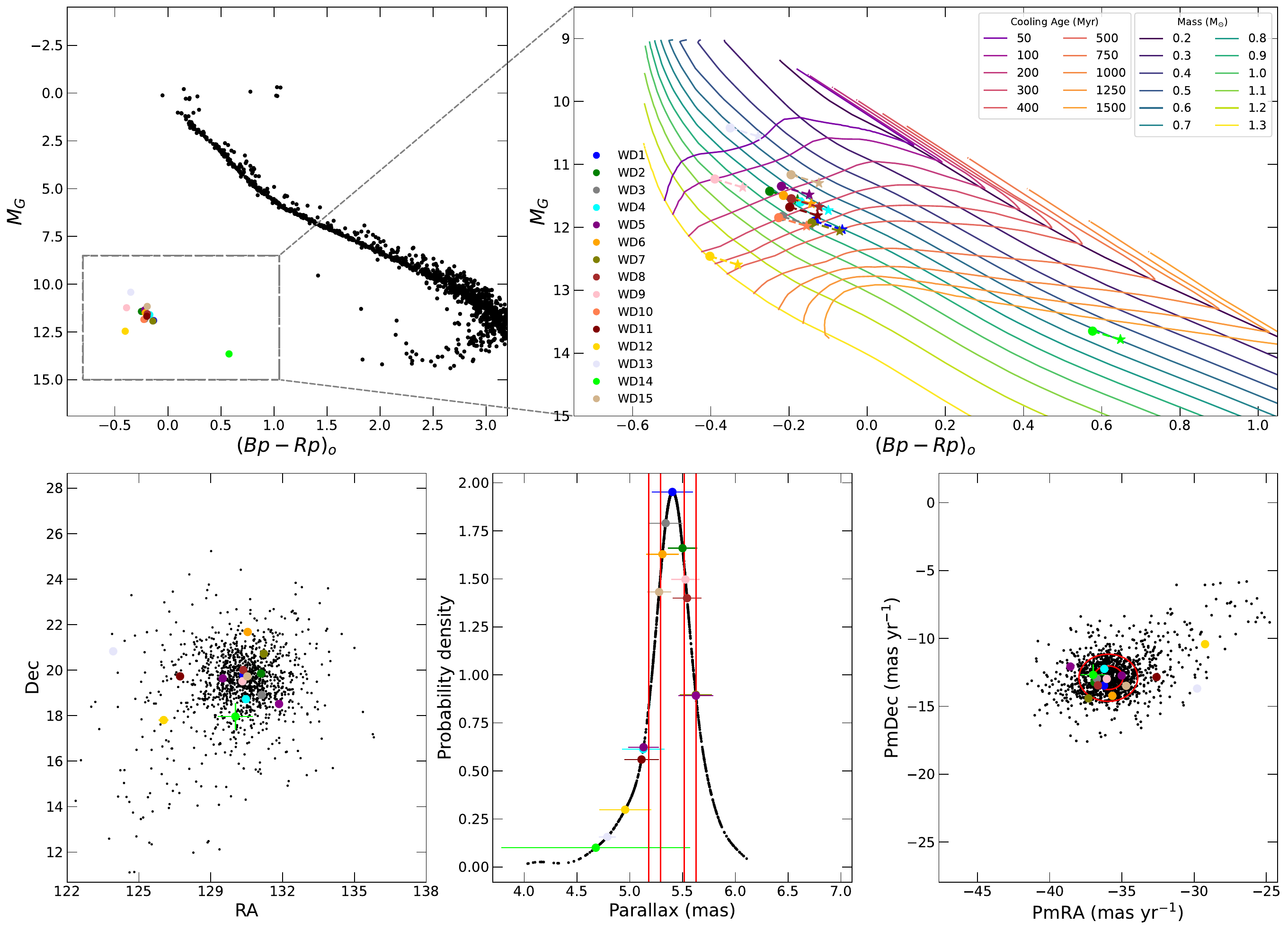}
    \caption{Same as Fig.~\ref{fig:CMD_alessi_13}, but for the Praesepe cluster.}
    \label{fig:CMD_praesepe}
\end{figure*}

\begin{figure*}[!tp]
    \centering
    \includegraphics[width=0.75\textwidth]{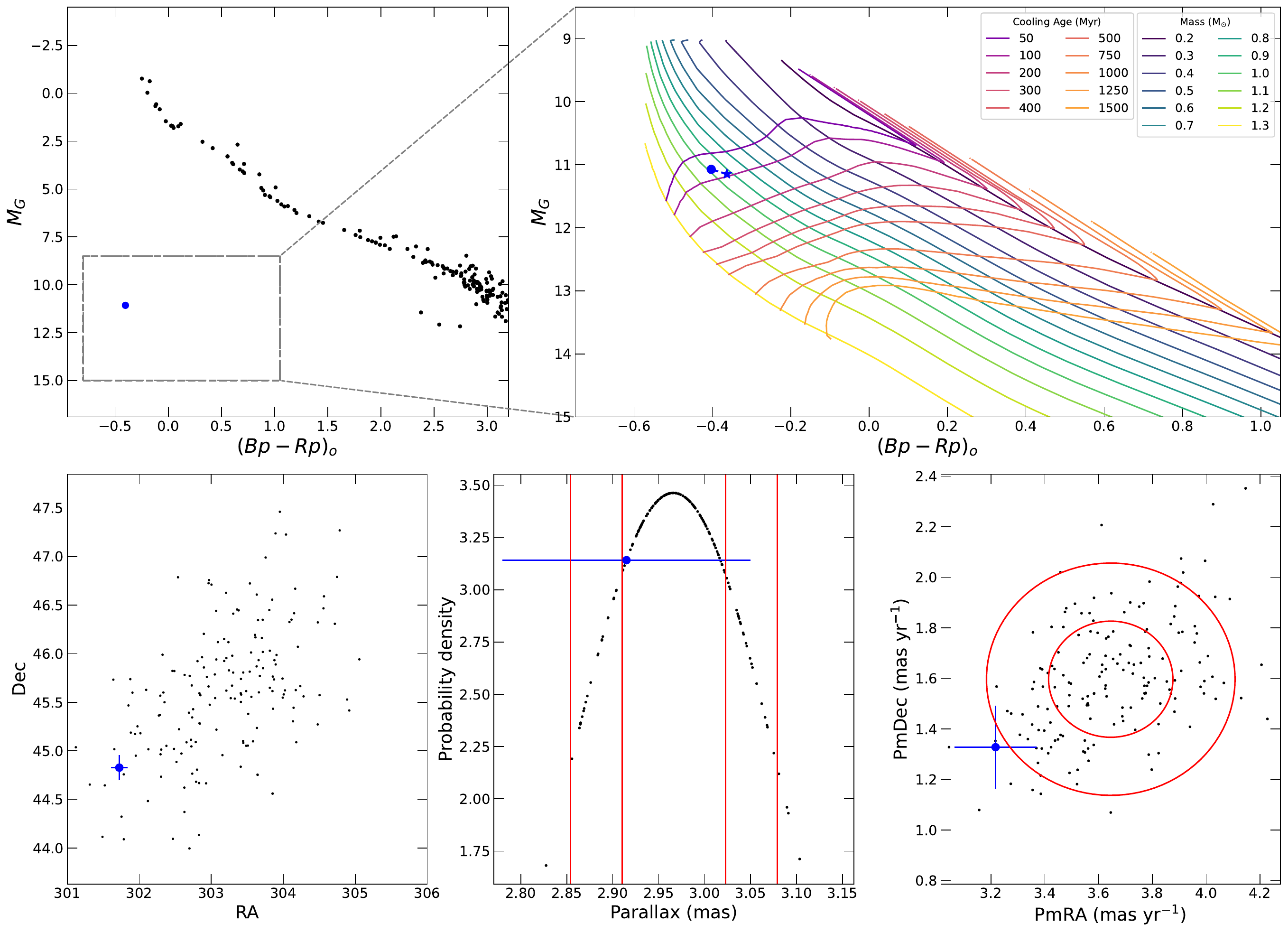}
    \caption{Same as Fig.~\ref{fig:CMD_alessi_13}, but for the RSG 5 cluster.}
    \label{fig:CMD_rsg_5}
\end{figure*}

\begin{figure*}[!tp]
    \centering
    \includegraphics[width=0.75\textwidth]{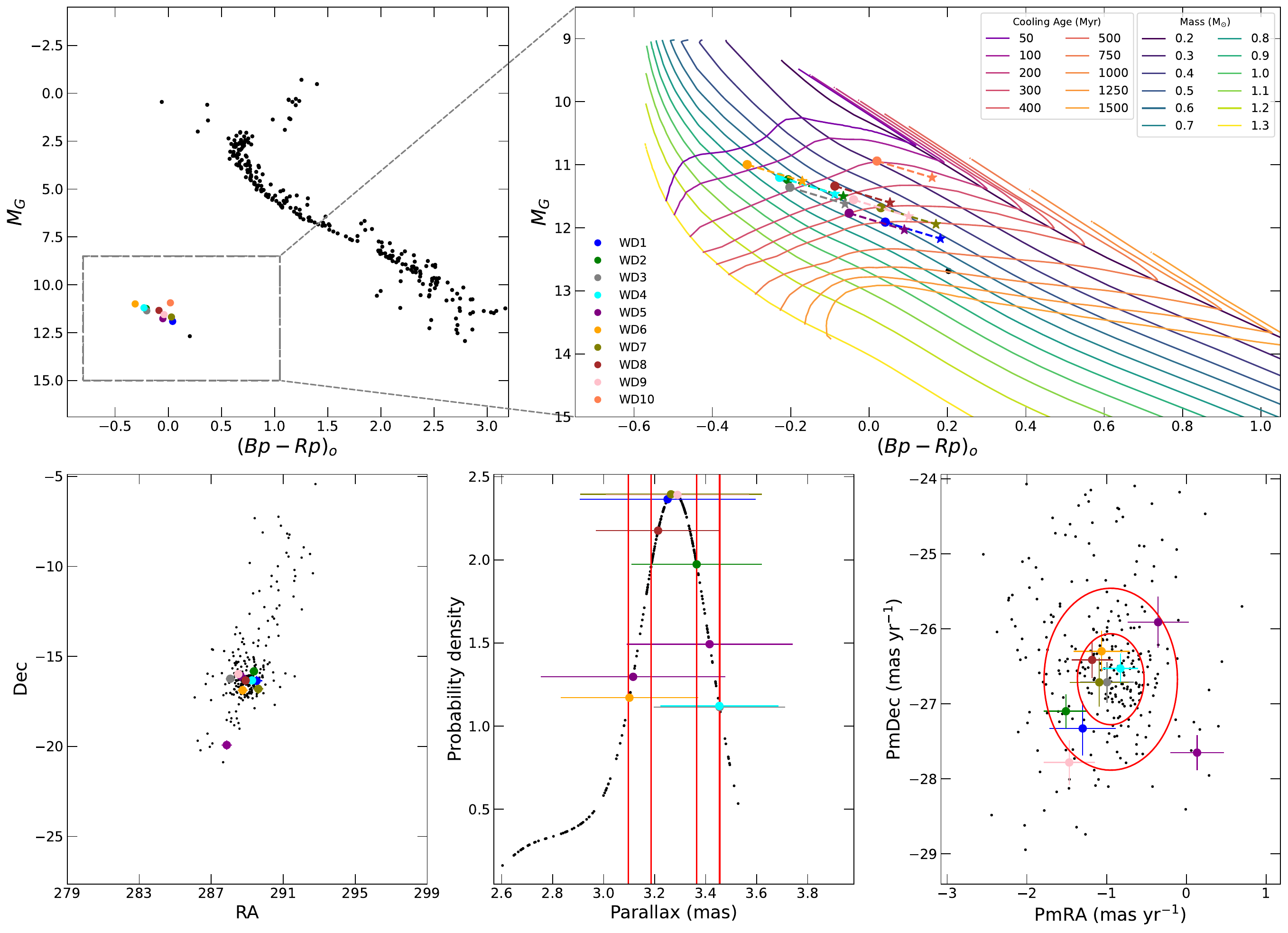}
    \caption{Same as Fig.~\ref{fig:CMD_alessi_13}, but for the Ruprecht 147 cluster.}
    \label{fig:CMD_ruprecht_147}
\end{figure*}

\begin{figure*}[!tp]
    \centering
    \includegraphics[width=0.75\textwidth]{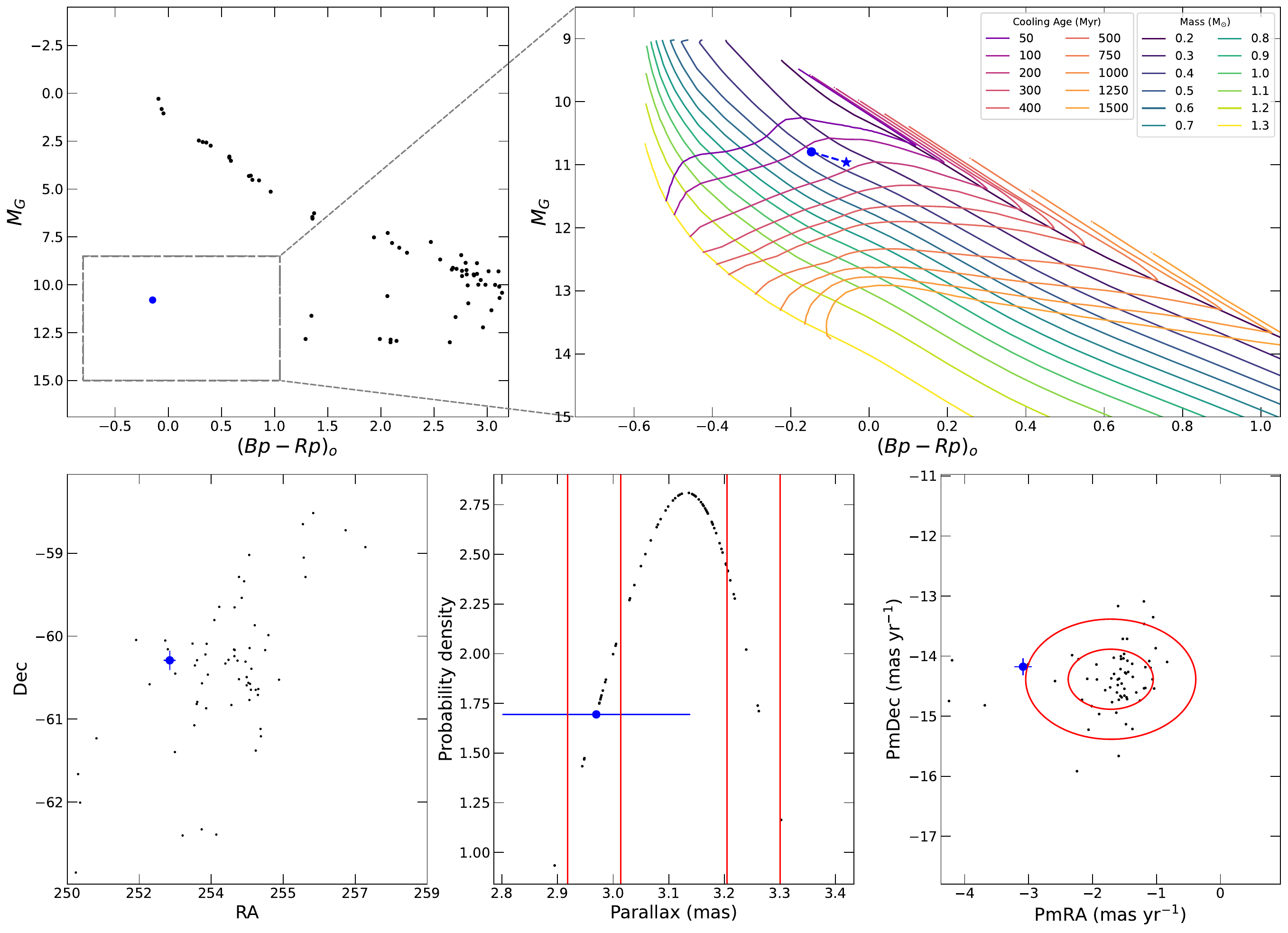}
    \caption{Same as Fig.~\ref{fig:CMD_alessi_13}, but for the UPK 624 cluster.}
    \label{fig:CMD_upk_624}
\end{figure*}

Many key figures and tables are included in the appendix. Table~\ref{tab:candidate_clusters} lists select properties for clusters with WD candidate members in the \hunt catalog. WD candidates are summarized in two appendix tables: those with existing literature spectra (Table~\ref{tab:candidates_literature}) and those we did not observe and that also lack literature spectra (Table~\ref{tab:candidates_notobserved}). Figs.~\ref{fig:spectrafits_a} and \ref{fig:spectrafits_b} show Balmer series fits for newly obtained non-magnetic DA WD spectra, including Stock 2 WD10 which is repeated from the main text (see Fig.~\ref{fig:spectrafit_stock2wd10}). Table~\ref{tab:wd_results_litsources} provides literature spectroscopic parameters for WDs with Gaia astrometric cluster membership support, while Table~\ref{tab:wd_results_nongaia} includes those that cannot be fairly assessed for membership with Gaia astrometry. Isochrone cluster age fits using the heuristic method are given in Figs.~\ref{fig:cluster_ages_1}, and \ref{fig:cluster_ages_2}, including ASCC 113, which is repeated from the main text (Fig.~\ref{fig:ASCC_113_age}). Determined cluster parameters, as well as those sourced from the literature, are summarized in Table~\ref{tab:cluster_ages_final}. Derived parameters for literature WDs with Gaia astrometric cluster membership support are provided in Table~\ref{tab:wd_results_litsources_derived}, while sources that cannot be fairly assessed with Gaia astrometry are listed in Table~\ref{tab:wd_results_nongaia_derived}. Table~\ref{tab:cummings_2018} examines the WD IFMR sources of \citet{2018ApJ...866...21C} and compares them to this work. Figs.~\ref{fig:corner_gaia_ifmr}, \ref{fig:corner_full_fit1_ifmr}, and \ref{fig:corner_full_fit2_ifmr} show corner plots for our IFMR fits. Finally, Figs.~\ref{fig:CMD_alessi_13} through \ref{fig:CMD_upk_624} show CMDs and astrometric distributions for each cluster with at least one WD candidate in our \hunt examination, as in Fig.~\ref{fig:CMD_NGC2516} of the main text.

\begin{table*}[ht]
\tiny
\centering
\caption{Properties of open clusters with candidate white dwarf members, selected from the \hunt catalog. N and N WDs indicate the total number of candidate member stars and white dwarfs, respectively. CST is the Cluster Significance Test, and class is the median homogeneity parameter. The final column provides select common alternative cluster names. $\dagger$ indicates cluster was reclassified as a moving group in \citet{2024A&A...686A..42H}.}
\label{tab:candidate_clusters}
\begin{tabular}{lccccccc}
\toprule
Cluster        & 
Distance & 
Age            & 
CST  & 
Class & 
N    & 
N WDs & 
Comments 
\\
& 
\multicolumn{1}{c}{$[\mathrm{pc}]$} &
\multicolumn{1}{c}{$[\mathrm{Myr}]$} &
\multicolumn{5}{c}{} 
\\
\midrule
Alessi 13 $\dagger$ &          104 &     $\asymnsmaller{25}{14}{13}$ & 18.8 &      0.97 &      167 &      1 &         \\
Alessi 22 $\dagger$ &          354 &  $\asymnsmaller{672}{473}{229}$ &  8.9 &      0.92 &       49 &      1 &         \\
Alessi 62 &          606 &  $\asymnsmaller{425}{336}{161}$ & 25.1 &      1.00 &      285 &      1 &         \\
Alessi 94 $\dagger$ &          553 &   $\asymnsmaller{184}{133}{76}$ & 12.3 &      0.99 &       86 &      2 &         \\
BH 99 &          440 &     $\asymnsmaller{45}{43}{19}$ & 32.4 &      0.99 &      643 &      1 &  vdB Hagen 99       \\
CWNU 41 &          729 &  $\asymnsmaller{629}{413}{234}$ &  8.1 &      0.87 &       44 &      1 &         \\
CWNU 515 $\dagger$ &          203 &      $\asymnsmaller{21}{17}{9}$ & 13.0 &      1.00 &       69 &      1 &         \\
CWNU 1012 $\dagger$ &          284 &   $\asymnsmaller{135}{102}{65}$ & 22.8 &      0.96 &      225 &      1 &         \\
CWNU 1066 $\dagger$ &           99 &     $\asymnsmaller{93}{88}{46}$ & 10.2 &      0.97 &       56 &      1 &         \\
CWNU 1095 $\dagger$ &          245 &  $\asymnsmaller{254}{239}{101}$ & 18.1 &      0.95 &      163 &      1 &         \\
Haffner 13 &          552 &       $\asymnsmaller{15}{8}{7}$ & 34.6 &      1.00 &      726 &      1 &         \\
HSC 242 $\dagger$ &          120 &   $\asymnsmaller{119}{146}{54}$ &  9.4 &      1.00 &       62 &      1 &         \\
HSC 381 $\dagger$ &          313 &     $\asymnsmaller{31}{20}{17}$ & 22.0 &      0.96 &      288 &      1 &         \\
HSC 448 $\dagger$ &          344 &    $\asymnsmaller{103}{67}{51}$ & 12.2 &      0.99 &      163 &      2 &         \\
HSC 452 $\dagger$ &          238 &   $\asymnsmaller{117}{108}{52}$ &  6.8 &      0.93 &       39 &      1 &         \\
HSC 601 $\dagger$ &          594 &     $\asymnsmaller{77}{49}{38}$ & 11.6 &      0.92 &      120 &      1 &         \\      
HSC 1155 $\dagger$ &          283 &   $\asymnsmaller{203}{197}{90}$ &  8.2 &      1.00 &       96 &      1 &         \\
HSC 1555 $\dagger$ &          173 &   $\asymnsmaller{128}{110}{59}$ &  6.2 &      0.95 &       74 &      2 &         \\
HSC 1630 $\dagger$ &          134 &   $\asymnsmaller{166}{239}{68}$ &  8.9 &      0.57 &       94 &      1 &         \\
HSC 1710 $\dagger$ &          137 &   $\asymnsmaller{125}{148}{72}$ &  6.5 &      0.81 &       34 &      1 &         \\
HSC 2139 $\dagger$ &          188 &       $\asymnsmaller{12}{7}{6}$ & 12.2 &      0.99 &       67 &      1 &         \\
HSC 2156 $\dagger$ &          176 &      $\asymnsmaller{19}{15}{9}$ &  6.4 &      0.76 &       41 &      2 &         \\
HSC 2263 $\dagger$ &          274 &     $\asymnsmaller{81}{55}{44}$ & 10.0 &      0.92 &       39 &      1 &         \\
HSC 2304 &         1070 &   $\asymnsmaller{187}{131}{75}$ & 12.9 &      1.00 &      116 &      1 &         \\
HSC 2609 $\dagger$ &          360 &    $\asymnsmaller{108}{81}{53}$ &  6.0 &      0.92 &       28 &      1 &         \\
HSC 2630 $\dagger$ &          180 &        $\asymnsmaller{8}{5}{4}$ &  8.2 &      0.98 &       29 &      1 &         \\
HSC 2971 $\dagger$ &          102 &  $\asymnsmaller{238}{168}{110}$ &  9.1 &      0.93 &       57 &      1 &         \\
Hyades     &           47 &  $\asymnsmaller{577}{413}{223}$ & 27.4 &      1.00 &      927 &     10 & Melotte 25 in \hunt   \\
IC 4665 &          343 &     $\asymnsmaller{33}{20}{14}$ & 24.1 &      0.99 &      308 &      1 &         \\
IC 4756 &          467 &  $\asymnsmaller{837}{502}{282}$ & 37.5 &      0.99 &      718 &      2 &         \\
LISC 3534 &          453 &   $\asymnsmaller{149}{109}{77}$ & 29.1 &      1.00 &      430 &      1 &         \\
LISC-III 3668 $\dagger$ &          574 &     $\asymnsmaller{29}{20}{12}$ & 26.0 &      0.91 &      356 &      1 &         \\
Mamajek 4 &          444 &  $\asymnsmaller{312}{224}{140}$ & 34.2 &      0.99 &      655 &      1 &         \\
Melotte 111 &           85 &  $\asymnsmaller{654}{433}{236}$ & 21.8 &      0.99 &      271 &      1 & Coma Berenices \\
NGC 752 &          434 & $\asymnsmaller{1{,}421}{913}{487}$ & 31.6 &      0.99 &      442 &      2 &         \\
NGC 1901 &          416 &  $\asymnsmaller{283}{225}{119}$ & 22.6 &      1.00 &      231 &      1 &         \\
NGC 2099 &         1402 &  $\asymnsmaller{486}{246}{165}$ & 99.0 &      1.00 &     2424 &      1 & M37      \\
NGC 2516 &          407 &   $\asymnsmaller{125}{127}{52}$ & 99.0 &      0.99 &     3784 &      6 &         \\
NGC 2547 &          382 &      $\asymnsmaller{22}{19}{9}$ & 32.4 &      1.00 &      675 &      1 &         \\
NGC 2548 &          747 &  $\asymnsmaller{379}{293}{142}$ & 99.0 &      1.00 &      778 &      1 &         \\
NGC 2682 &          837 & $\asymnsmaller{1{,}688}{945}{531}$ & 99.0 &      0.98 &     $1{,}844$ &      4 & M67      \\
NGC 3532 &          472 &   $\asymnsmaller{238}{218}{93}$ & 99.0 &      1.00 &     $3{,}414$ &     12 &         \\
NGC 5822 &          808 & $\asymnsmaller{1{,}040}{643}{330}$ & 99.0 &      1.00 &      804 &      1 &         \\
NGC 6475 &          276 &   $\asymnsmaller{171}{163}{59}$ & 99.0 &      1.00 &     $1{,}427$ &      1 &         \\
NGC 6633 &          390 &  $\asymnsmaller{472}{326}{172}$ & 31.5 &      1.00 &      515 &      1 &         \\
NGC 6991 &          557 &  $\asymnsmaller{989}{630}{345}$ & 29.1 &      0.94 &      395 &      3 &         \\
Platais 9 &          185 &     $\asymnsmaller{37}{29}{16}$ & 20.7 &      0.95 &      251 &      1 &         \\
Pleiades   &          135 &    $\asymnsmaller{122}{75}{66}$ & 99.0 &      1.00 &     $1{,}721$ &      1 & Melotte 22 in \hunt \\
Praesepe &          183 &  $\asymnsmaller{346}{275}{145}$ & 99.0 &      1.00 &     $1{,}314$ &     15 & NGC 2632 in \hunt, Beehive, M44 \\
RSG 5 &          334 &     $\asymnsmaller{23}{13}{12}$ & 18.2 &      1.00 &      170 &      1 &         \\
Ruprecht 147 &          303 &  $\asymnsmaller{889}{420}{319}$ & 24.4 &      0.93 &      279 &     10 & NGC 6774 \\
Stock 1 &          402 &   $\asymnsmaller{258}{178}{99}$ & 24.7 &      1.00 &      290 &      1 &         \\
Stock 2 &          370 &  $\asymnsmaller{290}{165}{129}$ & 99.0 &      0.99 &     $3{,}011$ &     20 &         \\
Stock 12 &          431 &    $\asymnsmaller{124}{78}{60}$ & 22.9 &      1.00 &      273 &      1 &         \\
Theia 172 &          465 &   $\asymnsmaller{121}{100}{60}$ & 26.5 &      1.00 &      322 &      1 &         \\
Theia 248 $\dagger$ &          477 &    $\asymnsmaller{112}{77}{45}$ & 16.9 &      0.97 &      153 &      1 &         \\
Theia 517 &          296 &    $\asymnsmaller{127}{95}{60}$ & 31.6 &      1.00 &      511 &      1 & M39, NGC 7092 \\
Theia 558 &          683 &   $\asymnsmaller{202}{135}{95}$ & 13.0 &      0.99 &       96 &      1 &         \\
Theia 817 $\dagger$ &          360 &  $\asymnsmaller{196}{150}{100}$ &  8.3 &      0.94 &      120 &      2 &         \\
Theia 1315 $\dagger$ &          432 &     $\asymnsmaller{42}{33}{18}$ &  7.1 &      0.97 &       86 &      1 &         \\
Turner 5 $\dagger$ &          412 &   $\asymnsmaller{164}{126}{72}$ & 16.9 &      0.98 &      116 &      1 &         \\
UPK 5 &          552 &     $\asymnsmaller{83}{44}{37}$ & 16.9 &      1.00 &      156 &      1 &         \\
UPK 303 $\dagger$ &          211 &    $\asymnsmaller{111}{94}{57}$ & 21.4 &      0.98 &      261 &      1 &         \\
UPK 552 $\dagger$ &          314 &    $\asymnsmaller{108}{86}{49}$ & 14.1 &      1.00 &       79 &      1 &         \\
UPK 624 $\dagger$ &          314 &     $\asymnsmaller{25}{14}{12}$ & 11.5 &      0.97 &       66 &      1 &        \\

\bottomrule
\end{tabular}
\end{table*}

\begin{table*}[ht]
\tiny
\centering
\caption{As in Table~\ref{tab:candidates_observed}, but for candidates with available literature spectra. The final column indicates whether it was included in the Gaia-based IFMR of \citet{2022ApJ...926L..24M}; if not, it indicates inclusion in the IFMR of \citet{2018ApJ...866...21C}. We also list the spectral type for non-DA WDs.}
\label{tab:candidates_literature}
\begin{tabular}{lcccccccccc}
\toprule
Name & 
Gaia DR3 Source ID &
RA & 
Dec & 
$G_\mathrm{obs}$ & 
$t_\mathrm{cool}$ &
$M_\mathrm{WD}$ &
$M_\mathrm{prog}$ &
$t_\mathrm{total}$ & 
$\tfrac{t_\mathrm{total}-\mathrm{Age}_\mathrm{cl}}{\sigma_\mathrm{Age,cl}}$ & 
Comments \\
      & 
      & 
[deg] & 
[deg] & 
[mag] & 
[Myr] & 
[$M_\odot$] & 
[$M_\odot$] & 
[Myr] & 
[$\sigma_\mathrm{Age,cl}$]  &  
\\
\midrule
Hyades WD1 & 3308403897837092992 & 70.100 & 13.979 & 14.91 & 198 & 0.81 & 3.20 & 564 & -0.1 & in \cite{2022ApJ...926L..24M} IFMR, DBA \\
Hyades WD2 & 39305036729495936 & 60.926 & 14.991 & 15.02 & 309 & 0.87 & 3.68 & 550 & -0.1 & in \cite{2022ApJ...926L..24M} IFMR \\
Hyades WD3 & 3302846072717868416 & 58.842 & 9.788  & 14.55 & 279 & 0.83 & 3.32 & 603 & 0.1  & in \cite{2022ApJ...926L..24M} IFMR \\
Hyades WD4 & 218783542413339648 & 58.001 & 34.124 & 15.18 & 368 & 0.88 & 3.75 & 585 & 0.0  & in \cite{2022ApJ...926L..24M} IFMR \\
Hyades WD6 & 48522895539002624 & 63.288 & 18.994 & 16.27 & 653 & 0.88 & 3.80 & 860 & 0.7  & DC              \\
Melotte 111 & 4008511467191955840 & 184.734 & 25.766 & 16.61 & 337 & 0.89 & 3.91 & 547 & -0.5 &                 \\
NGC 2099 & 3451205783698632704 & 88.109 & 32.624 & 19.15 & 1   & 0.68 & 2.09 & $1{,}232$ & 3.0 &                 \\
NGC 2516 WD1 & 5290767695648992128 & 119.705 & -60.806 & 19.10 & 63  & 0.93 & 4.34 & 218 & 0.7  & in \cite{2018ApJ...866...21C} IFMR \\
NGC 2516 WD2 & 5290834387897642624 & 119.474 & -60.480 & 19.12 & 56  & 0.77 & 2.92 & 530 & 3.2  & in \cite{2022ApJ...926L..24M} IFMR \\
NGC 2516 WD3 & 5290720695823013376 & 119.461 & -60.832 & 18.95 & 79  & 0.85 & 3.51 & 350 & 1.8  & in \cite{2018ApJ...866...21C} IFMR \\
NGC 2516 WD6 & 5290719287073728128 & 119.263 & -60.915 & 19.34 & 146 & 0.57 & 1.06 & $7{,}307$ & 56.6 & \\
NGC 2682 WD1 & 604917698073661952 & 132.833 & 11.811 & 18.74 & 0   & 1.00 & 5.33 & 91  & -3.0 & photometric nonmember\\
NGC 2682 WD3 & 604898490980773888  & 132.895 &  11.705 & 20.16 & 6     & 0.23 & —    & —         & —      & DB \\
NGC 3532 WD6 & 5340219811654824448 & 165.905 & -58.311 & 18.53 & 1   & 1.11 & 6.18 & 65  & -1.9 & in \cite{2018ApJ...866...21C} IFMR \\
NGC 3532 WD7 & 5338652084186678400 & 166.399 & -58.874 & 19.30 & 60  & 1.15 & 6.67 & 114 & -1.3 & in \cite{2018ApJ...866...21C} IFMR \\
NGC 3532 WD8 & 5338718261060841472 & 165.813 & -58.362 & 19.89 & 335 & 1.29 & 7.97 & 371 & 0.6  & in \cite{2018ApJ...866...21C} IFMR \\
Pleiades & 66697547870378368 & 58.047 & 24.930 & 16.59 & 56  & 1.03 & 5.70 & 134 & 0.2  & in \cite{2022ApJ...926L..24M} IFMR \\
Praesepe WD1 & 661311267210542080 & 129.947 & 19.770 & 18.37 & 385 & 0.84 & 3.38 & 683 & 1.2  & in \cite{2022ApJ...926L..24M} IFMR \\
Praesepe WD2 & 661353224747229184 & 130.718 & 19.853 & 17.86 & 182 & 0.82 & 3.27 & 518 & 0.6  & in \cite{2022ApJ...926L..24M} IFMR \\
Praesepe WD3 & 661010005319096192 & 130.742 & 18.910 & 18.32 & 298 & 0.93 & 4.38 & 448 & 0.4  & in \cite{2022ApJ...926L..24M} IFMR \\
Praesepe WD4 & 659494049367276544 & 130.055 & 18.724 & 18.18 & 266 & 0.76 & 2.86 & 767 & 1.5  & in \cite{2022ApJ...926L..24M} IFMR \\
Praesepe WD5 & 662798086105290112 & 129.042 & 19.638 & 17.93 & 180 & 0.74 & 2.67 & 792 & 1.6  & in \cite{2022ApJ...926L..24M} IFMR \\
Praesepe WD6 & 665139697978259200 & 130.131 & 21.678 & 18.00 & 216 & 0.79 & 3.06 & 630 & 1.0  & \\
Praesepe WD7 & 661841163095377024 & 130.841 & 20.725 & 18.31 & 390 & 0.86 & 3.54 & 656 & 1.1  & \\
Praesepe WD8 & 664325543977630464 & 129.940 & 20.004 & 17.96 & 239 & 0.78 & 2.99 & 683 & 1.2  & \\
Praesepe WD9 & 661270898815358720 & 129.902 & 19.512 & 17.65 & 101 & 1.00 & 5.40 & 189 & -1.1 & \\
Praesepe WD10 & 660178942032517760 & 131.508 & 18.513 & 18.23 & 297 & 0.95 & 4.66 & 427 & 0.3  & in \cite{2022ApJ...926L..24M} IFMR \\
Praesepe WD12 & 662152977721471488 & 126.448 & 17.805 & 19.12 & 353 & 1.29 & 7.97 & 389 & 0.2  & DC \\
Praesepe WD13 & 675767959626673152 & 124.234 & 20.829 & 17.15 & 27  & 0.62 & 1.54 & $2{,}464$ & 7.7  & DB \\
Praesepe WD15 & 661297901272035456 & 130.117 & 19.726 & 17.68 & 154 & 0.62 & 1.47 & $2{,}911$ & 9.3  & \\
Ruprecht 147 WD2 & 4184169822810795648 & 289.507 & -15.832 & 18.87 & 164 & 0.67 & 1.99 & $1{,}479$ & 1.4  & \\
Ruprecht 147 WD3 & 4088108859141437056 & 288.219 & -16.243 & 18.93 & 191 & 0.71 & 2.43 & $1{,}001$ & 0.3  & DB \\
Ruprecht 147 WD5 & 4184148073089506304 & 288.765 & -16.062 & 19.56 & 403 & 0.67 & 2.01 & $1{,}674$ & 1.9  & \\
Ruprecht 147 WD6 & 4087806832745520128 & 288.891 & -16.880 & 18.80 & 87  & 0.77 & 2.93 & 557 & -1.0 & \\
Ruprecht 147 WD8 & 4183937688413579648 & 289.057 & -16.337 & 19.07 & 240 & 0.53 & 0.86 & 12.845 & 28.5 & \\
Stock 2 WD1 & 506862078583709056 & 33.555 & 59.070 & 19.59 & 111 & 0.81 & 3.22 & 469 & 1.1  & in \cite{2022ApJ...926L..24M} IFMR \\
Stock 12 & 1992469104239732096 & 352.406 & 52.744 & 19.05 & 78  & 0.51 & 0.83 & $14{,}296$ & 181.8 & \\
Theia 517 & 2170776080281869056 & 321.749 & 48.749 & 19.16 & 238 & 1.00 & 5.36 & 329 & 2.1  & in \cite{2022ApJ...926L..24M} IFMR \\
\bottomrule
\end{tabular}
\end{table*}

\clearpage

\setlength{\tabcolsep}{1pt}
\renewcommand{\arraystretch}{0.9}
\setlength\LTcapwidth{\textwidth}
\begin{longtable}{>{\tiny}l
  >{\tiny\centering}c
  *{2}{>{\tiny\centering}p{1.5cm}}
  >{\tiny\centering}p{0.5cm}
  *{5}{>{\tiny\centering}p{1.3cm}}
  >{\tiny\centering\arraybackslash}p{2.0cm}}
\caption{%
  \begin{minipage}[t]{0.8\textwidth}
    \raggedright
    As in Table~\ref{tab:candidates_observed}, but for candidates without follow-up observations or literature spectra. The final column lists the primary reason each source was not prioritized for spectroscopy.
  \end{minipage}%
}
\label{tab:candidates_notobserved}\\
\toprule
Name & 
Gaia DR3 Source ID &
RA & 
Dec & 
$G_\mathrm{obs}$ & 
$t_\mathrm{cool}$ &
$M_\mathrm{WD}$ &
$M_\mathrm{prog}$ &
$t_\mathrm{total}$ & 
$\tfrac{t_\mathrm{total}-\mathrm{Age}_\mathrm{cl}}{\sigma_\mathrm{Age,cl}}$ & 
Comments \\
      & 
      & 
[deg] & 
[deg] & 
[mag] & 
[Myr] & 
[$M_\odot$] & 
[$M_\odot$] & 
[Myr] & 
[$\sigma_\mathrm{Age,cl}$]  &  
\\
\midrule
\endfirsthead

\multicolumn{11}{c}{\bfseries \tablename\ \thetable{}} \\[-0.5ex]
\multicolumn{11}{c}{(Continued)} \\  
\toprule
Name & 
Gaia DR3 Source ID &
RA & 
Dec & 
$G_\mathrm{obs}$ & 
$t_\mathrm{cool}$ &
$M_\mathrm{WD}$ &
$M_\mathrm{prog}$ &
$t_\mathrm{total}$ & 
$\tfrac{t_\mathrm{total}-\mathrm{Age}_\mathrm{cl}}{\sigma_\mathrm{Age,cl}}$ & 
Comments \\
      & 
      & 
[deg] & 
[deg] & 
[mag] & 
[Myr] & 
[$M_\odot$] & 
[$M_\odot$] & 
[Myr] & 
[$\sigma_\mathrm{Age,cl}$]  &  
\\
\midrule
\endhead 

\midrule
\endfoot

\bottomrule
\endlastfoot

Alessi 13 $\dagger$          & 4853382867764646912 &  53.256 & -38.674 & 17.36 & 697   & 0.76 & 2.86 & $1{,}199$ & 84.5   & (1) \\
Alessi 22 $\dagger$          & 2879428195013510144 & 355.656 &  36.327 & 19.34 & 151   & 0.74 & 2.72 & 730       & 0.1    & (3) \\
Alessi 62                    & 4519349757798439936 & 283.875 &  21.694 & 18.60 & 0     & 1.02 & 5.55 & 83        & -2.1   & (6) \\
Alessi 94 WD1 $\dagger$      & 391939027303287040  &   4.906 &  46.169 & 18.77 & 9     & 0.72 & 2.51 & 740       & 4.2    & (8) \\
Alessi 94 WD2 $\dagger$      & 386116254240723712  &   3.762 &  45.588 & 18.96 & 10    & 0.60 & 1.34 & $3{,}591$ & 25.6   & (1) \\
CWNU 41 $\dagger$            & 5798954758752816512 & 228.573 & -70.634 & 20.20 & 96    & 0.37 & —    & —         & —      & (2) \\
CWNU 515 $\dagger$           & 5248875855957835136 & 139.616 & -64.477 & 19.40 & $1{,}250$ & 1.03 & 5.64 & $1{,}330$ & 77.4 & (1) \\
CWNU 1012 $\dagger$          & 6831938232172511360 & 323.725 & -17.244 & 19.00 & 392   & 0.54 & 0.92 & $11{,}185$ & 108.4 & (1) \\
CWNU 1066 $\dagger$          & 2389169910242015104 & 357.321 & -21.775 & 18.49 & $2{,}419$ & 0.76 & 2.84 & $2{,}932$ & 32.4 & (1) \\
CWNU 1095 $\dagger$          &  851411295734572416 & 153.291 &  52.730 & 18.37 & 149   & 0.77 & 2.96 & 605       & 1.5    & (3) \\
Haffner 13                   & 5616599380898334336 & 111.658 & -25.274 & 20.20 & 111   & 0.89 & 3.89 & 323       & 38.6   & (1) \\
HSC 242 $\dagger$            & 4446494085800777984 & 248.220 &   9.517 & 19.60 & $2{,}482$ & 0.47 & —    & —         & —      & (2) \\
HSC 381 $\dagger$            & 4502736137087916032 & 269.554 &  17.354 & 16.84 & 3     & 0.65 & 1.83 & $1{,}575$ & 78.1   & (1) \\
HSC 448 WD1 $\dagger$        & 2678298278156118144 & 335.827 &   0.099 & 16.75 & 4     & 0.39 & —    & —         & —      & (2) \\
HSC 448 WD2 $\dagger$        & 2673038386327748224 & 327.202 &  -3.928 & 19.10 & 155   & 0.51 & —    & —         & —      & (2) \\
HSC 452 $\dagger$            & 4317953552994722048 & 293.818 &  14.219 & 19.61 & $1{,}077$ & 0.89 & 3.82 & $1{,}286$ & 10.8 & (1) \\
HSC 1155 $\dagger$           &  335525529520188032 &  42.242 &  38.972 & 18.13 & 8     & 0.52 & 0.83 & $13{,}669$ & 68.5 & (1) \\
HSC 1555 WD1 $\dagger$       & 661798660099237760  & 132.010 &  20.787 & 18.05 & $3{,}117$ & 0.76 & 2.86 & $3{,}618$ & 31.6 & (1) \\
HSC 1555 WD2 $\dagger$       & 612137052768974848  & 136.015 &  18.464 & 18.25 & $4{,}017$ & 0.79 & 3.05 & $4{,}435$ & 39.0 & (1) \\
HSC 1630 $\dagger$           & 3299641271199703680 &  67.310 &   9.623 & 17.03 & 353   & 0.94 & 4.47 & 496       & 1.4    & (3) \\
HSC 1710 $\dagger$           & 5092006917107465216 &  65.852 & -19.867 & 19.91 & $3{,}434$ & 0.66 & 1.84 & $4{,}942$ & 32.5 & (1) \\
HSC 2139 $\dagger$           & 5323610318430135680 & 133.697 & -52.980 & 18.75 & 615   & 0.68 & 2.13 & $1{,}784$ & 248.1 & (1) \\
HSC 2156 WD1 $\dagger$       & 5317454079808243456 & 132.027 & -54.199 & 17.28 & 51    & 0.97 & 4.98 & 159       & 9.3    & (1) \\
HSC 2156 WD2 $\dagger$       & 5322374634856765184 & 126.860 & -50.936 & 20.13 & $2{,}638$ & 0.83 & 3.33 & $2{,}956$ & 193.5 & (1) \\
HSC 2263 $\dagger$           & 5413780270582193664 & 153.297 & -46.000 & 20.41 & $1{,}366$ & 1.19 & 7.69 & $1{,}405$ & 24.1 & (1) \\
HSC 2304                     & 5257090272981848192 & 147.424 & -59.923 & 19.99 & 1     & 0.91 & 4.15 & 167       & -0.3   & (4) \\
HSC 2609 $\dagger$           & 5785132690708928768 & 216.913 & -79.364 & 19.68 & 417   & 0.50 & —    & —         & —      & (2) \\
HSC 2630 $\dagger$           & 5892219110301811200 & 213.299 & -57.186 & 18.38 & 264   & 0.59 & 1.20 & $5{,}176$ & $1{,}006.6$ & (1) \\
HSC 2971 $\dagger$           & 6020441063956859392 & 249.542 & -35.414 & 18.56 & $2{,}201$ & 0.85 & 3.46 & $2{,}477$ & 13.3 & (1) \\
Hyades WD5                   & 866166024521723776  & 110.619 &  22.203 & 16.23 & 168   & 0.69 & 2.22 & $1{,}212$ & 1.5    & (8) \\
Hyades WD7                   & 164297449857329408  &  62.176 &  29.358 & 16.95 & $1{,}052$ & 0.60 & 1.35 & $4{,}515$ & 9.5  & (1) \\
Hyades WD8                   & 3408058645321741568 &  77.018 &  20.265 & 18.22 & $2{,}254$ & 0.64 & 1.69 & $4{,}112$ & 8.6  & (1) \\
Hyades WD9                   & 3237581051061123712 &  81.836 &   5.383 & 19.38 & $2{,}960$ & 0.66 & 1.88 & $4{,}401$ & 9.3  & (1) \\
Hyades WD10                  & 3026376377878137088 &  89.591 &  -0.518 & 18.76 & $5{,}039$ & 0.68 & 2.09 & $6{,}263$ & 13.8 & (1) \\
IC 4665                      & 4486161063724137600 & 266.068 &   6.331 & 16.85 & 518   & 0.51 & —    & —         & —      & (2) \\
IC 4756 WD1                  & 4284010735661963648 & 279.352 &   5.505 & 16.78 & 171   & 0.77 & 2.94 & 635       & -0.7   & (4) \\
IC 4756 WD2                  & 4283928577215973120 & 279.737 &   5.045 & 19.16 & 4     & 0.33 & —    & —         & —      & (2) \\
LISC 3534                    & 403000732753623680  &  14.624 &  50.743 & 19.02 & 24    & 0.69 & 2.22 & 1,068     &  8.5   & (1) \\
LISC-III 3668 $\dagger$      & 5514623525712723072 & 121.989 & -48.822 & 17.98 & 107   & 0.66 & 1.85 & 1,634     & 81.1   & (1) \\
Mamajek 4                    & 6653882460188145152 & 273.967 & -52.692 & 17.25 & 54    & 0.74 & 2.68 & 658       &  1.5   & (8) \\
NGC 752 WD1                  & 322159621357863424  &  23.464 &  36.785 & 18.11 & 275   & 0.45 & —    & —         & —      & (2) \\
NGC 752 WD2                  & 342523646152426368  &  28.862 &  37.385 & 18.30 & 239   & 0.44 & —    & —         & —      & (2) \\
NGC 1901                     & 4659513404187412736 &  85.597 & -66.826 & 18.04 & 51    & 0.79 & 3.06 & 466       &  0.8   & (8) \\
NGC 2547                     & 5517879759465329024 & 119.765 & -47.778 & 18.73 & 382   & 0.72 & 2.46 & 1,160     & 58.5   & (1) \\
NGC 2548                     & 3064576401124133120 & 123.573 &  -5.814 & 18.43 & 10    & 0.64 & 1.64 & 2,022     &  5.6   & (4) \\
NGC 2682 WD2                 & 598655567036876928  & 132.618 &  10.905 & 17.92 & 15    & 0.80 & 3.12 & 407       & -2.4   & (5) \\
NGC 2682 WD4                 & 604721293514179712  & 133.605 &  11.642 & 19.83 & 113   & 0.30 & —    & —         & —      & (2) \\
NGC 3532 WD1                 & 5340220262646771712 & 166.040 & -58.329 & 18.26 & 2     & 0.73 & 2.55 & 699       &  2.1   & (7) \\
NGC 3532 WD3                 & 5337742307052922752 & 167.137 & -59.998 & 18.08 & 73    & 0.82 & 3.27 & 411       &  0.8   & (7) \\
NGC 3532 WD5                 & 5340257302454589696 & 166.094 & -57.818 & 18.29 & 230   & 0.80 & 3.15 & 611       &  1.7   & (7) \\
NGC 3532 WD11                & 5338641604462074240 & 167.043 & -58.988 & 18.47 & 165   & 0.70 & 2.28 & 1,134     &  4.1   & (7) \\
NGC 3532 WD12                & 5339400679820581248 & 167.449 & -58.742 & 20.03 & 271   & 0.41 & —    & —         & —      & (2) \\
NGC 5822                     & 5887666586717940224 & 226.164 & -54.407 & 19.16 & 7     & 0.36 & —    & —         & —      & (2) \\
NGC 6475                     & 4041649403352615936 & 267.513 & -34.225 & 18.89 & 366   & 0.56 & 1.02 & 8,184     & 49.1   & (1) \\
NGC 6633                     & 4477214475044842368 & 276.793 &   6.438 & 19.72 & 41    & 0.25 & —    & —         & —      & (2) \\
NGC 6991 WD1                 & 2166915179559503232 & 313.505 &  47.803 & 19.25 & 10    & 0.71 & 2.41 & 840       & -0.4   & (8) \\
NGC 6991 WD2                 & 2166829112734085248 & 313.419 &  47.296 & 19.89 & 117   & 0.40 & —    & —         & —      & (2) \\
NGC 6991 WD3                 & 2163824456685399168 & 314.173 &  46.972 & 18.21 & 463   & 0.50 & —    & —         & —      & (2) \\
Platais 9                    & 5429378904609527808 & 139.857 & -40.698 & 18.94 & 4,344 & 0.95 & 4.61 & 4,478     & 151.2  & (1) \\
Praesepe WD11                & 662998983199228032  & 127.166 &  19.729 & 18.31 & 271   & 0.84 & 3.40 & 571       &  0.8   & (8) \\
Praesepe WD14                & 659335994570713600  & 129.594 &  17.962 & 18.59 & 1,910 & 0.62 & 1.48 & 4,625     & 15.5   & (1) \\
Ruprecht 147 WD1             & 4183926006112672768 & 289.650 & -16.358 & 18.83 & 504   & 0.62 & 1.45 & 3,377     &  5.9   & (1) \\
Ruprecht 147 WD4             & 4183928888026931328 & 289.403 & -16.334 & 19.28 & 148   & 0.69 & 2.24 & 1,166     &  0.7   & (8) \\
Ruprecht 147 WD7             & 4183847562828165248 & 289.743 & -16.806 & 19.31 & 407   & 0.53 & 0.88 & 12,499    & 27.6   & (1) \\
Ruprecht 147 WD9             & 4184196073644880000 & 288.659 & -15.983 & 19.06 & 335   & 0.56 & 1.03 & 8,022     & 17.0   & (1) \\
Ruprecht 147 WD10            & 4083739606090532480 & 288.028 & -19.927 & 19.64 & 195   & 0.30 & —    & —         & —      & (2) \\
Stock 2 WD11                 & 507362012775415552  &  35.131 &  59.845 & 18.87 & 120   & 0.77 & 2.90 & 605       &  1.9   & (7) \\
Stock 2 WD12                 & 458923547711833216  &  36.754 &  58.524 & 19.12 & 123   & 0.68 & 2.13 & 1,284     &  6.0   & (1) \\
Stock 2 WD13                 & 464505424647551744  &  42.242 &  59.398 & 18.67 & 264   & 0.81 & 3.19 & 632       &  2.1   & (7) \\
Stock 2 WD14                 & 507277870080186624  &  34.359 &  59.741 & 19.05 & 194   & 0.56 & 0.99 & 8,990     & 52.8   & (1) \\
Stock 2 WD15                 & 507555904779576064  &  32.971 &  60.457 & 19.07 & 78    & 0.47 & —    & —         & —      & (2) \\
Stock 2 WD16                 & 508705826443985536  &  29.259 &  61.839 & 19.25 & 122   & 0.40 & —    & —         & —      & (2) \\
Stock 2 WD17                 & 507414067782288896  &  35.316 &  60.416 & 19.47 & 7     & 0.35 & —    & —         & —      & (2) \\
Stock 2 WD18                 & 506848643933335296  &  33.581 &  58.778 & 18.90 & 45    & 0.63 & 1.63 & 2,099     & 11.0   & (1) \\
Stock 2 WD19                 & 459783602023720704  &  44.158 &  56.437 & 18.69 & 39    & 0.63 & 1.57 & 2,322     & 12.3   & (1) \\
Stock 2 WD20                 & 507221863701989248  &  34.344 &  59.136 & 18.43 & 442   & 0.66 & 1.94 & 1,813     &  9.2   & (1) \\
Theia 172                    & 3114831641658036608 & 108.419 &   1.654 & 18.53 & 61    & 0.91 & 4.11 & 236       &  1.1   & (4) \\
Theia 558                    & 3324040907394753792 &  94.480 &   6.151 & —     & —     & >1.3 & —    & —         & —      & (3) \\
Theia 817 WD2 $\dagger$      & 4529222337115434240 & 276.349 &  21.664 & 19.34 & 31    & 0.60 & 1.32 & 3,769     & 23.8   & (1) \\
Theia 1315 $\dagger$         & 5249293056193304320 & 144.741 & -64.798 & 19.05 & 243   & 0.66 & 1.93 & 1,566     & 46.4   & (1) \\
Turner 5 $\dagger$           & 5433483136700686336 & 146.176 & -36.668 & 19.28 & 262   & 0.67 & 2.01 & 1,544     & 11.0   & (1) \\
UPK 5                        & 4098106821451715584 & 274.905 & -15.497 & 18.90 & 8     & 0.66 & 1.92 & 1,401     & 29.9   & (1) \\
UPK 552 $\dagger$            & 5364608495685924480 & 159.963 & -47.782 & 19.11 &1,316  & 0.56 & 1.00 & 9,543     &110.1   & (1) \\
UPK 624 $\dagger$            & 5914732847840333440 & 252.593 & -60.291 & 18.60 & 116   & 0.41 & —    & —         & —      & (2) \\

\midrule
\multicolumn{11}{@{}p{\textwidth}}{\footnotesize
(1) Total age substantially older than the \hunt mean cluster age. (2) Low-mass WD below the lower bound of the IFMR used for progenitor mass determination. (3) Poor-quality CMD inhibits precise age determination. (4) Significant astrometric uncertainty. (5) Relative age of cluster and source requires significant cooling delay for membership. (6) Very young WD whose parameters are particularly sensitive to the assumed reddening. (7) Lower-mass source in a cluster with many WD candidates needing spectroscopic follow-up. (8) Generic lower priority, highest priority amongst unobserved candidates.}\\
\bottomrule
\end{longtable}

\begin{table*}[ht]
\tiny
\caption{Spectroscopic parameters for Gaia-based astrometric member sources with spectra in literature. * indicates the cluster name was added for cases where the literature source name does not make the WD's cluster association clear. $\dagger$ indicates literature source shares a name with one from this work but is not the same source. $\ddagger$ indicates use of photometric log g estimate due to impact of magnetic field.}
\label{tab:wd_results_litsources}
\centering
\begin{tabular}{lllccl}
\toprule
Name (literature source) & 
Name (this work) & 
Gaia DR3 Source ID & 
$\log g$ & 
$T_\textrm{eff}$ & 
Spectrum Fit Source \\
  &
  &
  &
\multicolumn{1}{c}{$\mathrm{[cm\,s^{-2}]}$} &
\multicolumn{1}{c}{$\mathrm{[K]}$} &
\\
\midrule
ABDMG*, GD 50               & -            & 3251244858154433536           & $9.20\pm0.07$   & $42{,}700\pm800$    & \cite{2011ApJ...743..138G} \\
Alpha Per WD1              & -            & 439597809786357248            & $9.05\pm0.03$   & $41{,}600\pm200$    & \cite{2022ApJ...926L..24M} \\
Alpha Per WD2              &  -           & 244003693457188608            & $8.98\pm0.04$   & $46{,}200\pm300$    & \cite{2022ApJ...926L..24M} \\
Alpha Per WD5              &   -          & 1983126553936914816            & $8.84\pm0.05$   & $47{,}500\pm500$    & \cite{2022ApJ...926L..24M} \\
ASCC 47                    &    -         & 5529347562661865088            & $8.47\pm0.05$$\ddagger$   & $114{,}000\pm3{,}000$ & \cite{2020ApJ...901L..14C} \\
ASCC 113                   &     -        & 1871306874227157376            & $8.71\pm0.07$   & $25{,}400\pm300$    & \cite{2021ApJ...912..165R} \\
Hyades HS0400+1451             & Hyades WD2            & 39305036729495936            & $8.25\pm0.01$   & $14{,}620\pm60$     & \cite{2018ApJ...866...21C} \\
Hyades WD0348+339              & Hyades WD4            & 218783542413339648            & $8.31\pm0.05$   & $14{,}820\pm350$    & \cite{2018ApJ...866...21C} \\
Hyades WD0352+096              & Hyades WD3             & 3302846072717868416            & $8.30\pm0.05$   & $14{,}670\pm380$    & \cite{2018ApJ...866...21C} \\
Hyades WD0406+169     & -            & 45980377978968064            & $8.38\pm0.05$   & $15{,}810\pm290$    & \cite{2018ApJ...866...21C} \\
Hyades WD0421+162     & -            & 3313606340183243136            & $8.13\pm0.05$   & $20{,}010\pm320$    & \cite{2018ApJ...866...21C} \\
Hyades WD0425+168     & -            & 3313714023603261568            & $8.12\pm0.05$   & $25{,}130\pm380$    & \cite{2018ApJ...866...21C} \\
Hyades WD0431+126     & -            & 3306722607119077120            & $8.11\pm0.05$   & $21{,}890\pm350$    & \cite{2018ApJ...866...21C} \\
Hyades WD0438+108     & -            & 3294248609046258048            & $8.15\pm0.05$   & $27{,}540\pm400$    & \cite{2018ApJ...866...21C} \\
Hyades WD1$\dagger$            & -            & 560883558756079616            & $9.55\pm0.11$   & $26{,}400\pm200$    & \cite{2023ApJ...956L..41M} \\
M39                       & Theia 517            & 2170776080281869056            & $8.54\pm0.04$$\ddagger$   & $18{,}400\pm300$    & \cite{2020ApJ...901L..14C} \\
Melotte 111* J1218+2545   & Melotte 111            & 4008511467191955840            & $8.460\pm0.013$ & $16{,}099\pm224$    & \cite{2020ApJ...898...84K} \\
N752-WD1      & - & 342902771504611712 & $8.12\pm0.04$   & $15{,}500\pm250$  & \cite{2020NatAs...4.1102M} \\
NGC 2099* 0552+3236             & -            & 3451182182857026048            & $8.08\pm0.29$   & $76{,}000\pm4{,}640$  & \cite{2023A_A...678A..89W} \\
NGC 2516-1              & NGC 2516 WD1            & 5290767695648992128            & $8.47\pm0.04$   & $30{,}100\pm350$    & \cite{2018ApJ...866...21C} \\
NGC 2516-2               & NGC 2516 WD3             & 5290720695823013376            & $8.55\pm0.07$   & $35{,}500\pm550$    & \cite{2018ApJ...866...21C} \\
NGC 2516-3               & NGC 2516 WD6            & 5290719287073728128            & $8.46\pm0.04$ & $29{,}100\pm350$    & \cite{2018ApJ...866...21C} \\
NGC 2516-5              & NGC 2516 WD2            & 5290834387897642624            & $8.54\pm0.05$   & $32{,}200\pm400$    & \cite{2018ApJ...866...21C} \\
Pleiades* EGGR 25        & Pleiades           & 66697547870378368            & $8.66\pm0.03$   & $33{,}000\pm200$    & \cite{2022ApJ...926..132H} \\
Pleiades* Lan 532        & -            & 228579606900931968            & $8.61\pm0.05$   & $33{,}700\pm300$    & \cite{2022ApJ...926..132H} \\
Pleiades* WD 0518-105     & -            & 3014049448078210304            & $8.70\pm0.06$   & $33{,}700\pm400$    & \cite{2022ApJ...926..132H} \\
Prae WD0833+194               & Praesepe WD5            & 662798086105290112            & $8.325\pm0.042$ & $14{,}500\pm300$    & \cite{2018ApJ...866...21C} \\
Prae WD0836+199           & Praesepe WD1            & 661311267210542080            & $8.351\pm0.043$ & $14{,}900\pm300$    & \cite{2018ApJ...866...21C} \\
Prae WD0837+185               & Praesepe WD4            & 659494049367276544            & $8.413\pm0.046$ & $14{,}750\pm350$    & \cite{2018ApJ...866...21C} \\
Prae WD0840+190               & Praesepe WD3            & 661010005319096192            & $8.452\pm0.047$ & $14{,}800\pm400$    & \cite{2018ApJ...866...21C} \\
Prae WD0840+200               & Praesepe WD2            & 661353224747229184            & $8.226\pm0.042$ & $16{,}050\pm200$    & \cite{2018ApJ...866...21C} \\
Prae WD0843+184              & Praesepe WD10            & 660178942032517760            & $8.456\pm0.043$ & $14{,}850\pm300$    & \cite{2018ApJ...866...21C} \\
Praesepe* EGGR 61         & Praesepe WD15            & 661297901272035456            & $8.230\pm0.040$ & $17{,}200\pm200$    & \cite{2018ApJ...866...21C} \\
Praesepe* SDSS J083936.48+193043.6   & Praesepe WD9            & 661270898815358720            & $8.37\pm0.08$   & $21{,}334^{+1{,}332}_{-1{,}201}$ & \cite{2019A_A...628A..66L} \\
Praesepe* SDSS J083945.55+200015.66  & Praesepe WD8             & 664325543977630464            & $8.750\pm0.019$ & $16{,}550\pm140$    & \cite{2019MNRAS.486.2169K} \\
Praesepe* SDSS J084031.46+214042.7  & Praesepe WD6            & 665139697978259200            & $8.463\pm0.048$ & $17{,}996\pm284$    & \cite{2019MNRAS.482.5222T} \\
Praesepe* SDSS J084321.98+204330.2  & Praesepe WD7            & 661841163095377024            & $8.351\pm0.046$ & $14{,}731\pm553$    & \cite{2019MNRAS.482.5222T} \\
Praesepe* SDSS J092713.51+475659.6 & -     & 825640117568825472            & $8.389\pm0.028$ & $14{,}648\pm318$    & \cite{2019MNRAS.482.5222T} \\
Praesepe* US 2088         & -            & 1008837274655408128            & $8.502\pm0.040$ & $15{,}170\pm515$    & \cite{2019MNRAS.482.5222T} \\
NGC 3532-1               & NGC 3532 WD7            & 5338652084186678400            & $8.52\pm0.04$   & $23{,}100\pm300$    & \cite{2018ApJ...866...21C} \\
NGC 3532-5                & -            & 5338650984675000448            & $8.28\pm0.05$   & $27{,}700\pm350$    & \cite{2018ApJ...866...21C} \\
NGC 3532-9              & NGC 3532 WD6            & 5340219811654824448            & $8.18\pm0.04$   & $31{,}900\pm400$    & \cite{2018ApJ...866...21C} \\
NGC 3532-10              & NGC 3532 WD8             & 5338718261060841472            & $8.34\pm0.04$   & $26{,}300\pm350$    & \cite{2018ApJ...866...21C} \\
NGC 3532-J1106-584        & -          & 5340149103605412992            & $8.52\pm0.05$   & $20{,}200\pm300$    & \cite{2018ApJ...866...21C} \\
R147-WD01                 & Ruprecht 147 WD6            & 4087806832745520128            & $8.11\pm0.04$   & $17{,}850\pm250$    & \cite{2020NatAs...4.1102M} \\
R147-WD04                 &  -           &  4183919237232621056           & $8.07\pm0.04$   & $19{,}550\pm250$    & \cite{2020NatAs...4.1102M} \\
R147-WD07                 & Ruprecht 147 WD8             & 4183937688413579648            & $8.08\pm0.04$   & $15{,}900\pm200$    & \cite{2020NatAs...4.1102M} \\
R147-WD08                 &  -           &  4183978061110910592           & $8.11\pm0.05$   & $13{,}000\pm200$    & \cite{2020NatAs...4.1102M} \\
R147-WD10                 & Ruprecht 147 WD2            & 4184169822810795648            & $8.08\pm0.04$   & $17{,}950\pm250$    & \cite{2020NatAs...4.1102M} \\
Stock 2                   &  Stock 2 WD1            & 506862078583709056            & $8.58\pm0.05$   & $24{,}900\pm400$    & \cite{2021ApJ...912..165R} \\
Stock 12                  &  Stock 12           & 1992469104239732096            & $8.50\pm0.04$   & $31{,}600\pm200$    & \cite{2021ApJ...912..165R} \\
\bottomrule
\end{tabular}
\end{table*}

\begin{table*}[ht]
\tiny
\caption{Literature spectroscopic parameters for white dwarfs which cannot be properly assessed for astrometric cluster membership with Gaia.}
\label{tab:wd_results_nongaia}
\centering
\begin{tabular}{llccl}
\toprule
Name (literature source) &
Gaia DR3 Source ID &
\multicolumn{1}{c}{$\log g$} &
\multicolumn{1}{c}{$T_\textrm{eff}$} &
Spectrum fit source \\
  &
  &
\multicolumn{1}{c}{$\mathrm{[cm\,s^{-2}]}$} &
\multicolumn{1}{c}{$\mathrm{[K]}$} &
\\
\midrule
M67:WD6       & - & $7.989\pm0.104$ & $8{,}830\pm144$   & \cite{2018ApJ...867...62W,2021AJ....161..169C} \\
M67:WD9       & - & $7.918\pm0.059$ & $13{,}720\pm493$  & \cite{2018ApJ...867...62W,2021AJ....161..169C} \\
M67:WD14      & - & $8.030\pm0.059$ & $13{,}390\pm294$  & \cite{2018ApJ...867...62W,2021AJ....161..169C} \\
M67:WD16      & - & $8.035\pm0.083$ & $10{,}030\pm163$  & \cite{2018ApJ...867...62W,2021AJ....161..169C} \\
M67:WD17      & - & $8.036\pm0.060$ & $11{,}360\pm201$  & \cite{2018ApJ...867...62W,2021AJ....161..169C} \\
M67:WD25      & - & $7.977\pm0.045$ & $20{,}440\pm308$  & \cite{2018ApJ...867...62W,2021AJ....161..169C} \\
M67:WD26      & - & $7.934\pm0.072$ & $9{,}580\pm146$   & \cite{2018ApJ...867...62W,2021AJ....161..169C} \\
NGC 1039-WD15 & 340173367032619008 & $8.580\pm0.070$ & $25{,}900\pm500$  & \cite{2018ApJ...866...21C} \\
NGC 1039-WD17 & 337155723012394752 & $8.610\pm0.050$ & $26{,}050\pm350$  & \cite{2018ApJ...866...21C} \\
NGC 1039-WDS2 & 337044088221827456 & $8.460\pm0.040$ & $31{,}600\pm400$  & \cite{2018ApJ...866...21C} \\
NGC 2099-WD2  & - & $8.240\pm0.070$ & $22{,}200\pm650$  & \cite{2018ApJ...866...21C} \\
NGC 2099-WD5  & - & $8.210\pm0.010$ & $18{,}100\pm650$  & \cite{2018ApJ...866...21C} \\
NGC 2099-WD6  & - & $8.440\pm0.110$ & $16{,}700\pm750$  & \cite{2018ApJ...866...21C} \\
NGC 2099-WD9  & - & $7.950\pm0.140$ & $16{,}200\pm800$  & \cite{2018ApJ...866...21C} \\
NGC 2099-WD10 & - & $8.160\pm0.084$ & $19{,}250\pm500$  & \cite{2018ApJ...866...21C} \\
NGC 2099-WD13 & - & $8.526\pm0.126$ & $20{,}250\pm850$  & \cite{2018ApJ...866...21C} \\
NGC 2099-WD16 & - & $8.334\pm0.144$ & $17{,}150\pm850$  & \cite{2018ApJ...866...21C} \\
NGC 2099-WD17 & - & $8.571\pm0.154$ & $18{,}000\pm950$  & \cite{2018ApJ...866...21C} \\
NGC 2099-WD18 & - & $8.210\pm0.060$ & $24{,}900\pm600$  & \cite{2018ApJ...866...21C} \\
NGC 2099-WD21 & - & $8.370\pm0.110$ & $16{,}900\pm700$  & \cite{2018ApJ...866...21C} \\
NGC 2099-WD24 & - & $8.290\pm0.110$ & $18{,}700\pm700$  & \cite{2018ApJ...866...21C} \\
NGC 2099-WD25 & - & $8.110\pm0.060$ & $27{,}500\pm450$  & \cite{2018ApJ...866...21C} \\
NGC 2099-WD28 & - & $8.200\pm0.060$ & $22{,}000\pm400$  & \cite{2018ApJ...866...21C} \\
NGC 2168-LAWDS1 & - & $8.440\pm0.060$ & $33{,}500\pm450$  & \cite{2018ApJ...866...21C} \\
NGC 2168-LAWDS2 & - & $8.490\pm0.100$ & $33{,}400\pm600$  & \cite{2018ApJ...866...21C} \\
NGC 2168-LAWDS5 & 3426295183834581376 & $8.210\pm0.060$ & $52{,}700\pm900$  & \cite{2018ApJ...866...21C} \\
NGC 2168-LAWDS6 & 3426294977676135168 & $8.050\pm0.060$ & $57{,}300\pm1{,}000$ & \cite{2018ApJ...866...21C} \\
NGC 2168-LAWDS11 & - & $8.350\pm0.050$ & $19{,}900\pm350$  & \cite{2018ApJ...866...21C} \\
NGC 2168-LAWDS12 & - & $8.600\pm0.060$ & $34{,}200\pm500$  & \cite{2018ApJ...866...21C} \\
NGC 2168-LAWDS14 & - & $8.570\pm0.060$ & $30{,}500\pm450$  & \cite{2018ApJ...866...21C} \\
NGC 2168-LAWDS15 & - & $8.610\pm0.060$ & $30{,}100\pm400$  & \cite{2018ApJ...866...21C} \\
NGC 2168-LAWDS27 & - & $8.720\pm0.060$ & $30{,}700\pm400$  & \cite{2018ApJ...866...21C} \\
NGC 2168-LAWDS29 & 3426285249576179328 & $8.560\pm0.060$ & $33{,}500\pm450$  & \cite{2018ApJ...866...21C} \\
NGC 2168-LAWDS30 & - & $8.390\pm0.080$ & $29{,}700\pm500$  & \cite{2018ApJ...866...21C} \\
NGC 2287-2      & 2927203353930175232 & $8.450\pm0.050$ & $25{,}900\pm350$  & \cite{2018ApJ...866...21C} \\
NGC 2287-4      & 2927020766282196992 & $8.710\pm0.050$ & $26{,}500\pm350$  & \cite{2018ApJ...866...21C} \\
NGC 2287-5      & 2926996577021773696 & $8.440\pm0.040$ & $25{,}600\pm350$  & \cite{2018ApJ...866...21C} \\
NGC 2323-WD10   & 3051559974457298304 & $8.680\pm0.090$ & $52{,}800\pm1{,}350$ & \cite{2018ApJ...866...21C} \\
NGC 2323-WD11   & 3051568873629418624 & $8.690\pm0.070$ & $54{,}100\pm1{,}000$ & \cite{2018ApJ...866...21C} \\
NGC 3532-J1106-590 & 5338636244376571136 & $8.480\pm0.050$ & $21{,}100\pm350$  & \cite{2018ApJ...866...21C} \\
NGC 3532-J1107-584 & 5340148691289324416 & $8.590\pm0.050$ & $20{,}700\pm300$  & \cite{2018ApJ...866...21C} \\
NGC 6121 WD00    & - & $7.771\pm0.076$ & $20{,}900\pm500$  & \cite{2018ApJ...866...21C} \\
NGC 6121 WD04    & - & $7.776\pm0.074$ & $25{,}450\pm550$  & \cite{2018ApJ...866...21C} \\
NGC 6121 WD05    & - & $7.767\pm0.072$ & $28{,}850\pm500$  & \cite{2018ApJ...866...21C} \\
NGC 6121 WD06    & - & $7.903\pm0.069$ & $26{,}350\pm500$  & \cite{2018ApJ...866...21C} \\
NGC 6121 WD15    & - & $7.887\pm0.081$ & $24{,}600\pm600$  & \cite{2018ApJ...866...21C} \\
NGC 6121 WD20    & - & $7.792\pm0.084$ & $21{,}050\pm550$  & \cite{2018ApJ...866...21C} \\
NGC 6121 WD24    & - & $7.789\pm0.069$ & $26{,}250\pm500$  & \cite{2018ApJ...866...21C} \\
NGC 6819-6      & - & $7.944\pm0.051$ & $21{,}700\pm350$  & \cite{2018ApJ...866...21C} \\
NGC 7789-5      & - & $8.116\pm0.061$ & $31{,}700\pm450$ & \cite{2018ApJ...866...21C} \\
NGC 7789-8      & - & $8.114\pm0.074$ & $24{,}800\pm550$   & \cite{2018ApJ...866...21C} \\
NGC 7789-11     & - & $8.270\pm0.095$ & $20{,}500\pm650$    & \cite{2018ApJ...866...21C} \\
NGC 7789-14     & - & $7.987\pm0.144$ & $21{,}100\pm1{,}000$   & \cite{2018ApJ...866...21C} \\
\bottomrule
\end{tabular}
\end{table*}

\begin{table*}[ht]
\caption{Cluster age, reddening, and metallicity values are primarily derived using our heuristic iterative isochrone fitting method for most clusters in the \hunt catalog. Exceptions, along with the specific methodology applied to each cluster, are described in the text. For Stock 2, due to a substantial extended MSTO likely caused by differential reddening, we employ individual source reddening corrections rather than a single cluster-wide value. For clusters where ages are sourced from the literature, we also adopt the corresponding metallicity values, as these are required for progenitor mass determination. Reddening values are not included unless we determine them ourselves. Solar metallicity is taken to be 0.015, which is used to convert from Fe/H where necessary.}
\label{tab:cluster_ages_final}
\centering
\begin{tabular}{lcccc}
\toprule
Name &
Av &
Z &
Age &
Source
\\
 &
$\mathrm{[mag]}$ & 
& 
$\mathrm{[Myr]}$ &
\\
\midrule
ABDMG        & - & 0.015 & $133^{+15}_{-20}$ & \citet{2018ApJ...861L..13G}\\
ASCC 47      & - & 0.015 & $90\pm20$ & \citet{2020ApJ...901L..14C} \\
ASCC 113 & 0.128 & 0.017 & $282^{+24}_{-9}$ & this work \\
BH 99 & 0.132 & 0.018 & $76^{+10}_{-28}$ & this work \\
HSC 601 & 0.220 & 0.018 & $126^{+44}_{-22}$ & this work \\
Hyades & 0.000 & 0.024 & $646^{+85}_{-84}$ & this work \\
Alpha Persei & - & 0.0207 & $81\pm6$ & \citet{2021arXiv211004296H} \\
Melotte 111 & 0.000 & 0.014 & $617^{+11}_{-42}$ & this work \\
NGC 752 & 0.026 & 0.016 & $1{,}549^{+157}_{-156}$ & this work \\
NGC 1039 & 0.143 & 0.015 & $313^{+15}_{-12}$ & this work \\
NGC 2099 & 0.718 & 0.015 & $535^{+27}_{-43}$ & this work \\
NGC 2168 & 0.593 & 0.012 & $208^{+34}_{-21}$ & this work \\
NGC 2287 & 0.086 & 0.013 & $206^{+14}_{-16}$ & this work \\
NGC 2323 & 0.533 & 0.016 & $200^{+33}_{-38}$ & this work \\
NGC 2516 & 0.185 & 0.017 & $206^{+37}_{-12}$ & this work \\
NGC 2682 & 0.132 & 0.012 & $3{,}819^{+292}_{-86}$ & this work \\
NGC 3532 & 0.060 & 0.014 & $385^{+43}_{-33}$ & this work \\
NGC 6121 &   -    & 0.0012 & $10{,}200\pm1{,}000$ & \citet{2018ApJ...866...21C} \\
NGC 6819 &   -    & 0.014  & $2{,}430\pm150$    & \citet{2018ApJ...866...21C} \\
NGC 7789 &    -   & 0.0095 & $1{,}560\pm100$    & \citet{2018ApJ...866...21C} \\
Pleiades & - & 0.0147 & $128\pm7$ & \citet{2021arXiv211004296H} \\
Praesepe & 0.059 & 0.019 & $741^{+72}_{-30}$ & this work \\
Ruprecht 147 & 0.113 & 0.018 & $2{,}754^{+408}_{-124}$ & this work \\
Stock 1 & 0.388 & 0.016 & $447^{+19}_{-40}$ & this work \\
Stock 2 & - & 0.015 & $437^{+22}_{-16}$ & this work \\
Stock 12 & 0.292 & 0.015 & $234^{+49}_{-11}$ & this work \\
Theia 248 & 0.312 & 0.023 & $490^{+18}_{-14}$ & this work \\
Theia 517 & 0.015 & 0.015 & $407^{+32}_{-20}$ & this work \\
Theia 817 & 0.152 & 0.020 & $302^{+16}_{-25}$ & this work \\
UPK 303 & 0.081 & 0.015 & $263^{+31}_{-14}$ & this work \\
\bottomrule
\end{tabular}
\end{table*}

\begin{table*}[ht]
\tiny
\caption{Rederived parameters for Gaia-based astrometric member sources with spectra in literature. Names are from literature spectra sources as in Table~\ref{tab:wd_results_litsources}.}
\label{tab:wd_results_litsources_derived}
\centering
\begin{tabular}{lcccccl}
\toprule
Name & 
Mass & 
$t_\textrm{cool}$ 
& Radius & 
Initial Mass & 
Comments \\
  &
$\msun$ & 
$\mathrm{[Myr]}$ & 
$\mathrm{[km]}$ &
$\msun$ &
\\
\midrule
ABDMG*, GD 50              & $1.249\pm0.019$  & $71^{+18}_{-17}$          & $3{,}300\pm200$         & $6.5^{+1.2}_{-0.8}$& \\
Alpha Per WD1              & $1.201\pm0.012$  & $45^{+6}_{-5}$            & $3{,}780\pm110$         & $8.3^{+0.9}_{-0.8}$   &  \\
Alpha Per WD2              & $1.175\pm0.016$  & $13^{+5}_{-4}$            & $4{,}050\pm150$         & $6.2\pm0.3$          &  \\
Alpha Per WD5              & $1.11\pm0.02$    & $2.5^{+1.2}_{-0.2}$       & $4{,}600\pm200$         & $5.83^{+0.21}_{-0.18}$&  \\
ASCC 47                    & $1.01\pm0.02$    & $0.069^{+0.015}_{-0.012}$  & $6{,}700\pm300$         & $5.5^{+0.5}_{-0.4}$   & magnetic \\
ASCC 113                   & $1.06^{+0.02}_{-0.04}$ & $148\pm19$        & $5{,}200\pm300$         & $4.7^{+0.3}_{-0.4}$   &  \\
Hyades HS0400+1451              & $0.765\pm0.006$  & $339^{+6}_{-9}$           & $7{,}560\pm60$          & $3.5\pm0.3$          &  \\
Hyades WD0348+339              & $0.80\pm0.03$    & $350^{+40}_{-20}$         & $7{,}200\pm300$         & $3.5^{+0.4}_{-0.3}$   &  \\
Hyades WD0352+096              & $0.80\pm0.03$    & $360^{+40}_{-30}$         & $7{,}300\pm300$         & $3.6^{+0.4}_{-0.3}$   &  \\
Hyades WD0406+169     & $0.85\pm0.03$    & $330^{+30}_{-20}$         & $6{,}900\pm300$         & $3.4\pm0.3$          &  \\
Hyades WD0421+162     & $0.70\pm0.03$    & $93^{+16}_{-7}$           & $8{,}300\pm300$         & $2.84\pm0.14$         &  \\
Hyades WD0425+168     & $0.70\pm0.03$    & $28^{+7}_{-2}$            & $8{,}400\pm300$         & $2.73\pm0.12$         &  \\
Hyades WD0431+126     & $0.70\pm0.03$    & $60^{+10}_{-8}$           & $8{,}400\pm300$         & $2.78\pm0.13$         &  \\
Hyades WD0438+108     & $0.73\pm0.03$    & $18\pm3$                  & $8{,}300\pm300$         & $2.72^{+0.12}_{-0.11}$&  \\
Hyades WD1     & $1.317^{+0.015}_{-0.018}$ & $563^{+7}_{-20}$ & $2{,}200\pm300$     & $5.7^{+2.1}_{-1.4}$   &  \\
Melotte 111* J1218+2545                 & $0.902\pm0.008$  & $357^{+14}_{-13}$         & $6{,}440\pm70$          & $3.60^{+0.22}_{-0.09}$  &  \\
M39             & $0.96^{+0.02}_{-0.03}$ & $282^{+22}_{-18}$  & $6{,}000\pm200$          & $4.8^{+0.4}_{-0.5}$    & magnetic \\
N752-WD1      & $0.69^{+0.03}_{-0.02}$  & $237^{+12}_{-19}$      & $8{,}300\pm200$ & $1.97^{+0.18}_{-0.09}$  &  \\
NGC 2099* 0552+3236             & $0.77^{+0.15}_{-0.12}$ & $0.57\pm0.13$   & $9{,}000^{+3{,}000}_{-2{,}000}$ & $2.80^{+0.08}_{-0.05}$&  \\
NGC 2516-1               & $0.93\pm0.02$    & $39^{+7}_{-6}$            & $6{,}500\pm200$         & $4.29^{+0.14}_{-0.32}$  &  \\
NGC 2516-2               & $0.98\pm0.04$    & $20^{+9}_{-7}$            & $6{,}100\pm400$         & $4.12^{+0.12}_{-0.28}$  &  \\
NGC 2516-3               & $0.92\pm0.02$    & $45\pm7$                  & $6{,}500\pm200$         & $4.36^{+0.15}_{-0.33}$ &  \\
NGC 2516-5              & $0.97\pm0.03$    & $36^{+8}_{-7}$            & $6{,}100\pm300$         & $4.27^{+0.13}_{-0.31}$  &  \\
NGC 3532-1               & $0.95\pm0.02$    & $136^{+13}_{-10}$         & $6{,}200\pm200$         & $3.67^{+0.18}_{-0.22}$   &  \\
NGC 3532-5                & $0.81\pm0.03$    & $28^{+7}_{-5}$            & $7{,}500\pm300$         & $3.22^{+0.10}_{-0.13}$  &  \\
NGC 3532-9               & $0.75\pm0.02$    & $8.5^{+0.8}_{-0.5}$        & $8{,}100\pm200$         & $3.16^{+0.10}_{-0.12}$  &  \\
NGC 3532-10              & $0.84\pm0.02$    & $50\pm7$                  & $7{,}100\pm200$         & $3.29^{+0.12}_{-0.14}$  &  \\
NGC 3532-J1106-584 & $0.94\pm0.03$   & $208^{+21}_{-18}$         & $6{,}200^{+300}_{-200}$  & $4.2^{+0.3}_{-0.4}$    &  \\
Pleiades* EGGR 25                   & $1.041^{+0.014}_{-0.017}$ & $50^{+4}_{-5}$  & $5{,}500\pm150$         & $5.8^{+0.2}_{-0.3}$   &  \\
Pleiades* Lan 532        & $1.01\pm0.03$    & $37^{+8}_{-7}$            & $5{,}700^{+300}_{-200}$ & $5.5\pm0.2$          &  \\
Pleiades* WD 0518-105     & $1.06^{+0.02}_{-0.03}$ & $49^{+4}_{-7}$   & $5{,}300\pm300$         & $5.8^{+0.2}_{-0.3}$   &  \\
Prae WD0833+194               & $0.81\pm0.03$    & $390^{+30}_{-20}$         & $7{,}100\pm200$         & $3.27^{+0.13}_{-0.21}$  &  \\
Prae WD0836+199              & $0.83\pm0.03$    & $380^{+20}_{-30}$         & $7{,}000\pm200$         & $3.25^{+0.11}_{-0.21}$  &  \\
Prae WD0837+185              & $0.87\pm0.03$    & $430\pm30$                & $6{,}700\pm200$         & $3.41^{+0.16}_{-0.26}$  &  \\
Prae WD0840+190              & $0.90\pm0.03$    & $450^{+40}_{-30}$         & $6{,}500\pm200$         & $3.5^{+0.2}_{-0.3}$    &  \\
Prae WD0840+200              & $0.75\pm0.03$    & $244^{+24}_{-10}$         & $7{,}700\pm200$         & $2.91^{+0.08}_{-0.13}$  &  \\
Prae WD0843+184              & $0.90\pm0.03$    & $440^{+40}_{-20}$         & $6{,}500\pm200$         & $3.5^{+0.2}_{-0.3}$    &  \\
Praesepe* EGGR 61         & $0.76\pm0.03$    & $200^{+18}_{-9}$          & $7{,}700\pm200$         & $2.83^{+0.06}_{-0.12}$ &  \\
Praesepe* SDSS J083936.48+193043.6               & $0.85\pm0.05$    & $130^{+30}_{-20}$         & $6{,}900\pm400$         & $2.72^{+0.06}_{-0.11}$ &  \\
Praesepe* SDSS J083945.55+200015.66              & $1.067\pm0.007$  & $570^{+40}_{-20}$         & $5{,}030\pm90$          & $4.25\pm0.5$          & magnetic \\
Praesepe* SDSS J084031.46+214042.7              & $0.91\pm0.03$    & $260\pm20$                & $6{,}400^{+300}_{-200}$  & $2.95^{+0.08}_{-0.14}$  &  \\
Praesepe* SDSS J084321.98+204330.2               & $0.83\pm0.03$    & $400^{+50}_{-40}$         & $7{,}000\pm200$         & $3.29^{+0.18}_{-0.23}$  &  \\
Praesepe* SDSS J092713.51+475659.6 & $0.855\pm0.018$ & $400^{+30}_{-20}$ & $6{,}810\pm150$  & $3.36^{+0.15}_{-0.24}$  &  \\
Praesepe* US 2088         & $0.93\pm0.03$    & $450^{+50}_{-40}$         & $6{,}200\pm200$         & $3.56^{+0.2}_{-0.3}$   &  \\
R147-WD01          & $0.68\pm0.02$ & $145^{+7}_{-14}$  & $8{,}400\pm200$ & $1.59^{+0.02}_{-0.07}$ &  \\
R147-WD04     & $0.66\pm0.02$ & $91^{+11}_{-7}$   & $8{,}700\pm200$ & $1.58^{+0.02}_{-0.07}$ &  \\
R147-WD07           & $0.66\pm0.02$ & $201^{+17}_{-8}$  & $8{,}500\pm200$ & $1.60^{+0.03}_{-0.07}$ &  \\
R147-WD08     & $0.67\pm0.03$ & $400^{+20}_{-30}$ & $8{,}300\pm300$ & $1.64^{+0.03}_{-0.08}$ &  \\
R147-WD10          & $0.67\pm0.02$ & $133\pm11$ & $8{,}600\pm200$ & $1.59^{+0.02}_{-0.07}$ &  \\
Stock 2               & $0.99\pm0.03$    & $121^{+14}_{-13}$         & $7{,}300^{+300}_{-200}$  & $3.37^{+0.08}_{-0.09}$  &  \\
Stock 12                 & $0.95\pm0.02$    & $33\pm6$                  & $6{,}300\pm200$         & $3.98^{+0.10}_{-0.31}$   &  potentially magnetic\\
\bottomrule
\end{tabular}
\end{table*}

\begin{table*}[ht]
\tiny
\caption{Rederived parameters for literature cluster member white dwarfs which cannot be properly assessed for astrometric cluster membership with Gaia.}
\label{tab:wd_results_nongaia_derived}
\centering
\begin{tabular}{lccccccl}
\toprule
Name (literature source) &
Mass & 
$t_\textrm{cool}$ & 
Radius & 
Initial Mass &
Comments \\
  &
$\msun$ & 
$\mathrm{[Myr]}$ & 
$\mathrm{[km]}$ & 
$\msun$ &
\\
\midrule
M67:WD6       & $0.59^{+0.07}_{-0.06}$ & $1{,}840^{+130}_{-820}$ & $8{,}900\pm600$ & $1.65^{+0.04}_{-0.18}$ &  \\
M67:WD9       & $0.56\pm0.03$          & $260\pm30$           & $9{,}500\pm400$ & $1.371^{+0.011}_{-0.034}$ &  \\
M67:WD14      & $0.63^{+0.04}_{-0.03}$  & $340^{+20}_{-30}$      & $8{,}800^{+400}_{-300}$ & $1.381^{+0.012}_{-0.035}$ &  \\
M67:WD16      & $0.62^{+0.05}_{-0.04}$  & $760^{+80}_{-40}$      & $8{,}700\pm500$ & $1.438^{+0.017}_{-0.041}$ &  \\
M67:WD17      & $0.63\pm0.04$          & $540^{+40}_{-30}$      & $8{,}700^{+400}_{-300}$ & $1.407^{+0.013}_{-0.038}$ &  \\
M67:WD25      & $0.61\pm0.03$          & $59^{+8}_{-4}$         & $9{,}300\pm300$ & $1.347^{+0.010}_{-0.030}$ &  \\
M67:WD26      & $0.56\pm0.04$          & $700^{+40}_{-20}$      & $9{,}300\pm400$ & $1.433^{+0.014}_{-0.040}$ &  \\
NGC 1039-WD15 & $0.99\pm0.04$          & $106^{+18}_{-17}$      & $5{,}900^{+400}_{-300}$ & $3.94^{+0.16}_{-0.15}$ &  \\
NGC 1039-WD17 & $1.01\pm0.03$          & $111^{+13}_{-12}$      & $5{,}700\pm200$ & $3.97^{+0.13}_{-0.14}$ &  \\
NGC 1039-WDS2 & $0.92\pm0.02$          & $28^{+6}_{-5}$         & $6{,}500\pm200$ & $3.49^{+0.06}_{-0.07}$ &  \\
NGC 2099-WD2  & $0.77\pm0.44$          & $82^{+16}_{-14}$       & $7{,}700\pm400$ & $2.97^{+0.10}_{-0.07}$ &  \\
NGC 2099-WD5  & $0.746\pm0.006$         & $164^{+21}_{-19}$      & $7{,}810\pm60$  & $3.18^{+0.15}_{-0.09}$ &  \\
NGC 2099-WD6  & $0.89\pm0.07$          & $310^{+60}_{-50}$      & $6{,}600\pm600$ & $3.8^{+0.5}_{-0.3}$    &  \\
NGC 2099-WD9  & $0.58^{+0.08}_{-0.07}$  & $160\pm40$           & $9{,}300\pm900$ & $3.17^{+0.16}_{-0.13}$ &  \\
NGC 2099-WD10 & $0.71\pm0.05$          & $120^{+24}_{-20}$      & $8{,}100\pm500$ & $3.06^{+0.12}_{-0.08}$ &  \\
NGC 2099-WD13 & $0.95\pm0.08$          & $210^{+60}_{-40}$      & $6{,}100^{+700}_{-600}$ & $3.33^{+0.25}_{-0.16}$ &  \\
NGC 2099-WD16 & $0.82\pm0.09$          & $240^{+70}_{-50}$      & $7{,}100\pm800$ & $3.5^{+0.3}_{-0.2}$    &  \\
NGC 2099-WD17 & $0.97^{+0.09}_{-0.10}$ & $320^{+100}_{-80}$     & $5{,}900^{+800}_{-700}$ & $3.9^{+0.7}_{-0.4}$   &  \\
NGC 2099-WD18 & $0.75\pm0.04$          & $41^{+10}_{-7}$        & $7{,}900^{+400}_{-300}$ & $2.88^{+0.09}_{-0.06}$ &  \\
NGC 2099-WD21 & $0.84\pm0.07$          & $270^{+60}_{-40}$      & $6{,}900\pm600$ & $3.6^{+0.3}_{-0.2}$    &  \\
NGC 2099-WD24 & $0.80\pm0.07$          & $170^{+40}_{-30}$      & $7{,}400\pm600$ & $3.20^{+0.17}_{-0.12}$ &  \\
NGC 2099-WD25 & $0.70^{+0.04}_{-0.03}$  & $15.8^{+3.7}_{-1.3}$    & $8{,}500\pm400$ & $2.83^{+0.08}_{-0.05}$ &  \\
NGC 2099-WD28 & $0.75\pm0.04$          & $75^{+14}_{-12}$       & $7{,}900^{+400}_{-300}$ & $2.95^{+0.10}_{-0.07}$ &  \\
NGC 2168-LAWDS1 & $0.91\pm0.04$        & $16^{+6}_{-4}$         & $6{,}600\pm300$ & $4.01^{+0.18}_{-0.24}$ &  \\
NGC 2168-LAWDS2 & $0.94\pm0.06$        & $22^{+14}_{-9}$        & $6{,}400\pm500$ & $4.1^{+0.2}_{-0.3}$    &  \\
NGC 2168-LAWDS5 & $0.80\pm0.03$        & $1.49^{+0.08}_{-0.07}$   & $8{,}100\pm400$ & $3.90^{+0.16}_{-0.21}$ &  \\
NGC 2168-LAWDS6 & $0.72\pm0.03$        & $1.31^{+0.07}_{-0.06}$   & $9{,}300^{+500}_{-400}$ & $3.90^{+0.15}_{-0.21}$ &  \\
NGC 2168-LAWDS11 & $0.84\pm0.03$       & $158^{+15}_{-17}$       & $7{,}000\pm300$ & $7.2\pm1.7$           &  \\
NGC 2168-LAWDS12 & $1.01^{+0.03}_{-0.04}$ & $32^{+9}_{-8}$       & $5{,}800\pm300$ & $4.2^{+0.2}_{-0.3}$    &  \\
NGC 2168-LAWDS14 & $0.99\pm0.04$       & $53\pm11$              & $5{,}900\pm300$ & $4.4\pm0.03$          &  \\
NGC 2168-LAWDS15 & $1.01^{+0.03}_{-0.04}$ & $64^{+12}_{-11}$     & $5{,}700\pm300$ & $4.5^{+0.3}_{-0.4}$    &  \\
NGC 2168-LAWDS27 & $1.06^{+0.02}_{-0.03}$ & $76\pm8$             & $5{,}200\pm300$ & $4.7^{+0.3}_{-0.4}$    &  \\
NGC 2168-LAWDS29 & $0.98\pm0.04$        & $31^{+9}_{-8}$        & $6{,}000\pm300$ & $4.1^{+0.2}_{-0.3}$    &  \\
NGC 2168-LAWDS30 & $0.88\pm0.05$        & $30^{+12}_{-9}$       & $6{,}900\pm400$ & $4.1^{+0.2}_{-0.3}$    &  \\
NGC 2287-2      & $0.91\pm0.03$        & $76^{+11}_{-10}$       & $6{,}500\pm300$ & $4.7^{+0.3}_{-0.2}$    &  \\
NGC 2287-4      & $1.058^{+0.017}_{-0.023}$ & $129\pm12$       & $5{,}200\pm200$ & $5.9^{+0.6}_{-0.5}$    &  \\
NGC 2287-5      & $0.90\pm0.03$        & $77^{+9}_{-8}$         & $6{,}600\pm200$ & $4.7^{+0.3}_{-0.2}$    &  \\
NGC 2323-WD10   & $1.06^{+0.03}_{-0.04}$ & $1.48\pm0.02$       & $5{,}400\pm500$ & $4.0^{+0.3}_{-0.2}$    &  \\
NGC 2323-WD11   & $1.06^{+0.02}_{-0.03}$ & $1.30^{+0.13}_{-0.17}$ & $5{,}400\pm400$ & $4.0^{+0.3}_{-0.2}$  &  \\
NGC 3532-J1106-590 & $0.92\pm0.03$      & $169\pm16$            & $6{,}400\pm300$ & $3.9^{+0.2}_{-0.3}$     &  \\
NGC 3532-J1107-584 & $0.99\pm0.03$      & $220\pm20$            & $5{,}800\pm200$ & $4.3\pm0.04$           &  \\
NGC 6121 WD00    & $0.51^{+0.04}_{-0.03}$ & $35^{+6}_{-3}$     & $10{,}700^{+600}_{-500}$ & $0.89\pm0.02$     &  \\
NGC 6121 WD04    & $0.52^{+0.04}_{-0.03}$ & $16.9^{+1.1}_{-1.3}$ & $10{,}800^{+600}_{-500}$ & $0.89\pm0.02$     &  \\
NGC 6121 WD05    & $0.53\pm0.03$      & $10.5^{+0.8}_{-1.1}$   & $10{,}900\pm600$ & $0.89\pm0.02$         &  \\
NGC 6121 WD06    & $0.59^{+0.04}_{-0.03}$ & $15.5^{+1.4}_{-1.8}$ & $9{,}900\pm500$ & $0.89\pm0.02$         &  \\
NGC 6121 WD15    & $0.57\pm0.04$      & $21\pm2$              & $9{,}900\pm600$ & $0.89\pm0.02$         &  \\
NGC 6121 WD20    & $0.52\pm0.04$      & $39^{+4}_{-5}$         & $10{,}500\pm600$ & $0.89\pm0.02$        &  \\
NGC 6121 WD24    & $0.53\pm0.03$      & $15.1^{+1.3}_{-2.5}$    & $10{,}700\pm500$ & $0.89\pm0.02$        &  \\
NGC 6819-6      & $0.60\pm0.03$      & $39^{+9}_{-3}$         & $9{,}500\pm300$ & $1.58\pm0.03$         &  \\
NGC 7789-5      & $0.71\pm0.04$      & $8.2\pm0.5$            & $8{,}500\pm400$ & $1.76\pm0.04$ &  \\
NGC 7789-8      & $0.70\pm0.04$      & $29^{+10}_{-5}$        & $8{,}500^{+500}_{-400}$ & $1.76\pm0.04$ &  \\
NGC 7789-11     & $0.79\pm0.06$      & $120^{+30}_{-20}$      & $7{,}500\pm500$           & $1.80\pm0.05$ &  \\
NGC 7789-14     & $0.62\pm0.08$      & $53^{+21}_{-13}$        & $9{,}200^{+1{,}000}_{-900}$ & $1.77^{+0.05}_{-0.04}$  &  \\
\bottomrule
\end{tabular}
\end{table*}

\begin{table*}[ht]
\tiny
\caption{Comparison of this work with the \citet{2018ApJ...866...21C} IFMR sample. * indicates a source cannot be properly assessed for membership with Gaia but is included in our IFMR as a non-Gaia-based astrometric cluster member.}
\label{tab:cummings_2018}
\centering
\begin{tabular}{lccl}
\toprule
\citet{2018ApJ...866...21C} Name & 
Gaia DR3 Source ID & 
Name (this work) & 
Comments \\
\midrule
NGC 2099-WD10 & - & - & (2), * \\
NGC 2099-WD13 & - & - & (2), * \\
NGC 2099-WD16 & - & - & (2), * \\
NGC 2099-WD17 & - & - & (2), * \\
NGC 2099-WD2 & - & - & (2), * \\
NGC 2099-WD5 & - & - & (2), * \\
NGC 2099-WD6 & - & - & (2), * \\
NGC 2099-WD9 & - & - & (2), * \\
NGC 2099-WD18 & - & - & (2), * \\
NGC 2099-WD21 & - & - & (2), * \\
NGC 2099-WD24 & - & - & (2), * \\
NGC 2099-WD25 & - & - & (2), * \\
NGC 2099-WD28 & - & - & (2), * \\
NGC 2099-WD33 & - & - & (2), (8) \\
Pra WD0833+194 & 662798086105290112 & Praesepe WD5 & (1) \\
Pra WD0836+199 & 661311267210542080 & Praesepe WD1 & (1) \\
Pra WD0837+185 & 659494049367276544 & Praesepe WD4 & (1) \\
Pra WD0837+199 & 661297901272035456 & Praesepe WD15 & (1) \\
Pra WD0840+190 & 661010005319096192 & Praesepe WD3 & (1) \\
Pra WD0840+200 & 661353224747229184 & Praesepe WD2 & (1) \\
Pra WD0843+184 & 660178942032517760 & Praesepe WD10 & (1) \\
NGC 6121 WD00 & - & - & (2), * \\
NGC 6121 WD04 & - & - & (2), *  \\
NGC 6121 WD05 & - & - & (2), *  \\
NGC 6121 WD06 & - & - & (2), *  \\
NGC 6121 WD15 & - & - & (2), *  \\
NGC 6121 WD20 & - & - & (2), *  \\
NGC 6121 WD24 & - & - & (2), *  \\
NGC 6819-6 & - & - & (2), *  \\
NGC 7789-5 & - & - & (2), *  \\
NGC 7789-8 & - & - & (2), *  \\
NGC 7789-11 & - & - & (2), *  \\
NGC 7789-14 & - & - & (2), *  \\
NGC 1039-WD15 & 340173367032619008 & - & (3), * \\
NGC 1039-WD17 & 337155723012394752 & - & (3), * \\
NGC 1039-WDS2 & 337044088221827456 & - & (3), * \\
NGC 2168-LAWDS1 & - & - & (2), * \\
NGC 2168-LAWDS2 & - & - & (2), * \\
NGC 2168-LAWDS5 & 3426295183834581376 & - & (3), * \\
NGC 2168-LAWDS6 & 3426294977676135168 & - & (3), * \\
NGC 2168-LAWDS11 & - & - & (2), * \\
NGC 2168-LAWDS12 & - & - & (2), * \\
NGC 2168-LAWDS14 & - & - & (2), * \\
NGC 2168-LAWDS15 & - & - & (2), * \\
NGC 2168-LAWDS22 & 3426288857352638080 & - & (4) \\
NGC 2168-LAWDS27 & - & - & (2), * \\
NGC 2168-LAWDS29 & 3426285249576179328 & - & (5), * \\
NGC 2168-LAWDS30 & - & - & (2), * \\
NGC 2287-2 & 2927203353930175232 & - & (3), * \\
NGC 2287-4 & 2927020766282196992 & - & (5), * \\
NGC 2287-5 & 2926996577021773696 & - & (3), * \\
NGC 2323-WD10 & 3051559974457298304 & - & (5), * \\
NGC 2323-WD11 & 3051568873629418624 & - & (3), * \\
NGC 2516-1 & 5290767695648992128 & NGC 2516 WD1 & (1) \\
NGC 2516-2 & 5290720695823013376 & NGC 2516 WD3 & (1) \\
NGC 2516-3 & 5290719287073728128 & NGC 2516 WD6 & (1) \\
NGC 2516-5 & 5290834387897642624 & NGC 2516 WD2 & (1) \\
NGC 3532-1 & 5338652084186678400 & NGC 3532 WD7 & (1) \\
NGC 3532-5 & 5338650984675000448 & - & (1) \\
NGC 3532-9 & 5340219811654824448 & NGC 3532 WD6 & (1) \\
NGC 3532-10 & 5338718261060841472 & NGC 3532 WD8 & (1) \\
NGC 3532-J1106-584 & 5340149103605412992 & - & (1) \\
NGC 3532-J1106-590 & 5338636244376571136 & - & (6), * \\
NGC 3532-J1107-584 & 5340148691289324416 & - & (3), * \\
VPHASJ1103-5837 & 5338711045522811776 & - & (3), (4), (6) \\
Sirius B & 2947050466531873024 & - & (7) \\
Pleiades-LB 1497 & 66697547870378368 & Pleiades & (1) \\
GD 50 & 3251244858154433536 & - & (1) \\
PG 0136+251 & 292454841560140032 & - & (5) \\
Hyades HS0400+1451 & 39305036729495936 & Hyades WD2 & (1) \\
Hyades WD0348+339 & 218783542413339648 & Hyades WD4 & (1) \\
Hyades WD0352+096 & 3302846072717868416 & Hyades WD3 & (1) \\
Hyades WD0406+169 & 45980377978968064 & - & (1) \\
Hyades WD0421+162 & 3313606340183243136 & - & (1) \\
Hyades WD0425+168 & 3313714023603261568 & - & (1) \\
Hyades WD0431+126 & 3306722607119077120 & - & (1) \\
Hyades WD0437+138 & 3308403897837092992 & Hyades WD1 & (1), DBA \\
Hyades WD0438+108 & 3294248609046258048 & - & (1) \\
Hyades WD0625+415 & 958262527212954752 & - & (4) \\
Hyades WD0637+477 & 966561537900256512 & - & (4) \\
\bottomrule
\end{tabular}
\begin{minipage}{\linewidth}
\medskip
{\textbf{Notes}: (1) Cluster membership supported by Gaia astrometry. (2) Below Gaia magnitude limit. (3) Questionable cluster member, substantial Gaia astrometric error. (4) Gaia proper motion or parallax makes cluster membership unlikely. (5) Incomplete Gaia DR3 proper motion and/or parallax data. (6) Low astrometric fidelity. (7) Not associated with an open cluster. (8) Requires cooling delay from past merger history for cluster membership.}
\end{minipage}
\end{table*}

\begin{figure*}[!tp]
    \centering
    \includegraphics[width=0.75\textwidth]{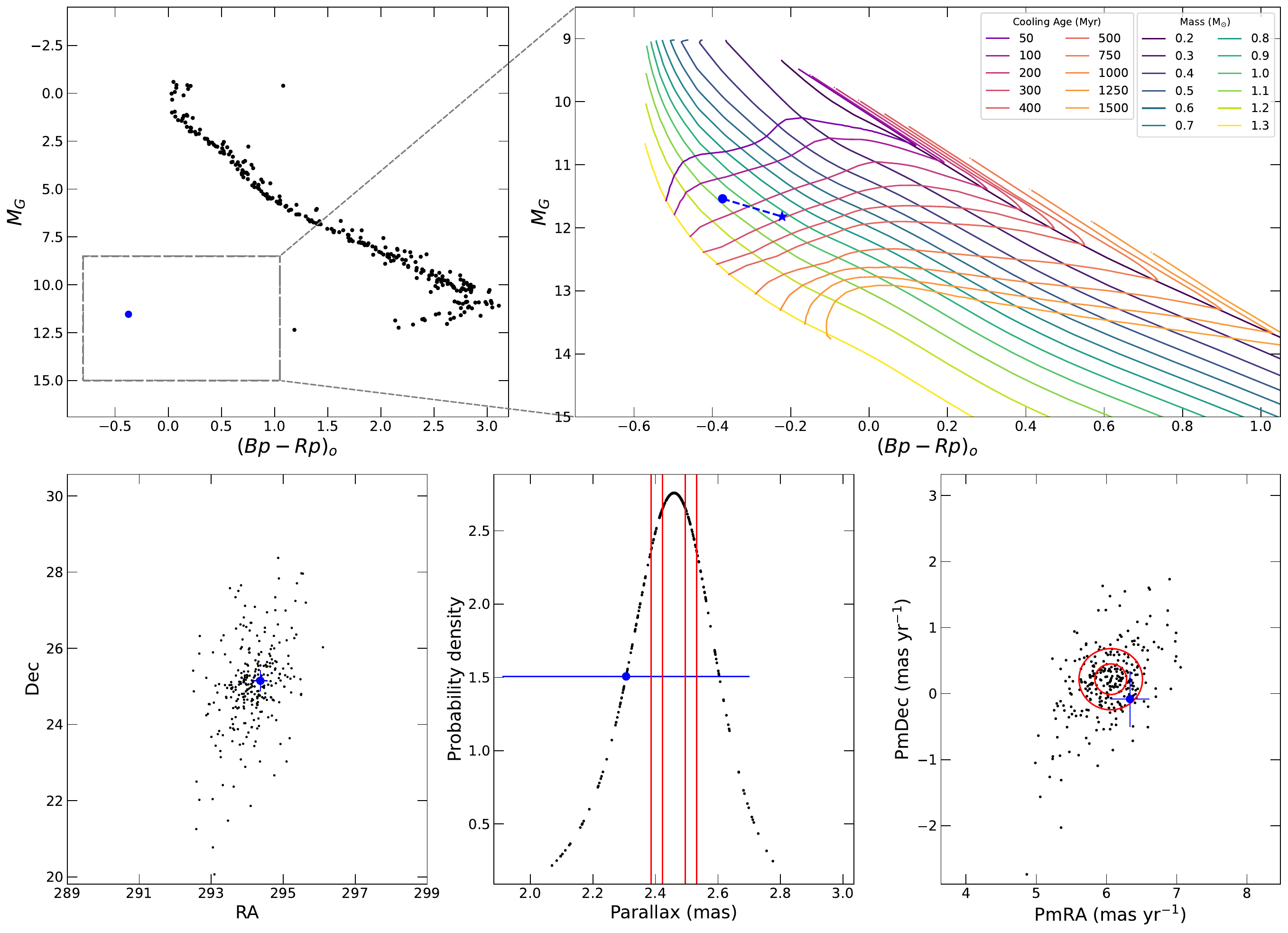}
    \caption{Same as Fig.~\ref{fig:CMD_alessi_13}, but for the Stock 1 cluster.}
    \label{fig:CMD_stock_1}
\end{figure*}

\begin{figure*}[!tp]
    \centering
    \includegraphics[width=0.75\textwidth]{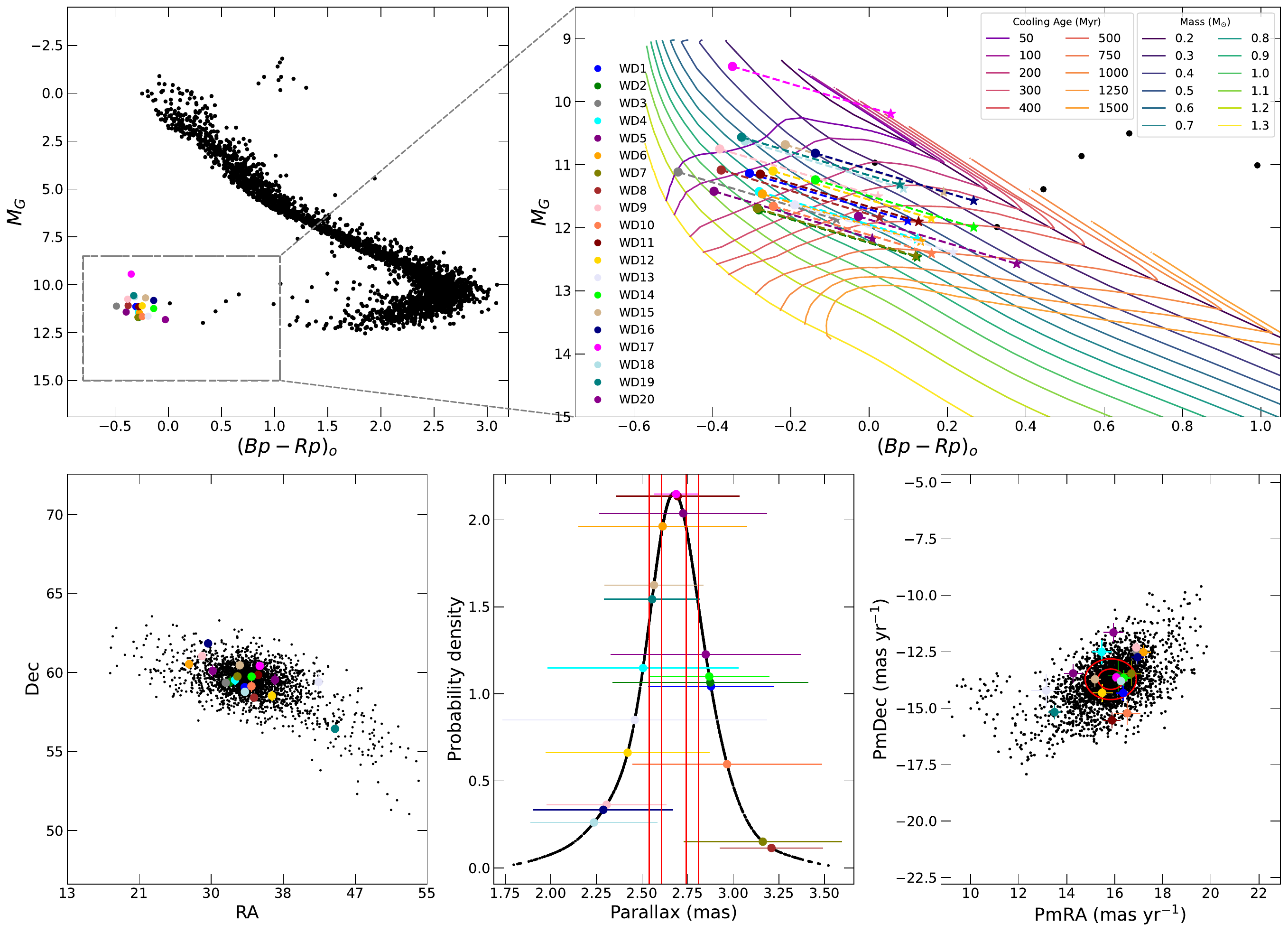}
    \caption{Same as Fig.~\ref{fig:CMD_alessi_13}, but for the Stock 2 cluster.}
    \label{fig:CMD_stock_2}
\end{figure*}

\begin{figure*}[!tp]
    \centering
    \includegraphics[width=0.75\textwidth]{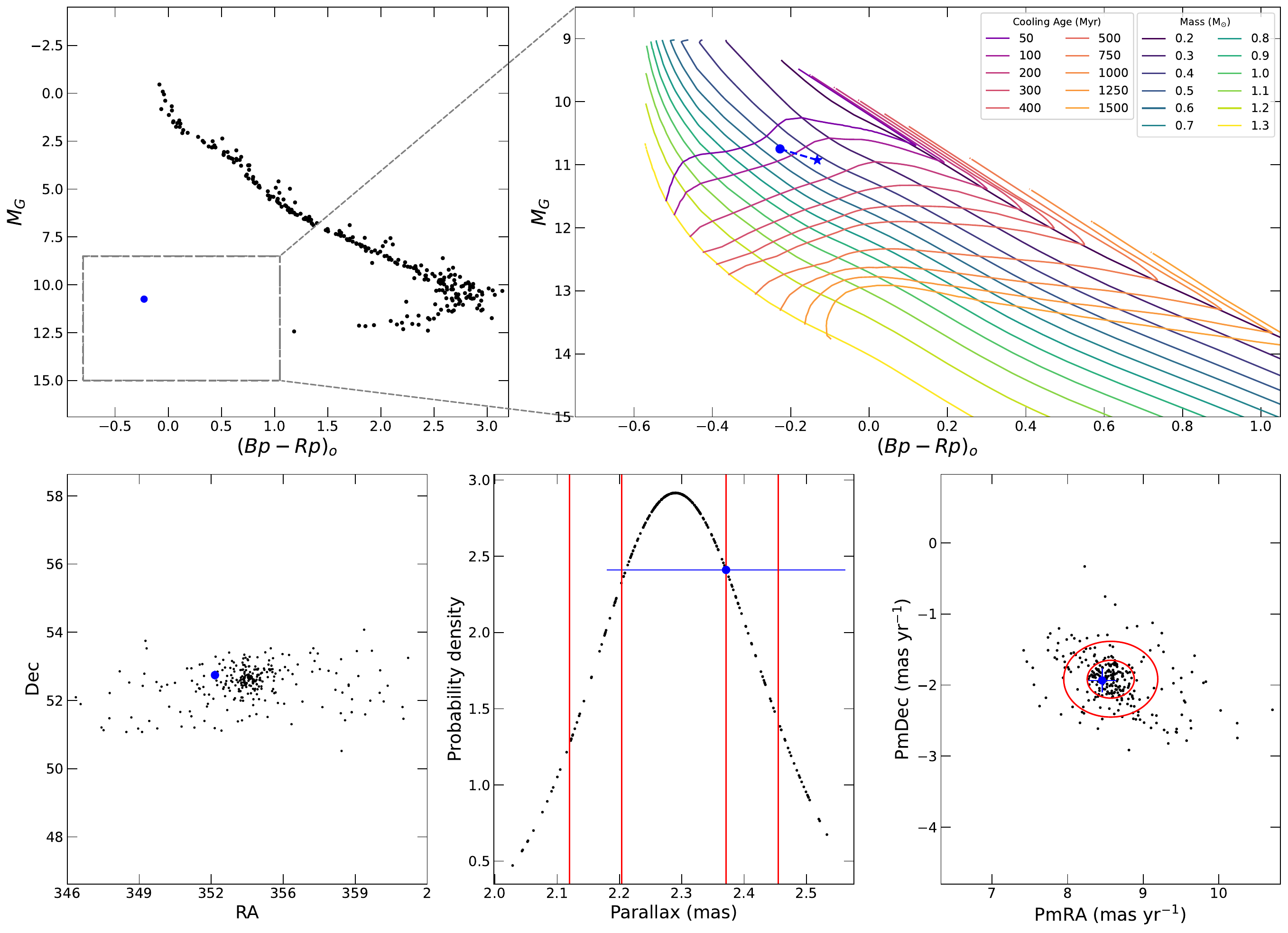}
    \caption{Same as Fig.~\ref{fig:CMD_alessi_13}, but for the Stock 12 cluster.}
    \label{fig:CMD_stock_12}
\end{figure*}

\begin{figure*}[!tp]
    \centering
    \includegraphics[width=0.75\textwidth]{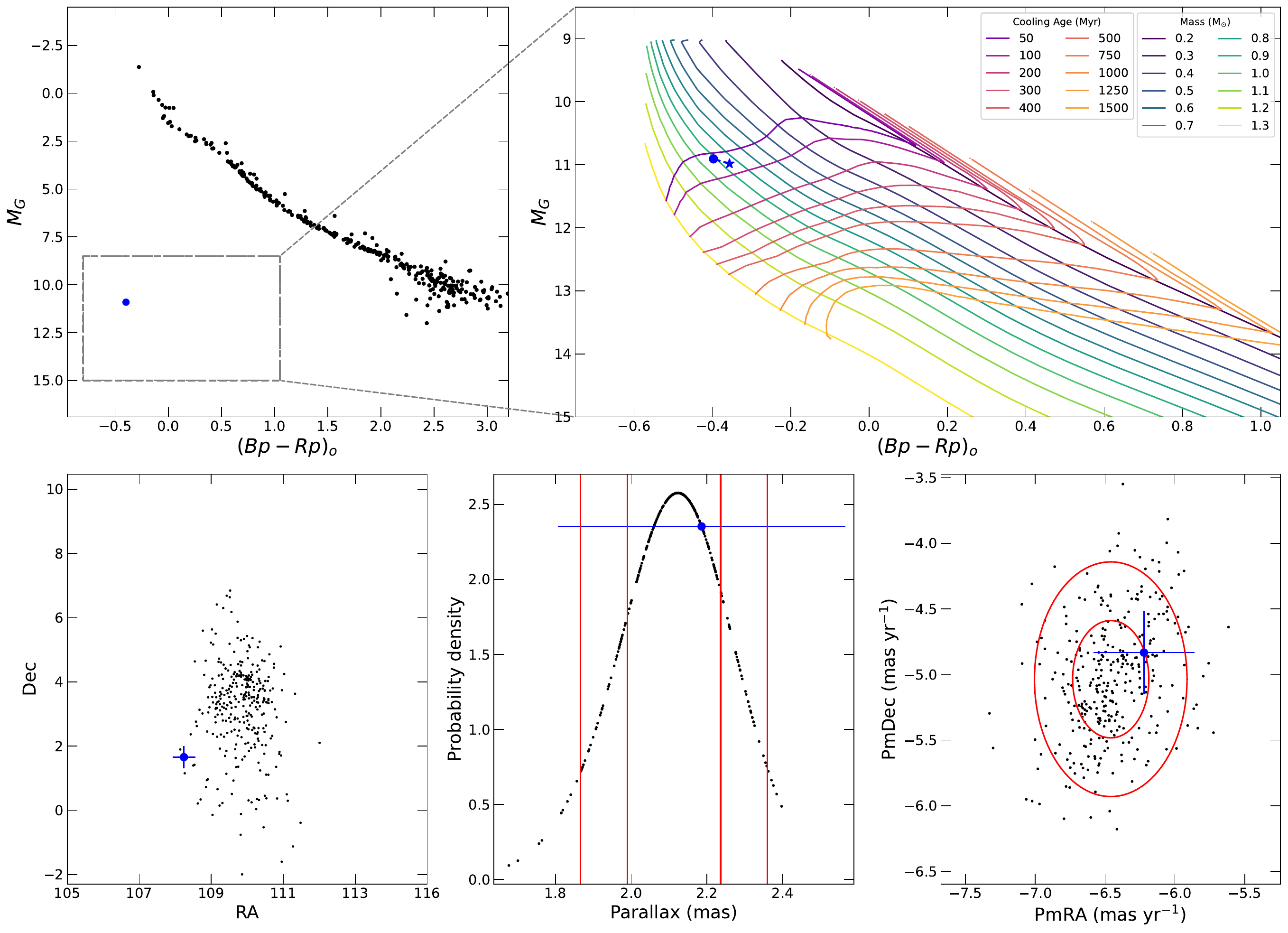}
    \caption{Same as Fig.~\ref{fig:CMD_alessi_13}, but for the Theia 172 cluster.}
    \label{fig:CMD_theia_172}
\end{figure*}

\begin{figure*}[!tp]
    \centering
    \includegraphics[width=0.75\textwidth]{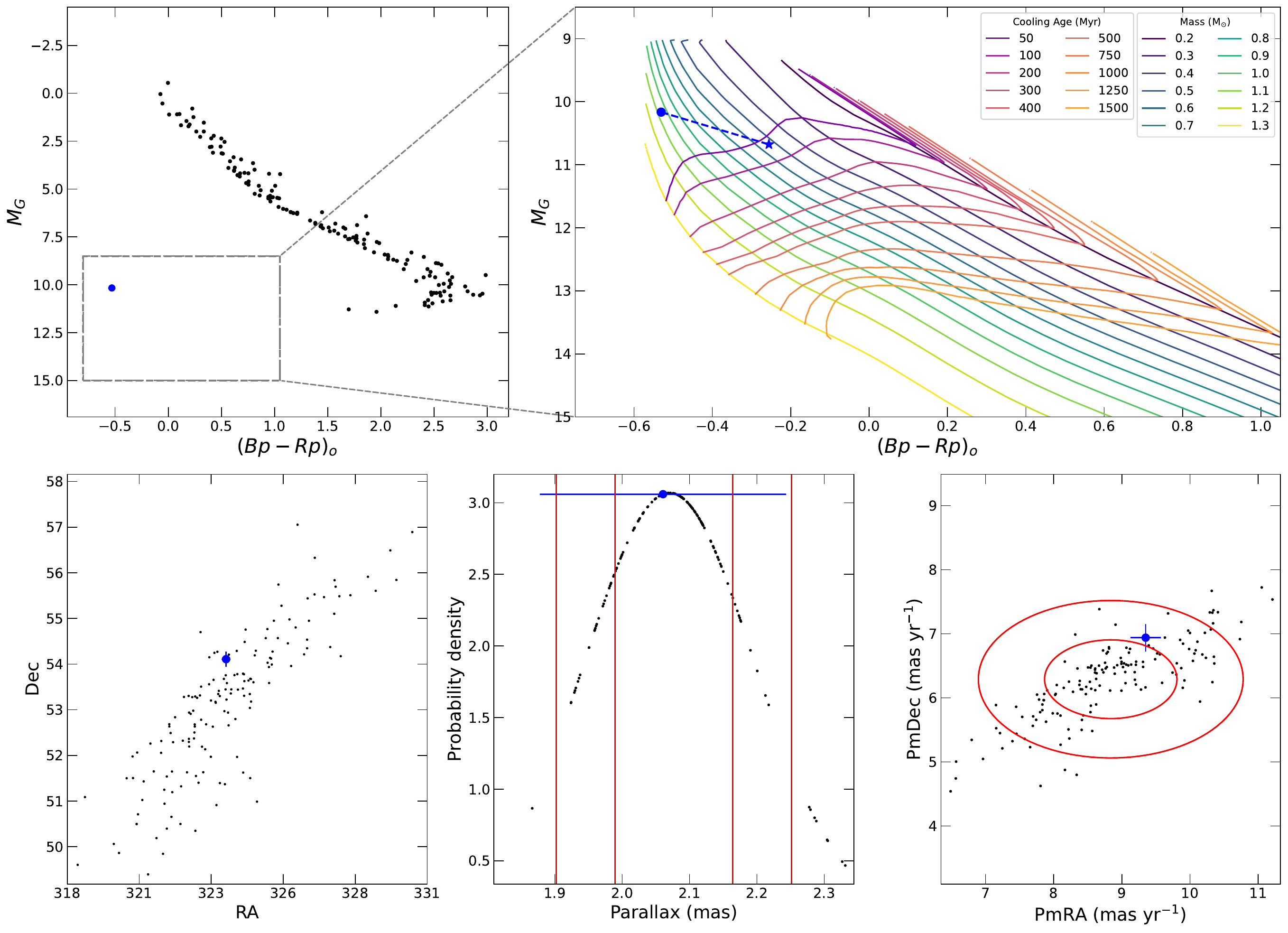}
    \caption{Same as Fig.~\ref{fig:CMD_alessi_13}, but for the Theia 248 cluster.}
    \label{fig:CMD_theia_248}
\end{figure*}

\begin{figure*}[!tp]
    \centering
    \includegraphics[width=0.75\textwidth]{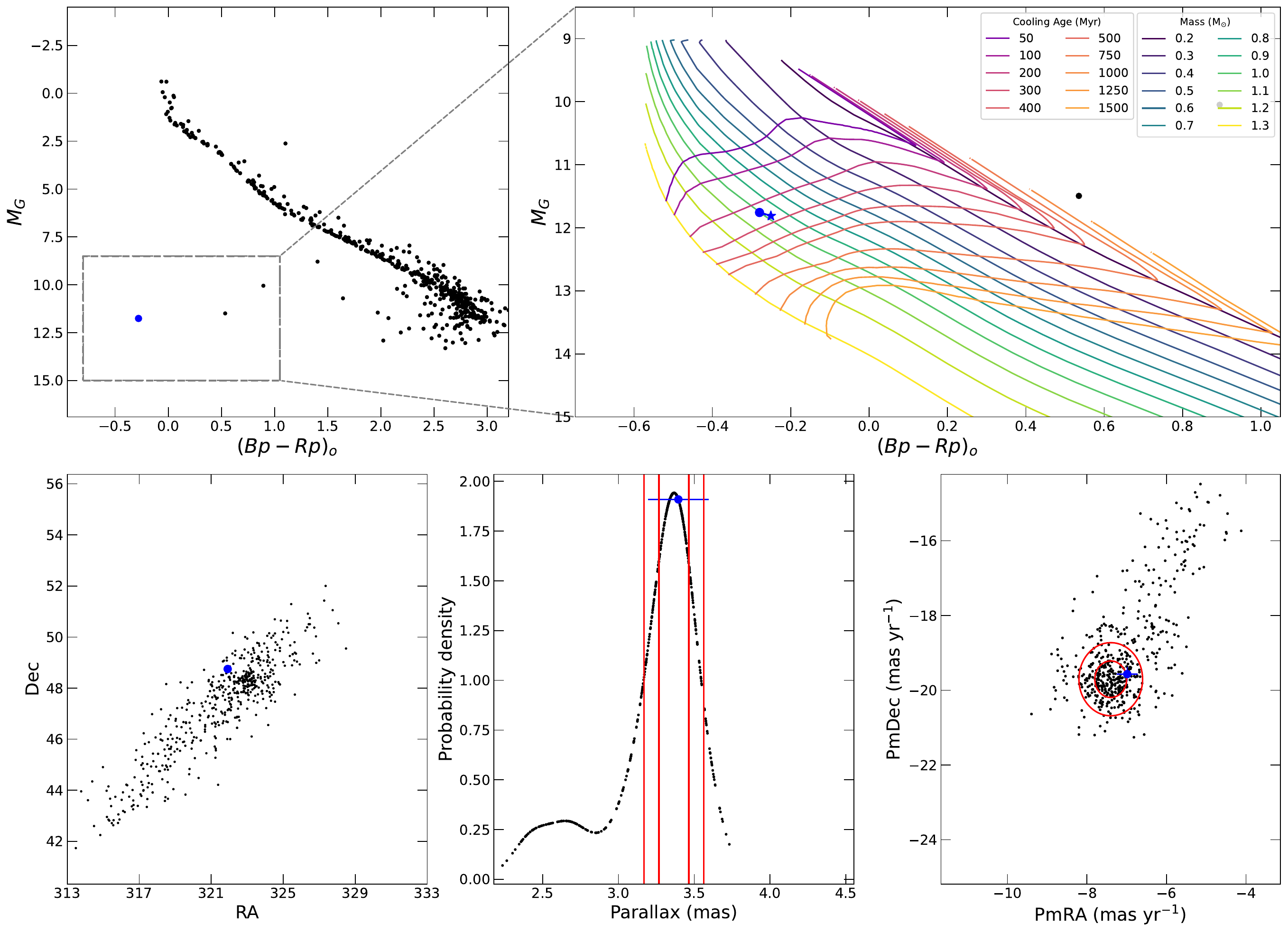}
    \caption{Same as Fig.~\ref{fig:CMD_alessi_13}, but for the Theia 517 cluster.}
    \label{fig:CMD_theia_517}
\end{figure*}

\begin{figure*}[!tp]
    \centering
    \includegraphics[width=0.75\textwidth]{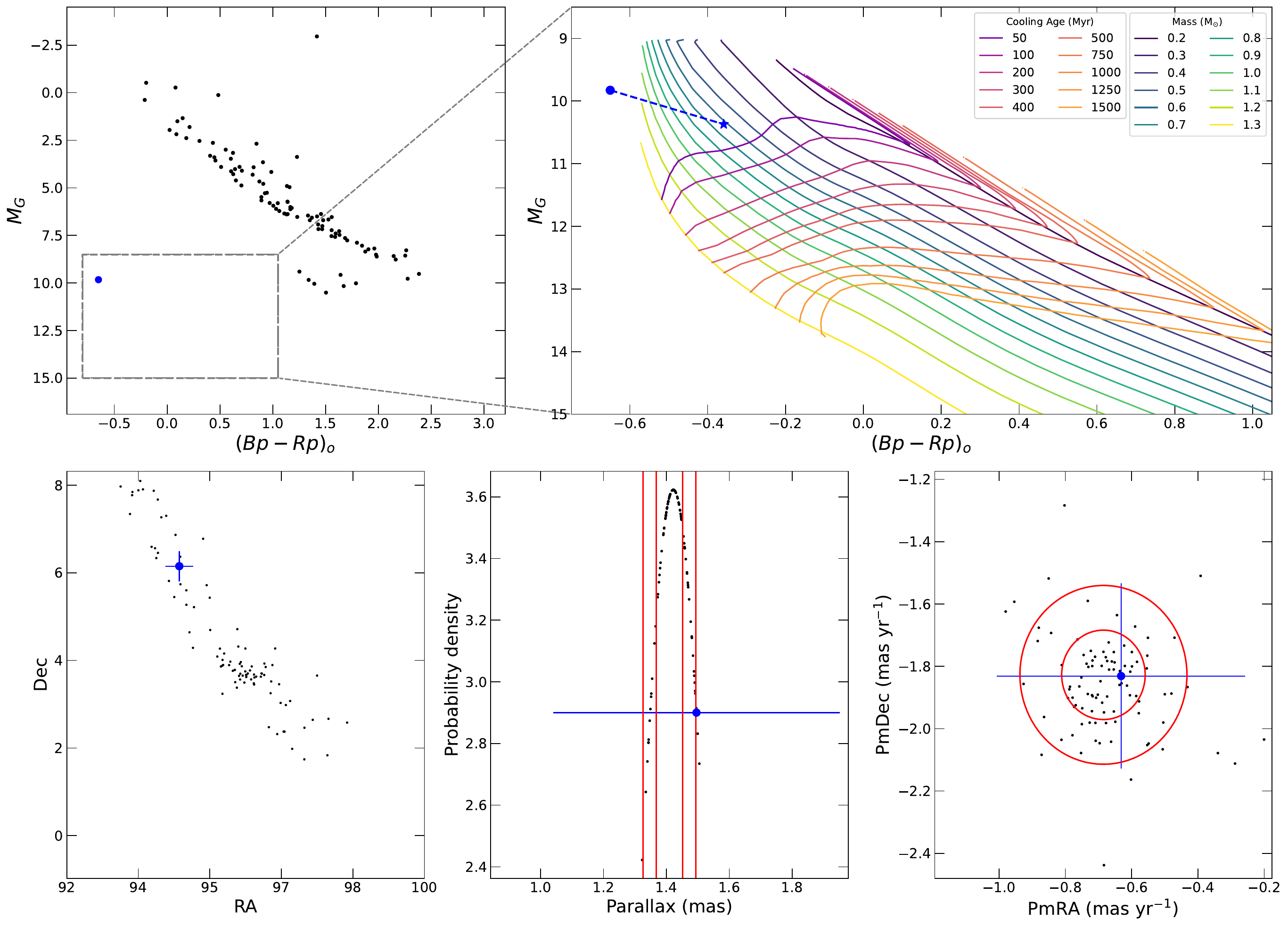}
    \caption{Same as Fig.~\ref{fig:CMD_alessi_13}, but for the Theia 558 cluster.}
    \label{fig:CMD_theia_558}
\end{figure*}

\begin{figure*}[!tp]
    \centering
    \includegraphics[width=0.75\textwidth]{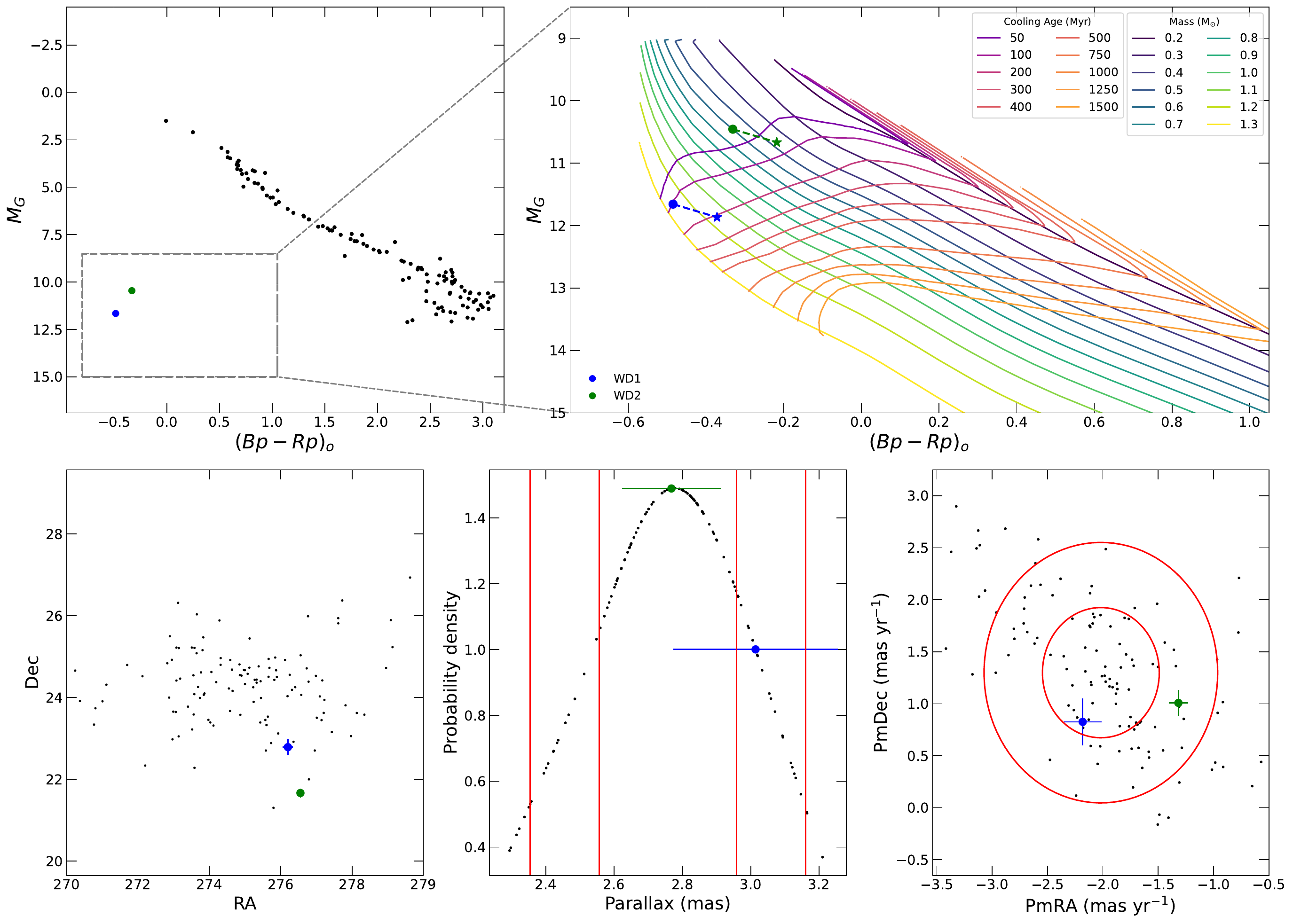}
    \caption{Same as Fig.~\ref{fig:CMD_alessi_13}, but for the Theia 817 cluster.}
    \label{fig:CMD_theia_817}
\end{figure*}

\begin{figure*}[!tp]
    \centering
    \includegraphics[width=0.75\textwidth]{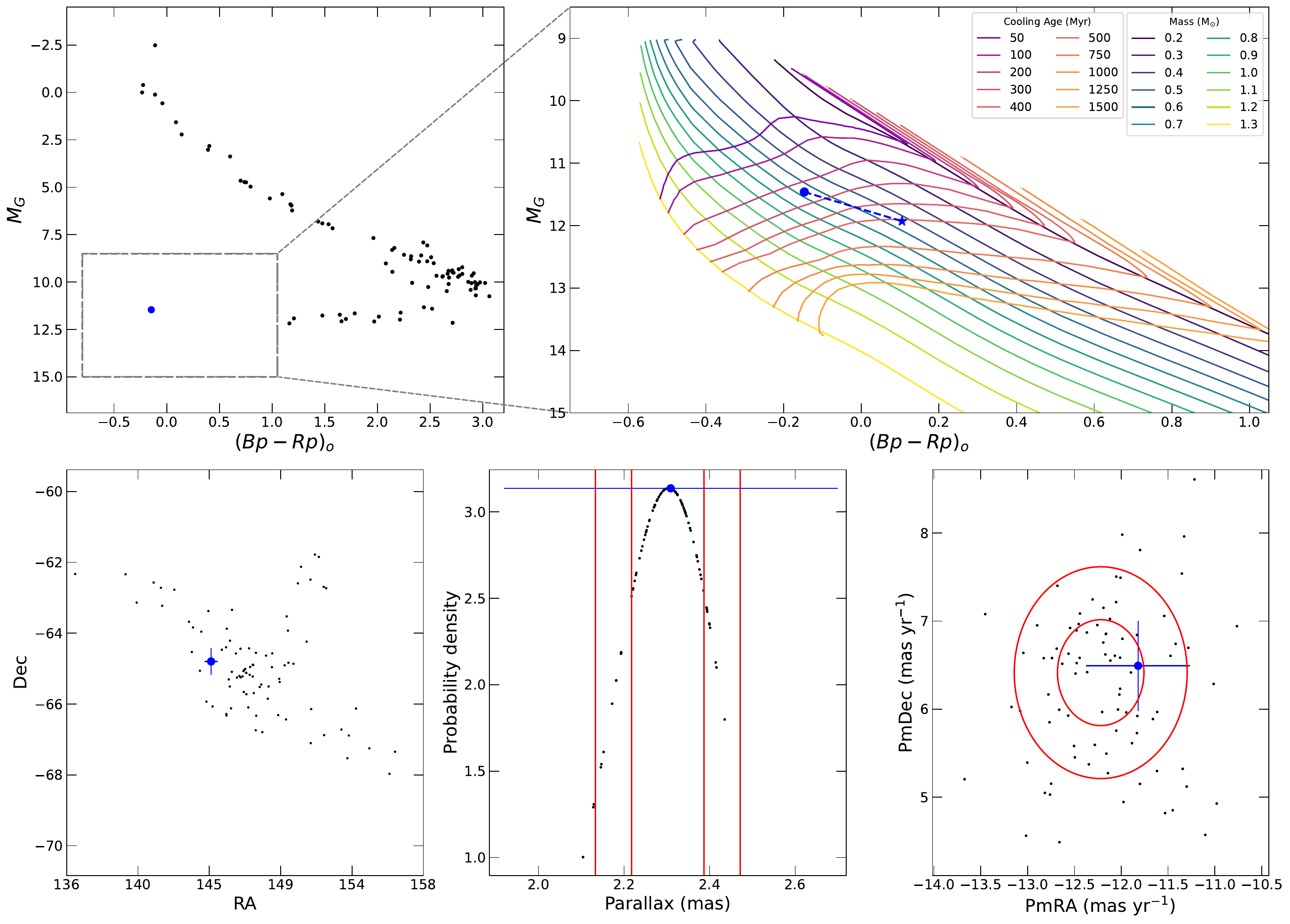}
    \caption{Same as Fig.~\ref{fig:CMD_alessi_13}, but for the Theia 1315 cluster.}
    \label{fig:CMD_theia_1315}
\end{figure*}

\begin{figure*}[!tp]
    \centering
    \includegraphics[width=0.75\textwidth]{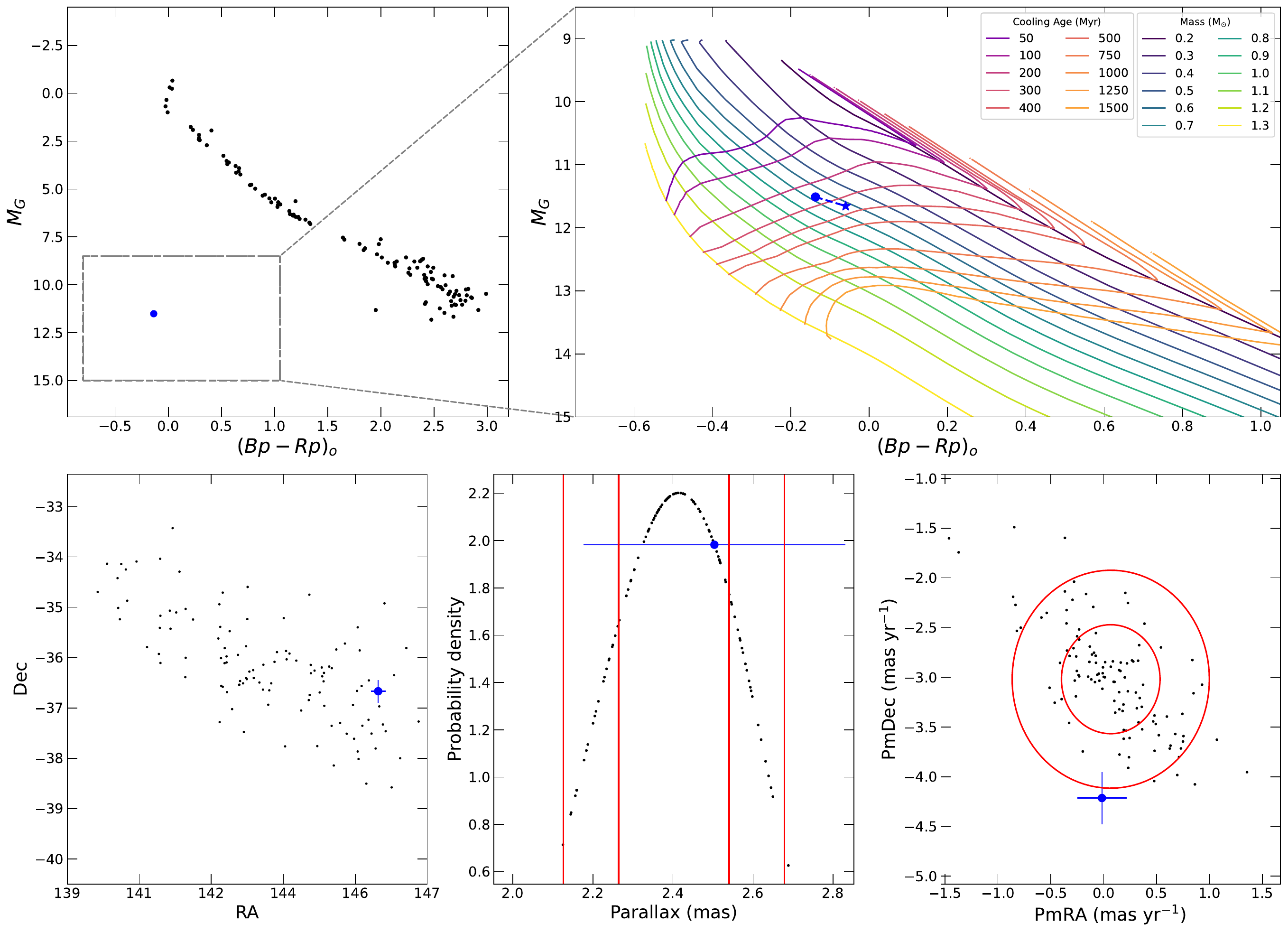}
    \caption{Same as Fig.~\ref{fig:CMD_alessi_13}, but for the Turner 5 cluster.}
    \label{fig:CMD_turner_5}
\end{figure*}

\begin{figure*}[!tp]
    \centering
    \includegraphics[width=0.75\textwidth]{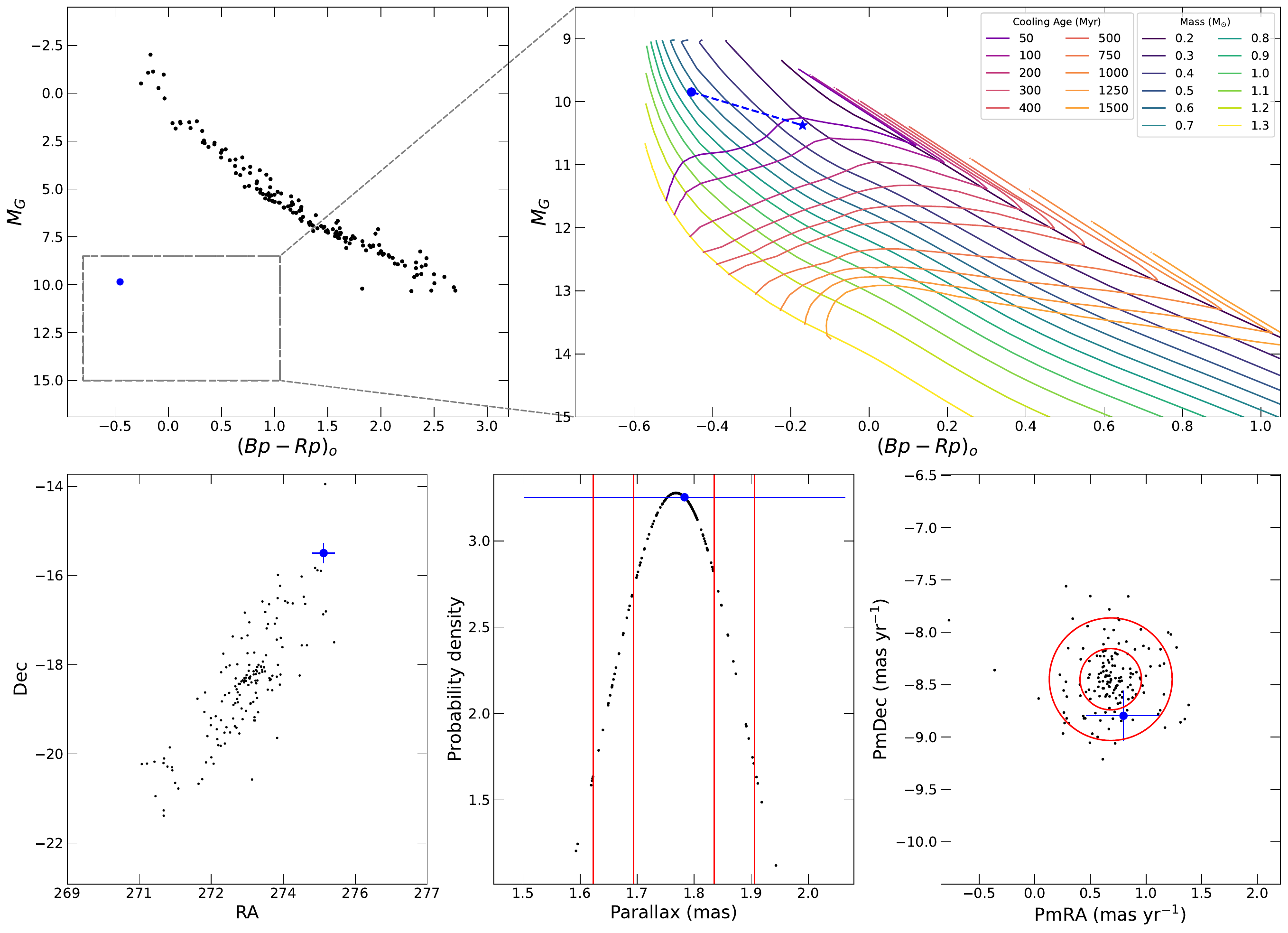}
    \caption{Same as Fig.~\ref{fig:CMD_alessi_13}, but for the UPK 5 cluster.}
    \label{fig:CMD_upk_5}
\end{figure*}

\begin{figure*}[!tp]
    \centering
    \includegraphics[width=0.75\textwidth]{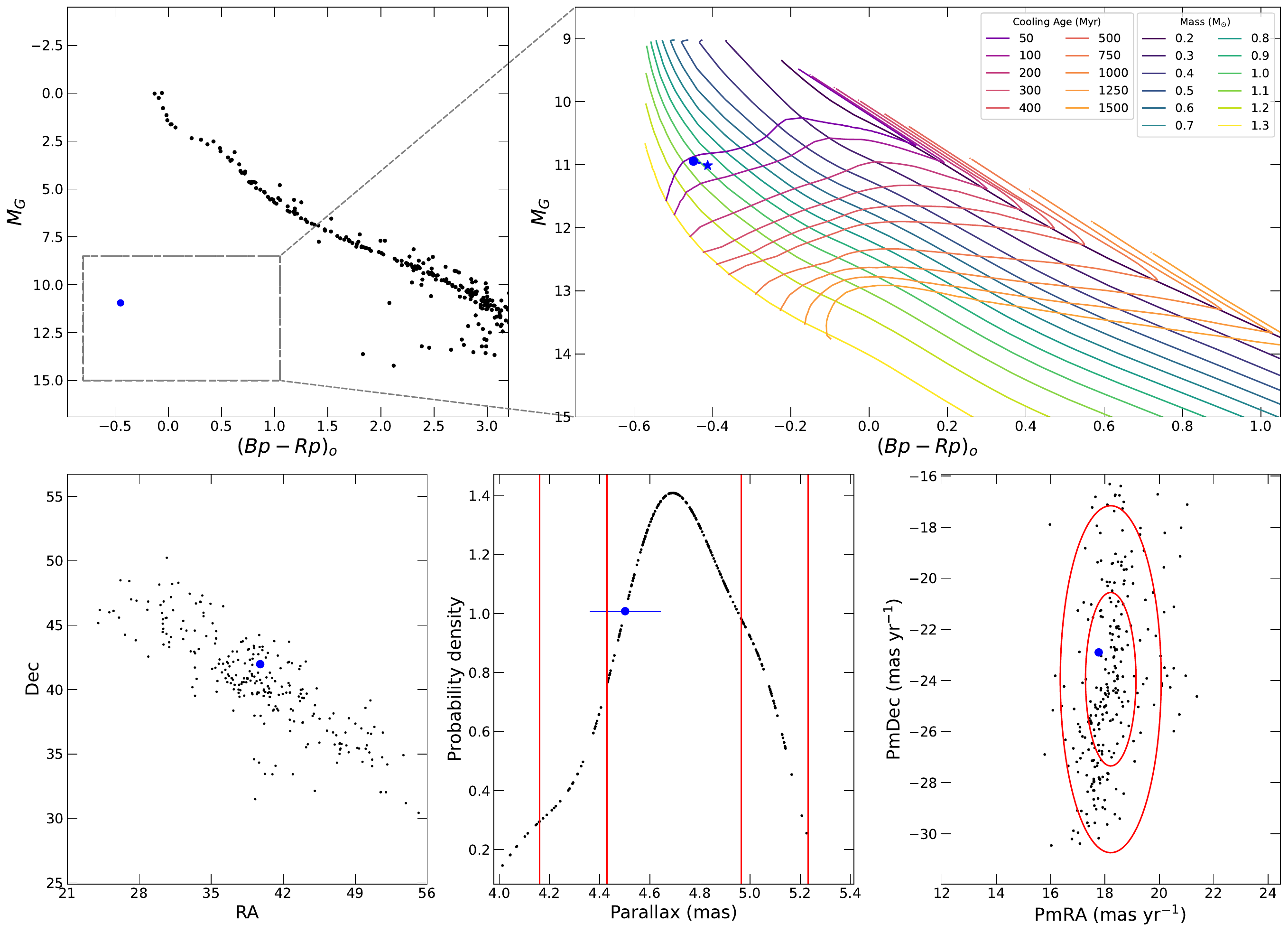}
    \caption{Same as Fig.~\ref{fig:CMD_alessi_13}, but for the UPK 303 cluster.}
    \label{fig:CMD_upk_303}
\end{figure*}

\begin{figure*}[!tp]
    \centering
    \includegraphics[width=0.75\textwidth]{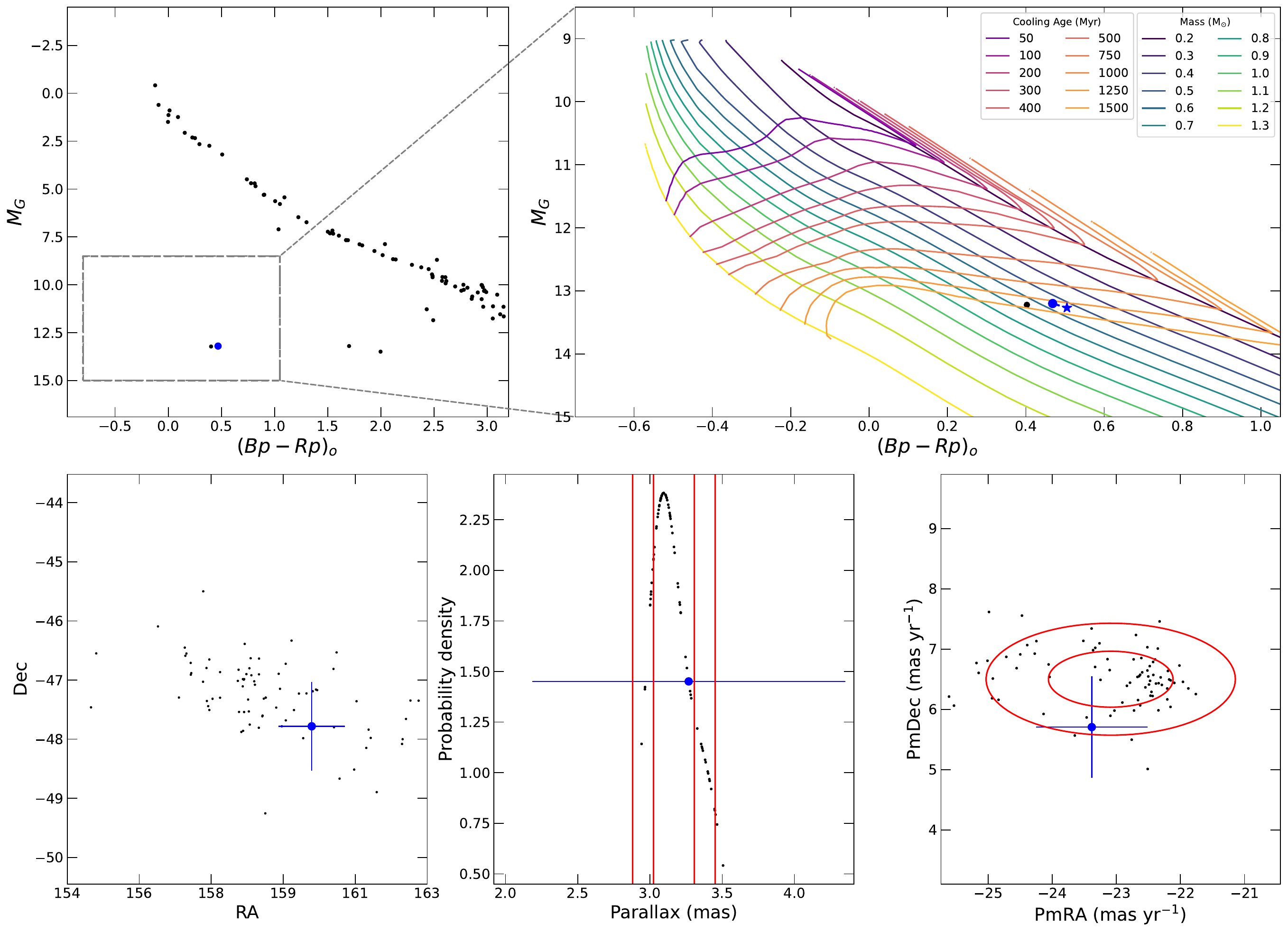}
    \caption{Same as Fig.~\ref{fig:CMD_alessi_13}, but for the UPK 552 cluster.}
    \label{fig:CMD_upk_552}
\end{figure*}

\end{document}